# A Survey of Stellar Families: Multiplicity of Solar-Type Stars

Deepak Raghavan[1,2], Harold A. McAlister[1], Todd J. Henry[1], David W. Latham[3], Geoffrey W. Marcy[4], Brian D. Mason[5], Douglas R. Gies[1], Russel J. White[1], Theo A. ten Brummelaar[6]

## ABSTRACT

We present the results of a comprehensive assessment of companions to solar-type stars. A sample of 454 stars, including the Sun, was selected from the *Hipparcos* catalog with $\pi > 40$ mas, $\sigma_\pi/\pi < 0.05$, $0.5 \leq B - V \leq 1.0$ ($\sim$ F6–K3), and constrained by absolute magnitude and color to exclude evolved stars. These criteria are equivalent to selecting all dwarf and subdwarf stars within 25 pc with $V$-band flux between 0.1 and 10 times that of the Sun, giving us a physical basis for the term "solar-type". New observational aspects of this work include surveys for (1) very close companions with long-baseline interferometry at the Center for High Angular Resolution Astronomy (CHARA) Array, (2) close companions with speckle interferometry, and (3) wide proper motion companions identified by blinking multi-epoch archival images. In addition, we include the results from extensive radial-velocity monitoring programs and evaluate companion information from various catalogs covering many different techniques. The results presented here include four new common proper motion companions discovered by blinking archival images. Additionally, the spectroscopic data searched reveal five new stellar companions. Our synthesis of results from many methods and sources results in a thorough evaluation of stellar and brown dwarf companions to nearby Sun-like stars.

[1]Center for High Angular Resolution Astronomy, Georgia State University, P.O. Box 3969, Atlanta, GA 30302-3969

[2]raghavan@chara.gsu.edu

[3]Harvard-Smithsonian Center for Astrophysics, 60 Garden Street, Cambridge, MA 02138

[4]Department of Astronomy, University of California, Berkeley, CA 94720-3411

[5]US Naval Observatory, 3450 Massachusetts Avenue NW, Washington DC 20392-5420

[6]The CHARA Array, Mount Wilson Observatory, Mount Wilson, CA 91023

The overall observed fractions of single, double, triple, and higher order systems are $56\% \pm 2\%$, $33\% \pm 2\%$, $8\% \pm 1\%$, and $3\% \pm 1\%$, respectively, counting all confirmed stellar and brown dwarf companions. If all candidate, i.e., unconfirmed, companions identified are found to be real, the percentages would change to $54\% \pm 2\%$, $34\% \pm 2\%$, $9\% \pm 2\%$, and $3\% \pm 1\%$, respectively. Our completeness analysis indicates that only a few undiscovered companions remain in this well-studied sample, implying that the majority ($54\% \pm 2\%$) of solar-type stars are single, in contrast to the results of prior multiplicity studies.

Our sample is large enough to enable a check of the multiplicity dependence on various physical parameters by analyzing appropriate subsamples. Bluer, more massive stars are seen as more likely to have companions than redder, less massive ones, consistent with the trend seen over the entire spectral range. Systems with larger interaction cross sections, i.e., those with more than two components or long orbital periods, are preferentially younger, suggesting that companions may be stripped over time by dynamical interactions. We confirm the planet-metallicity correlation (i.e., higher metallicity stars are more likely to host planets), but are unable to check it for brown dwarfs due to the paucity of such companions, implying that the brown dwarf desert extends over all separation regimes. We find no correlation between stellar companions and metallicity for $B - V < 0.625$, but among the redder subset, metal-poor stars ([Fe/H] $< -0.3$) are more likely to have companions with a $2.4\sigma$ significance.

The orbital-period distribution of companions is unimodal and roughly log-normal with a peak and median of about 300 years. The period-eccentricity relation shows the expected circularization for periods below 12 days, caused by tidal forces over the age of the Galaxy, followed by a roughly flat distribution. The mass-ratio distribution shows a preference for like-mass pairs, which occur more frequently in relatively close pairs. The fraction of planet hosts among single, binary, and multiple systems are statistically indistinguishable, suggesting that planets are as likely to form around single stars as they are around components of binary or multiple systems with sufficiently wide separations. This, along with the preference of long orbital periods among stellar systems, increases the space around stars conducive for planet formation, and perhaps life.

*Subject headings:* Stellar multiplicity - Binary stars - Solar-type stars - Solar neighborhood - Exoplanet systems - Brown dwarfs - Survey

## 1. Introduction

Our Sun is the only known star in the Universe that hosts a life-bearing planet. It is no surprise then that many astronomical programs are focused on developing a more comprehensive understanding of Sun-like stars. Paramount to this is an assessment of how often solar-type stars have stellar and substellar companions, and whether there any relationships between a star's tendency to have one or more of these companion types. This is the primary motivator of this effort, which is aimed at moving us a step closer to comprehending the availability of habitable space and the possibility of extraterrestrial life in the Universe. Additionally, statistical results from comprehensive multiplicity analyses can provide vital clues about star formation and evolution. The long-held belief that the majority of stars have stellar companions (e.g., Heintz 1969; Abt & Levy 1976; Duquennoy & Mayor 1991, hereafter DM91) has significant bearing on their formation mechanisms and needs to be tested as new observational data become available. Relationships identified between multiplicity and the physical parameters of stars such as mass (DM91, Henry & McCarthy 1990; Fischer & Marcy 1992; Mason et al. 1998a; Burgasser et al. 2003) and age (Mason et al. 1998b) also inform us about the formation and evolution of stars. Another valuable result of studying stellar multiplicity is the determination of dynamical masses of the component stars, enabling an observational basis for testing stellar structure and evolution models. Moreover, simply knowing whether a star is single or not is important for many astronomical efforts because calibration stars used in spectroscopic, photometric, and astrometric studies are often required to be single.

While many previous studies have investigated the multiplicity of solar-type stars (e.g, Abt & Levy 1976, DM91), an update is warranted for several reasons: (i) More accurate astrometry available from the *Hipparcos* catalog (Perryman & ESA 1997) enables the selection of a more accurate volume-limited sample, minimizing biases in the results. The DM91 sample was selected using the parallaxes in Gliese (1969) (hereafter referred to as CNS, i.e, the Catalogs of Nearby Stars, along with Gliese & Jahreiß (1979, 1991)), and we will show in § 2.3 that it was contaminated by parallax errors. (ii) The vast amount of attention garnered by nearby solar-type stars enables this study to build on this knowledge with targeted new observations, resulting in the most comprehensive assessment yet. (iii) The larger sample of this study compared to prior efforts not only increases the confidence in the results, but also enables analyses of various subsamples, resulting in an improved understanding of star formation and evolution. (iv) High-precision radial-velocity surveys (e.g., Marcy et al. 2004; Mayor et al. 2004; Butler et al. 2006) and high-contrast imaging efforts (e.g, Burgasser et al. 2003; Metchev & Hillenbrand 2009) since the prior studies yield observational results of substellar companions and enable us to better constrain our incompleteness analysis.

In § 2, we describe the sample selection process, present the resulting set of solar-type stars, discuss sample completeness and characteristics, and compare our sample to that of DM91, which represents the seminal work in this area. In § 3, we discuss the various sources and observational methods used to derive the list of companions. The synthesized multiplicity results of each system are presented in § 4, followed by a discussion of survey completeness, comparison with prior studies, and implications of our results in § 5.

## 2. The Sample of Solar-type Stars

The sample of stars studied in this work is comprised of 454 solar-type primary stars in the solar neighborhood (see Table 1), selected from the *Hipparcos* catalog. The distance limit for our sample is 25 pc from the Sun, corresponding to a *Hipparcos* parallax of $\pi \geq 40$ mas. Stars with parallax errors larger than 5% of the corresponding value are excluded from the sample (see § 2.1). Söderhjelm (1999) improved the precision and accuracy of several *Hipparcos* parallaxes for close binaries by combining *Hipparcos* and ground-based observations, and we utilize his results in place of the *Hipparcos* values, when available. To restrict the sample to solar-type stars, we require $0.5 \leq B - V \leq 1.0$, which roughly corresponds to spectral types F6–K3. To exclude significantly evolved stars while selecting all dwarfs and subdwarfs, we limit selection to a band within 2 magnitudes above and 1.5 magnitudes below the Cox (2000) main sequence. The larger band above the main sequence was chosen to minimize any biases against close binaries and multiple star systems. Figure 1 shows the sample and the selection criteria on an HR diagram. The two regions marked with an 'X' include no stars, and hence the above criteria are equivalent to selecting all dwarf stars with $V$-band fluxes of $10^{-1}$ to $10^{+1}$ that of the Sun, giving us a physical basis for the term "solar-type".

The above criteria select 462 stars from the *Hipparcos* catalog, nine of which are companions to other stars in the sample, yielding the final sample of 454 stars when the Sun is included. The *Hipparcos* astrometry was recently updated by (van Leeuwen 2007a,b), but our sample selection and much of the observational work were completed before they were available. However, the impact of these revisions on our survey is minimal. If we had selected our sample from van Leeuwen (2007b), 15 new stars would have been included and 18 sample stars would have been left out. Compared to the sample size, these differences are small and show no systematic effects. They are all due to parallax fluctuations within $2\sigma$ near the 40 mas limit.

Our volume limit yields a sufficiently large sample to produce robust multiplicity statistics and to enable the analyses of subsamples based on various physical characteristics.

Moreover, solar-type stars in this volume of space have been extensively studied over many decades, allowing our work to benefit from those prior results. We will show in § 2.2 that the *Hipparcos* catalog is not constrained by magnitude limits for solar-type stars within 25 pc, resulting in a complete volume-limited sample.

The 12 stars above the magnitude selection limit in Figure 1 are likely evolved, as suggested by their spectral classes in Table 2. The table also lists the three stars that are too faint to be included in our sample. Two of these, plotted in Figure 1 below the cutoff are, in fact, M-dwarfs with erroneous *Hipparcos* $B - V$ values, as confirmed by their spectral types listed in the table and magnitudes in the Tycho-2 Catalog (Høg et al. 2000). The third star below the cutoff, far too faint to be included in the figure, is a nearby white dwarf. Note that the selection band we used around the main sequence was chosen to select all dwarfs and subdwarfs. While the spectral types listed in Table 1 do not include any of luminosity class VI, the sample nevertheless contains 31 stars with [Fe/H] $\leq -0.50$ (see § 5.3.1 for a tabulation of the metallicity values, and Jao et al. (2008) for a discussion of subdwarf spectral types). The four sample stars farthest below the main sequence in Figure 1 are also the four lowest metallicity stars in the sample with [Fe/H] $\leq -0.95$, and no stars are excluded near this limit, indicating that the few subdwarfs within 25 pc are included in the sample.

## 2.1. Stars Excluded Due to Large Parallax Errors

While we chose the 5% error threshold to limit our sample to good-quality parallax measures, this raises the question of a potential bias, namely, could astrometric perturbations from close companions lead to relatively large errors in the *Hipparcos* astrometry, preferentially excluding these binaries from our study? Alternatively, the large errors could be due to the faintness of the sources, which would imply stars too far to fit within our volume-limit, so a closer inspection of these stars is warranted. Table 3 lists the 24 stars that were left out of the sample because their parallax errors were greater than 5% of the corresponding parallax value. Columns 13 and 14 list an alternate parallax and its corresponding error, estimated as identified by Column 15.

Five stars have parallaxes with better than 5% accuracy for a physically associated companion. One of these (HD 53706) is included in this work as a companion of HD 53705, while the remaining four are too distant to make the 40-mas cutoff. For an additional ten, our distance estimates using *Hipparcos* photometry yields parallaxes below 17 mas. While these estimates are crude, they nevertheless place these stars far beyond our distance limits, justifying their exclusion. It should be noted that these distance estimates assume that the stars are main-sequence dwarfs. If they are subdwarfs, their actual distances will be less than the photometric estimates. However, only 7% of our sample is comprised of subdwarfs, and so any effect of this error on these 24 stars is minimal. Two more entries in the table (HD 139460 and 139461) represent a physically bound pair, of which HD 139461 is itself a spectroscopic binary with an M-dwarf companion. The spectral type measures of these stars from Gray et al. (2003) and the consistent mass estimates from Tokovinin & Gorynya (2001) place this system beyond the 25 pc limit. This leaves only seven stars that could possibly lie within the distance limit of our sample. Two of these (HD 23439 and 217580) have single-lined spectroscopic (SB1) companions and a third (HD 212697) has a common proper motion (CPM) companion. The implied multiplicity fraction is consistent with our overall results, so the effect of this possible bias is minimal.

## 2.2. Sample Completeness

The *Hipparcos* catalog contains astrometry of unprecedented quality of over 118 000 stars. The faintest target that could be reached by the telescope was about $V = 12.4$ mag (Perryman et al. 1995), but the catalog is complete to significantly brighter limits. The *Hipparcos* target list was selected based on two criteria. About 52 000 "survey" stars were selected as a complete sample down to well-defined magnitude limits varying from 7.3–9.0 mag depending on spectral type and galactic latitude. Stars later than G5 were restricted to a limit brighter by 0.6 mag as compared to earlier type stars to preferentially select nearby A, F, and early G-type stars over distant red giants, which would otherwise have dominated the list (Turon et al. 1992a,b). The remaining 66 000 "faint" stars were selected down to about 12 mag from various proposals submitted to the *Hipparcos* Scientific Selection Committee (Turon et al. 1992b). Given that the current sample includes several stars fainter than the 7.3 mag limit to which the catalog is complete for all spectral types and galactic latitudes, a further analysis is warranted to ensure that our sample does not suffer from any brightness limits. Such limits, if present, would introduce the Malmquist bias, favoring the selection of binaries in our most distant bins.

First, we check the completeness of the *Hipparcos* catalog against the Tycho-2 Catalog, which is 99% complete down to $V = 11.0$ mag (Høg et al. 2000). Figure 2 shows the $V$ magnitude distributions of the two catalogs as well as the percentage of Tycho-2 stars that are also in the *Hipparcos* catalog. A few things are apparent from this figure. First, the linear slope on a logarithmic scale of the Tycho-2 catalog out to $V \sim 11$ confirms its completeness out to that limit. Similarly, the *Hipparcos* catalog is seen as complete out to about 7.5 mag, no surprise given the limits discussed above. Second, while the Tycho-2 distribution falls off rapidly beyond the peak, the *Hipparcos* distribution shows a moderated fall. This is somewhat due to the graded *Hipparcos* limits based on spectral types and galactic latitudes, but mainly because of the selection of more than half of the *Hipparcos* stars fainter than the completeness limits based on their scientific rationale. Presumably, most nearby solar-type stars would have made this cut, not only because of their interest to various scientific investigations but also because most of these had existing ground-based parallax measurements, which the *Hipparcos* mission could verify and improve.

Figure 3 shows the distance distribution of the *Hipparcos* stars conforming to our sample selection criteria, but expanding the distance limit out to 35 pc. The dotted curve marks the expected distribution for a complete sample extrapolated from the 15-pc sample. This extrapolation makes two assumptions. First, it assumes that the space density of nearby stars is uniform, which is true for our sample of stars within 25 pc. Second, it assumes that the 15 pc sample is complete. This is justifiable because our subsample within 15 pc includes only two stars fainter than the 7.3 mag limit to which the *Hipparcos* catalog is complete for all spectral types and galactic latitudes. One of these is within the *Hipparcos* magnitude limit for its galactic latitude, given the specific magnitude limit definitions in Turon et al. (1992b), and the other is just 0.14 magnitudes fainter than the limit. Furthermore, these two stars are among the most intrinsically faint stars in the sample (spectral types K2.5 V and K3 V) and close to the limit of 15 pc (both are 14 pc away). Thus, the chances of any stars being left out of the 15-pc sample due to brightness limits are slim. Given this, Figure 3 suggests that the actual 25-pc sample is also complete and does not suffer from brightness limits as the actual distribution matches the expected one fairly well. The deficiency in the 23–25 pc bin is within a $1\sigma$ deviation for Poisson statistics and is made up in the very next distance bin, suggesting that it may be due to statistical scatter alone. While the overall profile of the curve shows the effects of brightness limits creeping in, they occur beyond the 25-pc limit of our sample. We will revisit this topic in § 5.2 using the multiplicity data gathered by this effort to show that the percentage of single stars is statistically equivalent for various distance-limited subsamples, consistent with the expectations of a complete volume-limited sample.

### 2.3. Comparison with the DM91 Sample

The DM91 multiplicity survey was based on a volume-limited sample of 164 solar-type stars, selected with parallaxes of 45 mas or greater, spectral types F7–G9, and spectral classes IV-V, V, or VI. Selecting a sample subject to these same criteria from the *Hipparcos* catalog results in 148 primaries, but only 92 of these overlap with the DM91 sample. This indicates that 72 stars (44%) of the DM91 sample are now known to lie outside their selection criteria and 56 (38%) of the stars meeting their criteria were not included in their sample. The important question is whether these substantial differences introduce possible biases that would taint their multiplicity results. While the distribution of the CNS stars in Halbwachs et al. (2003) showed no systematic deviations from the main sequence when plotted using *Hipparcos* data, Figure 4 shows a preferential shift for the DM91 sample. The majority of stars that are now known to lie outside the DM91 criteria are elevated above the main sequence and even include evolved stars, demonstrating the need for an updated survey. Furthermore, a closer inspection of the *Hipparcos* stars matching the DM91 criteria reveals that several are erroneously included due to incorrect spectral type assignments in *Hipparcos*. Our sample selection, based on $B - V$ color, mitigates this problem.

## 3. Survey Methods and Multiplicity Results

This multiplicity survey of solar-type stars synthesizes companion information from various imaging and spectroscopic observations and augments them with a thorough examination of reported companions in the various catalogs and publications. The closest companions are best revealed through radial-velocity searches, and this work benefits from a complementary effort by Latham et al. (2010), which reports radial velocities and their analyses for 344 of our sample stars and 38 companions. Additionally, high-precision measures of radial velocities gathered as part of the California and Carnegie Planet Search (CCPS, described in Butler et al. 1996) program were investigated to identify stellar companions. As seen in Figure 5, the excellent coverage of these high-precision data allow us to not only identify stellar companions of all masses, but also help constrain our incompleteness analysis (see § 5.1). An astrometric search for companions too wide for radial-velocity campaigns and too close for traditional visual techniques such as speckle interferometry was conducted through a systematic survey for separated fringe packet (SFP) binaries using the Center for High Angular Resolution Astronomy (CHARA) Array (ten Brummelaar et al. 2005). This effort, which yielded no new companion detections, is discussed in Raghavan et al. (2010a). Augmenting comprehensive prior efforts with several new observations, we now have speckle interferometry observations for every target in the sample. Finally, a systematic search for the widest of companions was conducted by blinking multi-epoch archival images (see § 3.1).

Apart from the targeted observations mentioned above, our synthesis effort brings together information about previously known and suspected companions, which we evaluate individually. In § 3.2, we examine the astrometric companions discovered by the *Hippar-* *cos* mission, §3.3 assesses the CPM candidates in the Washington Double Star Catalog[1] (WDS) and the Fourth Catalog of Interferometric Measurements of Binary Stars[2] (hereafter INT4), §3.4 addresses the resolved-pair and photocentric-motion visual binaries cataloged in the Sixth Catalog of Orbits of Visual Binary Stars[3] (hereafter ORB6), and §3.5 reviews the CPM as well as spectroscopic binaries listed in the CNS. Surveys for eclipsing binaries are fruitful in identifying short-period systems but generally pertain to stars beyond our volume limit. The few eclipsing binaries in the sample are discussed in §3.6. Finally, brown dwarf companions are treated in §3.7, and published planetary companions to stars in this sample are covered in §3.8.

In assessing companions identified by the various methods, we assume that whenever such detections are consistent with one physical companion, only one companion exists. Sometimes this could be confirmed, as in the case of spectroscopic and visual orbits with the same orbital periods, but often, we relied solely on consistency. For example, does a slow linear drift in radial velocity over 20 years indicate a new companion, or is it measuring the effects of a known 250-year visual pair? The nature of hierarchical systems suggests that these two measurements are likely related to the same companion, and this is the assumption we make. Similarly, proper motion acceleration seen for one component of a visual or spectroscopic binary is assumed to not indicate an additional component. It is important to note that even if this approach misidentifies two different companions as one, a later rectification can only enhance the multiplicity order of binaries or higher-order systems, but will not affect the estimated percentage of single stars.

## 3.1. The Search for CPM Companions

Due to the proximity of the stars studied here, their proper motions are generally quite large (367 of the 453 stars have proper motions larger than $0\rlap{.}{''}2\,\mathrm{yr}^{-1}$). This, along with archival digitized images taken several years apart, facilitates the identification of CPM companions by blinking the images. The primary source of the archival images utilized in this effort is the multi-epoch STScI Digitized Sky Survey[4] (DSS). In some cases, when the time interval between the two DSS images was not sufficient to easily detect the proper motion of the primary star, SuperCOSMOS Sky Survey (Hambly et al. 2001, SSS) images were used for the earlier epoch, significantly increasing the apparent motion seen upon blinking. Figure 6 shows the epoch-distribution of the images blinked. The earlier epoch distribution is bimodal, with 266 frames from 1949–1957 for predominantly northern hemisphere stars and 187 frames from 1974–1989 for mostly southern hemisphere stars. The later epoch is more tightly constrained, with all 453 frames obtained during 1984–2000. The resulting time interval between the pairs of images blinked varies from 1–49 years, with well over half of the intervals exceeding 28 years. Figure 7 illustrates this method with the two images blinked for HD 9826 ($v$ And), allowing the identification of the CPM companion and the detection of two WDS entries as unrelated field stars with minimal proper motion. The effect of the target star's proper motion, while noticeable in these printed images, is readily apparent when they are blinked numerous times in quick succession. While we utilized image subtractions as an additional tool, our experience shows that a visual examination by blinking the images offers the most effective method for identifying candidate CPM companions. This technique is well suited for identifying companions with separations greater than $\sim 15$–$30''$ from the primary (depending on its brightness), as companions closer than this are often buried in the saturation around the bright primaries. However, in many instances, bright companions inside the saturation region can be identified by twin, comoving diffraction spikes.

---

[1] *http : //www.usno.navy.mil/USNO/astrometry/optical − IR − prod/wds/WDS/*

[2] *http : //www.usno.navy.mil/USNO/astrometry/optical − IR − prod/wds/int4*

[3] *http : //www.usno.navy.mil/USNO/astrometry/optical − IR − prod/wds/orb6*

[4] `http://stdatu.stsci.edu/cgi-bin/dss_form`

### 3.1.1. CPM Companions Identified

To enable an effective search out to a linear radius of about 10 000 AU from each primary, we blinked $22'$ square images for systems closer than 20 pc and $15'$ square images for the more distant systems. The larger images were blinked as four sub-images of $15'$ square from each corner, allowing for a closer inspection. Figure 8 shows the distribution of the linear projected separation searched around each primary, illustrating that most systems were effectively searched out to a separation of $\sim 10 000$ AU. While companions farther than this separation limit have been reported (e.g., Latham et al. 1991; Poveda et al. 1994), such wide companions are rare and are not often expected to survive dynamical interactions in clusters (Parker et al. 2009), so the range selected here enables an effective search of the region containing the vast majority of companions. Companions down to $R = 17$ mag could be readily identified in the images used, as evidenced by the refuted candidate companion to HD 141004, which has $R = 16.9$ and was clearly seen above the background. In comparison, the apparent $R$ magnitude of an M5 dwarf at 25 pc is 12.5 (Cox 2000), and the star-brown dwarf transition occurs at about $R = 19.0$ at 25 pc, estimated using data from Dahn et al. (2002). This implies that our search is sensitive to all but the lowest-mass stellar companions. Of the 453 image pairs blinked, 44 did not show a discernible motion of the primary and an additional 43 had marginally detectable motions. Hence, for 366 of the 453 targets investigated, this technique proved to be fully effective in identifying most CPM stellar companions, and somewhat effective for an additional 43.

The 366 primaries adequately inspected reveal 88 candidate companions in the vicinity of 79 primaries. These include a few previously reported CPM companions at wider separations than our systematic search, which we confirmed by blinking adequately large images. While CPM is necessary for a physical association, it is by no means sufficient. Accordingly, we confirm candidates as physical only if their (a) independently measured parallaxes and proper motions from the literature are consistent with the primaries' *Hipparcos* values, (b) existing or new photometry and/or spectral types indicate distances compatible with the primaries' *Hipparcos* parallaxes, (c) multiple resolved images show conclusive orbital motions, or (d) proximity to the primary and matching large proper motions (greater than $0\rlap{.}''5\,\mathrm{yr}^{-1}$) argue for a physical association. Table 4 lists every candidate identified by this method, grouped by the ones confirmed as physical and those refuted by follow-up work. Table 5 lists the candidates confirmed or refuted by their photometric distance estimates, derived as described in Henry et al. (2004), along with their spectral types, proper motions, and photometry. The separations and position angles listed in Table 4 are approximate measurements from the more recent of the images blinked, whose epoch is also noted in the table. These measurements are generally good to within a few arcseconds and degrees, respectively, except for measurements between twin diffraction spikes, which typically have larger uncertainties. The separations are repeated in Table 5 to identify the star whose distance is estimated. Overall, this method led to the identification of 70 CPM companions, including four new discoveries. The new companions are identified in the archival images in Figure 9. In addition, while many of the entries in Table 4 were previously suspected CPM companions, we confirm their definitive physical association as documented by Column 5 of the table.

### 3.1.2. Linear Motions of Field Stars

In addition to identifying CPM companion candidates, we used the blink results to recognize many double star entries in catalogs such as the WDS as chance alignments of unrelated field stars. These stars show minimal motion upon blinking as is characteristic of distant field stars, or in a few cases, show motions in different directions than their primaries. Additionally, some close WDS pairs within the primary's saturation region were identified as field stars because their WDS measures demonstrated a linear change in relative separations, reflecting the differential proper motion between the two stars (see Figure 12 in Raghavan et al. (2009) for an example). Table 6 lists the 298 WDS entries for the stars of this study that were identified as optical doubles because they do not share the primary's proper motion. All data except the primary name in Column 4 are from the WDS, and the astrometry listed in Columns 6–8 is for the most recent observation in the catalog as of 2008 July.

## 3.2. *Hipparcos* Double Stars

In addition to producing unprecedented astrometry, the *Hipparcos* mission identified many stellar companions, which we evaluate next for our sample stars. The primary identification of companions in the *Hipparcos* catalog is reported in field H59, with further details in the *Double and Multiple Systems Annex*. Potential companions are identified as one of five types by various multiplicity flags: C, G, O, and X, which are discussed in the subsections below, and V (photocentric movement due to the variability of one or more components), which was not assigned to any star of this survey. In addition to the multiplicity flag, field H61 of the main catalog lists an 'S' for suspected non-single stars, based on the astrometric fit obtained, although a satisfactory double star solution could not be obtained. Some of these correspond to the 'X' entries of field H59. Also, field H52 contains a 'D' (duplicity-induced variability) identifying photometric variability presumably caused by a companion. While H52 and H61 contribute 16 possible companions that are not identified by field H59, we do not consider them reliable companion detection indicators for this work unless there is an independent confirmation. There are several additional flags in the catalog that can imply a companion, such as H10, which contains a reference flag for components of double or multiple systems, and H62, which is a component designation for double or multiple systems. However, these indicators overlap with the flags discussed above in all cases, and are not discussed further because they do not contribute any new companions. The following subsections treat each of the companion indicators considered and describe the methods of this survey in evaluating their verity.

### 3.2.1. *Component Solutions*

These companions, identified by a 'C' in field H59, represent double stars resolved by *Hipparcos* as separated components that could be modeled as single stars, usually with an assumed common parallax. For these systems, field H61 gives an indication of the reliability, with a quality of A (good), B (fair), C (poor), or D (uncertain). Among the sample studied here, 62 stars were flagged as *Hipparcos* component solutions, 59 of which have independent supporting evidence. Of the remaining three, two (HD 64606 and 148704) are refuted by this effort (see § 4.3) and one (HD 111312) remains a candidate binary. Table 7 lists all entries of this type identified by their HD and HIP names and the companion ID as in the *Annex*, along with the quality flag (field H61), the final status used in this effort, and the reason for this conclusion.

### 3.2.2. *Accelerating Proper Motions*

The *Hipparcos* catalog identifies many suspected companions because the proper motion of a presumed single star required higher-order terms to obtain an acceptable fit. The implied curvature in the proper motions of these stars was interpreted as the effect of an unseen companion, and these stars were accordingly flagged with a value of 'G' in field H59. In addition, significant differences between the *Hipparcos* proper motions, measured over approximately three years, and those from the Tycho-2 catalog, which contains the average of ground-based measures over a much longer time interval, can indicate the presence of an unseen companion (e.g., Makarov & Kaplan 2005; Frankowski et al. 2007, and references therein). This is because the *Hipparcos* measurements would be sensitive to deviations due to companions with periods of about 10 years, while the Tycho-2 measurements would average out the effects of such orbital motions and represent the transverse motion of the system. On the other hand, orbits of many decades to a few centuries will affect the Tycho-2 measurements, but hardly influence the *Hipparcos* data. Makarov & Kaplan (2005) followed this approach in identifying companions based on a $3.5\sigma$ difference between the proper motion in either coordinate from these two catalogs and estimated a minimum companion mass. Frankowski et al. (2007) developed a $\chi^2$ test to minimize false identification of companions, and identified 3 565 proper-motion binaries with a greater than 99.99% confidence level, which they estimate to be an order of magnitude better than that of Makarov & Kaplan (2005).

Table 8 lists the 91 stars from the current sample that either have a 'G' designation in *Hipparcos* field H59 ($N = 18$), or have a greater than $3\sigma$ difference between the *Hipparcos* and Tycho-2 proper motions ($N = 73$). The significance of the proper motion difference, listed in Column 8 of the table, is computed as the root-sum-squared of the difference in each axis divided by the corresponding larger error. Of the 18 entries in the table that have the *Hipparcos* 'G' designation, 16 have independent confirmation of a nearby companion or a greater than $3\sigma$ difference between the *Hipparcos* and Tycho-2 proper motions. The remaining two are retained as candidates because they meet the statistical criteria of Frankowski et al. (2007). Of the 73 stars with greater than a $3\sigma$ difference between the *Hipparcos* and Tycho-2 proper motions (but without a *Hipparcos* 'G' designation), 68 are considered phys-ical because they have independent confirmations of a nearby companion ($N = 65$) or meet the Frankowski et al. (2007) $\chi^2$ test ($N = 3$). The remaining five are retained as candidates for further investigation.

### 3.2.3. Orbital Solutions

Nineteen of the targets studied in this work have orbital solutions in the *Hipparcos* catalog, for which the *Hipparcos* data enabled the determination of at least one of the orbital parameters. Despite this evidence, we investigated each claim and found that three of them could in fact be refuted based on more recent studies, and one is retained as a candidate. Table 9 lists these stars along with their orbital periods from *Hipparcos*, the final status for our multiplicity statistics, and a corresponding reason for this conclusion.

### 3.2.4. Stochastic Solutions

Three of the sample stars have a stochastic solution designation in *Hipparcos*, implying that they had neither a satisfactory single nor double-star solution. The three stars are HD 21175, 200525, and 224930, all of which are confirmed to have companions (see § 4.3 for more details on the first two and § 3.4 for the third).

## 3.3. Speckle Interferometric Searches and the WDS and INT4 Catalogs

Building on comprehensive prior efforts (e.g., McAlister 1978a,b; McAlister et al. 1987, 1993; Hartkopf & McAlister 1984; Mason et al. 1998b; Hartkopf et al. 2008; Horch et al. 2008), we were able perform targeted speckle observations of 21 stars to ensure that every target on the list has at least one observation using this technique (Mason et al. 2010). While limited by $\Delta$mag $\lesssim 3$, speckle interferometry offers an efficient way of splitting pairs down to the diffraction limits of ground-based telescopes. This, along with the systematic searches for visual pairs for centuries, has resulted in the recording of many measures for each pair in the WDS, enabling us to definitively confirm many suspected companions as physical associations and show others to be unrelated field stars. The primary catalogs utilized by us for this purpose are the WDS and the INT4 as of 2008 July. The WDS lists measurements of pairs with epoch, separation, and position angle from techniques such as visual micrometry, speckle interferometry, and long-baseline optical interferometry (LBOI). The INT4 documents the results from high-resolution searches. While there is a significant overlap between these sources, the WDS includes historical observations which may not be of high precision, but nevertheless are very useful in significantly increasing the time baseline of measurements. On the other hand, the INT4 is useful in determining the status of many pairs, not only based on the measures of resolved pairs, but also because it includes the null results along with detection limits. The results from these techniques and catalogs are discussed in several sections. The linear motions of field stars were covered in § 3.1.2. Various other results are discussed throughout this work, including the evaluation of many *Hipparcos* pairs in § 3.2, and visual orbits derived from those measures in § 3.4. The WDS pairs that could be confirmed or refuted via photometric distance estimates are discussed next. Table 5 includes 23 confirmed and two refuted WDS pairs that were also seen upon blinking the archival images, and Table 10 lists the 12 additional WDS pairs that were confirmed using photometric distance estimates. Finally, the individual evaluation of other WDS entries and their corresponding results are discussed in § 4.3.

## 3.4. Visual Binaries in the ORB6 Catalog

The ORB6 catalog contains orbital solutions from the measurements of either relative separations between resolved pairs or photocentric motions of unresolved pairs. Orbits based on the explicit measurements of relative separations are coded with grades 1 (definitive) to 5 (indeterminate), those determined by visibility modulations measured by LBOI are coded with grade 8, and photocentric-motion orbits are flagged as grade 9. Table 11 lists the 98 orbits from this catalog for the stars of our sample, separated into the resolved and unresolved pairs. The first four columns identify the pair and are taken from the ORB6 catalog. Columns 5 & 6 from the WDS show the number of measurements of the pair and the time-span of these observations. Column 7 lists the orbital period from the ORB6 in days (d) or years (y), and is useful in matching spectroscopic orbits. Columns 8 & 9 list the orbit's grade and reference from the ORB6, and Column 10 identifies whether the orbit has a corresponding single-lined (1) or double-lined (2) spectroscopic orbit. Finally, Column 11 lists the status we adopt for the pair's physical association.

Orbits of grade 8 are robust solutions, typically combining spectroscopic and interferometric observations for short-period orbits. The quality of the other resolved-pair orbits degrades from 1–5, due to diminishing orbital coverage provided by measurements of the pair. Accordingly, there is a strong correlation between an orbit's grade and its period. While orbital solutions of grades 4–5 can potentially undergo significant revisions, the components are likely to be physically bound. Our investigation of the 72 individual systems found supporting evidence in all but one instance. We confirm physical associations due to one or more of the following factors: the existence of matching spectroscopic orbits, the many WDS measurements demonstrating not only CPM but also curved orbital motion, and in the case of very long-period orbits, the WDS measures showing CPM and linear motions consistent with a small arc of the orbit. The lone exception, HD 32923, despite an orbit of grade 3 in ORB6, was refuted by follow-up investigations (see § 4.3). We also investigated each of the 26 photocentric orbits in Table 11 and found all but four to be physical associations, confirmed by matching spectroscopic orbits or definitive photocentric orbits presented in the references indicated. We refute three claims of photocentric orbits, flagged as 'NO' in Column 11, because their expected radial-velocity modulations are not seen by high-precision measurements (Abt & Biggs 1972; Nidever et al. 2002; Gontcharov 2006; Latham et al. 2010). One orbit, noted as 'MAY' in Column 11, is retained as a candidate for further investigation.

## 3.5. Companions Listed in the CNS

While the astrometry of the CNS catalog has been superseded by *Hipparcos* (as discussed in § 2.3), the CNS also identified stellar components as either CPM pairs or ones with definitive or suggestive radial-velocity variations. For the stars studied here, the CNS listed 148 CPM companions and 50 suspected companions due to radial-velocity variations. Our investigations refuted six of the claimed CPM companions, and all remaining pairs were confirmed as physical due to supporting evidence from one or more of the methods described by this work. While most of the companions listed in this source are also present in the other sources checked, two wide companions (15′ from HD 63077 and 20′ from 137763) were identified solely based on their CNS entries, both of which were subsequently verified by blinking large-enough images and further confirmed by matching proper motions and parallaxes. The CNS, however, was not as accurate on claims of radial-velocity variations. The results of modern surveys have enabled us to refute 23 of these 50, while three more (HD 20010, 23484, 90839) remain candidates (see § 4.3). Table 12 lists the refuted companion claims. Columns 2 and 3 identify the specific companion claim (spectroscopic binary or radial-velocity variations). Column 4 lists the references we used to conclude that this star does not show any evidence of radial-velocity variations in high-precision, long-term campaigns and/or from independent measures over an extended time period.

### 3.6. Eclipsing Binaries

Most eclipsing binaries are more distant than our limit of 25 pc. Our search of the All Sky Automated Survey[5] for variable stars and Malkov et al. (2006), a catalog of 6 330 eclipsing binaries, reveals only three potential companions to the stars of this study: HD 123, 9770, and 133640. While the latter two are real, HD 123 was later refuted by Griffin (1999) (see § 4.3). For HD 9770, Cutispoto et al. (1997) present an eclipsing light curve and a corresponding 11.4-hour orbit. HD 133640 is listed in Malkov et al. (2006) with a period of 6.4 hours, which matches that of the double-lined spectroscopic orbit.

### 3.7. Brown Dwarf Companions

Searches for field brown dwarfs in infrared surveys, particularly the 2MASS database (Skrutskie et al. 2006), over the last ten or so years have identified many such substellar objects (e.g., Kirkpatrick et al. 1999, 2000). These have been augmented by high-resolution, high-contrast searches for substellar companions to nearby solar-type stars (e.g, Liu et al. 2002; Potter et al. 2002; Luhman et al. 2007). While these efforts have revealed several hundred brown dwarfs, a far greater tally would be expected if the stellar mass function extends into the sub-stellar regime. The deficiency of field and companion brown dwarfs has given rise to the term "brown dwarf desert". To identify brown dwarf companions to the stars in our sample, we searched the publications and also used the Dwarf Archives[6], an online source that maintains a comprehensive list of known brown dwarfs along with their spectral types and astrometric information. A search of the contents of this catalog as of 2009 November revealed 17 entries within half a degree of 15 sample stars. Eight of these entries are confirmed companions to seven stars (HD 3651, 79096, 97334, 130948, 137107, 190406, and 206860), at separations of $0\farcs9$–$200''$ from their primaries. Further details of these substellar objects are included in the tables and figures discussed in § 4.2, § 5.3.1, and § 5.3.8. A widely separated brown dwarf, about $27'$ from HD 145958, shows consistent proper motion and is hence retained as a candidate (see § 4.3). The remaining eight entries, representing brown dwarfs $225''$–$30'$ from their primaries (HD 53927, 86728, 90508, 130948, 136202, 161198, 202751, and HIP 91605), were refuted as physical companions because they are 2–3 magnitudes fainter than expected for their spectral type at the primary's distance, or because their proper motion and/or parallax measures from the Dwarf Archives or from Faherty et al. (2009) were significantly different than the corresponding primary's value. These eight chance alignments of unassociated brown dwarfs are consistent with the expectation of random alignments of the 453 sample stars and the 752 brown dwarfs in the Dwarf Archives, assuming isotropic distributions, which is valid for these nearby objects. Finally, none of the substellar companions to our sample stars in the exoplanet catalogs has a minimum mass of over 13 $M_J$, and only three of 52 have minimum masses greater than 5 $M_J$, consistent with the expectations of the brown dwarf desert.

---

[5]*http : //archive.princeton.edu/ asas/*

[6]*http : //spider.ipac.caltech.edu/staff/davy/ARCHIVE/index.shtml*

### 3.8. Planetary Companions

To complete our exhaustive search for companions, we extracted information on planetary companions from the two regularly updated online sources, namely, the Extrasolar Planets Encyclopedia[7] and the CCPS program's website[8]. As of 2009 October, 52 planetary companions are known to orbit 35 stars of our sample, excluding the Sun. Their details are listed in Table 13 and the relevant discussions are covered in §5. As discussed in §4.3, two of the planetary candidates, one each around HD 143761 and 217107, might, in fact, be stellar companions. Every exoplanet belonging to our sample was discovered by radial-velocity measurements and one (HD 189733b) has complementary photometric observations of transiting events as well (Bouchy et al. 2005).

### 4. Synthesis of Results

### 4.1. Nomenclature

We follow the Washington Multiplicity Catalog[9] (WMC) standards for naming stellar and brown dwarf companions as prescribed in Hartkopf & Mason (2004) and use the actual WDS designations when available. This hierarchical nomenclature designates components using a combination of uppercase and lowercase alphabets, and in the case of exceptionally complex systems, numbers as well. It is best understood by following the illustration of a fictitious example in Figure 10, which is adapted from a similar example in Hartkopf & Mason (2004) and described in the figure's caption. While this is not a perfect method, it adequately handles our evolving knowledge of components while capturing the hierarchical relationship between them. While the WMC standards were intended to cover companions of all types, a different de-facto standard has evolved for naming planetary companions, including some brown dwarf companions discovered by the planet search teams. This method attaches a lowercase alphabetic suffix, separated by a blank, to the host star's name for each substellar companion, starting with 'b' and incremented alphabetically in discovery sequence. Due to the wide acceptance of this standard, we follow this lowercase letter designation for the planetary companions.

---

[7]$http : //exoplanet.eu/$

[8]$http : //exoplanets.org/$

[9]$http : //www.usno.navy.mil/USNO/astrometry/optical − IR − prod/wds/wmc$

## 4.2. Multiplicity Results for Each System

Table 13 lists each star of our sample along with its stellar and planetary companions, if any. This table only includes confirmed or candidate companions, leaving out components that are now known to be physically unrelated. Such refuted components are discussed in the preceding sections and in § 4.3. The stars of the current sample are listed in order of right ascension, immediately followed in subsequent lines by the stellar and planetary companions of that system. Note that six stars in this sample (identified by superscript 'a' in Column 6) are actually not the primary components of their systems, but are still listed first within their group. Their primaries (component A, F0–K2) are not included in our sample because they are not *Hipparcos* stars or do not satisfy our $B - V$ color limits. Stellar companions are listed in order of proximity to the primary and planetary companions are listed directly under the star they orbit, also sequenced by proximity to the star. For ease of readability, the right ascension and declination (Columns 1 & 2) are only listed at the beginning of each group and correspond to the sample star. Column 3 lists the HD number of the star, when available, and Column 4 lists an alternate name of the star or the name of the planet. In the first line of each system, Column 5 lists an 'N' if this system has specific notes in § 4.3. For systems without notes, the companions listed, or lack thereof, are fully explained in the preceding sections and do not require any further comments.

For the first line of each system, Column 6 identifies the component designation of the sample star according to the nomenclature described above, unless it has no companions, in which case the column is empty. For stellar companions, this column identifies the pair whose details are listed in Columns 7–15. The stellar components of the pair are each named according to the WMC standards. The components are separated by a comma, which is suppressed when both components are represented by uppercase alphabets (e.g., AB; AB,C; and Aa,Ab are valid pair designations). In some instances, a single component may represent more than one star. For example, consider a visual binary, the secondary of which is itself a single-lined spectroscopic binary. This system will have the sample star listed first, followed by two pairs: AB representing the measurements of the visual binary, and Ba,Bb with details of the spectroscopic orbit. The B component of the AB pair hence identifies the spectroscopic binary and represents its center of light as observed by the visual observations. For planetary companions, this column is left empty. Column 7 lists the orbital period of the pair, when available from spectroscopic and/or visual orbit solutions, in hours (h), days (d), or years (y). Column 8 lists the angular separation measured between the components in arcseconds. It is the most recent measure from the WDS or other published references, or is our approximate measure from the archival images for new CPM companions. Column 9 lists the semi-major axis from orbital solutions when Column 8 is empty, or the projected linear separation corresponding to the angular separation in Column 8, in AU. Column 10 lists the status of companionship used for the multiplicity statistics derived here – 'Y' indicates a confirmed companion and 'M' implies an unconfirmed candidate. The next five columns list additional information about the techniques used to identify each companion as follows.

Column 11 corresponds to visual orbits and contains an 'O' for robust orbits of grade 1, 2, 3, or 8; 'P' for preliminary orbits of grade 4 or 5; and 'U' for unresolved photocentric-motion orbits (see § 3.4 and § 4.3 for details). Column 12 identifies spectroscopic companions as a '1' for single-lined orbits, '2' for double-lined orbits, and 'V' for radial-velocity variations indicating a companion, but without an orbital solution (see Latham et al. (2010) and § 4.3). Column 13 identifies CPM companions as close pairs with matching proper motions ('M'), pairs with evidence of orbital motion ('O') companions with matching proper motions and photometric distances ('P', see Tables 5 and 10), close pairs with published evidence of companionship ('R', see § 4.3), companions with matching proper motions and spectral type identifications that are consistent with the primary's distance ('S', see § 4.3), or pairs with independently-measured matching proper motions and trigonometric parallaxes ('T'). Column 14 identifies unresolved companions other than spectroscopic or visual ones such as eclipsing binaries ('E', see § 3.6), companions indicated by an overluminous star ('L', see § 4.3), or implied by proper motion accelerations ('M', see § 3.2.2). Finally, Column 15 identifies companions seen by CHARA LBOI as SFP ('S', Raghavan et al. 2010a) or as visibility-modulation binaries ('V', Raghavan et al. 2009, 2010b).

To identify the proficiency of various techniques in revealing companions and to illustrate the overlap among the various methods, Table 14 lists the number of confirmed companions found as visual binaries (VB, including resolved-pair and photocentric-motion binaries), spectroscopic binaries (SB, including single-lined, i.e., SB1, and double-lined, i.e., SB2, orbits, and radial-velocity variations without an orbit), CPM pairs (CP), and other techniques (OT, representing mostly accelerating proper motions, but including the few eclipsing binaries and over-luminous objects indicating unseen companions). Column 2 shows the total number of companions found by each of the above methods, Column 3 lists the number of companions detected uniquely by each method, and the remaining columns show the overlap of companion detections by more than one method.

### 4.3. Notes on Individual Systems

Following are the notes on individual systems which are marked with an 'N' in Column 5 of Table 13. These systems require explanations about confirmed, refuted, or candidate companions beyond what is covered in the previous sections.

**HD 123**: *Triple.* This system is composed of a 107-year visual-orbit pair, the secondary of which was shown to be the more massive component from absolute astrometry (Griffin 1999), suggesting that it itself was an unresolved binary. Brettman et al. (1983) reported a periodic variation in the component's brightness over roughly a 1-day period, which Griffin (1999) later disproved based on *Hipparcos* photometry and instead showed it to be a spectroscopic binary with a 47.7-day period, estimating component masses of 0.98, 0.95, and 0.22 $M_\odot$ for the three components.

**HD 1237**: *Binary, one planet.* In a systematic search for faint companions to planet hosts, Chauvin et al. (2006) discovered a CPM companion to this star using VLT NACO adaptive optics and demonstrated orbital motion. Chauvin et al. (2007) characterized the companion as M4±1V.

**HD 3651**: *Binary, one planet.* Luhman et al. (2007) reported the discovery of a T7.5 ± 0.5 companion 43″ away from this planet-host star using Spitzer IRAC images, and confirmed CPM using 2MASS images. The brown dwarf's infrared colors are consistent with the distance to the primary, confirming companionship. They estimate the companion's mass as 0.051 ± 0.014 $M_\odot$ and age as 7 ± 3 Gyr by comparing luminosity with evolutionary tracks. This was the first substellar object imaged around an exoplanet host. An additional component listed in the WDS is clearly optical (see Table 6).

**HD 4391**: *Triple.* The WDS lists three measurements of a companion with separations ranging 10″.0–16″.6 over 98 years, and we discovered an additional companion 49″ away by blinking archival images (see § 3.1.1). Our $VRI$ images, obtained in 2007 July and October at the CTIO 0.9-m telescope, clearly reveal both companions, confirming CPM. The closer companion was saturated in all but one $V$-band image, but the differential photometry extracted from this image allowed confirmation as a companion (see Table 10). For the newly-discovered wider companion, our absolute $VRI$ photometry along with 2MASS $JHK_S$ magnitudes confirmed a physical association (see Table 5).

**HD 4628**: *Single,* candidate *Binary.* Heintz & Borgman (1984) detected a companion, $2\overset{''}{.}7$ away, on 11 exposures over two nights, but did not see the companion on 164 other plates or on multiple visual checks with a micrometer. Their two observations about 25 days apart show evidence of variation in the companion's brightness by about 1 magnitude. Heintz (1994) notes an acceleration in proper motion for the primary and speculates that this might be caused by the companion reported earlier. Roberts et al. (2005) did not detect a companion using Adaptive Optics (AO) down to $\Delta I \lesssim 10$ and note that only a white dwarf companion could have escaped detection, while the flaring companion as seen by Heintz should have been detected. Moreover, the *Hipparcos* and Tycho-2 proper motions of HD 4628 match to within $2\sigma$, and the INT4 lists several null results with speckle interferometry and adaptive optics. BDM and DR observed this target using the KPNO 4-m telescope in 2008 June, and while the separation was too wide and the $\Delta$m too large for speckle observations, the finder TV showed a faint source about $5''$ away at about 230°. Could this be the companion seen by Heintz after about 30 years of orbital motion? While a possibility, follow-up observations by Elliott Horch two days later with the WIYN 3.6m telescope on KPNO failed to identify the source. Additionally, 5-second exposures in $VRI$ taken by TJH at the CTIO 0.9-m telescope in 2008 June also failed to identify any companion, although saturation around the primary could hide the companion in these images. Nidever et al. (2002) report that the primary shows no variations in its radial-velocity measurements. At this time, we do not have sufficient information to confirm or refute this companion, although, chances of a physical companion appear slim. The wider component listed in the WDS is clearly optical (see Table 6).

**HD 4676**: *Binary.* Boden et al. (1999) presented a visual orbit based on LBOI observations for this 14-day SB2 and derived component masses of $1.223 \pm 0.021$ M$_\odot$ and $1.170 \pm 0.018$ M$_\odot$. Earlier, Nadal et al. (1979) had speculated on the presence of a third companion based on temporal changes in the spectroscopic orbital elements. While this suspected companion has been mentioned in subsequent literature (Fekel 1981; Tokovinin et al. 2006), it was refuted by Boden et al. (1999) based on imaging and spectroscopic evidence. Two additional components listed in the WDS are clearly optical (see Table 6).

**HD 9826**: *Binary, three planets.* Lowrance et al. (2002) discovered an M4.5V companion, $55''$ away from planet host $\upsilon$ And and confirmed its physical association by demonstrating CPM and showing that its spectral type is consistent with its magnitudes at the primary's distance. Two additional components listed in the WDS are clearly optical (see Table 6).

**HD 13445**: *Binary, one planet.* This planet-host star with a 4 M$_J$ planet exhibits a longterm trend in radial velocity, consistent with a stellar companion beyond 20 AU (Queloz et al. 2000). A significant difference between the *Hipparcos* and Tycho-2 proper motions (see

Table 8) also suggests a nearby unseen companion. Later work has resolved this companion and demonstrated orbital motion (Lagrange et al. 2006). The companion was initially misidentified as a T dwarf (Els et al. 2001) and later shown to be a white dwarf based on spectroscopy (Mugrauer & Neuhäuser 2005) and a dynamical analysis of astrometry and radial velocities (Lagrange et al. 2006).

**HD 20010**: *Binary,* candidate *Triple.* The secondary of a 5″ CPM pair with a preliminary visual orbit is listed in the CNS as "RV-Var?". Eggen (1956) mentions that there is a strong evidence of variability, quoting van den Bos (1928), but this reference could not be found. With insufficient evidence to confirm or refute a physical association, the additional companion is retained as a candidate.

**HD 20807**: *Binary.* The wide CPM companion, HD 20766, lies 309″ away and is confirmed by matching proper motions and parallax. Additionally, the WDS lists a single speckle interferometry measure of a companion in 1978, 0″046 away at 11° (Bonneau et al. 1980). However, Bonneau et al. failed to resolve the companion in 1979 and da Silva & Foy (1987) mention that the 1978 measure was in fact an artifact in the diffraction pattern of the telescope spider.

**HD 21175**: *Binary.* While this companion only has three ground-based measurements, they span more than 50 years and are consistent with a bound pair. The *Hipparcos* catalog also identifies this star as a suspected binary because the astrometry did not adequately fit either a single or binary solution (H59 = 'X'). Söderhjelm (1999) presents a visual orbit combining *Hipparcos* and ground-based measures, confirming a physical association.

**HD 22049**: *Single, one planet.* This is $\epsilon$ Eri, the well-studied exoplanet host. The WDS lists a single speckle resolution of a potential stellar companion, 0″048 away (Blazit et al. 1977), significantly closer than the planet. However, 13 other attempts by speckle and AO have failed to resolve the companion (e.g., McAlister 1978a; Hartkopf & McAlister 1984; Oppenheimer et al. 2001). Presumably, the Blazit measure is spurious. The WDS lists 10 additional components, all of which were confirmed as optical by blinking archival images (see Table 6).

**HD 23484**: *Single,* candidate *Binary.* The CNS lists this as "RV-Var", but no radial-velocity data could be found in modern surveys. Catalogs (Abt & Biggs 1972; Duflot et al. 1995; Gontcharov 2006) list velocities with RMS scatter of about 3 km s$^{-1}$, but this could be due to measurement errors or zero-point variances. This candidate companion is retained as a candidate.

**HD 24496**: *Binary.* The two measurements with $\rho = 2″6$–$2″7$ and $\theta = 254°$–$256°$ listed in the WDS are by Wulff Heintz, nine years apart and consistent with a bound pair. The first measure is based on observations over three nights and the second on observations over two additional nights. Given the quality of the observations ($\Delta$m = 4–5 measured) and the reasonably high proper motion of the primary, this is likely a physical companion, but one that could use new measurements.

**HD 25680**: *Binary*. A companion $0\rlap{.}''2$ away was discovered by McAlister et al. (1993) with speckle interferometry and confirmed by the same technique by Hartkopf et al. (2008). These measures show evidence of orbital motion, and given the $0\rlap{.}''2\,\mathrm{yr}^{-1}$ proper motion of the primary and an elapsed time of 15 years between them, this companion can be confirmed as physical. Due to the constant radial velocity of the primary (Latham et al. 2010), this might be close to a face-on orbit. The WDS lists another potential companion $177''$ away, which we also identified by blinking archival images. This candidate (HIP 19075) was however refuted based on its significantly different proper motion in *Hipparcos* from the corresponding value of the primary (see Table 4). The two additional WDS entries are clearly optical (see Table 6).

**HD 26491**: *Binary*. A comparison of *Hipparcos* and Tycho-2 proper motions shows a significant difference suggesting a companion (see §3.2.2), which is confirmed by radial-velocity variations (Jones et al. 2002). A preliminary spectroscopic orbit exists (H. Jones 2008, private communication).

**HD 32923**: *Single*. The WDS lists 19 measurements at roughly $0\rlap{.}''1$ separation over 76 years, and Eggen (1956) even derived two preliminary visual orbits from these measures. However, Heintz & Borgman (1984) suggest that this is likely spurious and show that the observations are not consistent with orbital motion of any period. Three additional speckle observations exist since the Heintz & Borgman publication, from 1984–1987 (Tokovinin 1985; Tokovinin & Ismailov 1988; McAlister et al. 1993), but there are 17 null detections listed in the INT4 by speckle interferometry as well as by AO. This star has a stable radial velocity in Nidever et al. (2002) and Latham et al. (2010). It appears that these multiple, but sporadic, measures are spurious.

**HD 35296**: *Binary*. DM91 noted the primary of a $12'$ CPM pair as "SB", but one that was not confirmed by their work. Modern measures (Nidever et al. 2002; Latham et al. 2010) show this star to have a stable radial velocity, refuting the earlier claim.

**HD 36705**: *Quadruple*. The WDS lists two measures of this $10''$ pair (AB Dor AB), separated by 69 years and consistent with a bound pair. The first observation, by Rossiter (1955), measured a $\Delta$m $\sim 6$, explaining the lack of many more observations. Close et al. (2005) recovered this pair with AO at the VLT, and it is also seen in $VRI$ images obtained by TJH in 2008 September at the CTIO 0.9-m telescope. While the photometric distance estimate varies from the primary's *Hipparcos* distance by $7\sigma$ (see Table 10), the $V$ mag-nitude from Rossiter (1955) is likely approximate. Given the high proper motion of the primary, the consistent measures over 79 years indicate a physical association. The 2MASS colors indicate an M-dwarf with a V magnitude estimate of about 12.0, in good agreement with the measure of Rossiter (1955) and consistent with the primary's *Hipparcos* distance. High-contrast AO efforts have split each of these components into binaries themselves. The primary was identified by *Hipparcos* as showing accelerating proper motion, indicating an unseen companion, and this is supported by the significant difference between *Hipparcos* and Tycho-2 proper motions (see § 3.2.2). The suspected companion has since been revealed by VLBI (Guirado et al. 1997), resolved by AO (Close et al. 2005), and confirmed as a physical association by photometry and spectroscopy (Close et al. 2005, 2007; Boccaletti et al. 2008, and references therein). Close et al. (2005) also split the secondary into a 0″.070 pair, which was later confirmed by Janson et al. (2007) who measured it 66.1 mas away at 238°.5.

**HD 40397**: *Triple.* The five measures in the WDS for AB between 1902 and 1932 are consistent with a bound pair. The measured $\Delta m \sim 7$ makes this a difficult target for classical techniques and out of the reach of speckle interferometry. Given that more than 70 years have passed since the latest measure, this is a good candidate for follow-up AO observations. This pair also has a wide CPM companion, NLTT 15867, which was confirmed by photometric distance estimates (see Table 10), and an additional WDS component, which is clearly optical (see Table 6).

**HD 43834**: *Binary.* Eggenberger et al. (2007) detected a companion, 3″ away, three times over three years with AO at the VLT, demonstrating CPM and showing a hint of orbital motion. They also mention a linear trend in CORALIE radial velocities consistent with this companion, confirming a physical association, and estimate the companion to be M3.5–M6.5 with a mass of $0.14 \pm 0.01$ M$_\odot$.

**HD 45270**: *Single,* candidate *Binary.* The WDS lists three measurements spanning 43 years of a $\Delta m \sim 4$ companion separated by about 16″, which are consistent with a bound pair. Curiously, no additional measurements exist. This companion was listed in the *Hipparcos* input catalog, but not resolved by *Hipparcos*. 2MASS lists a source near this candidate companion, but it is clearly not the same star because its infrared colors are more than three magnitudes fainter than the visual magnitude of 10.6 from the *Hipparcos* input catalog. No additional information was found on this companion and hence it is retained as a candidate.

**HD 48189**: *Binary.* The WDS lists 19 measurements over 105 years that are consistent with a bound pair. During this time, the separation has closed in from about 3″ to about 0″.3 and the position angle has changed by about 15°. Given the small projected separation of 6–60 AU, one might expect a greater change in position angle as evidence of orbital motion. The change of only $15°$ indicates that the semimajor axis is larger than the observed separations, perhaps due to a high inclination. While a more robust confirmation is not available, the primary has moved about $9''$ during the measures, and the companion seems to be moving along with it, indicating a physical association.

**HIP 36357**: *Triple,* candidate *Quadruple.* The primary of this system, HD 58946, lies about $13'$ away, and its physical association is confirmed by matching parallax and proper motion. The primary has a closer companion, about $3''$ away, confirmed by greater than $11\sigma$ difference between *Hipparcos* and Tycho-2 proper motions (see Table 8) and five measurements in the WDS over 25 years which demonstrate CPM and orbital motion. Additionally, proper motion variations suggest an unseen companion to HIP 36357 as well, but one we could not confirm (see Table 8). Two wider WDS components are clearly optical (see Table 6).

**HD 64606**: *Binary.* For the primary of an SB1 pair, the WDS lists two measures of a $\Delta m \sim 4$ component separated by $4''.9$, one each from *Hipparcos* and Tycho. The *Hipparcos* solution is flagged as "poor" quality, and there is no independent confirmation of this companion. TJH observed this star using the CTIO 0.9-m telescope in 2008 September and obtained 1-second exposure images in $VRI$. No source was found at the expected position in those images, whereas a companion of $\Delta m \sim 4$ should easily have been seen above the background. While the SB1 pair is real, this astrometric detection is refuted.

**HD 65907**: *Triple.* LHS 1960 is a companion to this star, separated by about $60''$, and confirmed by photometric distance estimates (see Table 5). The WDS lists four measures of an additional companion to LHS 1960, observed 1930–1983, indicating that this component itself is a $3''$ CPM binary. No further evidence of companionship could be found, but given the high proper motion of the primary and the four consistent measurements by four different telescopes over 53 years, this system can be confirmed as a triple.

**HD 68257**: *Quintuple,* candidate *Sextuple.* The three brightest roughly solar-type components ($\zeta$ Cancri A, B, and C) are supported by over 1000 visual measurements each, corresponding to two visual orbits. Component C has been noted to have an irregular motion for most of its history and was identified as an SB1 with an orbit of $6302 \pm 59$ days (Griffin 2000), consistent with earlier astrometric orbits. However, earlier efforts (Heintz 1996) had noted a mass ratio for the C component binary of about 1, and with C being a G0 star, the non-detection was puzzling and attributed to the companion being a white dwarf or itself a binary. Hutchings et al. (2000) finally observed this pair (Ca,Cb) via AO observations at infrared wavelengths, designated Cb as an M2 dwarf based on its infrared colors, and argued on the basis of prior mass-ratio estimates that it itself is an unresolved binary (Cb1,Cb2). Richichi (2000) confirmed the presence of Cb via lunar occultation measures. While she could not confirm its binary nature, her K-band photometry supported the binary M-dwarf hypothesis, for which she determined an upper-limit for projected separation of 20-30 mas. Further, Richichi reported the potential discovery of a sixth component in this system (Cb3). While seen just above her detection limit and hence retained as a candidate for this work, she nonetheless confirmed its presence by three independent data analysis methods and excluded it from being the unresolved companion Cb2 noted above due to its larger separation of at least 1.6 AU from the lunar occultations. She tentatively identified Cb3 as an M2–M4 dwarf. In addition to all this, we identified a potential wide companion, $372''$ away at $107°$, which was later refuted (see Table 4), as were six additional WDS components, which are clearly optical (see Table 6).

**HD 72760**: *Binary.* This companion was suggested by a significant difference in *Hipparcos* and Tycho-2 proper motions. Recently, Metchev & Hillenbrand (2009) resolved the companion in a Palomar/Keck AO survey, confirmed companionship based on color and magnitude measurements, and estimated the companion's mass as 0.13 $M_\odot$.

**HD 73350**: *Single.* The WDS lists a B component $60''$ away and a C component about $10''$ from B. While the DSS images were taken over just a two-year interval, the SSS image provides a longer time baseline and helps confirm component B (HD 73351) as a field star (see Table 6). Component C is a CPM companion of B based on three consistent measures separated by over 100 years, and hence also physically unassociated with our sample star.

**HD 73752**: *Binary.* The CNS lists the primary of the $1''3$ visual binary as "SB" and notes that there are suspected perturbations in its proper motion. The reference detailing the perturbations (Hirst 1943) presents a 35-year inner orbit, which is noted as very preliminary with several different orbits equally permissible. The author also states that systematic effects alone may explain the residuals. His 214-year outer orbit was later revised to 145 years by Heintz (1968), who also points out that the observed range of radial velocities could be ascribed to scatter. Adopting a parallax of $0''058$, he derived a mass-sum of 1.1 $M_\odot$, and noted that at least one component must be over-luminous. If we adopt the HIP parallax of 50.2 mas, we get a mass-sum of 1.9 $M_\odot$, so the components are likely not over-luminous. Radial-velocity catalogs (Abt & Biggs 1972; Gontcharov 2006) list velocities in the range 40–48 $km\,s^{-1}$, but the differences could be due to zero-point offsets between observers. The early claim of a possible companion is not supported by subsequent observations, which in fact question it. While the visual binary is real, the third component is refuted. An additional wide component listed in the WDS is clearly optical (see Table 6).

**HD 75767**: *Quadruple.* Tokovinin et al. (2006) reported the discovery of a wide $\Delta m = 4.3$ companion to a 10.3-day SB1 binary with NACO AO and confirmed CPM using a partial resolution in 2MASS images. This companion was independently discovered by Fuhrmann et al. (2005), who obtained two observations four years apart, demonstrating CPM, and confirmed companionship by showing consistent radial velocity with the primary. Their spectra also enabled them to identify the companion itself as a double-lined binary, as evidenced by its H-alpha emission and near-infrared absorption lines appearing as pairs with an offset of about 21 km s$^{-1}$. Using composite-spectrum analysis, they derived spectral types of M3 and M4. Blinking archival images revealed a possible fifth companion 385″ away, and its photometric distance estimate matches the primary's *Hipparcos* value within 2$\sigma$. However, the Lépine & Shara (2005) proper motions of the two stars are significantly different, indicating that this might be a comoving star perhaps born out of the same cloud as HD 75767, but one that is not gravitationally bound to it.

**HD 79096**: *Quadruple.* Wilson et al. (2001) discovered an L8V companion (Gl 337C) 43″ from the SB2VB pair from 2MASS images. The two images, separated by 2.5 years, demonstrated CPM. They also showed that the magnitudes are consistent with the primary's distance to within 1$\sigma$, confirming companionship. Burgasser et al. (2005) later resolved Gl 337C as a nearly equal-magnitude binary separated by 0″53 ± 0″03 at 291° ± 8° using Lick natural guide star AO. Companionship was confirmed based on proximity and CPM, which was demonstrated by the absence of a source in 2MASS images at the expected position of a background star. Three other components listed in the WDS are clearly optical (see Table 6).

**HD 86728**: *Binary,* candidate *Triple.* Gizis et al. (2000) identified a wide CPM companion from 2MASS and confirmed it with a spectral type classification of M6.5. However, based on it being over-luminous ($M_K = 8.19$ using 2MASS magnitudes and *Hipparcos* parallax versus 9.60 for its spectral type) and having high activity (emission observed twice), they argued that it is an unresolved equal-mass binary, or even possibly a triple. We could not find any follow-up work confirming or refuting this claim, so while this system is confirmed as a binary, we retain the third component as a candidate. Additionally, the Dwarf Archives lists a brown dwarf about 25′ away from the primary, but its photometric distance estimate suggests a distant field object.

**HD 90839**: *Binary,* candidate *Triple.* The CNS lists the primary of a wide CPM pair as "SB?" and the secondary (HD 237903, GJ 394) as "RV-Var". The primary is a constant velocity star (see Table 12), refuting the CNS claim, but the modern surveys did not observe the secondary. DM91 listed this companion with a constant velocity of 8.24–8.62 km s$^{-1}$ over 700 days. Heintz (1981) listed velocities of 7.7–8.4 km s$^{-1}$ over four days but noted that the coverage was too weak to definitively show velocity variations. He also noted that the spectrum had emission features. Wilson (1967) listed a velocity of 7.8 km s$^{-1}$ over three plates with a range of 7.7 km s$^{-1}$ and standard deviation of 3.1 km s$^{-1}$. Radial-velocity catalogs (Abt & Biggs 1972; Duflot et al. 1995) list values that range over many $\mathrm{km\,s^{-1}}$, but this could be due to zero-point differences between observers, and these catalogs do not note any variation. While the wide binary is confirmed based on matching parallax and proper motion and the primary's SB claim is refuted, the possible radial-velocity variation of the secondary is inconclusive and hence retained as a candidate. An additional WDS component is clearly optical (see Table 6).

**HD 97334**: *Triple.* Kirkpatrick et al. (2001) discovered an L4.5V CPM companion (Gl 417B) 90″ away at 245° from the primary using 2MASS images and confirmed a physical association by demonstrating CPM and consistent parallaxes. Bouy et al. (2003) later resolved this brown dwarf into a binary (0″.070 ± 0″.0028 at 79°.6 ± 1°.2) using HST WFPC2. While companionship of this pair has not been established conclusively, proximity argues for a physical association. Three additional WDS components are clearly optical (see Table 6).

**HD 98230**: *Quadruple.* $\xi$ UMa is a quadruple system composed of a 60-year visual binary, the primary of which is a SB1VB and the secondary is an SB1. Mason et al. (1995) reported a possible fifth companion detected via speckle interferometry near the secondary. While the single detection reported is quite convincing, this companion has never again been seen, despite multiple attempts. Our efforts with CHARA, while limited to $\Delta K \lesssim 2.5$, also failed to resolve any additional components. Given only one measure and about a dozen null results with the same technique, this new companion is likely a chance alignment of an unrelated star. An additional wide component listed in the WDS is clearly optical (see Table 6).

**HD 100180**: *Binary,* candidate *Triple.* The primary of the 15″ CPM binary (see Table 5) has two speckle interferometry measurements of a close companion in the WDS, observed 0″.035 away at 6°.8 in 2001 and 0″.122 at 355°.8 in 2004. One of the two attempts by BDM and DR at the KPNO 4-m telescope in 2008 June resulted in an "uncertain" measure of 0″.218 at 14°.6. Given the 0″.378 yr$^{-1}$ proper motion of the primary, these measures are consistent with a bound pair, but further observations are warranted to obtain a definitive confirmation, especially given the constant radial velocity reported by Nidever et al. (2002).

**HD 100623**: *Binary.* The WDS lists only a single measure of this large $\Delta$m companion discovered by Luyten in 1960. While the proximity and large magnitude difference make follow-up observations difficult, Henry et al. (2002) obtained spectra of this 15th magnitude companion and showed that it is a DC or DQ white-dwarf, not an M-dwarf as reported in the CNS. The second observation confirms CPM, and the spectral type and photometry are consistent with a physical association.

**HD 102365**: *Binary.* The companion, discovered by Luyten in 1960, is LHS 313 and has a proper motion that matches the primary's $1\rlap{.}''6\,\mathrm{yr}^{-1}$. Hawley et al. (1996) identified the companion as M4V, which was recovered by 2MASS at a similar position angle and separation as Luyten's observation. Its infrared colors are consistent with an M4 dwarf at the primary's distance.

**HD 103095**: *Single.* The CNS and DM91 listed a companion with separation $2''$ at $175°$. DM91 noted that the companion was flaring with magnitudes of 8.5–12 and also mentioned that it was normally not seen. The INT4 lists four null measurements with speckle interferometry and there are no radial-velocity variations. Three attempts by BDM and DR in 2008 June at the KPNO 4-m telescope failed to identify a companion. Recently, Schaefer et al. (2000) have shown that the brightness enhancements observed are likely due to superflares on the stellar surface rather than due to a companion.

**HD 109358**: *Single.* The WDS lists a single speckle measure of a companion $0\rlap{.}''1$ away (Bonneau & Foy 1980), but the INT4 has over 20 null speckle detections. BDM and DR failed to resolve the suspected companion in 2008 June at the KPNO 4-m telescope. Given the mention of telescope artifacts as being responsible for some detections by this observer (da Silva & Foy 1987), we side with the many null detections, including one by the same observer. Additionally, the CNS listed this star as "SB" and Abt & Levy (1976) presented a preliminary orbital solution. However, those claims were subsequently refuted (Morbey & Griffin 1987). The wide AB pair in the WDS is clearly optical, as seen on blinking the archival images (see Table 6).

**HD 111312**: *Binary,* candidate *Triple.* This is a 2.7-year SB2. The WDS lists a single speckle measure with a separation of $0\rlap{.}''089$ at $90°\!.6$ in 2001 and the pair was seen again in 2006 with a separation of $0\rlap{.}''050$ at $44°\!.6$ by BDM. These measurements are consistent with the spectroscopic binary and more observations are needed to develop a visual orbit. The WDS lists an additional companion, $2\rlap{.}''7$ away with $\Delta m \sim 4$ based on *Hipparcos* and Tycho measures. The *Hipparcos* solution is flagged as "poor" quality, and there is no independent confirmation of this companion. Its orbital period, if real, would be too long to affect the velocities obtained over some 7 years. With no conclusive evidence to confirm or refute this companion, it is retained as a candidate requiring further observations.

**HD 112758**: *Triple.* This is a triple system with an inner SB1 pair and a wider visual component which was first resolved by van den Bos in 1945 and then again in 1960 with $\Delta m \sim 5$. McAlister et al. (1987) recovered this visual companion in 1983, and the three observations show evidence of orbital motion. The McAlister et al. measurement with speckle interferometry implies $\Delta m \lesssim 3$, suggesting that the companion may be variable. BDM and DR attempted to resolve this companion at the KPNO 4-m telescope in 2008 June, but could not see it, perhaps because of the large magnitude difference.

**HD 113449**: *Binary.* *Hipparcos* presents a photocentric orbit for this star with a period of about 231 days. Moore & Paddock (1950) noted this star as a radial-velocity variable, Gaidos et al. (2000) mentioned that the velocity changed by 20 km s$^{-1}$ over 10 months, and variations are also seen in the measurements reported in Latham et al. (2010). However, no definitive orbit exists. The companion was resolved at the Palomar 200-inch telescope with aperture masking in 2007 January, $35.65 \pm 0.6$ mas away at $225°\!.2 \pm 0°\!.5$ with $\Delta H \sim 1.6$, and confirmed at the Keck telescope more than a year later (M. Ireland 2008, private communication). With consistent astrometric and spectroscopic evidence, this is a bound pair.

**HD 120136**: *Binary, one planet.* $\tau$ Boo hosts a 4.13 M$_J$ minimum-mass planet in a 3-day orbit. Additionally, 56 observations over 170 years in the WDS confirm a stellar companion based on CPM and orbital motion, which has a preliminary orbital solution (see Table 11). Wright et al. (2007) and references therein mention a longterm drift in radial velocity consistent with this visual companion.

**HD 120780**: *Triple.* The WDS lists two measures of a $6''$ pair 51 years apart and consistent with a bound system, but follow-up observations have been difficult due to a magnitude difference of $\sim 5.5$. We obtained $I$-band images in 2006 July and 2007 June at the CTIO 0.9-m telescope. The companion was seen at both epochs about $5''\!.6$ away at $89°$ with $\Delta I \sim 3.3$. These three observations demonstrate CPM with a fast-moving primary $(0''\!.6\,\mathrm{yr}^{-1})$, and, in fact, hint at orbital motion, confirming companionship. Additionally, *Hipparcos* identifies this star as an accelerating proper-motion binary, and the Tycho-2 proper motion differs from the *Hipparcos* value by a $19\sigma$ significance (see Table 8). While the Tycho-2 proper motion, averaged over about 100 years, is no doubt affected by the wide pair mentioned above, whose orbital period could be about 1000 years, the *Hipparcos* observations are over some three years and indicate a closer companion.

**HD 124850**: *Single,* candidate *Binary.* The ORB6 catalog lists a photocentric orbit for this star with a period of 55 years and an inclination of $60°$. The corresponding preliminary orbit was recently presented by Gontcharov & Kiyaeva (2010) by combining *Hipparcos* data with astrometric ground-based observational catalogs, but we do not find the motion convincing enough to confirm the companion. This star was not included in the Nidever et al. (2002) or Latham et al. (2010) survey, and radial-velocity catalogs (Abt & Biggs 1972; Duflot et al. 1995; de Medeiros & Mayor 1999; Gontcharov 2006) do not indicate variations. BDM and DR could not resolve any companion via speckle interferometry on the KPNO 4-m telescope in 2008 June. With inconclusive evidence to confirm or refute this companion, it is retained as a candidate.

**HD 125276**: *Single,* candidate *Binary.* The *Hipparcos* and Tycho-2 proper motions

differ by greater than a $3\sigma$ significance, suggesting an unseen companion (see Table 8). These proper motion variations could be due to a WDS and CNS companion, separated by $3''$–$8''$ over four measures between 1891 and 1936. Some of these measures indicate a $\Delta m \sim 8$, which might explain the several non-detections also included in the WDS. An attempt by BDM and DR at the KPNO 4-m telescope in 2008 June could not detect this companion. While the pair is too wide and too high in contrast for detection via speckle, the finder TV, sensitive to faint companions, did not reveal any source at the expected position. Without conclusive evidence to confirm or refute this companion, it is retained as a candidate. An additional WDS companion is clearly a background star as seen by blinking archival images (see Table 6).

**HD 125455**: *Binary*. This companion was discovered by Kuiper in 1937 and has measurements in 1960 and 1987 that are consistent with a bound system. The companion is LHS 2895 with a proper motion that matches the primary's, and its 2MASS colors indicate a late M-dwarf at approximately the primary's distance.

**HD 128620**: *Triple*. This is the closest known star system, $\alpha$ Centauri, which is composed of an SB2VB pair and a wide companion, Proxima Centauri, about $2°$ away. While the angular separation to Proxima is extreme for bound systems, it translates to a linear projected separation of $10\,000$ AU, which is well within the limits of gravitationally bound pairs. Wertheimer & Laughlin (2006) used kinematic and radial-velocity data to show that Proxima Centauri is bound to $\alpha$ Centauri. A possible new substellar companion to Proxima Centauri was reported by Schultz et al. (1998) $0\rlap{.}''34$ away using the HST FOS as a coronagraphic camera. In a follow-up effort, Golimowski & Schroeder (1998) used HST WFPC2 to show that the FOS feature seen was likely an instrumental effect and exclude any stellar or substellar companion within $0\rlap{.}''09$–$0\rlap{.}''85$ of Proxima Centauri.

**HD 130948**: *Triple*. Potter et al. (2002) discovered a pair of brown dwarf companions using AO on the Gemini North 8-m telescope. They demonstrated CPM with observations over seven months and confirmed companionship based on their infrared colors, spectral-type of dL2 $\pm$ 2 and a consistent age with the primary of less than 0.8 Gyr derived by comparing their position on an HR diagram with theoretical models. They also noted that the relative youth is consistent with the high X-ray activity, Li abundance, and fast rotation. The Dwarf Archives lists an additional brown dwarf $523''$ away from the primary, but its photometric distance estimate suggests a distant field object. An additional CNS claim of an SB companion was refuted by modern surveys (see Table 12).

**HD 135204**: *Binary*. A companion, $0\rlap{.}''1$ away, is listed in the WDS and CNS and confirmed by the 17 measurements in the WDS over 82 years which not only demonstrate CPM, but also orbital motion.

**HD 137107**: *Triple.* Kirkpatrick et al. (2001) discovered a wide L8V companion (Gl 584C) to the SB2VB binary using 2MASS images and confirmed the physical association with additional measures and spectroscopy. Two additional WDS components are clearly optical (see Table 6).

**HD 137763**: *Quadruple.* The primary of a 52″ CPM binary is itself SB2, and also has a wide companion about 20′ away that was first mentioned by CNS and confirmed by a spectral-type identification of M4.5 and a distance estimate of 21.6 ± 1.9 pc (Reid et al. 1995). Another wide WDS component is clearly optical (see Table 6).

**HD 140901**: *Binary.* The WDS lists seven measures for this companion from 1897 to 1960 which are consistent with a bound system. We obtained $VRI$ images at the CTIO 0.9-m telescope in 2006 July and 2007 July, which reveal the companion at the expected position, confirming CPM. The magnitude difference of over six makes photometry difficult, but given the large matching proper motion and proximity, this companion can be confirmed as physical. An additional WDS component is clearly optical, as seen when blinking the archival images (see Table 6).

**HD 141272**: *Binary.* The WDS lists four micrometer observations of this pair over 56 years that are consistent with a bound system. Eisenbeiss et al. (2007) confirm companionship using photometry and spectroscopy, and derive an estimated mass for the dM3 ± 0.5 companion of $0.26^{+0.07}_{-0.06}$ M$_\odot$.

**HD 143761**: *Single planet-host or binary with no known planets.* ρ CrB definitely has a companion, but it is not clear whether it is planetary or stellar in nature. *Hipparcos* identified a photocentric orbit with a period of 78 days, exactly twice that of the planetary companion reported by Noyes et al. (1997). Gatewood et al. (2001) used *Hipparcos* and ground-based observations to conclude that the photocentric orbit is of the same period as the planet, and in fact the "planet" is an M-dwarf companion with a mass estimate of 0.14 M$_\odot$ in a nearly face-on orbit. Bender et al. (2005) failed to identify such a companion using high-resolution infrared spectroscopy, and placed an upper limit on the companion's mass of 0.11–0.15 M$_\odot$. Baines et al. (2008) attempted to resolve this question with LBOI observations at the CHARA Array, and could not settle the issue once again. While interferometric visibilities did not perfectly fit a single-star solution, additional data are required for a definitive conclusion. This system has a stellar or planetary companion, but not both. Further observations are warranted. A WDS component is clearly optical (see Table 6).

**HD 144284**: *Binary.* Mazeh et al. (2002) presented a 3-day SB2 orbit for this star using infrared spectroscopy to measure the faint companion, deriving a mass ratio of 0.380 ± 0.013. Mayor & Mazeh (1987) had identified this system as a possible triple based on a 1.7 km s$^{-1}$ variation in the velocity semiamplitude between their solution and that of Luyten (1936). While the velocity semiamplitude does seem to vary for the different orbital solutions presented for this pair (Luyten 1936; Mazeh et al. 2002, DM91) and the Latham et al. (2010) SB1 orbital solution has residuals of up to 2 km s$^{-1}$ on each side, there is no obvious periodic pattern or longterm drift over the 4.8 years of velocity coverage. The most recent velocity measure of this star in Latham et al. (2010) is from 1990, so additional observations are warranted.

**HD 144579**: *Binary.* The proper motion of the primary is 0$\farcs$574 yr$^{-1}$ and of the CPM candidate is 0$\farcs$550 yr$^{-1}$ from the LSPM. The companion's distance estimate has a large uncertainty and differs from the primary's *Hipparcos* value by 1.5$\sigma$ (see Table 5). Given the proximity of these two stars in the sky, the very large and similar proper motions, and similar distances, this appears to be a physical companion, as it has been previously recognized (DM91,CNS). The differences in the proper motions might indicate that the companion or the primary has a close unresolved companion and warrants further observations.

**HD 145958**: *Binary,* candidate *Triple.* The primary of a 4″ visual binary has two additional possible companions, one of which was refuted by this effort and the other remains a candidate. The WDS lists a nearby companion, 0$\farcs$2 away, detected by HAM in 1983. The INT4 catalog lists this as a weak detection and possibly spurious, and includes a null detection using the same technique. BDM and DR failed to resolve a companion at the KPNO 4-m telescope in 2008 June. Nidever et al. (2002) identifies this as a constant velocity star. Evidence seems to be mounting against this candidate companion, which is considered refuted for this work. Separately, the Dwarf Archives includes a T6 object about 27′ away from this star. Looper et al. (2007) discovered this T dwarf in the 2MASS survey, obtained spectra, typing it as T6, and estimated its proper motion and distance, but did not suggest an association with HD 145958. Their proper motion and distance estimates are similar to the primary's corresponding values from van Leeuwen (2007b). While the projected linear separation is very large at about 40 000 AU, this could be a loosely bound companion to HD 145958, and is retained as a candidate. An additional wide component listed in the WDS is clearly optical (see Table 6).

**HD 146361**: *Quintuple.* See Raghavan et al. (2009) for a comprehensive treatment of all components of this system.

**HD 147776**: *Binary,* candidate *Triple.* The WDS lists three candidate companions, but the details actually correspond to four stars. The $\Delta$m $\sim$ 4 component 103″ away at 281° is clearly a field star, as seen by blinking the archival images (see Table 6). Three additional companions were reported by Sinachopoulos (1988) – 6$\farcs$4 separation at 173° with $\Delta$m $\sim$ 3, 9$\farcs$7 separation at 14°, and 71$\farcs$9 separation at 28°. The latter two components do not have differential-magnitude measurements. Sinachopoulos measured these pairs using a 1.5-m telescope by combining 4–16 exposures of a few seconds each. The wide companion 72″ away should have been seen in the DSS images, but no stellar source was seen at the expected position. The closest star to this position in the 1995 DSS image is 83″ away at 15° and is clearly a field star. The other two sources seen by them would be buried in the saturation around the primary in the DSS images, so we obtained $VRI$ frames in 2008 May and August at the CTIO 0.9-m telescope. The images clearly show a faint companion about 9″ away at 19°. This is likely the 9″.7 companion seen by Sinachopoulos (1988), exhibiting CPM, and given the proximity, is likely physical. The closest source seen by Sinachopoulos is not detected in the CTIO images and remains a candidate. Additionally, the CNS lists a companion for this star 3″ away at 281° in 1909. This is likely the same as the component measured by Burnham as listed in the WDS, which was seen 103″ away at 281° in 1909 and as discussed above, is clearly optical.

**HD 148704**: *Binary*. This is a 32-day SB2 binary for which *Hipparcos* and Tycho identified another companion 4″.1 away at 221°. Our CTIO 0.9-m images obtained in 2008 October do not reveal any companion at the expected position, while a $\Delta$m $\sim$ 3 companion as indicated by *Hipparcos* should have been seen above the tail of the primary's PSF. However, given the proper motion of the primary, a field star would have moved closer and possibly could be buried within the primary's PSF. Gray et al. (2006) list the spectral type of the companion as G9V and its coordinates imply a separation of 2″.4 at 55°, the exact position where a field star would be fifteen years since the *Hipparcos* measure. The Gray et al. spectral type, along with the Tycho-2 $V = 10.5$ imply a significantly larger distance to this star compared to the primary, enabling us to refute this candidate.

**HD 149806**: *Binary*. This companion was first reported by Rossiter (1955) 5″.9 away at 22° and has two additional measurements in the WDS over the next 54 years, which are consistent with a bound system. While the photometric distance estimate is not a good match (see Table 10), the $R$ magnitude listed is likely approximate. Given the moderate proper motion of the primary, the consistent measures over 54 years indicate a physical association. The 2MASS colors indicate an M-dwarf with a V magnitude estimate of about 12, in fair agreement with the measure of Rossiter (1955). BDM and DR attempted to observe this companion at the KPNO 4-m telescope in 2008 June. While the separation and $\Delta$m are too large to be resolved using speckle, the finder image at the telescope showed a source at the expected position with a $\Delta$m similar to those of prior observations.

**HD 153557**: *Triple*. The WDS lists 17 measurements over 95 years with separations of 1″.9–4″.9, which are consistent with a bound system, and given the 0″.3 yr$^{-1}$ proper motion of the primary, imply a physical association. This pair also has a wider companion, HD 153525, about $2'$ away, which is confirmed by matching proper motion and parallax.

**HD 165341**: *Binary.* CNS lists component A of an 88-year SB2VB as a possible binary with a period of about 17 years, but this is inconsistent with modern measurements (Latham et al. 2010). Heintz (1988) presented a revised orbit of the SB2 and excluded the possibility of any additional companions with periods below 55 years, stating that the once suspected velocity variation of A is disallowed by the more precise recent measurements. The WDS lists 15 more components for this star, all of which are clearly optical (see Table 6).

**HD 165908**: *Binary,* candidate *Triple.* This is a 56-year VBO. Additionally, the WDS lists one speckle measure with a separation of $0\rlap{.}''228$ at $50\rlap{.}°2$ from Scardia et al. (2008), who list this new discovery as "faint". They also resolved the known VB companion about $1''$ away, and noted it as "very faint". In the absence of additional measures that can help confirm CPM, this close companion is retained as a candidate. Five other components listed in the WDS are clearly optical (see Table 6).

**HD 178428**: *Binary.* The primary of a 22-day SB1 has a single 1987 measure listed in the WDS with a separation of $0\rlap{.}''2$. However, the INT4 lists six null results and attempts by BDM and DR at the KPNO 4-m telescope in 2008 June once again failed to reveal any visual companion. A wider component listed in the WDS is clearly optical (see Table 6).

**HD 186408**: *Triple, one planet.* This close companion to 16 Cyg A was first resolved $3''$ away by Turner et al. (2001) with AO at the Mount Wilson Observatory and confirmed by Patience et al. (2002), who demonstrated CPM and measured infrared magnitudes consistent with the primary's distance. Four velocity measures over 25 years show a slow drift (Latham et al. 2010), consistent with this companion. This system also has a wide companion, 16 Cyg B, which is a planet host. The WDS lists an additional source, $16''$ away from 16 Cyg B, but Patience et al. (2002) measured the infrared magnitudes of this candidate, demonstrating that it is a background star. This is the only planetary system in this study with more than two stars.

**HD 190067**: *Binary.* This companion was discovered by Turner et al. (2001) with AO at the Mount Wilson Observatory, but the single-epoch measure with no color information does not allow confirmation of a physical association. BDM and DR observed this star at the KPNO 4-m telescope in 2008 June. While the separation and $\Delta$m are too large for speckle observations, a stellar source was seen at the expected position, confirming CPM, and given the proximity to a large proper motion ($0\rlap{.}''6\,\mathrm{yr}^{-1}$) primary, the physical association of this companion is very likely.

**HD 190406**: *Binary.* Liu et al. (2002) discovered a faint companion $0\rlap{.}''8$ from this star with AO at the Gemini North and Keck II telescopes and confirmed a physical association by demonstrating CPM, consistent spectroscopy, and longterm radial-velocity trend. They determined a spectral type for the companion of L4.5 $\pm$ 1.5, estimated its mass to be 55–78 $M_J$ and age as 1–3 Gyr. This is the first substellar object imaged so close to a solar-type star and indicates that brown dwarfs can exist in extrasolar systems at positions comparable to the gas giants in our solar system. Eight additional WDS components are clearly optical (see Table 6).

**HD 191499**: *Binary.* The WDS lists 51 measurements of this companion between 1782 and 2003, which are consistent with a bound system. There is little evidence of orbital motion during the roughly 200 years of observations, possibly because the companion is near apastron in an eccentric orbit or the orbit is highly inclined. *Hipparcos* and Tycho-2 proper motions differ by 5.6$\sigma$, suggesting some orbital motion (see Table 8). The photometric distance estimate is not a very good match (see Table 10), but photometry would be tricky for this close pair as indicated by the large uncertainties of the 2MASS magnitudes. Given the evidence of consistent WDS measures, proper motion differences between *Hipparcos* and Tycho-2, and similar distance estimates, this pair likely has a physical association. An additional wide WDS component is clearly optical (see Table 6).

**HD 195564**: *Binary.* The WDS lists 16 measures over 110 years that are consistent with a bound pair. While proximity to the primary and $\Delta$m $\sim$ 5 (from WDS) make photometry of the companion difficult, the proximity and CPM implied by the measures argue for a physical association. An additional wide WDS component is clearly optical (see Table 6).

**HD 200525**: *Triple.* The CNS and *Hipparcos* identified the closer pair as a possible binary (stochastic solution) and Goldin & Makarov (2006) derived a photocentric orbit using the *Hipparcos* intermediate astrometry data. Their orbital solutions using data from the two independent *Hipparcos* reduction methods, Fundamental Astronomy by Space Techniques (FAST) and the Northern Data Analysis Consortium (NDAC), are consistent. They tested their orbit determination method satisfactorily against 235 known binaries and derived a better than 99% confidence level based on simulations. The WDS lists four measurements from 1898 to 1932, during which time the separation reduced from about 1″ to 0.″16. The respective position angles of the measurements are consistent with a high-inclination orbit. Given these independent measurements leading to similar results, we conclude that while the orbital elements may be preliminary, this companion is physically bound. For the wider component, the WDS has five measures over 88 years that are also consistent with a bound pair. The companion is NLTT 50542 with a proper motion that matches the primary's, and the notes in the catalog recognize this component as gravitationally bound. 2MASS magnitudes have a large error, but are consistent with a mid-K dwarf companion at approximately the primary's distance.

**HD 200560**: *Binary.* The WDS has seven measures of a companion with separations $2\rlap{.}''8$–$3\rlap{.}''3$ over 28 years, suggesting a bound system. This is especially significant given the primary's large proper motion of $0\rlap{.}''4\,\mathrm{yr}^{-1}$. The companion, GJ 816.1B, is recognized in the CNS as bound, although no conclusive evidence is presented. The 2MASS photometry has large errors, and hence is not very useful. The *Hipparcos* and Tycho-2 proper motions are different by about $8\sigma$, providing evidence of orbital motion and lending credibility to a physical association. The WDS lists this pair as the CD component of the B3V binary HD 200595 AB, but there is clearly no physical association between HD 200560 and HD 200595 as seen by blinking archival images. An additional wide WDS component is also clearly optical (see Table 6).

**HD 202275**: *Binary.* This is a 5.7-year SB2VB. Tokovinin et al. (2006) gives an additional orbit with a period of 5.7 days, which is in fact the former orbit listed with an incorrect unit (A. Tokovinin 2007, private communication). This system is a binary with mass estimates of 1.2 $\mathrm{M}_\odot$ and 1.1 $\mathrm{M}_\odot$ by Pourbaix (2000). An additional wide WDS component is clearly optical (see Table 6).

**HD 206860**: *Binary.* Luhman et al. (2007) reported the discovery of a T2.5 $\pm$ 0.5 companion using Spitzer IRAC images and confirmed CPM using 2MASS images. The infrared colors are consistent with the distance to the primary, confirming companionship. By comparing the luminosity with evolutionary tracks, they estimate the companion's mass as 0.021 $\pm$ 0.009 $\mathrm{M}_\odot$ and age as 0.3 $\pm$ 0.2 Gyr. An additional, potentially wide companion was identified $591''$ away by blinking archival images but refuted based on its photometric distance estimate (see Table 5).

**HD 215648**: *Binary.* A companion, $11''$ away, is listed in the WDS and CNS and confirmed by the 23 measurements in the WDS over 179 years which not only demonstrate CPM, but also orbital motion. A wider WDS component is clearly optical (see Table 6).

**HD 217107**: *Single star with two planets or a binary with one planet.* The WDS lists two measurements, fifteen years apart, of a companion $0\rlap{.}''3$ away from this star, which also hosts two planets. These speckle interferometry detections could however not be confirmed by the same technique on at least three other occasions, indicating that this pair might have a large or varying $\Delta$m. Interestingly, the farther planet is one of the most widely separated planets reported, at least 5 AU from the star. Vogt et al. (2005) present orbital solutions with periods of 7–9 years, but mention that it could be three times larger. Wright et al. (2009) present an updated orbit with $P = 11.5 \pm 0.5$ years and $a = 5.27 \pm 0.36$ AU. At the 20-pc distance to the star, these separations are consistent with the speckle observations. Given the inconsistent measures, if we assume a $\Delta V$ near the speckle limit of about 3, the companion to the G8 IV-V primary (Gray et al. 2003) could be an early M-dwarf. The mass-sum of such a binary is consistent with the Wright et al. (2009) orbital elements. Vogt et al. (2005) note that an AO image obtained with the Keck telescope did not reveal any stars beyond 0″.1 from the primary, and Chauvin et al. (2006) confirm this null result with VLT and CFHT AO observations. The M-dwarf companion would also imply a significantly larger velocity semi-amplitude for the primary, but that possibility is not excluded by the orbital plot in Wright et al. (2009). While it appears that this "planetary" companion could be a star, further observations are warranted.

**HD 220140**: *Triple*. The WDS has five measurements over 100 years for the closer visual companion at separations of about 10″ that are consistent with a bound pair. The companion is NLTT 56532 with a proper motion matching that of the primary and 2MASS colors consistent with an early M-dwarf at approximately the primary's distance. The wide CPM companion, 16′ away, was first identified by Lépine & Shara (2005) and confirmed by Makarov et al. (2007) who show that the companion's trigonometric parallax is consistent with the primary's *Hipparcos* value. Their *BVRI* photometry along with 2MASS near-infrared magnitudes show that this star is over-luminous in the $K_s$ band, confirming its suspected pre-main-sequence status, and enabling an age estimate of 12–20 Myr.

## 5.   Analysis & Discussion

### 5.1.   Survey Sensitivities and Completeness

Table 15 enumerates the coverage of the sample by various systematic searches and Figure 11 identifies the detection limits of the methods. The shortest-period systems are comprehensively covered by the radial-velocity surveys. The CCPS effort is sensitive down to planetary masses for periods of some 10 years and the Latham et al. (2010) survey is capable of identifying $K_1 \geq 1\ \mathrm{km\,s^{-1}}$ with periods below 30 years[10]. The CHARA SFP detection space covers $\Delta K \leq 2.5$ for separations of 10–80 mas. These, and the following angular separations are converted into linear separations assuming a distance of 20 pc (the median distance of our sample) and translated into periods assuming a mass-sum of 1.5 $\mathrm{M_\odot}$. Speckle interferometry, typically done on 4-m class telescopes, is sensitive to $\Delta V \leq 3$ and separations 30 mas to 2″. AO and other visual searches for high-contrast binaries are done using telescopes of up to 10-m apertures and can detect $\Delta V \leq 2$ at 50 mas separation, improving to about 10 mag for 2″– 10″. While AO observations are not part of our systematic efforts, their results from other searches (e.g., Turner et al. 2001; Liu et al. 2002; Luhman & Jayawardhana 2002; Potter et al. 2002; Roberts et al. 2005; Chauvin et al. 2006; Lagrange et al. 2006; Metchev & Hillenbrand 2009) are nevertheless included in our multiplicity results. Unfortunately, many of these studies only report companion detections, leaving the list of targets searched, and the corresponding null results, unpublished. Nevertheless, we were able to identify the results of AO searches around 82 of our sample stars from studies that published their target lists (Luhman & Jayawardhana 2002; Chauvin et al. 2006; Metchev & Hillenbrand 2009), or for which we could obtain unpublished null results (Turner et al. 2001), and their results are included in the last row of Table 15. Finally, the widest companions are detected by blinking archival images, capable of identifying roughly equal-magnitude binaries as close as $5''$ as twin diffraction spikes. At wider separations beyond $15''$–$30''$ (depending on the particular image), this technique can reveal companions with $R \leq 17$, i.e., all but the lowest mass stars. Figure 11 also shows the overlap of the actual pairs from our results with estimates of masses and periods as described in § 5.3.1 and § 5.3.3, respectively, highlighting the thorough coverage by the various techniques and validating the detection-limit estimates. This, along with the robust overlap of the detection space by multiple complementary techniques, suggests that our tally of companions is comprehensive and nearly complete. Next, we estimate the number of stellar and sub-stellar companions missed by this survey.

---

[10]We assume $e = 0.25$, $i = 90°$, and mass-sum = 1.5 $\mathrm{M_\odot}$ (the average value for all our pairs) in generating the curves plotted, which are truncated at the period limits mentioned above for the respective surveys.

As seen in Column 4 of Table 15, our efforts added only a few new companions, underscoring the collective comprehensiveness of prior efforts. Moreover, the techniques employed are capable of detecting stellar companions of nearly all masses. We can thus estimate the number of stellar companions missed by simply extrapolating the number of new companions found by each technique (Column 4 of the table) to the full sample, accounting for their incomplete coverage. So, for example, the four new CPM companions found around 409 stars imply that one undetected companion might exist in the 45 stars not searched by this technique. Similarly, the radial-velocity techniques listed in the first two rows of Table 15 imply two spectroscopic companions missing from our tally. While these spectroscopic techniques are not sensitive to companions in nearly face-on orbits, the CCPS survey, in particular, should see a velocity trend from stellar companions in all but perfectly face-on orbits, which are highly improbable. Moreover, the dearth of short-period systems in general, and low-mass ones in particular, as seen in Figure 11, suggests that we are not missing many such companions. At longer periods, companions in face-on orbits should be detected by the visual techniques. The two new companions reported by AO searches (last row of Table 15) suggest that nine high-contrast companions might exist near the 371 stars not surveyed by these efforts. However, given that only six of at least 22 AO companions in our results were discovered by the references included in Table 15, most of these companions are expected to be counted in our results, and any further missing AO companions are expected to be covered by the estimates performed below.

Considering missing low-mass companions, Figure 11 identifies two possible gaps in the coverage of the parameter space for potential companions: the low-mass region between RV and AO and then again between AO and CPM. While these gaps suggest a few missing companions, they are nevertheless populated by some actual detections because of other techniques such as *Hipparcos* double stars, whose companions are investigated and included in our tally, but whose detection regions are not included in the plot. The dearth of low-mass companions was first established by radial-velocity surveys for $P \lesssim 5$ years, and extended to all separations by high-contrast visual searches (e.g., Gizis et al. 2001; Kirkpatrick et al. 2001; Lowrance et al. 2002; Burgasser et al. 2003; McCarthy & Zuckerman 2004; Neuhäuser & Guenther 2004; Grether & Lineweaver 2006; Kraus et al. 2006; Allen et al. 2007; Lépine & Bongiorno 2007; Looper et al. 2007; Metchev & Hillenbrand 2009). While many of these searches specifically targeted brown dwarf companions, they were more sensitive to, and often reported, the discovery of low-mass stellar companions as well. Additionally, a specific search for low-mass companions out to separations of 10 000 AU around solar-type stars within 10 pc using 2MASS infrared images resulted in no new detections (D. Looper 2009, private communication). Consistent with these findings, our results include only only eight brown dwarf companions, found by high-resolution techniques such as coronagraphy, AO, and space-based observations (e.g. Potter et al. 2002; Bouy et al. 2003; Burgasser et al. 2005; Luhman et al. 2007). Moreover, while our CPM efforts using archival images are not sensitive to brown dwarfs, widely-separated substellar companions to nearby stars have effectively been identified using 2MASS infrared images (Kirkpatrick et al. 2001; Lowrance et al. 2002; Looper et al. 2007). While these efforts readily present new detections, null results often remain unpublished. Nevertheless, these searches using 2MASS infrared images have largely yielded null detections (D. Kirkpatrick 2009, private communication), suggesting a real shortage of low-mass companions. As noted above, the CCPS survey is specifically suited to identifying relatively long-period low-mass companions, or those in nearly face-on orbits, as low-amplitude temporal drifts in the radial-velocity measurements of the host stars. Our search of the CCPS data revealed 21 stars which have such drifts with RMS deviation from a uniform velocity of less than 0.1 km s$^{-1}$ over some 10 years. Fourteen of these have known stellar companions, many of which are close enough to cause the radial-velocity variations seen. The remaining seven stars do not have any known companions, implying undetected stellar or sub-stellar companions to presumed single stars. While some of these could turn out to be planets in long-period orbits, they could all be new stellar or brown dwarf companions. Added to the one missing CPM companion and two missing spectroscopic companions discussed above, we estimate that our survey might be missing about 10 companions to presumed single stars.

The extrapolation of new companions used above to estimate the number of companions missed by our survey is valid only if the stars not observed by a particular technique have, in general, been historically covered in equal measure as those observed by that technique. If the stars not surveyed by a given technique fall into a set of generally neglected stars, the approach used above would underestimate missing companions. We address this assumption next for the Latham et al. (2010) and CPM surveys, as these are the primary techniques yielding new discoveries from observations of a subset of the overall sample. The speckle and *Hipparcos* searches exhaustively covered the sample, and hence no extrapolation is necessary. We exclude the CCPS survey from this analysis because of the inherent bias in its sample selection, as it avoids known close binaries. The percentage of single stars among the 344 primaries observed by Latham et al. (2010) is $55\% \pm 4\%$, compared to $59\% \pm 7\%$ for the 109 stars not observed by this survey. These percentages agree to well within $1\sigma$, indicating no deficit of historical attention for the stars excluded from the Latham et al. survey. Similarly, the fraction of single stars among the 409 stars explored by the CPM technique is $54\% \pm 3\%$, compared to $75\% \pm 13\%$ for the 44 unexplored stars, an agreement to $1.6\sigma$. While still consistent within expected statistical deviations, this is a larger departure than seen for the radial velocity survey. Moreover, stars not investigated for CPM companions exhibit low proper motion in archival images, a factor that could have contributed to their exclusion from some companion search efforts. However, even if we adopt the fraction of single stars from the surveyed subsets, this analysis suggests that $54\% - 55\%$ of Sun-like stars are likely to be single, consistent with the above estimate of about 10 missing companions in our survey.

Finally, we try another approach to estimate the number of missing companions. Rather than extrapolate new companions, we apply the fraction of detected companions found by each technique (Column 3 of Table 15) to the unsurveyed stars and estimate the number that would apply to presumed single stars. For example, the 70 companions discovered among 409 stars searched for CPM companions imply eight undiscovered companions in the 44 not searched. Given that 75% of these 44 stars do not have any observed companions from any technique, six of these eight missing companions may apply to presumed single stars. Because of the systematic reason for the exclusion of these 44 stars, i.e., these are not conducive to CPM searches by any efforts relying on archival images, this may be a better estimate than the one missing companion estimated by extrapolating new companions. However, some of these stars may be the same as the ones for which companions are suggested by the other subsets of the incompletion analysis, suggesting that about 10–15 companions may be missing from our results. Hence, the various approaches of estimating the incompletion of our survey yield consistent results, resulting in our estimate that about 13 presumed single stars may have yet undiscovered companions. Note that the extrapolation of total companions discovered, as discussed above for CPM, could also be applied to the radial velocity and AO surveys listed in Table 15. However, there is no systematic reason for excluding targets from the Latham et al. (2010) survey or the AO surveys (other than declination limits, which by themselves do not suggest any bias impacting companion fractions). The missing companions implied by these extrapolations, 12 from the radial-velocity survey and 27 from the AO survey, are largely expected to have been discovered by other efforts using these techniques, as suggested by consistent multiplicity fractions among the surveyed and unobserved subsets discussed above. The consistent multiplicity results for various subsamples with distance and declination limits, covered in the next section, further improve confidence in our estimates of missing companions for the substantially larger sample.

## 5.2. Overall Multiplicity Statistics

Table 16 lists our overall multiplicity results for solar-type stars in the solar neighborhood and facilitates a comparison with the DM91 results. Our overall observed fractions of single, double, triple, and higher order systems are $56\% \pm 2\%$, $33\% \pm 2\%$, $8\% \pm 1\%$, and $3\% \pm 1\%$, respectively, counting all confirmed stellar and brown dwarf companions. If all candidate, i.e., unconfirmed, companions identified are found to be real, the percentages would change to $54\% \pm 2\%$, $34\% \pm 2\%$, $9\% \pm 2\%$, and $3\% \pm 1\%$, respectively. Based on the completeness analysis discussed in § 5.1, we conclude that $54\% \pm 2\%$ of solar-type stars are single, i.e, without stellar or brown dwarf companions[11].

We now return to the question about the completeness of our volume-limited sample, explored in § 2.2. As pointed out, if our sample suffered from *Hipparcos* magnitude limits, we would expect the percentage of single stars to drop with increasing distance, because a magnitude-limited sample would favor the inclusion of spatially unresolved binaries at farther distances. However, the multiplicity fractions for the 15, 20, and 23 pc subsamples and that of the full 25 pc sample, shown in the second section of Table 16, are all statistically equivalent. This is consistent with the expectations for a complete, volume-limited sample, enabling us to conclude that our sample does not suffer from any magnitude limitations.

---

[11]These uncertainties are estimated using a bootstrap resampling analysis of 10 000 iterations. In each iteration, a random set of 454 stars is selected from our actual sample of 454 stars such that some stars may be selected more than once, while others may be excluded. This is the effect of the selection process, which does not disqualify a star from further random selections in a sample simply because of a prior random selection. The fractions of single, binary, triple, and higher order multiple systems are then computed for each sample. The distribution of these parameters for all 10 000 samples is roughly Gaussian with a peak corresponding to the value of the actual sample and a standard deviation representing its uncertainty.

Given the longstanding appreciation of the DM91 survey, we now compare our multiplicity fractions with their results, reproduced in the third section of Table 16. The comparison reveals several interesting facts. First, the percentage of triple and quadruple systems in the current study is roughly double that of DM91. Our results show that 25% of non-single stars are higher order multiples, compared to 13% in DM91, confirming their prediction that additional multiple systems were likely to be detected among nearby solar-type stars. Second, while the percentage of multiple systems has doubled, it has come solely at the expense of binary systems. It is indeed remarkable that despite almost twenty years of monitoring since DM91, the observed percentage of single stars before any incompleteness corrections has remained virtually unchanged – from 56.7% in DM91 to 56.4% in our study. Third, candidate companions have a smaller influence on our multiplicity statistics compared to DM91 – 2% versus 7%. DM91 considered radial-velocity variations with $P(\chi^2) < 0.01$ as non-random variations suggesting unseen companions. While this metric has been subsequently used as an indication of companionship (Nordström et al. 2004), it alone is not reliable for this purpose (Latham et al. 2010). In contrast, in the current study, we individually inspect each candidate and use multiple methods and observations to confirm or refute the companions whenever possible, leaving us with fewer candidates. Fourth, DM91 made significant incompleteness corrections, estimating that only 43% of solar-type stars have companions with mass ratio $q > 0.1$. Further, they assume that many low-mass companions were missed, concluding that only one-third of solar-type stars are truly single, i.e., without companions with masses greater than 10 $M_J$. The significant detection gaps discussed in DM91 in the spectroscopic, visual, and low-mass regimes have been effectively plugged by our effort, as shown in Figure 11, resulting in our modest incompleteness correction discussed in the previous section. We thus believe that DM91 significantly overestimated the number of low-mass companions missed by their survey.

However, there are some systematic differences between our sample and that of DM91, whose possible effects we investigate next. The last section in Table 16 includes results from various subsamples to facilitate this analysis. The first two lines present the statistics from the 106 systems that are common to the two studies. While the DM91 results for this subsample are consistent with their overall results, the updated results for this subset show a 7% reduction in the percentage of single stars, a $1\sigma$ deviation. A check of the individual results indicates that while three of the DM91 companions to these stars have since been refuted, 17 new companions have been discovered to 15 of these stars. We explored several avenues to investigate a systematic cause for this difference, but failed to identify any such factor. First, the third row in this group lists the statistics for 10 000 randomly selected subsets of 106 stars from our sample, showing that their results are consistent with our overall numbers and suggesting that the difference for the particular set of 106 stars overlapping with DM91 is just statistical scatter. Second, while our sample is all-sky, the DM91 sample was declination limited to north of $-15°$. To check whether stars in the southern hemisphere have been studied less completely, resulting in a greater percentage of presumed single stars, we checked the subsample of our study with declination north of $-15°$. The results, listed in the fourth row of this section in the table, indicate no such bias. A third factor could be the difference in the color ranges of the two samples. While our study includes stars with $0.5 \leq B - V \leq 1.0$, roughly corresponding to spectral types F6–K3, DM91 selected a narrower F7–G9 range. As later spectral types are known to have fewer companions (see § 5.3.2), this might be a factor. However, as shown in the last row of the table, a subsample of matching spectral types does not show a significant deviation from our overall results.

Finally, we examine whether the DM91 sample represents a particularly well-studied set of stars when compared to the non-DM91 stars in our sample. Of the 106 stars common to both surveys, 104 (98%) were observed for CPM companions. In comparison, $88\% \pm 5\%$ of the 347 sample stars not in the DM91 study were observed for CPM companions. The difference might be due the preferential selection of high proper-motion stars for ground-based parallax measurements that governed the selection of the DM91 sample, compared to the more comprehensive *Hipparcos* catalog used by this effort. Nevertheless, this difference might suggest moderately better coverage of the DM91 sample, but the companions possibly missed due to this effect were estimated in the previous section and are included in our results. Comparing the coverage of these subsamples by the Latham et al. (2010) survey, we find that 97% of the 106 common stars were observed by the Latham et al. survey, compared to 69% of the non-DM91 stars. Further analysis confirms that this is simply due to the declination limits of the DM91 ($\geq -15°$) and Latham et al. ($\geq -40°$) efforts, compared to our all-sky sample. Limiting the comparison to stars north of the DM91 declination limit, we find identical coverage of 97% by the Latham et al. (2010) effort for both subsets, i.e., the 106 common stars and the 201 non-DM91 stars. As discussed above, the declination-limited subsample yields multiplicity fractions consistent with our overall results, confirming that our comprehensive synthesis of other efforts adequately covers the southern-hemisphere stars. Apparently, the differences in the particular subsample common to the DM91 and our survey are simply due to statistical scatter.

## 5.3. Orbital Elements and Multiplicity Dependence on Physical Parameters

### 5.3.1. Sources and Estimation of Physical Parameters

The analyses described in the following subsections require the estimation of physical parameters for the components of binaries and multiple systems, which are extracted from publications or estimated as described here. Tables 17 and 18 list the physical parameters for the sample stars and companions, respectively, and their corresponding references. Masses for the primary stars are listed from dynamical estimates, when available. For other stars, the analyses below use interpolated mass estimates from spectral types using the relations in Cox (2000). For the companions, mass estimates and spectral types are extracted from the various discovery or characterization publications, when available. Otherwise, they are estimated using mass-ratios for double-lined spectroscopic binaries, or from multi-color photometry from catalogs, or using the $\Delta$mag measures in the WDS along with the primary's spectral type. Metallicity and chromospheric activity estimates of the primary are adopted for all components of the system.

### 5.3.2. Multiplicity by Spectral Type and Color

Figure 12 shows the multiplicity fraction for stars and brown dwarfs. Most O-type stars seem to form in binary or multiple systems, with an estimated lower limit of 75% in clusters and associations having companions (Mason et al. 1998a, 2009). Studies of OB-associations also show that over 70% of B and A type stars have companions (Shatsky & Tokovinin 2002; Kobulnicky & Fryer 2007; Kouwenhoven et al. 2007). In sharp contrast, M-dwarfs have companions in significantly fewer numbers, with estimates ranging from 11% for companions 14–825 AU away (Reid & Gizis 1997) to 34%–42% (Henry & McCarthy 1990; Fischer & Marcy 1992). Finally, estimates for the lowest-mass stars and brown dwarfs suggest that only 10–30% have companions (Burgasser et al. 2003; Siegler et al. 2005; Allen et al. 2007; Maxted et al. 2008; Joergens 2008). Our results for F6–K3 stars are consistent with this overall trend, as seen by the thick solid-lines for the incompleteness-corrected fraction. Moreover, the thick dashed lines for two subsamples of our study show that this overall trend is present even within the range of solar-type stars. $50\% \pm 4\%$ of the blue subsample $(0.5 \leq B - V \leq 0.625,$ F6–G2, $N = 131)$ have companions, compared to only $41\% \pm 3\%$ for the red subsample $(0.625 < B - V \leq 1.0,$ G2–K3, $N = 323)$.

### 5.3.3. Period Distribution

Figure 13 shows the period distribution of all 259 confirmed pairs, with an identification of the technique used to discover and/or characterize the system. To provide context, the axis at the top shows the semimajor axis corresponding to the period on the x-axis assuming a mass sum of 1.5 $M_\odot$, the average value of all the confirmed pairs. When period estimates are not available from spectroscopic or visual orbits, we estimate them as follows. For CPM companions with separation measurements, we estimate semimajor axes using the statistical relation $\log a'' = \log \rho'' + 0.13$ from DM91, where $a$ is the angular semimajor axis and $\rho$ is the projected angular separation, both in arcseconds. This, along with mass estimates as described in § 5.3.1 and Newton's generalization of Kepler's Third Law yields the period. For the remaining few unresolved pairs, we assume periods of 30–200 years for radial-velocity variables and 10–25 years for proper motion accelerations. The period distribution follows a roughly log-normal Gaussian profile with a mean of $\log P = 5.03$ and $\sigma_{\log P} = 2.28$, where P is in days. This average period is equivalent to 293 years, somewhat larger than Pluto's orbital period around the Sun. The median of the period distribution is 252 years, similar to the Gaussian peak. This compares with corrected mean and median values of 180 years from DM91. The larger value of the current survey is a result of more robust companion information for wide CPM companions. The similarity of the overall profile with the incompleteness-corrected DM91 plot suggests that most companions they estimated as missed have now been found. The shading in the figure shows the expected trend – the shortest-period systems are spectroscopic, followed by combined spectroscopic/visual orbits, then by visual binaries, and finally by CPM pairs. The robust overlap between the various techniques in all but the longest period bins underscores the absence of significant detection gaps in companion space and supports our earlier statements about the completeness of this survey. Binaries with periods longer than $\log P = 8$ are rare, and only 10 of the 259 confirmed pairs (4%) have estimated separations larger than 10 000 AU. Although separations wider than this limit were not searched comprehensively, Figure 8 shows that separations of up to 14 000 AU were searched for some systems, and 56% of the primaries were searched beyond 10 000 AU limit. The drop in the number of systems with companions thus appears to occur within our search space and hence is likely real. While binaries with separations $\sim 0.1$ pc have been reported (Latham et al. 1991), they are rare and unlikely to survive dynamical interactions in clusters (Parker et al. 2009). The few such extremely wide binaries in our sample might have formed outside clusters, which are the exception rather than the rule (e.g., Lada & Lada 2003). Looking at a closer limit, Lépine & Bongiorno (2007) concluded that at least 9.5% of *Hipparcos* stars within 100 pc have distant stellar companions with projected separations greater than 1 000 AU. Our results for the *Hipparcos* solar-type stars within 25 pc show that 11.5% have companions with projected separations greater than 1 000 AU, consistent with Lépine & Bongiorno's conclusions and suggesting that their companion search effort was fairly comprehensive.

### 5.3.4. Period-Eccentricity Relationship

Figure 14 shows the period-eccentricity relationship for the 127 pairs with estimates of these parameters from visual and/or spectroscopic orbital solutions. Pairs with orbital periods longer than the circularization limit of about 12 days have a roughly flat eccentricity distribution out to $e \sim 0.6$, independent of period, as shown by Figure 15. The deficiency of high eccentricities is likely due to dynamical interactions for short-period systems and the lack of measurements for long-period systems. The flat eccentricity distributions in Figure 15 are in contrast to the DM91 results, which showed a normal distribution for periods below 1000 days, and an $f(e) = 2e$ distribution for longer-period systems. Pairs with periods below 12 days are circularized, with one notable exception. The 7-day SB2 pair HD 45088 Aa-Ab seems to have an unusually high eccentricity of $0.1471 \pm 0.0034$ for its short period, and the longer 600-year AB orbit has an eccentricity of 0.25. A possible explanation is that this system is relatively young and hence not yet circularized. The $\log R'_{HK}$ of the primary of $-4.266$ (Gray et al. 2003) is among the highest for the stars of this sample, suggesting relative youth, which is consistent with a high rotational-velocity and emission features in its spectra (Mishenina et al. 2008). However, the chromospheric activity and high rotation can also be explained by tidal interactions between the components of the short-period binary. An alternate explanation of the high eccentricity is the Kozai mechanism (Kozai 1962), which causes periodic oscillations in the eccentricity and inclination of inner orbit due to tidal forces from the wide companion. Another similar system, HD 223778, has an outer orbit with an large eccentricity of 0.55, and an inner orbit with a small eccentricity of $0.0174 \pm 0.0035$. While low, the inner orbit's eccentricity is different from zero by a $5\sigma$ significance, suggesting that the Kozai mechanism could be at play in this system as well. In another observational support of the Kozai mechanism, Figure 14 shows that the upper-eccentricity envelope is dominated by components of triple systems, as also observed by DM91.

### 5.3.5. Mass-Ratio Distribution

Figure 16 shows the distribution of mass ratios ($M_2/M_1$) for binaries, pairs in higher-order multiple systems, and composite-mass pairs in multiple systems (see the figure caption for an example). These plots exclude 11 systems with components whose masses could not be estimated by the process described in § 5.3.1. The figure shows a roughly flat distribution for mass ratios 0.2–0.95. Binaries show a deficiency of low-mass companions, and Figure 17 illustrates the lack of low-mass, short-period companions. This parameter space is effectively covered by the CCPS and Latham et al. (2010) radial-velocity studies (see Figure 11), so the deficiency of such companions appears to be real, suggesting that short-period systems prefer higher mass ratios. The fraction of short-period ($P < 100$ days) systems increases from 0% for mass ratios below 0.2, to 4% for mass-ratios below 0.45, to 8% for mass ratios 0.45–0.9, and to 16% for mass ratios above 0.9. Figure 16 shows that the preference for like-mass pairs applies to binaries and pairs in higher-order multiples (such as Aa,Ab in a Aa,Ab,B triple), but does not apply when the aggregate mass of a pair is compared with the third component (e.g., mass of Aa+Ab, compared to mass of B). Consistent with the above observations, like-mass pairs ($M_2/M_1 > 0.95$) prefer relatively short periods – only six of 27 such pairs have periods longer than some 200 years and none have periods longer than about 1 000 years, corresponding to about 115 AU. These results agree with predictions from hydrodynamical simulations (Bate et al. 2002), which show that gas around a protobinary preferentially accretes on to the lower-mass component. This is a consequence of the lower-mass component sweeping a larger space, thereby aggregating more mass until the masses are roughly equal. These results contradict DM91's conclusions, which showed no preference for like-mass pairs. Figure 17 also shows that the majority of short-period pairs belong to triple systems. Of the 16 systems with periods below 100 days, seven are binaries and nine are the inner pairs of triple systems. These observations support predictions from hydrodynamical simulations, which suggest that many short-period pairs could have formed at wider separations and migrated closer as a result of dynamical interactions in unstable multiple systems or orbital decays due to gas accretion and/or the interaction of a binary with its circumbinary disk (Bate et al. 2002).

### 5.3.6. Multiplicity by Chromospheric Activity and Age

Multiplicity studies of young stars (Ghez et al. 1997; Kouwenhoven et al. 2007), nearby solar-type stars (Mason et al. 1998b), and aging stars in globular clusters (Sollima et al. 2007) show an anti-correlation between age and the fraction of stars with companions. To check this trend within our sample, we use chromospheric emission as measured by $\log R'_{HK}$, an age indicator (Henry et al. 1996; Mason et al. 1998b; Mamajek & Hillenbrand 2008). We avoid the potential contamination from tidally-induced high activity in short-period binaries by excluding systems with orbital periods less than 12 days. The overall multiplicity fraction of the active ($\log R'_{HK} \geq -4.75$) and inactive ($\log R'_{HK} < -4.75$) subsamples are statistically equivalent. $40\% \pm 3\%$ of low activity (older) stars have companions, compared to $44\% \pm 4\%$ of high activity (younger) stars. Interestingly, there appears to be a relationship between age and the interaction cross-sections of systems, i.e., systems with more stellar components or those with long orbital periods are preferentially younger. Figure 18 shows the relationship between orbital period, chromospheric activity, and multiplicity order. Not surprisingly, all systems with $P < 12$ days are active, presumably because of tidal interactions. These systems are excluded from any age-related analyses. Among systems that have no orbital periods shorter than 12 days, the active subset has a larger fraction of higher order multiples, i.e., 26% of the active systems are triples or higher order multiples compared to only 17% of low-activity systems. Consistent with this trend, only 32% of binaries are active, compared to 44% of higher order multiples. Also, the fraction of active single stars is 33%, virtually identical to the fraction for binaries. Thus, it appears that binaries tend to remain intact after dynamical interactions, but higher order multiples get disrupted, assuming that the formation rates of higher order multiples have not changed over time. Similarly, long period systems also tend to erode with age. For systems with $\log P < 8$ (i.e., periods below some 275 000 years or, assuming a mass-sum of 1.5 $M_\odot$, semimajor axes less than 5 000 AU), only $33\% \pm 4\%$ are in the active subset. In comparison, $69\% \pm 23\%$ of systems with longer periods are active. Comparing these distributions using the KS test shows consistent results, but the sample sizes in many cases are too small to yield definitive answers. In summary, our results suggest that systems with more than two stellar components or those with long orbital periods tend to preferentially lose companions with age, likely due to dynamical interactions.

### 5.3.7. Multiplicity by Metallicity

Figure 19 shows the fraction of stars with stellar, brown dwarf, and planetary companions as a function of metallicity. The planet-metallicity correlation, first demonstrated by Fischer & Valenti (2005), is clearly seen in our volume-limited sample as well. Due to the dearth of brown dwarfs, testing whether similar correlations exist for these more massive substellar companions is a challenge. While none of the seven stars of our sample with confirmed brown dwarf companions has [Fe/H] $< -0.3$, the steady rise with metallicity seen for planets is not evident for brown dwarfs. While the fraction of stars with planets rises from 2% at $-0.3 \leq$ [Fe/H] $< -0.2$ to 40% for $0.2 \leq$ [Fe/H] $< 0.3$, the corresponding numbers for brown dwarf companions are constrained between 0%–3%, and are all statistically equivalent due to the small numbers. To generate a larger sample, we looked at all brown dwarfs in the Dwarf Archives with known stellar companions. Twenty-two such stars were found, and 13 of them have metallicity measurements. All but one of these stars have metallicity values between $-0.4$ and 0.3, and the exception has [Fe/H] $= -0.69$. With such small numbers and the above exception, we are not able to conclude anything specific about the correlation between a star's metallicity and its tendency to have brown dwarf companions, although the data hint a possible relationship.

Figure 19 illustrates two additional points. First, the population of solar-type stars in the solar neighborhood is metal-poor compared to the Sun. Seventy percent of the 414 stars with metallicity measurements are less abundant in heavy elements than the Sun. Second, while stars with [Fe/H] $> -0.3$ show no correlation between metallicity and the fraction of stars with stellar companions, stars with metallicities below this limit show a greater tendency to have stellar companions, which appears to increase with decreasing metallicity. While the implied anti-correlation for the individual bins in this region is not statistically significant and consistent with a flat distribution[12], the aggregate differences on either side of this limit are significant. Using Poisson uncertainties, our data suggest that $57\% \pm 9\%$ of stars with [Fe/H] $< -0.3$ have companions, compared to $39\% \pm 3\%$ of the higher metallicity stars. This result is in contrast to several recent multiplicity studies of K and M subdwarfs, which find a lower companion fraction for subdwarfs when compared to corresponding main-sequence stars (Riaz et al. 2008; Jao et al. 2009; Lodieu et al. 2009). However, these surveys targeted known subdwarfs, which are often initially identified by their sub-main sequence location on the H-R diagram, potentially excluding some binaries because they are elevated on the diagram. On the other hand, based on an analysis of short-period binaries, Grether & Lineweaver (2007) concluded that low-metallicity stars are more likely to have companions and further pointed out that this tendency is limited to relatively red stars ($B - V > 0.75$). Our work confirms these conclusions. Splitting the analysis into red and blue subsets on either side of $B-V = 0.625$ (the limit used in § 5.3.2), we find no relationship between metallicity and the tendency to have stellar companions for $B - V < 0.625$. For the redder subset, however, the relationship shown in Figure 19 is not only repeated, but further accentuated. While the anti-correlation across the individual metallicity bins is still statistically insignificant, the aggregate percentage of stars with stellar companions is $63\% \pm 11\%$ for [Fe/H] $< -0.3$ and $35\% \pm 4\%$ for the higher metallicity subset. Although this interesting result is somewhat tentative, it may indicate that lower-metallicity clouds are more likely to fragment and form binary stars, as suggested by numerical simulations (Bate et al. 2003; Bate 2005; Machida et al. 2009). Because older-population stars are relatively metal-poor, these differences could appear to contradict the conclusion in § 5.3.6, but there is no relationship between metallicity and chromospheric activity for our sample of stars.

### 5.3.8. The Hierarchy of Multiple Systems

Figures 20–22 show the hierarchy of triple systems, Figure 23 that of quadruple systems, and Figure 24 that of higher order systems. Candidate companions are connected via dotted lines and confirmed companions via solid lines. The figures identify the relationship between the hierarchical pairs and the period or separation as appropriate. When available, the

---

[12]Furthermore, all four stars with [Fe/H] $< -0.9$, excluded from the plot, are single.

components' spectral types from Table 1 and mass estimates from Tables 17 and 18 are also listed. The hierarchical order of multiple systems is clearly seen in our results. All confirmed multiple systems, excluding two outliers discussed below, have period ratios between any pair and the next level up ranging 23–33 000. The corresponding semimajor axis ratios are 9–1 900. The outlier with the largest ratio is HD 133640, whose period and semimajor axis ratios of 251 000 and 5 300 are driven by the very short period of 6.4 hours for the inner pair. The outer pair is separated by about 2″ and has a visual orbit of 246 years. The exception with the lowest ratio can be more informative about dynamical interactions among components. HD 4391 has a period ratio of five and a semimajor axis ratio of three. While this seems to be low-enough to raise questions about the dynamical stability of this system, it is composed of two CPM pairs with separations of 17″ and 49″, the wider of which was discovered by this effort. The estimates of semimajor axes from observed separations using the statistical relationship discussed in § 5.3.3, and the resulting periods, are rough estimates useful for overall statistical analyses, but not accurate for individual systems. Hence, these ratios for this system are not meaningful and the observed projected separations can correspond to a whole host of stable orbital configurations.

The hierarchical relationships in the 33 confirmed triple systems suggests a strong preference for the tertiary component to orbit the more massive component. Twenty-five (76%) of these systems are like HD 40397 (G7+M4, M5) or HD 18143 (K2+K9, M7), where the plus symbol connects the close pair, and the comma delineates the wide component. Only eight (24%) of the systems have the two lower-mass components in a tight orbit, in orbit around a more distant primary (e.g. HD 65907: F9.5, M0+M5 and HD 101177: F9.5, K3+M2). Notably, both brown dwarf pairs in triple systems (HD 97334 and 130948; the third, HD 79096 is a quadruple system) are part of this minority, i.e., tight brown dwarf pairs are preferred rather than two brown dwarfs orbiting their host star in a planetary-system formation, consistent with the findings of Burgasser et al. (2005). These results might indicate a selection effect that the lower-mass components of binaries are less-often searched for companions. Alternatively, they may imply that the more massive cloud preferentially fragments in the formation of triple systems when the resulting components are stellar in nature.

All but two of the 13 systems with four or more components are based on a double-double relationship. Nested three level hierarchies for quadruple systems such as those for HD 9770 and HD 137763 appear rare. In instances when these systems are formed solely through fragmentation, these observables suggest that formation processes favor the fragmentation of both first-level clouds rather than the successive fragmentation of one cloud while its counterpart stays intact. In instances when the wider relationships are established via capture, these results suggest a binary-binary combination is more probable and stable than a triple-single combination.

### 5.3.9. Multiplicity in Exoplanet Systems

Contrary to earlier expectations, recent studies (e.g., Raghavan et al. 2006; Mugrauer & Neuhäuser 2009) have shown that planetary systems are quite common among binaries and multiple systems. In a comprehensive search for stellar companions to the then known exoplanet hosts, Raghavan et al. (2009) concluded that even against selection effects, as many as 23% (30 of 131) of the exoplanet systems also had stellar companions. A recent report by Mugrauer & Neuhäuser (2009) showed that 17% (43 of 250) of exoplanet hosts were members of binary or multiple systems. In comparison, 30% (11 of 36) exoplanet systems of this study have stellar companions. This represents the largest percentage of stellar companions in any sample yet of exoplanet systems, likely due to the thoroughness of companion detection in this sample of nearby Sun-like stars. This is however still smaller than the 46% of stars having companions in the overall sample, presumably because all the exoplanets discovered to-date are from surveys that avoid known spectroscopic binaries. One key question is whether binary and multiple systems are equally likely to form planets as are single stars. Our results show that $9\% \pm 2\%$ of the single stars have planets, compared to $7\% \pm 2\%$ of binaries and $3\% \pm 3\%$ of triples. These fractions are statistically equivalent, suggesting that single stars and stars with companions are equally likely to harbor planets. Moreover, while sufficiently short-period binaries will disrupt protoplanetary disks, hampering planet formation around either star (e.g., see Desidera & Barbieri 2007), Figure 13 shows that many binaries in this study have sufficiently long periods to foster planet formation around each stellar component.

## 6. Conclusion

We present the results of a comprehensive evaluation of the multiplicity of solar-type stars in the solar neighborhood. Our sample of 454 stars, including the Sun, is selected from the *Hipparcos* catalog based on the following criteria: $\pi_{\mathrm{trig}} > 40$ mas, $\sigma_\pi/\pi < 0.05$, $0.5 \leq B-V \leq 1.0$, and positioned on an HR diagram within a band extending 1.5 magnitudes below and 2 magnitudes above the main sequence. These criteria are equivalent to selecting all dwarf and subdwarf stars within 25 pc with $V$-band fluxes 0.1–10 times that of the Sun, providing a physical basis for the term "solar-type". Our analyses show that this selection is not magnitude limited and hence is a complete volume-limited sample (see § 2.2), minimizing selection effects. This work is an update to the seminal effort of DM91, utilizing a larger and more accurate sample. Despite DM91's intentions of selecting a complete volume-limited sample from Gliese (1969), updated *Hipparcos* results show that as many as 72 of their 164 stars are now known to lie outside their selection criteria and an additional 56 stars meeting their selection criteria were left out (see § 2.3), justifying this follow-up effort. See § 2 for details on our sample selection and analyses.

Our results are based on targeted new observations to augment the vast amount of data available from extensive multiplicity studies of these stars using many different techniques, resulting in the most complete survey of companions yet (§ 5.1). Complementary publications (Latham et al. 2010; Raghavan et al. 2010a) report on systematic searches for spectroscopic companions, including four new detections, and LBOI searches for separated fringe packet binaries, respectively. The null result of the LBOI search shows that the gap between short-period spectroscopic systems and relatively long-period visual systems has been bridged due to long-standing spectroscopic campaigns and higher resolution visual techniques. The widest of companions were searched by blinking archival images, resulting in the identification of four new companions (§ 3.1). Targeted speckle interferometry observations augmented comprehensive historical programs, ensuring that every target in our sample was inspected by this high resolution technique. Additionally, we have individually evaluated companion entries in various catalogs including *Hipparcos* (§ 3.2), WDS (§ 3.3), and CNS (§ 3.5) for stellar companions, the Dwarf Archives for brown dwarf companions (§ 3.7), and the exoplanet catalogs for planetary companions (§ 3.8). The CCPS data, obtained for planet searches, have been utilized here to identify new stellar companions, but yielded only one new companion, enhancing the robustness of our completeness analysis (§ 5.1).

The observed percentages of single, double, triple, and higher order systems, including stellar and brown dwarf components, are $56\%\pm2\% : 33\%\pm2\% : 8\%\pm1\% : 3\%\pm1\%$, and our incompleteness analysis suggests that $54\%\pm2\%$ of solar-type stars in the solar neighborhood are single. This reverses previous expectations that only 43% of solar-type stars lack companions with masses greater than 0.1 $M_\odot$ and only 33% are without companions more massive than 10 $M_J$ (DM91). Our results double the percentage of triple and quadruple systems as compared to DM91, confirming their prediction that these additional companions were missed in their survey. Remarkably, the percentage of single stars is essentially unchanged since DM91 despite the comprehensive monitoring of nearby Sun-like stars, suggesting that their incompleteness analysis significantly overestimated missed companions.

Consistent with the overall trend of decreasing multiplicity ratios with increasing spectral types, we find that $50\% \pm 4\%$ of the F6–G2 subsample have companions, compared to $41\% \pm 3\%$ of the G2–K3 subsample (§ 5.3.2). The period distribution is unimodal and log-normal, with a median period of about 300 years (§ 5.3.3). While the profile of this distribution is similar to that of DM91, the median value in the prior study was 180 years, highlighting the improved completeness of wide pairs in the current survey. We are unable to reproduce the period-eccentricity relationships in DM91 for either the short-period or long-period systems, but instead find no relationship between these parameters beyond the circularization for $P < 12$ days (§ 5.3.4). In another departure from DM91, our mass-ratio distribution shows a preference for like-mass pairs, which prefer relatively short orbital periods (§ 5.3.5). Our data also suggest that stars might lose companions with age, especially those with higher cross sections, i.e., systems with more than two components or with long orbital periods, suggesting that some companions are stripped away over time by dynamical interactions (§ 5.3.6). We confirm the planet-metallicity correlation and observe that most known brown dwarfs orbit relatively metal-rich stars, although the scarcity of such companions prevents us from making definitive conclusions. We see no relationship between a star's metallicity and its tendency to have stellar companions for $B - V < 0.625$. However, a preliminary but intriguing result is that stars redder than this limit are more likely to have companions when they are relatively metal poor (§ 5.3.7). Finally, our results show that planets are as likely to form around single stars as they are around components of binaries or multiple star systems (§ 5.3.9), at least when they are sufficiently widely-separated, increasing the real estate thought available for planets, and perhaps life.

We wish to thank Andy Boden, Bill Cochran, Richard Gray, Roger Griffin, Bill Hartkopf, Artie Hatzes, Elliott Horch, Mike Ireland, Hugh Jones, Davy Kirkpatrick, Dagny Looper, Dimitri Pourbaix, Sam Quinn, Andrei Tokovinin, and Nils Turner for their helpful insights on the nature of some specific systems, and in some instances, for making specific observations at our request and sharing their unpublished results. We would also like to acknowledge the insightful comments from an anonymous referee, which resulted in a more robust completion analysis of this survey, improving the confidence in our results. Research at the CHARA Array is supported by the College of Arts and Sciences at Georgia State University and by the National Science Foundation through NSF grants AST-0606958 and AST-0908253. Some of the photometric and spectroscopic observations reported here were carried out under the auspices of the SMARTS (Small and Moderate Aperture Research Telescope System) Consortium, which operates several small telescopes at CTIO. We thank RECONS group members, particularly, Jennifer Winters for reducing photometry data, and Wei-Chun Jao for his insights with some of the analysis. This research has made use of the SIMBAD literature database, operated at CDS, Strasbourg, France, NASA's Astrophysics Data System, and the Washington Double Star Catalog maintained by the U.S. Naval Observatory. This effort used multi-epoch images from the Digitized Sky Survey, which was produced at the Space Telescope Science Institute under U.S. Government grant NAG W-2166, and from the SuperCOSMOS Sky Survey. This publication also made use of data products from the Two Micron All Sky Survey (2MASS), which is a joint project of the University of Massachusetts and the Infrared Processing and Analysis Center/California Institute of Technology, funded

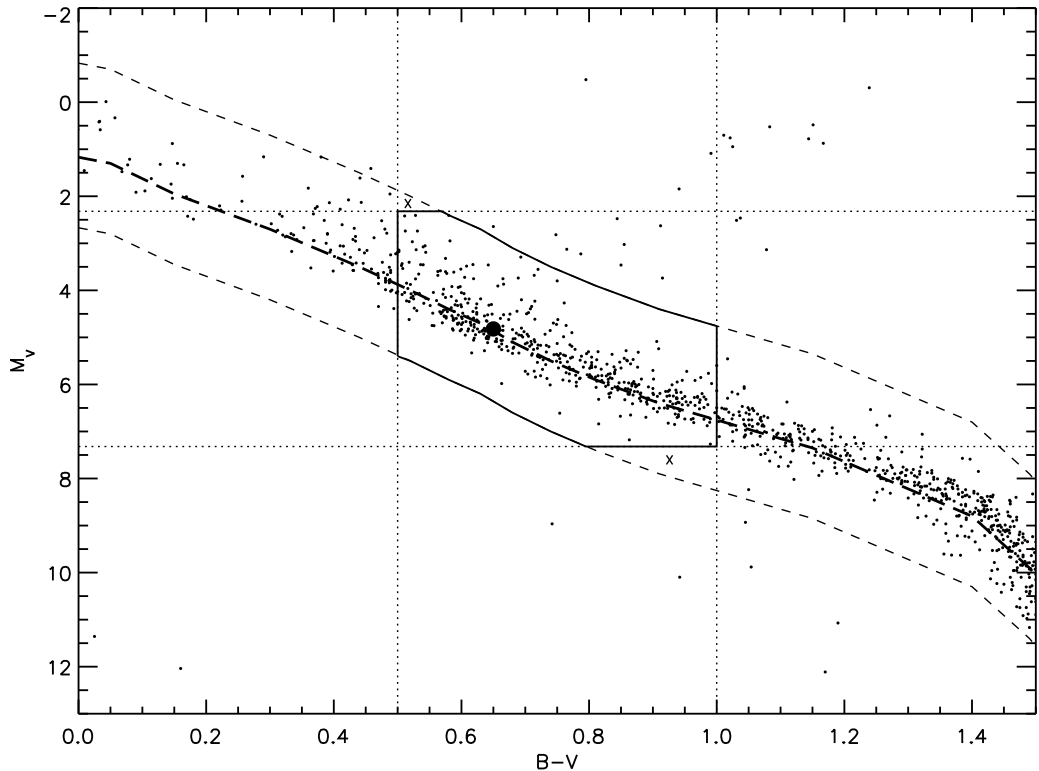

Fig. 1.— Volume-limited sample of solar-type stars from *Hipparcos*. The dots represent *Hipparcos* stars within our distance limit of 25 pc with parallax errors less than 5%. The long dashed line is the main sequence from Cox (2000), and the dashed lines 2 magnitudes above and 1.5 magnitudes below it identify the selection boundaries to limit spectral classes to IV, V, and VI. The dotted vertical lines mark the $B - V$ color limits of the sample, and the dotted horizontal lines mark the magnitudes corresponding to one-tenth and ten times the $V$-band flux of the Sun. The two triangles marked by 'X' indicate regions without corresponding stars, confirming that the color-based selection is equivalent to a selection based on luminosity. The solid outline encloses the final sample of 462 stars (including nine companions to other sample stars), and the large filled circle marks the position of the Sun.

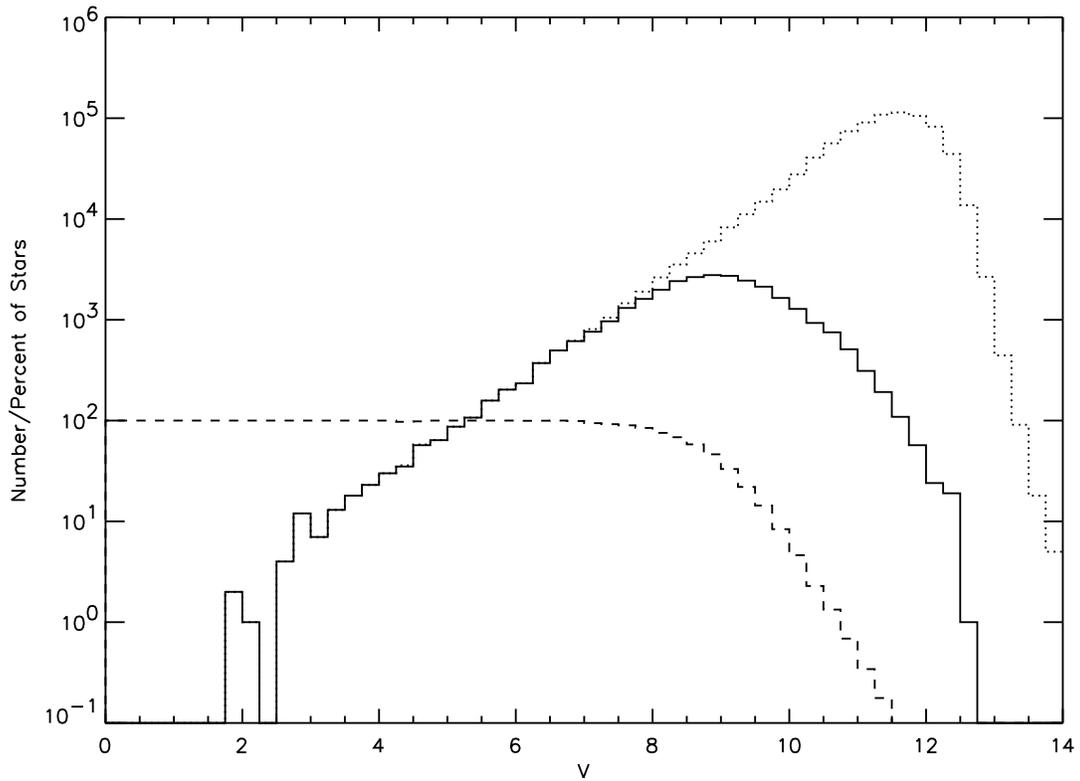

Fig. 2.— The *V* magnitude distributions of the *Hipparcos* and Tycho-2 catalogs. The solid line shows the *Hipparcos* distribution and the dotted line shows the Tycho-2 distribution. The number of stars is plotted on a logarithmic scale. The dashed line shows the percentage of Tycho-2 stars in each magnitude bin that are also included in the *Hipparcos* catalog and can be interpreted as the completeness of the *Hipparcos* catalog.

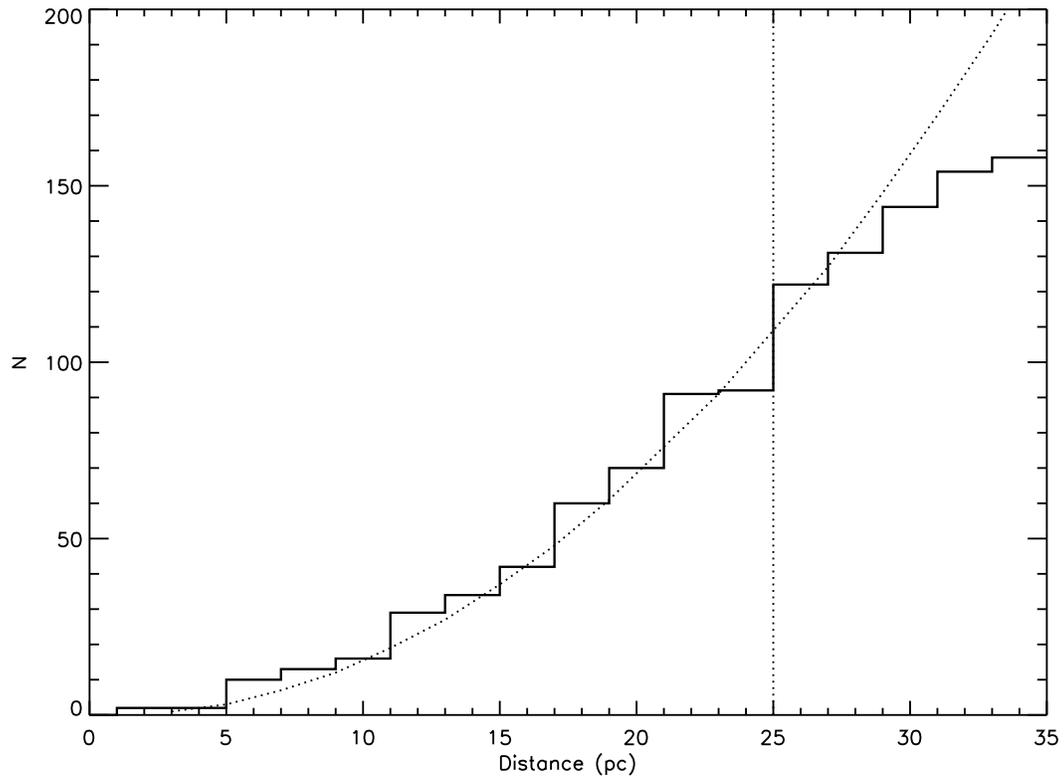

Fig. 3.— Distance distribution of the *Hipparcos* stars meeting our sample selection criteria, but expanding the distance limit out to 35 pc. The dotted curve shows the expected distribution of a complete sample, assuming completeness out to 15 pc. The vertical dotted line marks the distance limit of our sample.

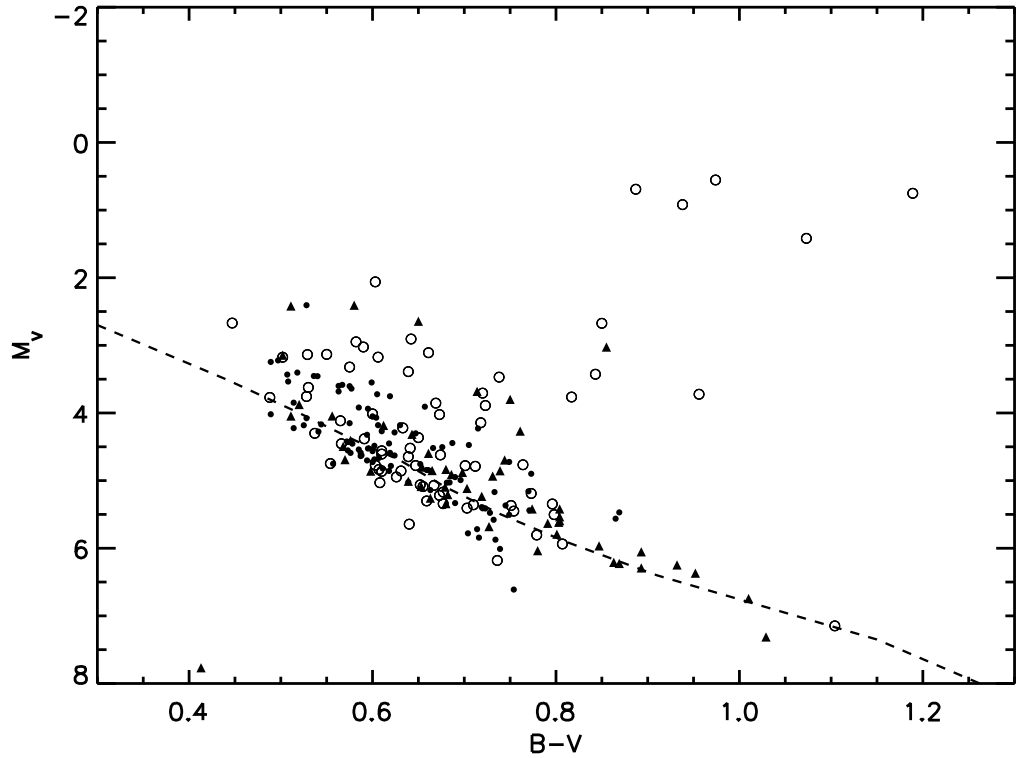

Fig. 4.— HR diagram for the DM91 sample from the CNS and *Hipparcos* catalogs. All points are plotted using magnitudes and parallaxes from *Hipparcos*. Filled circles represent DM91 sample stars that still match their criteria based on the *Hipparcos* data, while open circles identify DM91 sample stars that no longer meet their criteria. Filled triangles represent stars that were not part of the DM91 sample, but which meet their criteria based on *Hipparcos* data. The dashed line is the main sequence from Cox (2000).

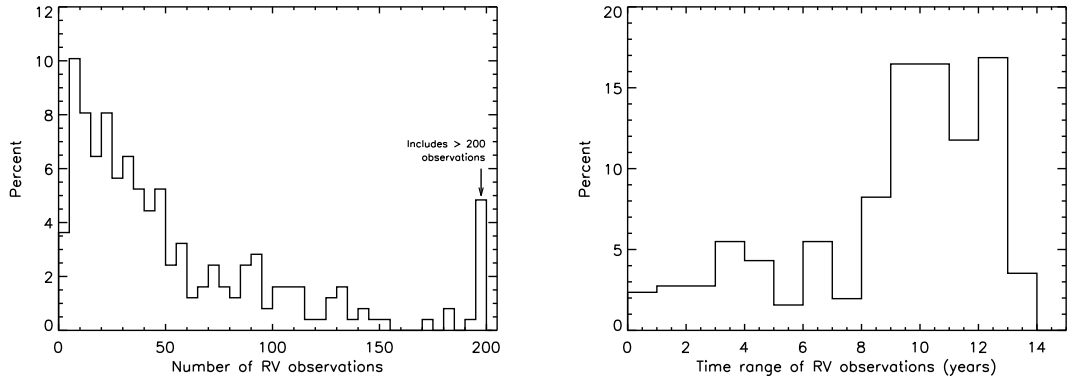

Fig. 5.— The distributions of the number of radial velocity measurements per star (left) and their time span (right) in the CCPS survey.

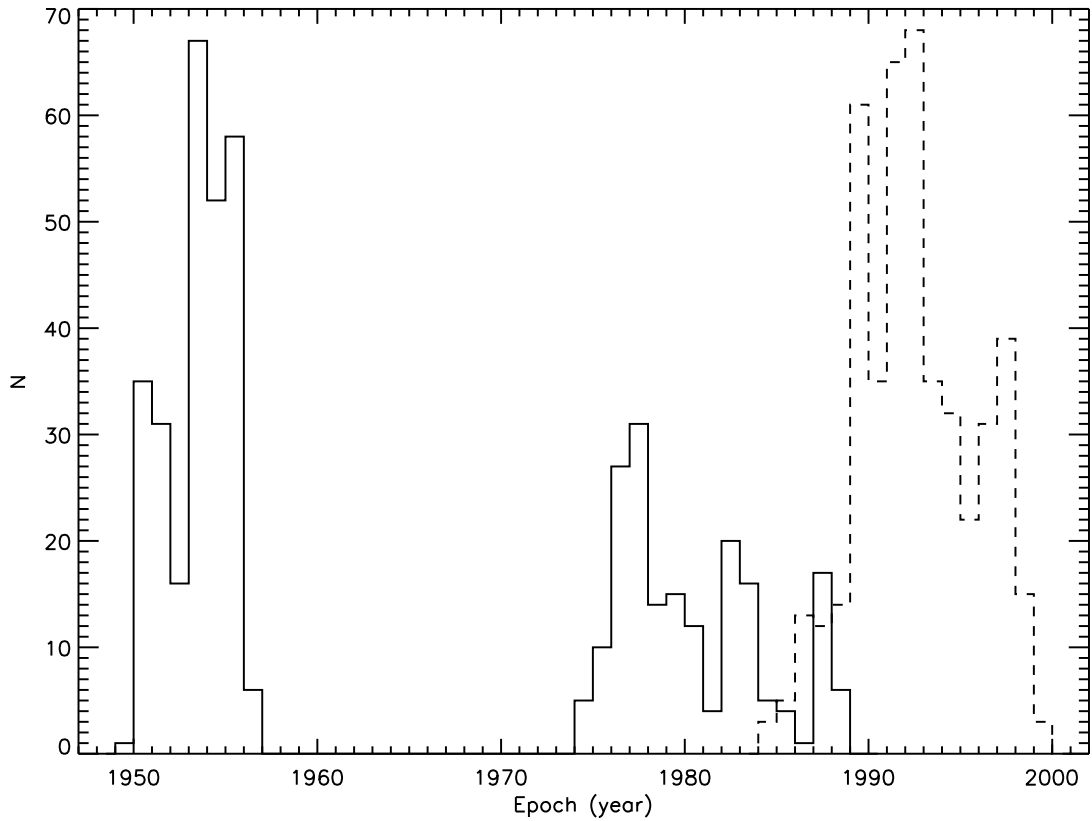

Fig. 6.— Epoch distribution of the images blinked to identify CPM companions. A pair of images from the DSS and/or SSS were blinked for each target, with the earlier epoch identified by the solid line and the later epoch by the dashed line.

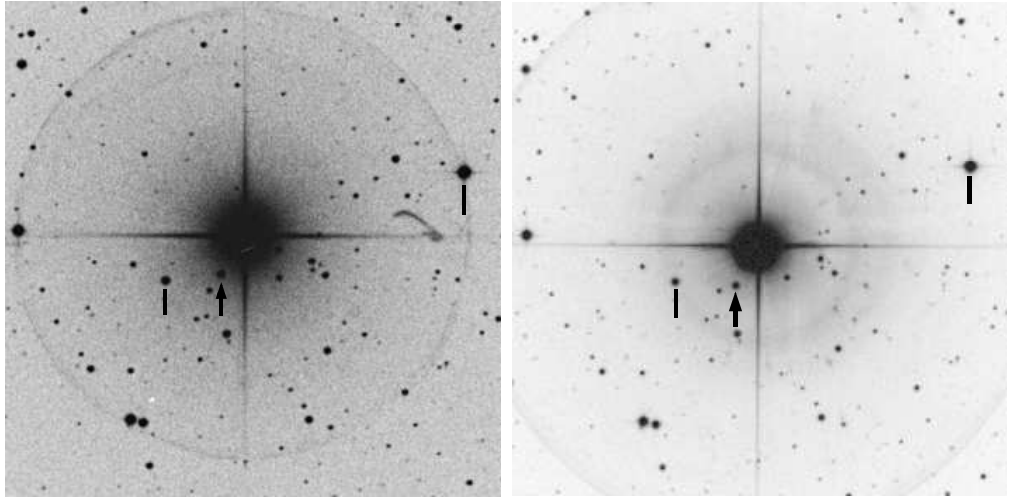

Fig. 7.— Example of the images blinked to identify CPM companions. The bright star at the center of these 10′ square images from the DSS is HD 9826 (υ And). North is up and east is to the left. The epoch of the left image is 1953.71 and for the right image is 1989.77. The arrow marks the CPM companion at a separation of 56″ and the lines identify WDS entries that are field stars. The primary's proper motion is $\mu_\alpha \cos\delta$ = -0″.173 yr$^{-1}$ and $\mu_\delta$ = -0″.381 yr$^{-1}$ from *Hipparcos*.

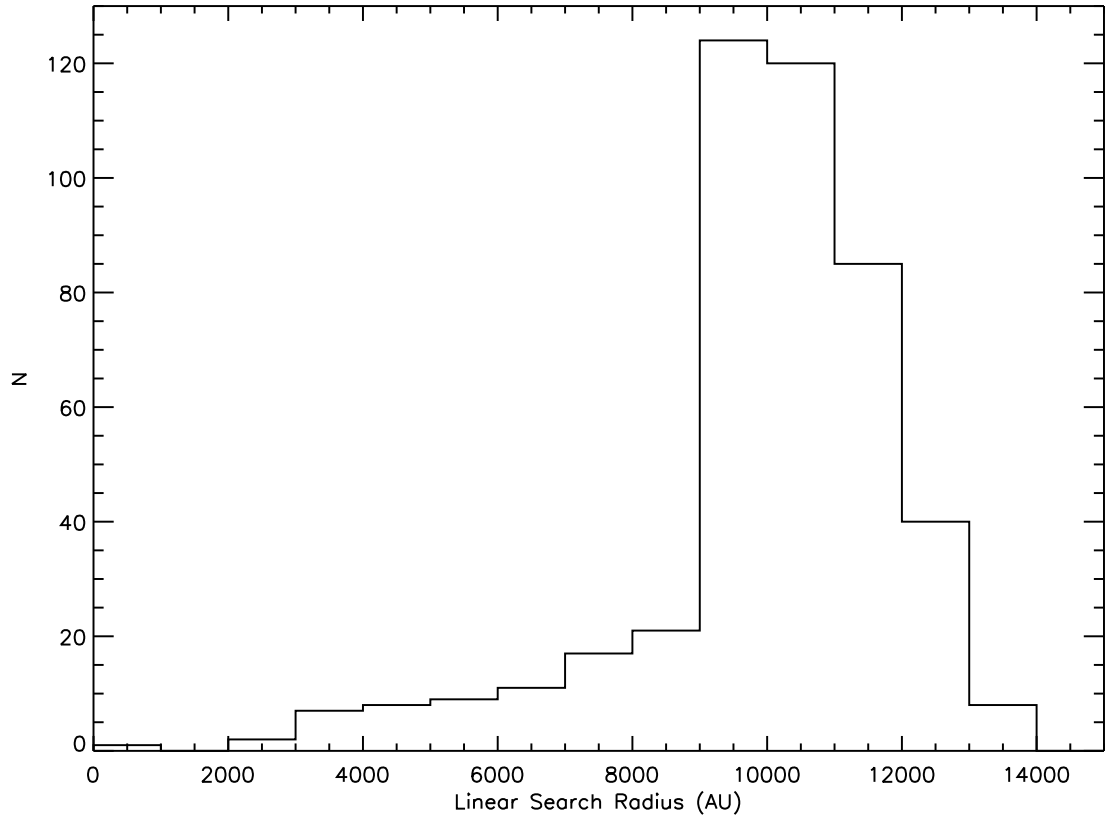

Fig. 8.— The distribution of the linear projected separation from the primary searched for CPM companions.

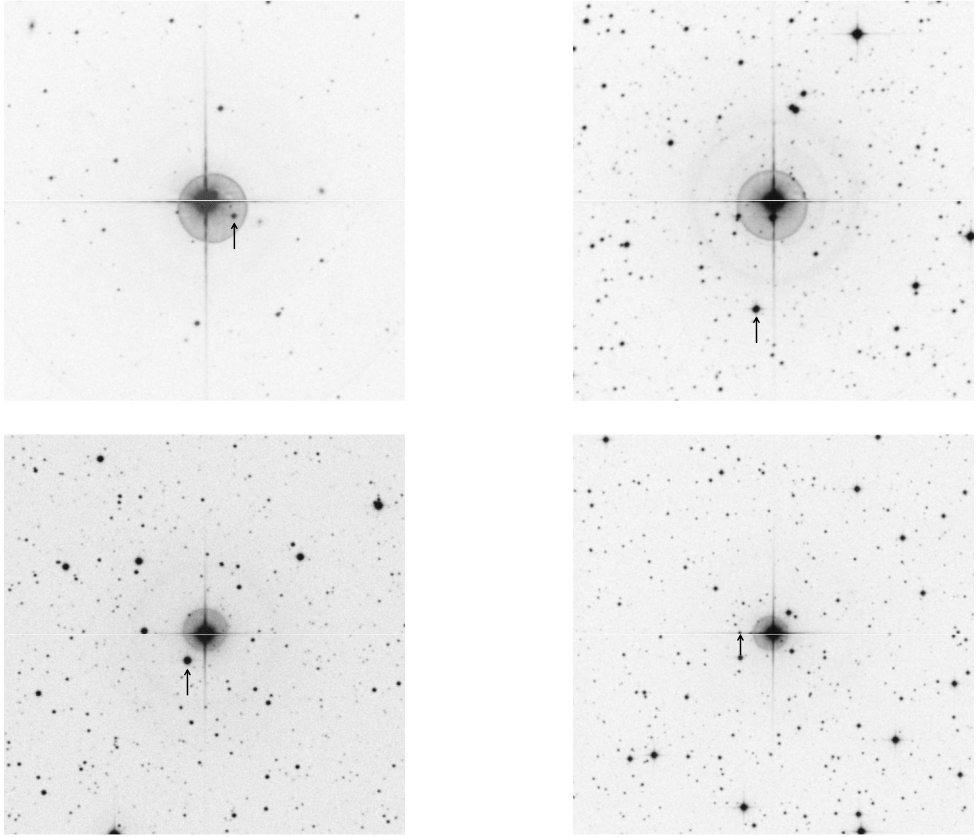

Fig. 9.— Archival images showing the new companions discovered. The primaries starting at the top-left image, going clockwise, are HD 4391, 43162, 218868, and 157347. In each image, the companion is marked by the arrow and the primary is the bright star at the center. The images are 10′ on each side and oriented with North up and East to the left.

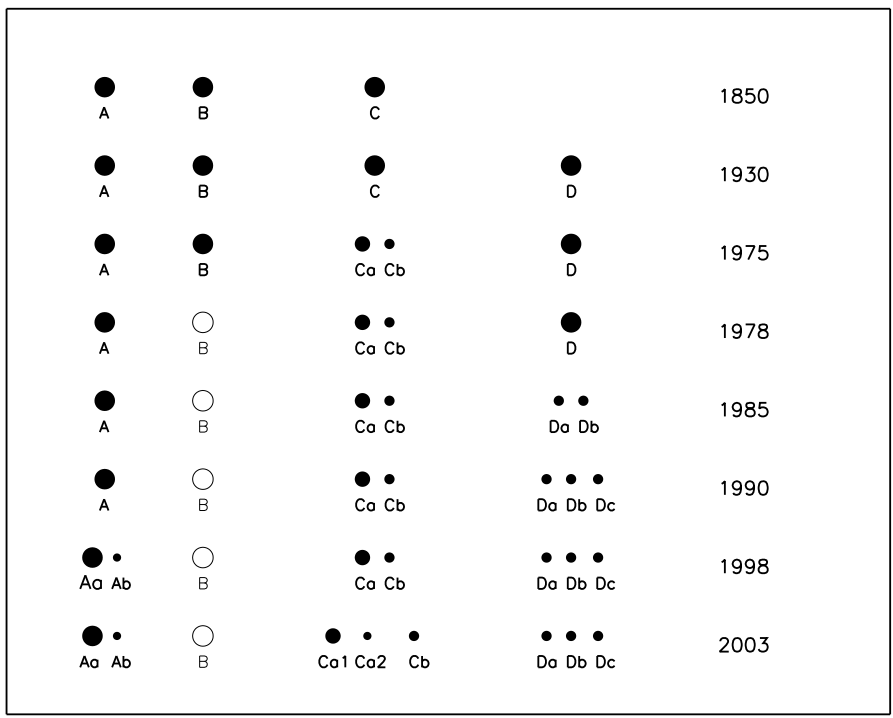

Fig. 10.— Companion nomenclature according to the WMC standards for a fictitious system that grows more complex over time. The following hypothetical events took place as components of this system were discovered. 1850: A telescopic inspection of this star reveals three visual components. The primary is designated as A and the companions are labeled B and C, respectively. 1930: A wide CPM companion is discovered and labeled D. 1975: Component C is split by speckle interferometry, and the components are named Ca and Cb, while C refers to the pair. 1978: A new measure of the AB pair shows that it is optical rather than physical, but the components names previously assigned are retained. 1985: Component D is split by speckle interferometry, and the components are named Da and Db. 1990: An additional speckle component is found in D and named Dc; 1998: AO reveals a faint source near A, which is identified as a brown dwarf companion by follow-up spectroscopy, splitting A into Aa, the stellar primary, and Ab, the brown dwarf. 2003: High-precision radial velocity measurements reveal a brown dwarf companion to Ca, splitting it into Ca1, the star, Ca2, the brown dwarf.

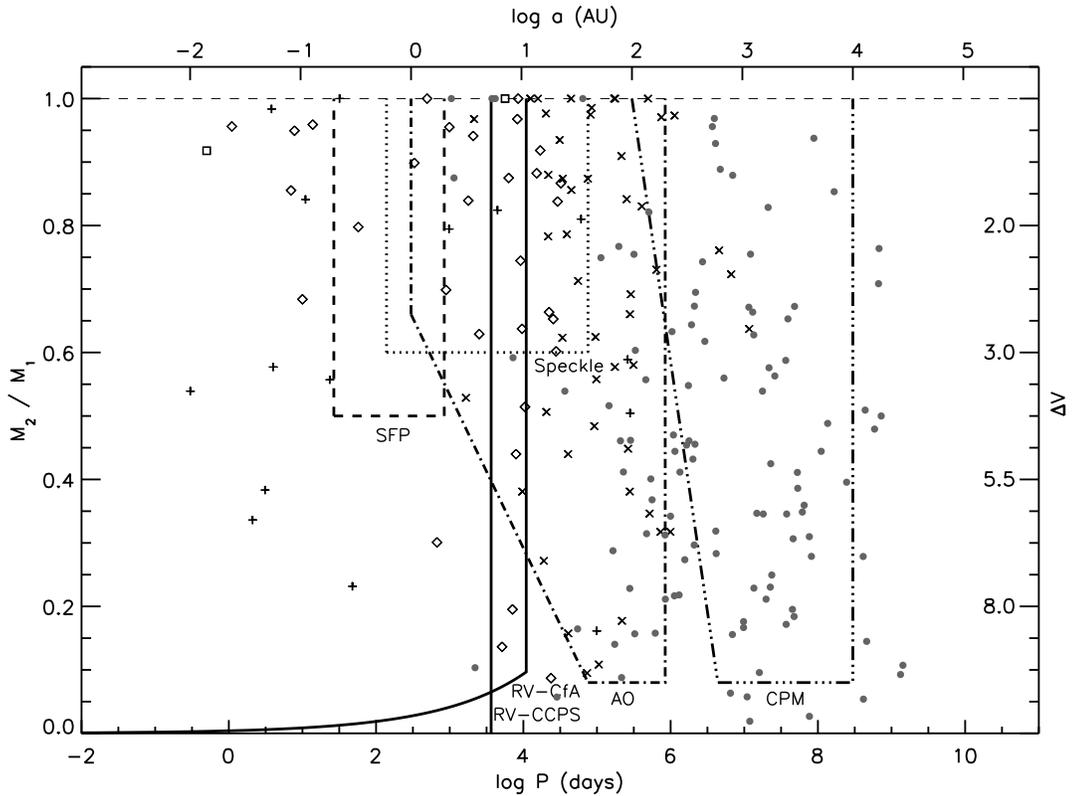

Fig. 11.— Detection limits of the various companion search methods by period and mass ratio. For convenience, the periods are converted to semi-major axes for our average mass-sum of 1.5 M$_\odot$ and shown on the top, and the mass ratios are converted to approximate $\Delta V$ values for primaries G0–K0 and shown on the right. The solid vertical line at $\log P$ just above 3.5 (10 years) is the limit for spectroscopic companions from the CCPS efforts, which are capable of detecting any companions to the left of it. The solid curve turning vertical at about $\log P = 4$ (30 years) marks the limit of the Latham et al. (2010) spectroscopic survey. The dashed lines show the detection region of the CHARA SFP search, the dotted lines mark the boundary of speckle interferometric searches, the dash-dot lines demarcate the AO detection space, and the dash-triple-dot lines delineate the CPM search region. The assumptions for detection limits for each technique are outlined in the text. The over-plotted points show the actual pairs found with estimates of periods and component masses. Plus signs identify spectroscopic orbits, crosses mark visual orbits, open diamonds show combined spectroscopic-visual solutions, filled circles identify CPM pairs and open squares show unresolved companions such as eclipsing binaries or overluminous sources.

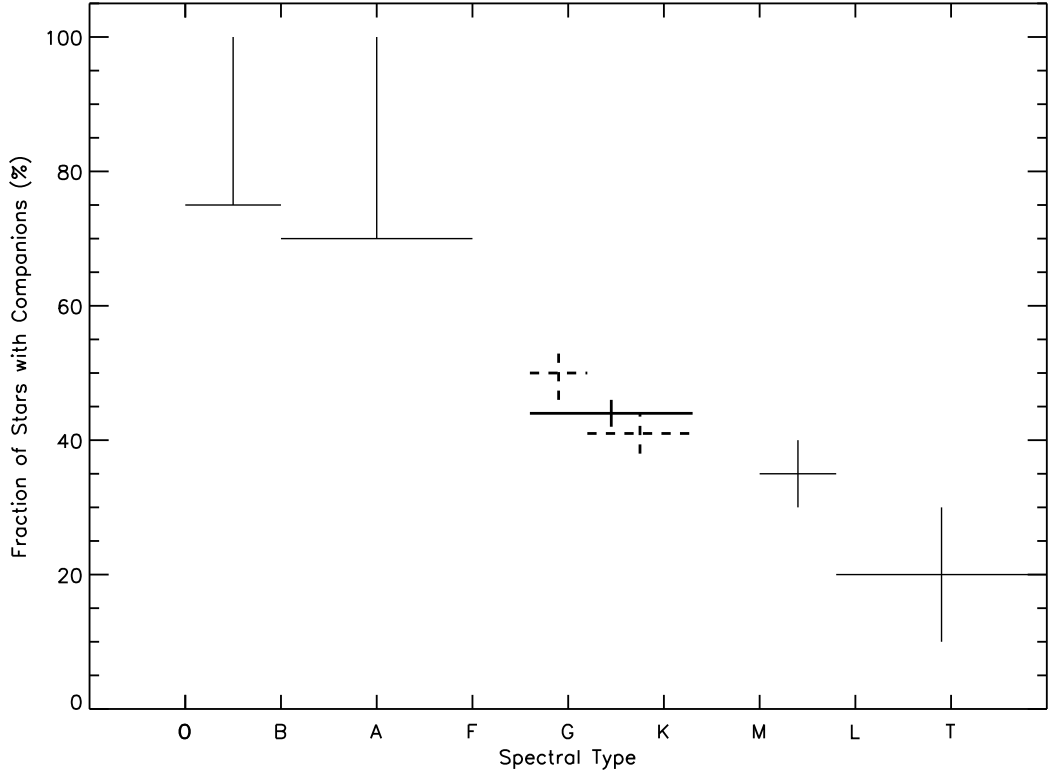

Fig. 12.— Multiplicity statistics by spectral type. The thin solid lines represent stars and brown dwarfs beyond the spectral range of this study, and their sources are listed in the text. For the FGK stars studied here, the thick dashed lines show our observed multiplicity fractions, i.e the percentage of stars with confirmed stellar or brown dwarf companions, for spectral types F6–G2 and G2–K3. The thick solid lines show the incompleteness-adjusted fraction for the entire F6–K3 sample. The uncertainties of the multiplicity fractions are estimated by bootstrap analysis as explained in §5.2.

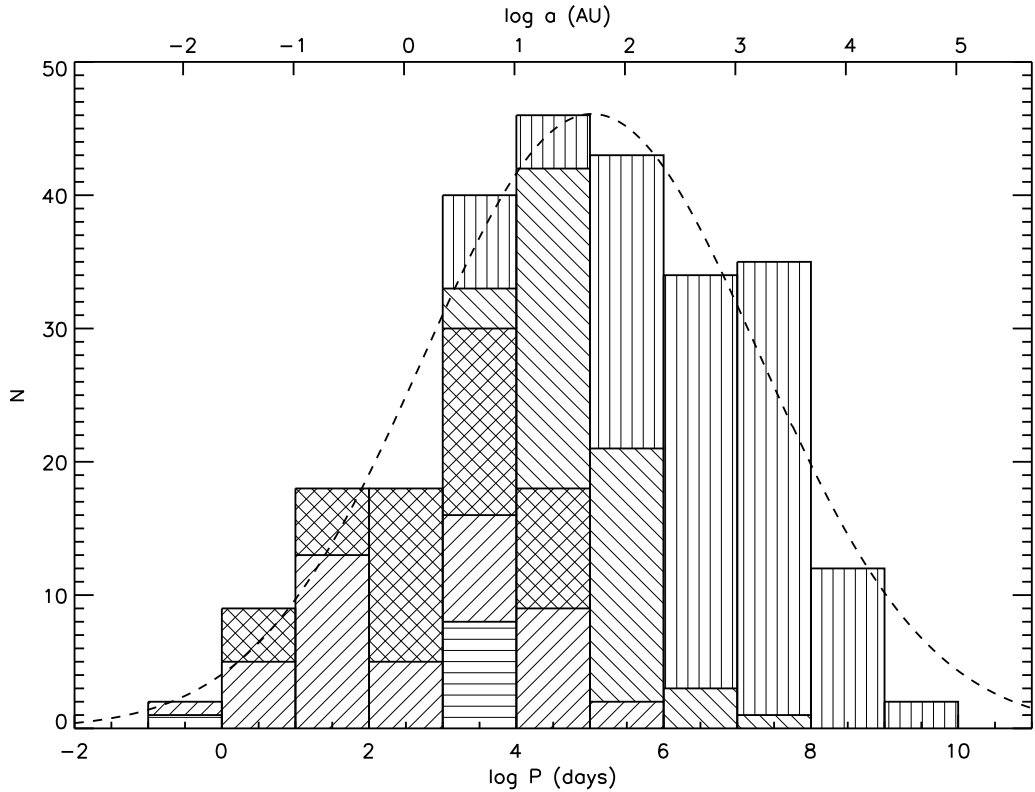

Fig. 13.— Period distribution for the 259 confirmed companions. The data are plotted by the companion detection method. Unresolved companions such as proper motion accelerations are identified by horizontal line shading, spectroscopic binaries by positively sloped lines, visual binaries by negatively sloped lines, companions found by both spectroscopic and visual techniques by crosshatching, and CPM pairs by vertical lines. The semimajor axes shown in AU at the top correspond to the periods on the x-axis for a system with a mass-sum of $1.5$ $M_\odot$, the average value for all the pairs. The dashed curve shows a Gaussian fit to the distribution, with a peak at $\log P = 5.03$ and standard deviation of $\sigma_{\log P} = 2.28$.

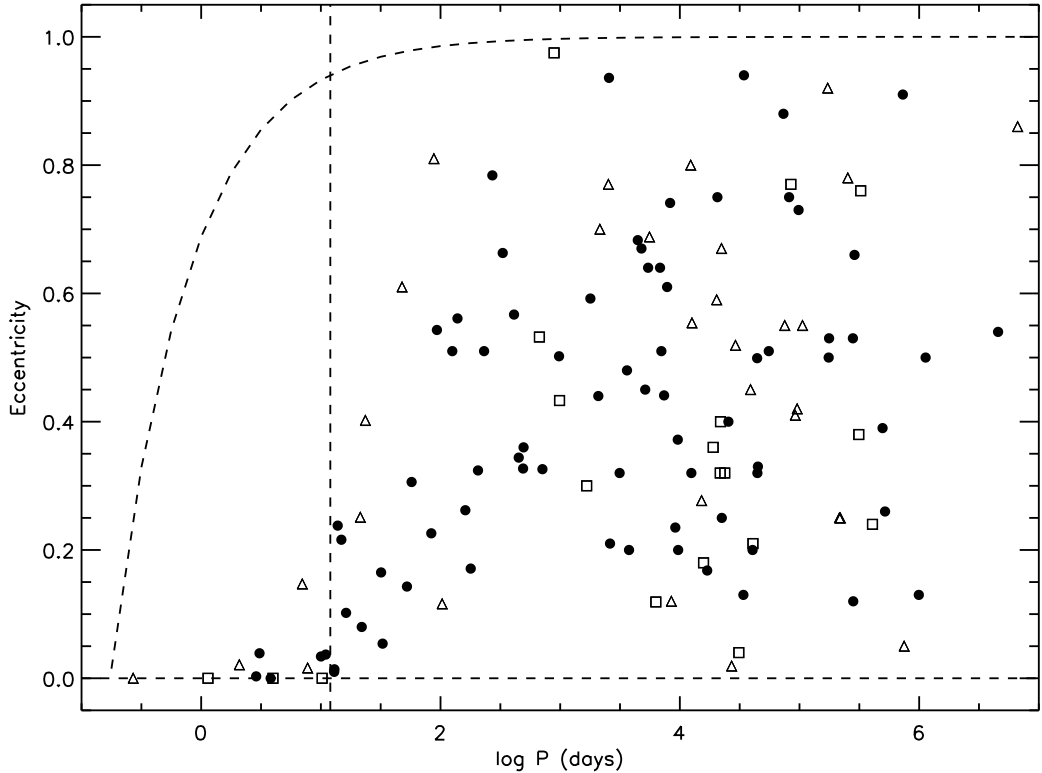

Fig. 14.— Period-eccentricity relationship for the 127 pairs with estimates of those parameters from visual and/or spectroscopic orbital solutions. Components of binaries are plotted as filled circles, of triples as open triangles, and of quadruple systems as open squares. The horizontal dashed line marks a zero-eccentricity limit and the vertical dashed line marks the 12-day period, which roughly corresponds to the circularization period for this population of stars. The exceptions with notable eccentricities to the left of this line are discussed in the text. The dashed curve represents a boundary, to the left of which pairs approaching periastron will pass within 1.5 R$_\odot$ and are hence likely to collide. The relation is derived assuming a mass-sum of 1.5 M$_\odot$, the average value for all the pairs.

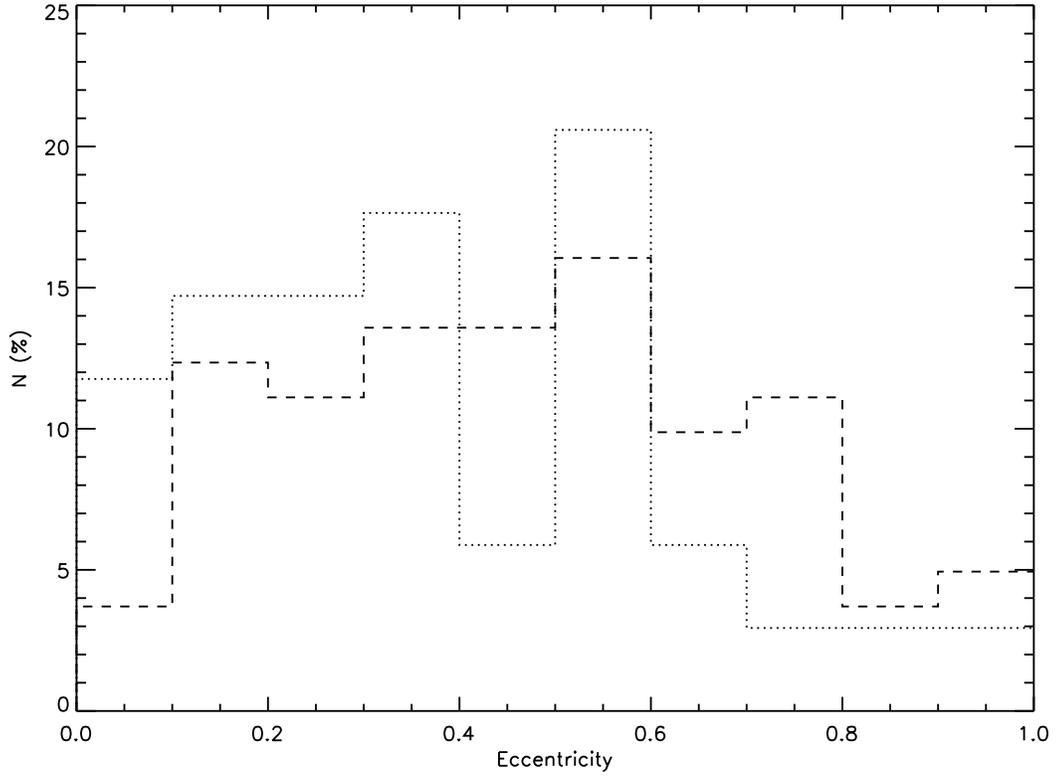

Fig. 15.— Eccentricity distribution for the 117 systems with periods longer than the 12-day circularization limit with estimated eccentricities from visual or spectroscopic solutions. The dotted line represents the 35 systems with periods below 1000 days, and the dashed line represents the 82 systems with periods longer than 1000 days.

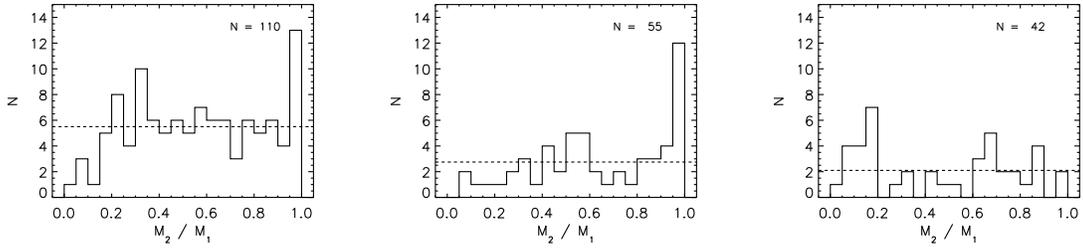

Fig. 16.— Mass-ratio distribution for binaries (left), pairs in higher-order multiple systems (middle), and composite-mass pairs in multiple systems (right). For example, in a triple system composed of a spectroscopic binary Aa,Ab and a visual binary AB, the $M_{Aa}$ to $M_{Ab}$ ratio is included in the middle panel and the $M_{(Aa+Ab)}$ to $M_B$ ratio is included in the right panel. The dashed lines mark the average value for each plot.

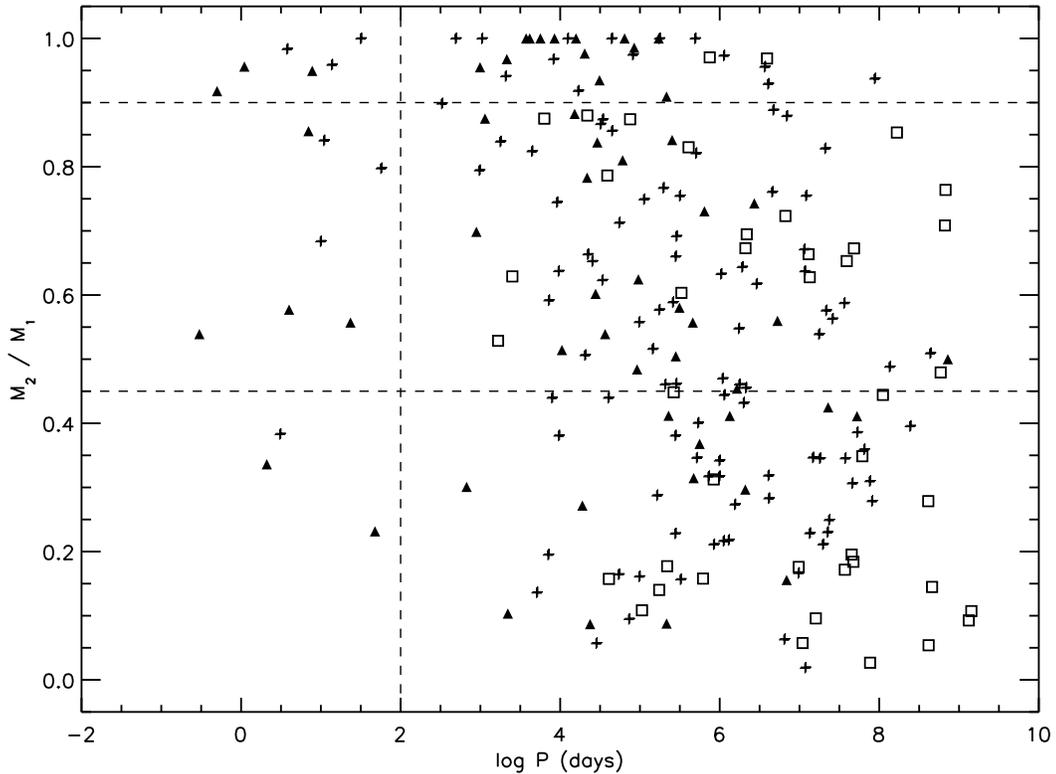

Fig. 17.— Mass-ratio – period relation. Plus signs represent binaries, filled triangles identify pairs in multiple systems, and open squares indicate composite-mass hierarchical pairs in multiples (see caption of Figure 16 for an example). The dotted lines are drawn to mark subdivisions for analysis, as described in the text.

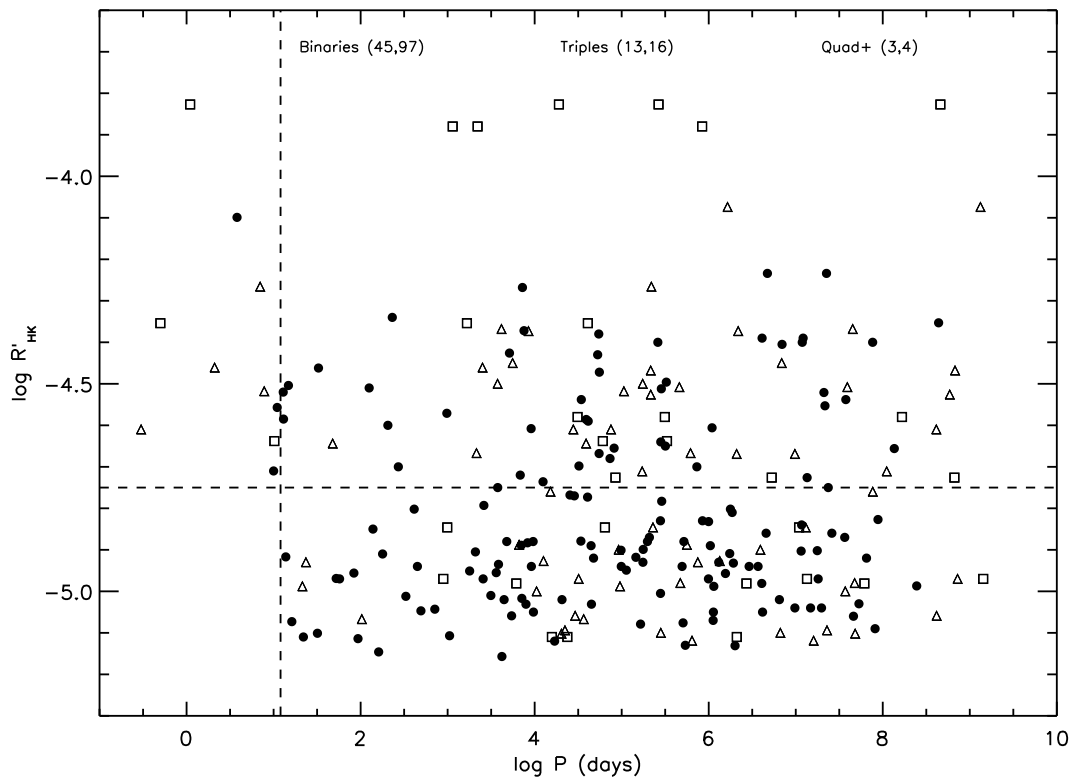

Fig. 18.— Plot of log $R'_{\rm HK}$ and orbital period for binaries (filled circles), pairs in triple systems (open triangles), and in quadruple and higher order systems (open squares). The horizontal dashed line delineates what we consider to be active stars (above the line) from inactive stars. The vertical dashed line at $P = 12$ days marks the tidal orbit-circularization limit. Systems to its left are likely to have enhanced activity due to tidal interactions independent of age. The numbers in parentheses at the top show the counts of active and inactive systems for binaries, triple and quadruple or higher order systems, excluding systems with periods less than 12 days. Because the plot includes each pair with its orbital period and the log $R'_{\rm HK}$ of the primary, triples and higher-order multiples have more than one point along horizontal lines.

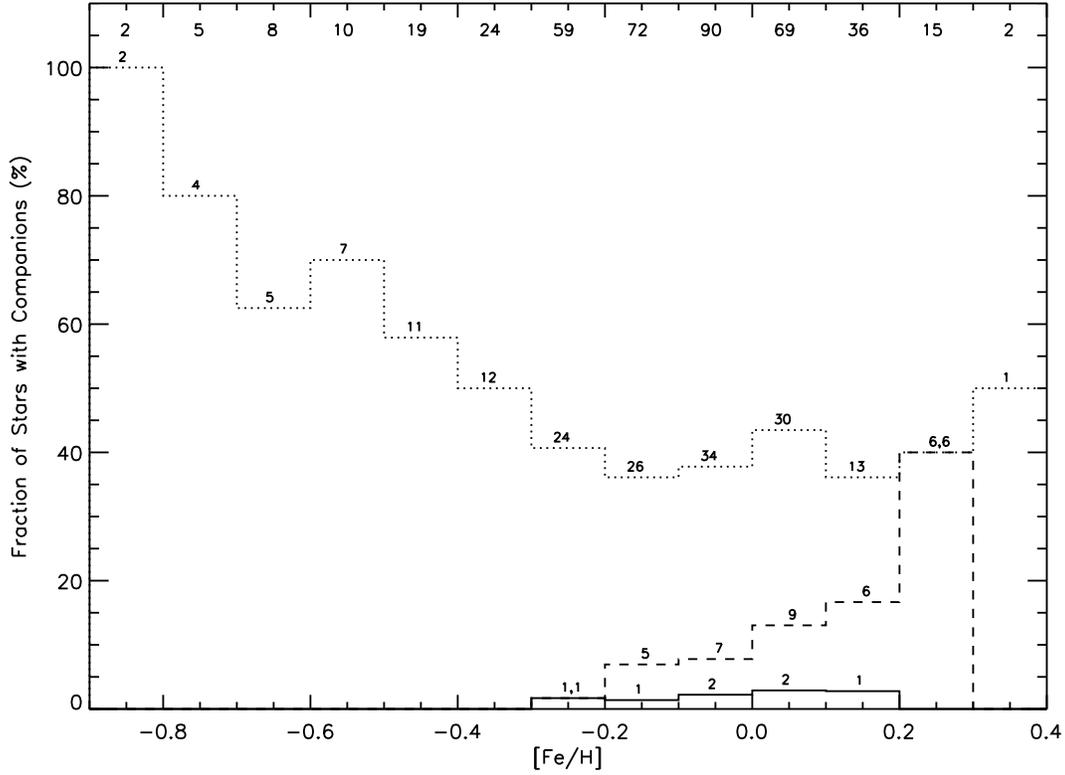

Fig. 19.— The fraction of stars with stellar (dotted line), brown dwarf (solid line), and planetary (dashed line) companions as a function of metallicity. The numbers at the top of the plot are the total number of stars in each 0.1 dex metallicity bin and the counts immediately above the histograms are the number of stars with the corresponding type of companion. Four stars with [Fe/H] < −0.9 are excluded from this plot, each of which is single with no known stellar or substellar companions.

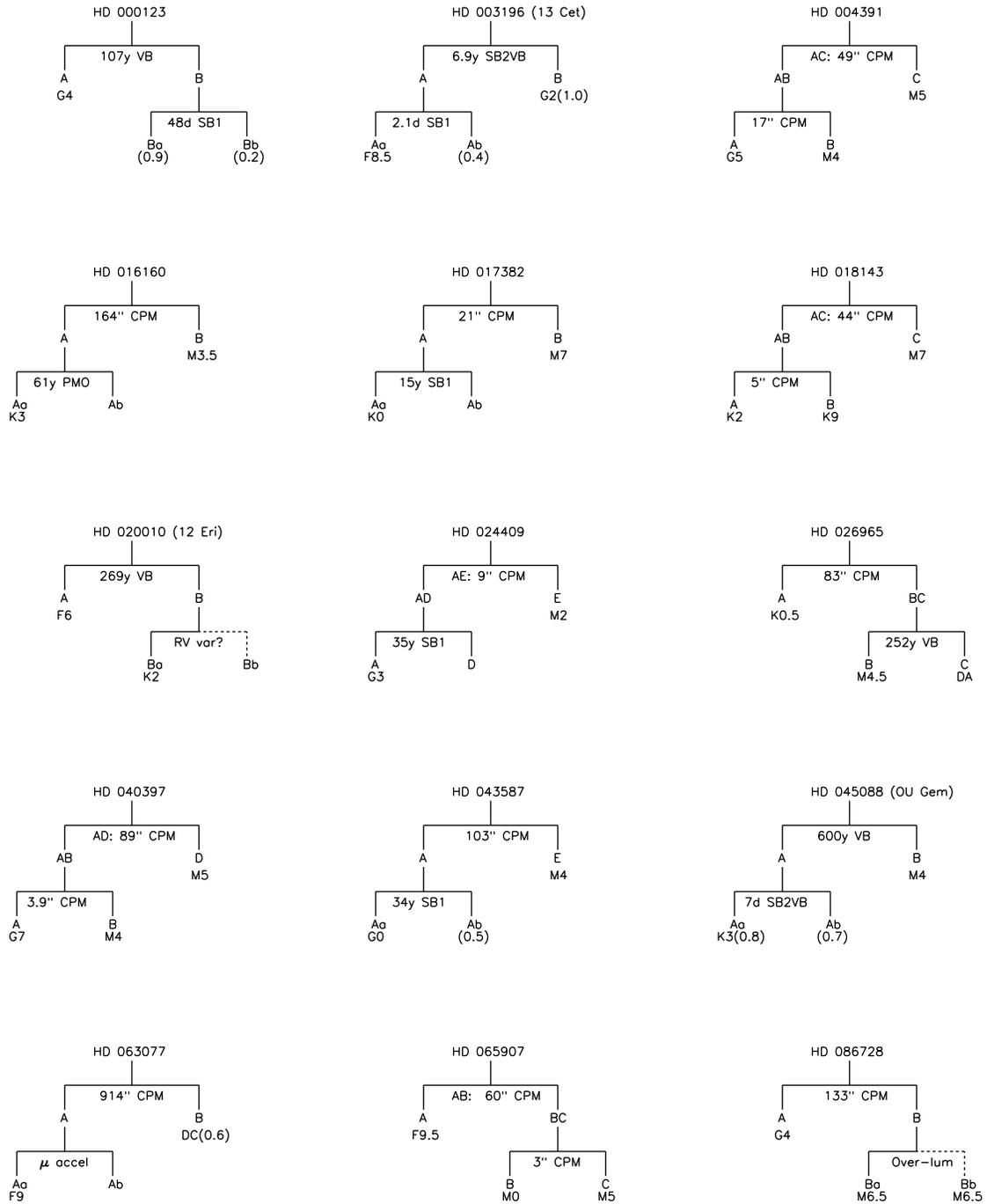

Fig. 20.— Mobile diagrams of triple systems (1 of 3). Solid lines connect confirmed companions and dashed lines connect candidate companions. Spectral types and/or mass estimates, in parenthesis in solar-mass units, are listed below each component, when available. See Tables 17 and 18 for the methods and references of the mass estimates.

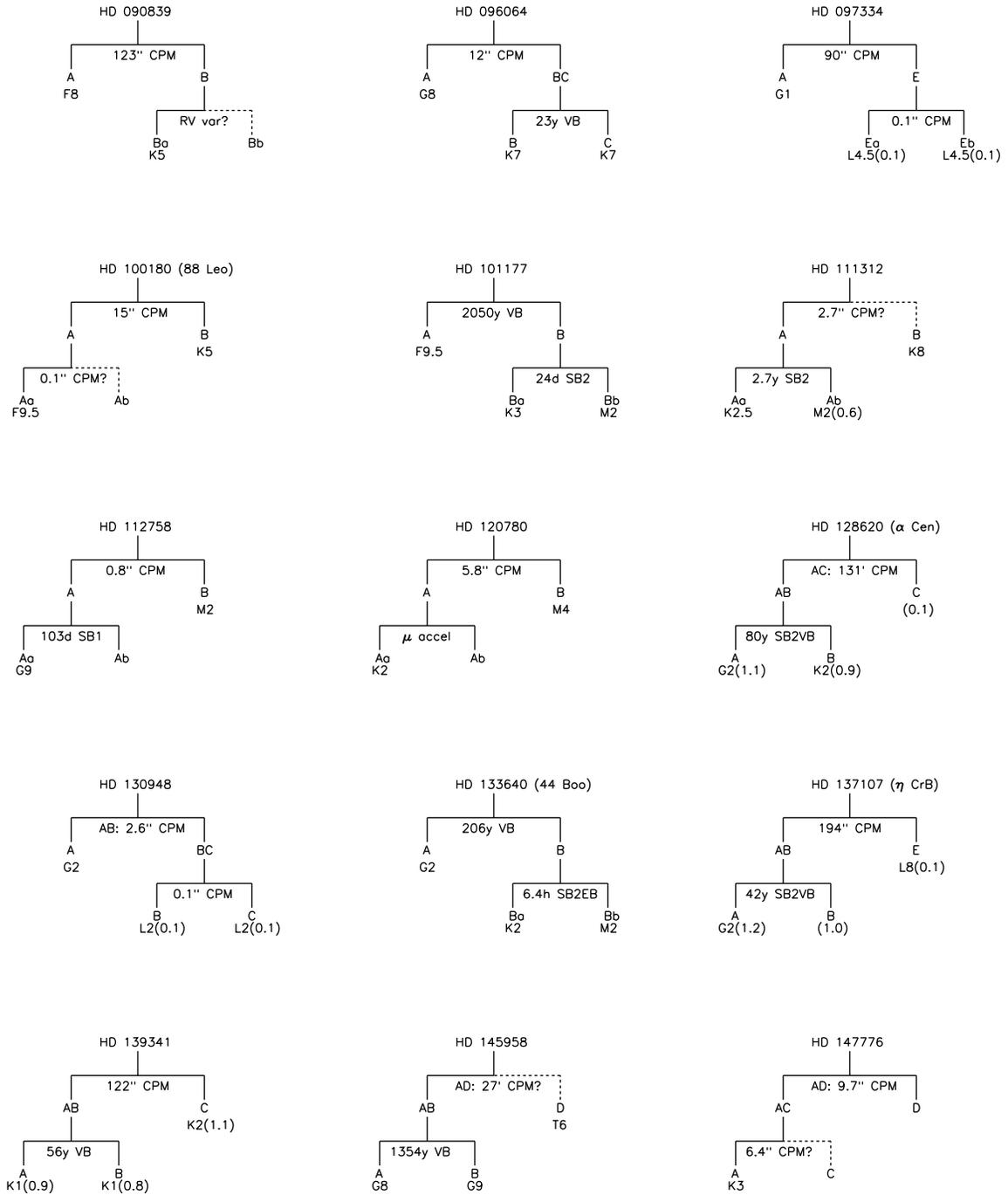

Fig. 21.— Mobile diagrams of triple systems (2 of 3). See caption under Figure 20 for a description of the format.

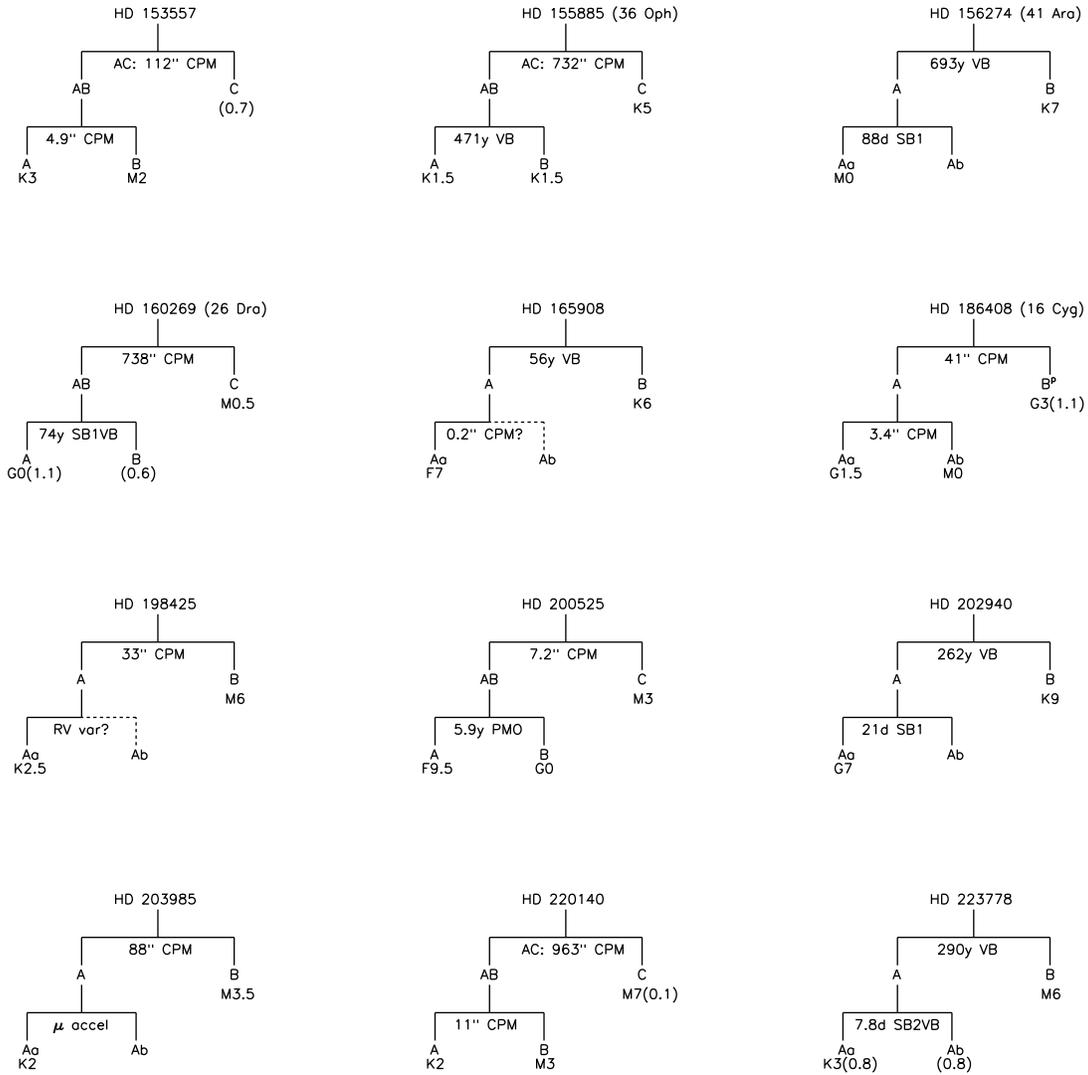

Fig. 22.— Mobile diagrams of triple systems (3 of 3). See caption under Figure 20 for a description of the format. The 'p' superscript for HD 186408 B indicates that it is a planet-host star.

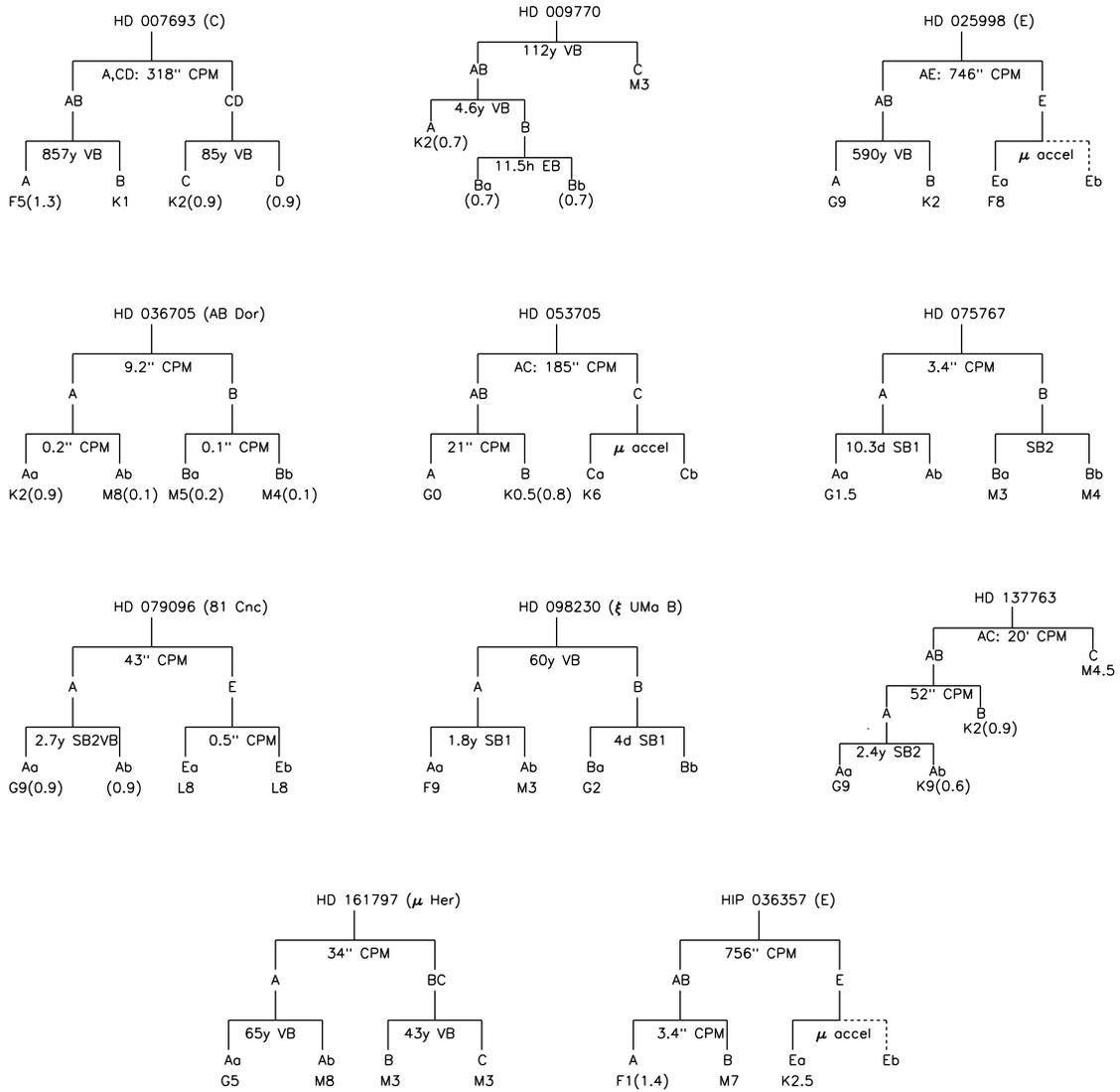

Fig. 23.— Mobile diagrams of quadruple systems. See caption under Figure 20 for a description of the format.

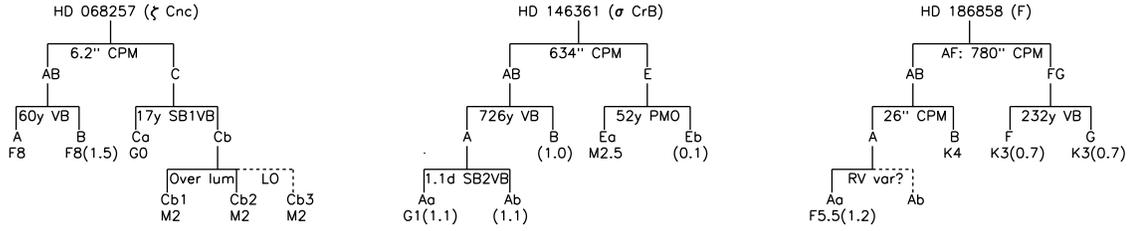

Fig. 24.— Mobile diagrams of quintuple and higher-order systems. See caption under Figure 20 for a description of the format.

Table 1.   Volume-limited Sample of 454 Solar-type Stars

| R.A. (J2000.0) (1) | Decl. (J2000.0) (2) | HIP Name (3) | HD Name (4) | Hipparcos $V$ (5) | Hipparcos $B-V$ (6) | Hipparcos $\pi$ (mas) (7) | Hipparcos $\sigma_\pi$ (mas) (8) | van Leeuwen (2007b) $\pi$ (mas) (9) | van Leeuwen (2007b) $\sigma_\pi$ (mas) (10) | Spec Type (11) | Ref (12) |
|---|---|---|---|---|---|---|---|---|---|---|---|
| . . . | . . . | Sun | . . . | . . . | 0.650 | . . . | . . . | . . . | . . . | G2 V | . . . |
| 00 02 10.16 | +27 04 56.1 | 000171 | 224930 | 5.80 | 0.690 | 80.63 | 3.03 | 82.17 | 2.23 | G3 V | 1 |
| 00 06 15.81 | +58 26 12.2 | 000518 | 000123 | 5.98 | 0.687 | 49.30 | 1.05 | 46.56 | 0.65 | G5 V | 1 |
| 00 06 36.78 | +29 01 17.4 | 000544 | 000166 | 6.07 | 0.752 | 72.98 | 0.75 | 73.15 | 0.56 | K0 V | 1 |
| 00 12 50.25 | −57 54 45.4 | 001031 | 000870 | 7.22 | 0.775 | 49.18 | 0.78 | 49.53 | 0.58 | K0 V | 2 |
| 00 16 12.68 | −79 51 04.3 | 001292 | 001237 | 6.59 | 0.749 | 56.76 | 0.53 | 57.15 | 0.31 | G8.5 V | 2 |
| 00 16 53.89 | −52 39 04.1 | 001349 | 001273 | 6.84 | 0.655 | 43.45 | 1.19 | 44.25 | 0.62 | G5 V | 2 |
| 00 18 41.87 | −08 03 10.8 | 001499 | 001461 | 6.47 | 0.674 | 42.67 | 0.85 | 43.02 | 0.51 | G0 V | 1 |
| 00 20 00.41 | +38 13 38.6 | 001598 | 001562 | 6.97 | 0.640 | 40.25 | 0.81 | 40.33 | 0.59 | G0 | 1 |
| 00 20 04.26 | −64 52 29.2 | 001599 | 001581 | 4.23 | 0.576 | 116.38 | 0.64 | 116.46 | 0.16 | F9.5 V | 2 |
| 00 22 51.79 | −12 12 34.0 | 001803 | 001835 | 6.39 | 0.659 | 49.05 | 0.91 | 47.93 | 0.53 | G5 V | 2 |
| 00 24 25.93 | −27 01 36.4 | 001936 | 002025 | 7.92 | 0.940 | 55.53 | 1.08 | 54.87 | 0.86 | K3 V | 2 |
| 00 25 45.07 | −77 15 15.3 | 002021 | 002151 | 2.82 | 0.618 | 133.78 | 0.51 | 134.07 | 0.11 | G0 V | 2 |
| 00 35 14.88 | −03 35 34.2 | 002762 | 003196 | 5.20 | 0.567 | 47.51 | 1.15 | 47.05 | 0.67 | F8 V... | 1 |
| 00 37 20.70 | −24 46 02.2 | 002941 | 003443 | 5.57 | 0.715 | 64.38 | 1.40 | 64.93 | 1.85 | G7 V | 2 |
| 00 39 21.81 | +21 15 01.7 | 003093 | 003651 | 5.88 | 0.850 | 90.03 | 0.72 | 90.42 | 0.32 | K0 V | 1 |
| 00 40 49.27 | +40 11 13.8 | 003206 | 003765 | 7.36 | 0.937 | 57.90 | 0.98 | 57.71 | 0.80 | K2 V | 1 |
| 00 44 39.27 | −65 38 58.3 | 003497 | 004308 | 6.55 | 0.655 | 45.76 | 0.56 | 45.34 | 0.32 | G6 V | 2 |
| 00 45 04.89 | +01 47 07.9 | 003535 | 004256 | 8.03 | 0.983 | 45.43 | 0.95 | 46.37 | 0.62 | K2 V | 1 |
| 00 45 45.59 | −47 33 07.2 | 003583 | 004391 | 5.80 | 0.635 | 66.92 | 0.73 | 65.97 | 0.39 | G5 V | 2 |
| 00 48 22.98 | +05 16 50.2 | 003765 | 004628 | 5.74 | 0.890 | 134.04 | 0.86 | 134.14 | 0.51 | K2.5 V | 2 |
| 00 48 58.71 | +16 56 26.3 | 003810 | 004676 | 5.07 | 0.502 | 41.80 | 0.75 | 42.64 | 0.27 | F8 V... | 1 |
| 00 49 06.29 | +57 48 54.7 | 003821 | 004614 | 3.46 | 0.587 | 167.99 | 0.62 | 167.98 | 0.48 | G0 V | 1 |
| 00 49 26.77 | −23 12 44.9 | 003850 | 004747 | 7.15 | 0.769 | 53.09 | 1.02 | 53.51 | 0.53 | G9 V | 2 |
| 00 49 46.48 | +70 26 58.1 | 003876 | 004635 | 7.75 | 0.900 | 46.47 | 0.70 | 46.23 | 0.53 | K0 | 1 |
| 00 50 07.59 | −10 38 39.6 | 003909 | 004813 | 5.17 | 0.514 | 64.69 | 1.03 | 63.48 | 0.35 | F7 IV-V | 1 |
| 00 51 10.85 | −05 02 21.4 | 003979 | 004915 | 6.98 | 0.663 | 45.27 | 0.97 | 46.47 | 0.66 | G0 | 1 |
| 00 53 01.13 | −30 21 24.9 | 004148 | 005133 | 7.15 | 0.936 | 71.01 | 0.78 | 70.56 | 0.61 | K2.5 V | 2 |
| 00 53 04.20 | +61 07 26.3 | 004151 | 005015 | 4.80 | 0.540 | 53.85 | 0.60 | 53.35 | 0.33 | F8 V | 1 |
| 01 08 16.39 | +54 55 13.2 | 005336 | 006582 | 5.17 | 0.704 | 132.40 | 0.60 | 132.38 | 0.82 | K1 V | 3 |
| 01 15 00.99 | −68 49 08.1 | 005842 | 007693 | 7.22 | 1.000 | 47.36 | 1.25 | 46.20 | 0.82 | K2+ V | 2 |
| 01 15 11.12 | −45 31 54.0 | 005862 | 007570 | 4.97 | 0.571 | 66.43 | 0.64 | 66.16 | 0.24 | F9 V | 2 |
| 01 16 29.25 | +42 56 21.9 | 005944 | 007590 | 6.59 | 0.594 | 42.30 | 0.75 | 43.11 | 0.45 | G0- V | 3 |
| 01 21 59.12 | +76 42 37.0 | 006379 | 007924 | 7.17 | 0.826 | 59.46 | 0.59 | 59.49 | 0.46 | K0 | 1 |
| 01 29 04.90 | +21 43 23.4 | 006917 | 008997 | 7.74 | 0.966 | 43.16 | 0.93 | 42.13 | 0.68 | K2.5 V | 3 |
| 01 33 15.81 | −24 10 40.7 | 007235 | 009540 | 6.97 | 0.766 | 51.27 | 0.88 | 52.49 | 0.46 | G8.5 V | 2 |
| 01 34 33.26 | +68 56 53.3 | 007339 | 009407 | 6.52 | 0.686 | 47.65 | 0.60 | 48.41 | 0.40 | G6.5 V | 3 |
| 01 35 01.01 | −29 54 37.2 | 007372 | 009770 | 7.11 | 0.909 | 42.29 | 1.47 | 46.24 | 3.07 | K2 V | 2 |
| 01 36 47.84 | +41 24 19.7 | 007513 | 009826 | 4.10 | 0.536 | 74.25 | 0.72 | 74.12 | 0.19 | F8 V | 1 |
| 01 37 35.47 | −06 45 37.5 | 007576 | 010008 | 7.66 | 0.797 | 42.35 | 0.96 | 41.75 | 0.74 | G9 V | 3 |
| 01 39 36.02 | +45 52 40.0 | 007734 | 010086 | 6.60 | 0.690 | 46.73 | 0.80 | 46.79 | 0.60 | G5 V | 3 |
| 01 39 47.54 | −56 11 47.0 | 007751 | 010360 | 5.76 | 0.880 | 122.75 | 1.41 | 127.84 | 2.19 | K2 V | 2 |

Table 1—Continued

| R.A. (J2000.0) (1) | Decl. (J2000.0) (2) | HIP Name (3) | HD Name (4) | *Hipparcos* V (5) | *Hipparcos* B − V (6) | *Hipparcos* π (mas) (7) | *Hipparcos* σ_π (mas) (8) | van Leeuwen (2007b) π (mas) (9) | van Leeuwen (2007b) σ_π (mas) (10) | Spec Type (11) | Ref (12) |
|---|---|---|---|---|---|---|---|---|---|---|---|
| 01 41 47.14 | +42 36 48.1 | 007918 | 010307 | 4.96 | 0.618 | 79.09 | 0.83 | 78.50 | 0.54 | G1 V | 3 |
| 01 42 29.32 | −53 44 27.0 | 007978 | 010647 | 5.52 | 0.551 | 57.63 | 0.64 | 57.36 | 0.25 | F9 V | 2 |
| 01 42 29.76 | +20 16 06.6 | 007981 | 010476 | 5.24 | 0.836 | 133.91 | 0.91 | 132.76 | 0.50 | K0 V | 3 |
| 01 44 04.08 | −15 56 14.9 | 008102 | 010700 | 3.49 | 0.727 | 274.17 | 0.80 | 273.96 | 0.17 | G8.5 V | 2 |
| 01 47 44.83 | +63 51 09.0 | 008362 | 010780 | 5.63 | 0.804 | 100.24 | 0.68 | 99.33 | 0.53 | G9 V | 3 |
| 01 59 06.63 | +33 12 34.9 | 009269 | 012051 | 7.14 | 0.773 | 40.74 | 0.90 | 40.03 | 0.58 | G9 V | 3 |
| 02 06 30.24 | +24 20 02.4 | 009829 | 012846 | 6.89 | 0.662 | 43.14 | 0.94 | 43.91 | 0.57 | G2 V- | 3 |
| 02 10 25.93 | −50 49 25.4 | 010138 | 013445 | 6.12 | 0.812 | 91.63 | 0.61 | 92.74 | 0.32 | K1 V | 2 |
| 02 17 03.23 | +34 13 27.2 | 010644 | 013974 | 4.84 | 0.607 | 92.20 | 0.84 | 92.73 | 0.39 | G0 V | 1 |
| 02 18 01.44 | +01 45 28.1 | 010723 | 014214 | 5.60 | 0.588 | 40.04 | 0.92 | 41.06 | 0.49 | G0 IV- | 3 |
| 02 18 58.50 | −25 56 44.5 | 010798 | 014412 | 6.33 | 0.724 | 78.88 | 0.72 | 78.93 | 0.35 | G8 V | 2 |
| 02 22 32.55 | −23 48 58.8 | 011072 | 014802 | 5.19 | 0.608 | 45.60 | 0.82 | 45.53 | 0.82 | G0 V | 2 |
| 02 36 04.89 | +06 53 12.7 | 012114 | 016160 | 5.79 | 0.918 | 138.72 | 1.04 | 139.27 | 0.45 | K3 V | 3 |
| 02 36 41.76 | −03 09 22.1 | 012158 | 016287 | 8.10 | 0.944 | 41.09 | 1.25 | 41.44 | 0.97 | K2.5 V | 3 |
| 02 40 12.42 | −09 27 10.3 | 012444 | 016673 | 5.79 | 0.524 | 46.42 | 0.82 | 45.96 | 0.41 | F8 V | 3 |
| 02 41 14.00 | −00 41 44.4 | 012530 | 016765 | 5.72 | 0.511 | 46.24 | 1.31 | 44.27 | 0.84 | F7 V | 3 |
| 02 42 14.92 | +40 11 38.2 | 012623 | 016739 | 4.91 | 0.582 | 40.52 | 1.25 | 41.34 | 0.43 | F9 IV-V | 3 |
| 02 42 33.47 | −50 48 01.1 | 012653 | 017051 | 5.40 | 0.561 | 58.00 | 0.55 | 58.25 | 0.22 | F9 V | 2 |
| 02 44 11.99 | +49 13 42.4 | 012777 | 016895 | 4.10 | 0.514 | 89.03 | 0.79 | 89.87 | 0.22 | F7 V | 1 |
| 02 48 09.14 | +27 04 07.1 | 013081 | 017382 | 7.56 | 0.820 | 44.71 | 1.15 | 40.59 | 1.28 | K0 V | 3 |
| 02 52 32.13 | −12 46 11.0 | 013402 | 017925 | 6.05 | 0.862 | 96.33 | 0.77 | 96.60 | 0.40 | K1.5 V | 2 |
| 02 55 39.06 | +26 52 23.6 | 013642 | 018143 | 7.52 | 0.953 | 43.71 | 1.26 | 42.57 | 0.84 | K2 IV | 3 |
| 03 00 02.81 | +07 44 59.1 | 013976 | 018632 | 7.97 | 0.926 | 42.66 | 1.22 | 41.00 | 1.12 | K2.5 V | 3 |
| 03 02 26.03 | +26 36 33.3 | 014150 | 018803 | 6.62 | 0.696 | 47.25 | 0.89 | 48.45 | 0.47 | G6 V | 3 |
| 03 04 09.64 | +61 42 21.0 | 014286 | 018757 | 6.64 | 0.634 | 43.74 | 0.84 | 41.27 | 0.58 | G1.5 V | 3 |
| 03 09 04.02 | +49 36 47.8 | 014632 | 019373 | 4.05 | 0.595 | 94.93 | 0.67 | 94.87 | 0.23 | F9.5 V | 3 |
| 03 12 04.53 | −28 59 15.4 | 014879 | 020010 | 3.80 | 0.543 | 70.86 | 0.67 | 70.24 | 0.45 | F6 V | 2 |
| 03 12 46.44 | −01 11 46.0 | 014954 | 019994 | 5.07 | 0.575 | 44.69 | 0.75 | 44.29 | 0.28 | F8.5 V | 3 |
| 03 14 47.23 | +08 58 50.9 | 015099 | 020165 | 7.83 | 0.861 | 44.96 | 1.09 | 44.15 | 0.83 | K1 V | 3 |
| 03 15 06.39 | −45 39 53.4 | 015131 | 020407 | 6.75 | 0.586 | 41.05 | 0.59 | 41.34 | 0.40 | G5 V | 2 |
| 03 18 12.82 | −62 30 22.9 | 015371 | 020807 | 5.24 | 0.600 | 82.79 | 0.53 | 83.11 | 0.19 | G0 V | 2 |
| 03 19 01.89 | −02 50 35.5 | 015442 | 020619 | 7.05 | 0.655 | 40.52 | 0.98 | 39.65 | 0.74 | G2 V | 3 |
| 03 19 21.70 | +03 22 12.7 | 015457 | 020630 | 4.84 | 0.681 | 109.18 | 0.78 | 109.41 | 0.27 | G5 Vvar | 1 |
| 03 19 55.65 | −43 04 11.2 | 015510 | 020794 | 4.26 | 0.711 | 165.02 | 0.55 | 165.47 | 0.19 | G8 V | 2 |
| 03 21 54.76 | +52 19 53.4 | 015673 | 232781 | 9.05 | 0.990 | 44.03 | 1.24 | 42.81 | 1.03 | K3.5 V | 3 |
| 03 23 35.26 | −40 04 35.0 | 015799 | 021175 | 6.90 | 0.840 | 58.53 | 1.04 | 57.40 | 0.67 | K1 V | 2 |
| 03 32 55.84 | −09 27 29.7 | 016537 | 022049 | 3.72 | 0.881 | 310.75 | 0.85 | 310.94 | 0.16 | K2 V | 3 |
| 03 36 52.38 | +00 24 06.0 | 016852 | 022484 | 4.29 | 0.575 | 72.89 | 0.78 | 71.62 | 0.54 | F9 V | 1 |
| 03 40 22.06 | −03 13 01.1 | 017147 | 022879 | 6.68 | 0.554 | 41.07 | 0.86 | 39.12 | 0.56 | F9 V | 1 |
| 03 43 55.34 | −19 06 39.2 | 017420 | 023356 | 7.10 | 0.927 | 71.17 | 0.91 | 71.69 | 0.67 | K2.5 V | 2 |
| 03 44 09.17 | −38 16 54.4 | 017439 | 023484 | 6.99 | 0.870 | 61.63 | 0.67 | 62.39 | 0.52 | K2 V | 2 |
| 03 54 28.03 | +16 36 57.8 | 018267 | 024496 | 6.81 | 0.719 | 48.36 | 1.02 | 48.95 | 0.70 | G7 V | 3 |

Table 1—Continued

| R.A. (J2000.0) (1) | Decl. (J2000.0) (2) | HIP Name (3) | HD Name (4) | *Hipparcos* $V$ (5) | $B-V$ (6) | *Hipparcos* $\pi$ (mas) (7) | $\sigma_\pi$ (mas) (8) | van Leeuwen (2007b) $\pi$ (mas) (9) | $\sigma_\pi$ (mas) (10) | Spec Type (11) | Ref (12) |
|---|---|---|---|---|---|---|---|---|---|---|---|
| 03 55 03.84 | +61 10 00.5 | 018324 | 024238 | 7.84 | 0.831 | 46.95 | 0.95 | 47.59 | 0.84 | K2 V | 3 |
| 03 56 11.52 | +59 38 30.8 | 018413 | 024409 | 6.53 | 0.698 | 46.74 | 0.96 | 45.49 | 0.62 | G3 V | 3 |
| 04 02 36.74 | −00 16 08.1 | 018859 | 025457 | 5.38 | 0.516 | 52.00 | 0.75 | 53.10 | 0.32 | F7 V | 3 |
| 04 03 15.00 | +35 16 23.8 | 018915 | 025329 | 8.51 | 0.863 | 54.14 | 1.08 | 54.68 | 0.92 | K3 Vp | 3 |
| 04 05 20.26 | +22 00 32.1 | 019076 | 025680 | 5.90 | 0.620 | 59.79 | 0.84 | 59.04 | 0.33 | G1 V | 3 |
| 04 07 21.54 | −64 13 20.2 | 019233 | 026491 | 6.37 | 0.636 | 43.12 | 0.50 | 42.32 | 0.28 | G1 V | 2 |
| 04 08 36.62 | +38 02 23.0 | 019335 | 025998 | 5.52 | 0.520 | 46.87 | 0.77 | 47.63 | 0.26 | F8 V | 3 |
| 04 09 35.04 | +69 32 29.0 | 019422 | 025665 | 7.70 | 0.952 | 54.17 | 0.79 | 53.33 | 0.71 | K2.5 V | 3 |
| 04 15 16.32 | −07 39 10.3 | 019849 | 026965 | 4.43 | 0.820 | 198.24 | 0.84 | 200.62 | 0.23 | K1 V | 3 |
| 04 15 28.80 | +06 11 12.7 | 019859 | 026923 | 6.32 | 0.570 | 47.20 | 1.08 | 46.88 | 0.47 | G0 IV-V | 3 |
| 04 43 35.44 | +27 41 14.6 | 021988 | 029883 | 8.00 | 0.907 | 44.74 | 0.99 | 45.61 | 0.81 | K5 III | 1 |
| 04 45 38.58 | −50 04 27.2 | 022122 | 030501 | 7.58 | 0.875 | 48.90 | 0.64 | 47.93 | 0.45 | K2 V | 2 |
| 04 47 36.29 | −16 56 04.0 | 022263 | 030495 | 5.49 | 0.632 | 75.10 | 0.80 | 75.32 | 0.36 | G1.5 V | 2 |
| 04 49 52.33 | −35 06 27.5 | 022451 | 030876 | 7.49 | 0.901 | 55.59 | 0.71 | 56.35 | 0.48 | K2 V | 1 |
| 05 02 17.06 | −56 04 49.9 | 023437 | 032778 | 7.02 | 0.636 | 44.94 | 0.58 | 44.48 | 0.36 | G7 V | 2 |
| 05 05 30.66 | −57 28 21.7 | 023693 | 033262 | 4.71 | 0.526 | 85.83 | 0.46 | 85.87 | 0.18 | F9 V | 2 |
| 05 06 42.22 | +14 26 46.4 | 023786 | 032850 | 7.74 | 0.804 | 41.70 | 1.14 | 42.24 | 0.92 | G9 V | 3 |
| 05 07 27.01 | +18 38 42.2 | 023835 | 032923 | 4.91 | 0.657 | 63.02 | 0.93 | 64.79 | 0.33 | G1 V | 3 |
| 05 18 50.47 | −18 07 48.2 | 024786 | 034721 | 5.96 | 0.572 | 40.11 | 0.76 | 39.96 | 0.40 | F9- V | 2 |
| 05 19 08.47 | +40 05 56.6 | 024813 | 034411 | 4.69 | 0.630 | 79.08 | 0.90 | 79.17 | 0.28 | G1 V | 3 |
| 05 22 33.53 | +79 13 52.1 | 025110 | 033564 | 5.08 | 0.506 | 47.66 | 0.52 | 47.88 | 0.21 | F7 V | 3 |
| 05 22 37.49 | +02 36 11.5 | 025119 | 035112 | 7.76 | 0.980 | 50.24 | 1.52 | 49.44 | 1.17 | K2.5 V | 3 |
| 05 24 25.46 | +17 23 00.7 | 025278 | 035296 | 5.00 | 0.544 | 68.19 | 0.94 | 69.51 | 0.38 | F8 V | 3 |
| 05 26 14.74 | −32 30 17.2 | 025421 | 035854 | 7.70 | 0.946 | 55.76 | 0.76 | 56.27 | 0.61 | K3- V | 2 |
| 05 27 39.35 | −60 24 57.6 | 025544 | 036435 | 6.99 | 0.755 | 51.10 | 0.52 | 52.08 | 0.45 | G9 V | 2 |
| 05 28 44.83 | −65 26 54.9 | 025647 | 036705 | 6.88 | 0.830 | 66.92 | 0.54 | 65.93 | 0.57 | K2 V | 3 |
| 05 36 56.85 | −47 57 52.9 | 026373 | 037572 | 7.95 | 0.845 | 41.90 | 1.74 | 39.82 | 1.36 | K1.5 V | 2 |
| 05 37 09.89 | −80 28 08.8 | 026394 | 039091 | 5.65 | 0.600 | 54.92 | 0.45 | 54.60 | 0.21 | G0 V | 3 |
| 05 38 11.86 | +51 26 44.7 | 026505 | 037008 | 7.74 | 0.834 | 48.72 | 1.00 | 49.60 | 0.72 | K1 V | 3 |
| 05 41 20.34 | +53 28 51.8 | 026779 | 037394 | 6.21 | 0.840 | 81.69 | 0.83 | 81.45 | 0.54 | K0 V | 3 |
| 05 46 01.89 | +37 17 04.7 | 027207 | 038230 | 7.34 | 0.833 | 48.60 | 1.03 | 45.76 | 0.76 | K0 V | 3 |
| 05 48 34.94 | −04 05 40.7 | 027435 | 038858 | 5.97 | 0.639 | 64.25 | 1.19 | 65.89 | 0.41 | G2 V | 3 |
| 05 54 04.24 | −60 01 24.5 | 027887 | 040307 | 7.17 | 0.935 | 77.95 | 0.53 | 76.95 | 0.37 | K2.5 V | 2 |
| 05 54 22.98 | +20 16 34.2 | 027913 | 039587 | 4.39 | 0.594 | 115.43 | 1.08 | 115.43 | 0.27 | G0 IV-V | 3 |
| 05 54 30.16 | −19 42 15.7 | 027922 | 039855 | 7.51 | 0.700 | 43.86 | 1.19 | 42.43 | 0.99 | G8 V | 2 |
| 05 58 21.54 | −04 39 02.4 | 028267 | 040397 | 6.99 | 0.720 | 43.10 | 0.93 | 42.46 | 0.63 | G7 V | 3 |
| 06 06 40.48 | +15 32 31.6 | 028954 | 041593 | 6.76 | 0.814 | 64.71 | 0.91 | 65.48 | 0.67 | G9 V | 3 |
| 06 10 14.47 | −74 45 11.0 | 029271 | 043834 | 5.08 | 0.714 | 98.54 | 0.45 | 98.06 | 0.14 | G7 V | 2 |
| 06 12 00.57 | +06 46 59.1 | 029432 | 042618 | 6.85 | 0.642 | 43.26 | 0.87 | 42.55 | 0.55 | G3 V | 3 |
| 06 13 12.50 | +10 37 37.7 | 029525 | 042807 | 6.43 | 0.663 | 55.20 | 0.96 | 55.71 | 0.44 | G5 V | 3 |
| 06 13 45.30 | −23 51 43.0 | 029568 | 043162 | 6.37 | 0.713 | 59.90 | 0.75 | 59.80 | 0.49 | G6.5 V | 2 |
| 06 17 16.14 | +05 06 00.4 | 029860 | 043587 | 5.70 | 0.610 | 51.76 | 0.78 | 51.95 | 0.40 | G0 V | 3 |

Table 1—Continued

| R.A. (J2000.0) (1) | Decl. (J2000.0) (2) | HIP Name (3) | HD Name (4) | *Hipparcos* V (5) | B − V (6) | *Hipparcos* π (mas) (7) | σ_π (mas) (8) | van Leeuwen (2007b) π (mas) (9) | σ_π (mas) (10) | Spec Type (11) | Ref (12) |
|---|---|---|---|---|---|---|---|---|---|---|---|
| 06 22 30.94 | −60 13 07.2 | 030314 | 045270 | 6.53 | 0.614 | 42.56 | 0.49 | 42.05 | 0.27 | G0 Vp | 2 |
| 06 24 43.88 | −28 46 48.4 | 030503 | 045184 | 6.37 | 0.626 | 45.38 | 0.63 | 45.70 | 0.40 | G1.5 V | 2 |
| 06 26 10.25 | +18 45 24.8 | 030630 | 045088 | 6.78 | 0.938 | 68.20 | 1.10 | 67.89 | 1.53 | K3 V | 3 |
| 06 38 00.36 | −61 32 00.2 | 031711 | 048189 | 6.15 | 0.624 | 46.15 | 0.64 | 46.96 | 0.81 | G1 V | 2 |
| 06 46 05.05 | +32 33 20.4 | 032423 | 263175 | 8.80 | 0.964 | 40.02 | 1.22 | 38.11 | 1.01 | K3 V | 3 |
| 06 46 14.15 | +79 33 53.3 | 032439 | 046588 | 5.44 | 0.525 | 56.02 | 0.53 | 55.95 | 0.27 | F8 V | 3 |
| 06 46 44.34 | +43 34 38.7 | 032480 | 048682 | 5.24 | 0.575 | 60.56 | 0.73 | 59.82 | 0.30 | F9 V | 3 |
| 06 55 18.67 | +25 22 32.5 | 033277 | 050692 | 5.74 | 0.573 | 57.89 | 0.90 | 58.00 | 0.41 | G0 V | 3 |
| 06 58 11.75 | +22 28 33.2 | 033537 | 051419 | 6.94 | 0.620 | 41.25 | 0.88 | 40.60 | 0.53 | G5 V | 3 |
| 06 59 59.66 | −61 20 10.3 | 033690 | 053143 | 6.81 | 0.786 | 54.33 | 0.54 | 54.57 | 0.34 | K0 IV-V | 2 |
| 07 01 13.74 | −25 56 55.4 | 033817 | 052698 | 6.71 | 0.882 | 68.42 | 0.72 | 68.27 | 0.61 | K1 V | 2 |
| 07 01 38.59 | +48 22 43.2 | 033852 | 051866 | 7.98 | 0.986 | 48.96 | 0.97 | 49.79 | 0.84 | K3 V | 3 |
| 07 03 30.46 | +29 20 13.5 | 034017 | 052711 | 5.93 | 0.595 | 52.37 | 0.84 | 52.27 | 0.41 | G0 V | 3 |
| 07 03 57.32 | −43 36 28.9 | 034065 | 053705 | 5.56 | 0.624 | 61.54 | 1.05 | 60.55 | 1.04 | G0 V | 2 |
| 07 08 04.24 | +29 50 04.2 | 034414 | 053927 | 8.32 | 0.907 | 44.92 | 1.43 | 44.93 | 0.97 | K2.5 V | 3 |
| 07 09 35.39 | +25 43 43.1 | 034567 | 054371 | 7.09 | 0.700 | 40.68 | 1.02 | 39.73 | 0.54 | G6 V | 3 |
| 07 15 50.14 | +47 14 23.9 | 035136 | 055575 | 5.54 | 0.576 | 59.31 | 0.69 | 59.20 | 0.33 | F9 V | 3 |
| 07 17 29.56 | −46 58 45.3 | 035296 | 057095 | 6.70 | 0.975 | 67.69 | 0.86 | 68.52 | 0.56 | K2.5 V | 2 |
| 07 27 25.47 | −51 24 09.4 | 036210 | 059468 | 6.72 | 0.694 | 44.43 | 0.53 | 44.10 | 0.36 | G6.5 V | 2 |
| 07 29 01.77 | +31 59 37.8 | 036357 | . . . | 7.73 | 0.923 | 56.98 | 1.24 | 56.63 | 0.93 | K2.5 V | 3 |
| 07 30 42.51 | −37 20 21.7 | 036515 | 059967 | 6.66 | 0.641 | 45.93 | 0.58 | 45.84 | 0.37 | G2 V | 2 |
| 07 33 00.58 | +37 01 47.4 | 036704 | 059747 | 7.68 | 0.863 | 50.80 | 1.29 | 50.60 | 0.94 | K1 V | 3 |
| 07 34 26.17 | −06 53 48.0 | 036827 | 060491 | 8.16 | 0.900 | 40.32 | 1.26 | 40.73 | 1.00 | K2.5 V | 3 |
| 07 39 59.33 | −03 35 51.0 | 037349 | 061606 | 7.18 | 0.891 | 70.44 | 0.94 | 70.37 | 0.64 | K3- V | 2 |
| 07 45 35.02 | −34 10 20.5 | 037853 | 063077 | 5.36 | 0.589 | 65.79 | 0.56 | 65.75 | 0.51 | F9 V | 2 |
| 07 49 55.06 | +27 21 47.4 | 038228 | 063433 | 6.90 | 0.682 | 45.84 | 0.89 | 45.45 | 0.53 | G5 V | 3 |
| 07 51 46.30 | −13 53 52.9 | 038382 | 064096 | 5.16 | 0.600 | 59.98 | 0.95 | 60.59 | 0.59 | G0 V | 2 |
| 07 54 34.18 | −01 24 44.1 | 038625 | 064606 | 7.43 | 0.739 | 52.01 | 1.85 | 49.78 | 1.85 | K0 V | 3 |
| 07 54 54.07 | +19 14 10.8 | 038657 | 064468 | 7.76 | 0.950 | 50.05 | 1.05 | 48.33 | 0.86 | K2.5 V | 3 |
| 07 56 17.23 | +80 15 55.9 | 038784 | 062613 | 6.55 | 0.719 | 58.67 | 0.57 | 58.17 | 0.36 | G8 V | 1 |
| 07 57 46.91 | −60 18 11.1 | 038908 | 065907 | 5.59 | 0.573 | 61.76 | 0.51 | 61.71 | 0.21 | F9.5 V | 2 |
| 07 59 33.93 | +20 50 38.0 | 039064 | 065430 | 7.68 | 0.833 | 43.21 | 0.96 | 42.15 | 0.71 | K0 V | 3 |
| 08 00 32.13 | +29 12 44.5 | 039157 | 065583 | 6.97 | 0.716 | 59.52 | 0.77 | 59.64 | 0.56 | K0 V | 3 |
| 08 02 31.19 | −66 01 15.4 | 039342 | 067199 | 7.18 | 0.872 | 57.88 | 0.58 | 57.76 | 0.41 | K2 V | 2 |
| 08 07 45.86 | +21 34 54.5 | 039780 | 067228 | 5.30 | 0.642 | 42.86 | 0.97 | 42.94 | 0.30 | G2 IV | 3 |
| 08 11 38.64 | +32 27 25.7 | 040118 | 068017 | 6.78 | 0.679 | 46.05 | 0.92 | 45.90 | 0.55 | G3 V | 3 |
| 08 12 12.73 | +17 38 52.0 | 040167 | 068257 | 4.67 | 0.531 | 41.10[a] | 0.90[a] | 39.87 | 0.82 | F8 V | 3 |
| 08 18 23.95 | −12 37 55.8 | 040693 | 069830 | 5.95 | 0.754 | 79.48 | 0.77 | 80.04 | 0.35 | G8+ V | 2 |
| 08 19 19.05 | +01 20 19.9 | 040774 | . . . | 8.35 | 0.901 | 42.89 | 1.32 | 43.61 | 1.26 | G5 | 1 |
| 08 27 36.79 | +45 39 10.8 | 041484 | 071148 | 6.32 | 0.624 | 45.89 | 0.84 | 44.94 | 0.46 | G1 V | 3 |
| 08 32 51.50 | −31 30 03.1 | 041926 | 072673 | 6.38 | 0.780 | 82.15 | 0.66 | 81.91 | 0.46 | G9 V | 2 |
| 08 34 31.65 | −00 43 33.8 | 042074 | 072760 | 7.32 | 0.791 | 45.95 | 1.01 | 47.31 | 0.72 | K0- V | 3 |

Table 1—Continued

| R.A. (J2000.0) (1) | Decl. (J2000.0) (2) | HIP Name (3) | HD Name (4) | *Hipparcos* | | *Hipparcos* | | van Leeuwen (2007b) | | Spec Type (11) | Ref (12) |
|---|---|---|---|---|---|---|---|---|---|---|---|
| | | | | V (5) | B − V (6) | π (mas) (7) | σ_π (mas) (8) | π (mas) (9) | σ_π (mas) (10) | | |
| 08 37 50.29 | −06 48 24.8 | 042333 | 073350 | 6.74 | 0.655 | 42.32 | 1.04 | 41.71 | 0.70 | G5 V | 3 |
| 08 39 07.90 | −22 39 42.8 | 042430 | 073752 | 5.05 | 0.720 | 50.20 | 0.98 | 51.55 | 0.63 | G5 IV | 2 |
| 08 39 11.70 | +65 01 15.3 | 042438 | 072905 | 5.63 | 0.618 | 70.07 | 0.71 | 69.66 | 0.37 | G1.5 Vb | 1 |
| 08 39 50.79 | +11 31 21.6 | 042499 | 073667 | 7.61 | 0.832 | 53.98 | 1.04 | 55.13 | 0.71 | K2 V | 3 |
| 08 42 07.52 | −42 55 46.0 | 042697 | 074385 | 8.11 | 0.904 | 44.73 | 0.79 | 43.73 | 0.55 | K2+ V | 2 |
| 08 43 18.03 | −38 52 56.6 | 042808 | 074576 | 6.58 | 0.917 | 89.78 | 0.56 | 89.76 | 0.37 | K2.5 V | 2 |
| 08 52 16.39 | +08 03 46.5 | 043557 | 075767 | 6.57 | 0.640 | 41.42 | 1.19 | 41.64 | 1.03 | G1.5 V | 3 |
| 08 52 35.81 | +28 19 50.9 | 043587 | 075732 | 5.96 | 0.869 | 79.80 | 0.84 | 81.03 | 0.75 | K0 IV-V | 3 |
| 08 54 17.95 | −05 26 04.1 | 043726 | 076151 | 6.01 | 0.661 | 58.50 | 0.88 | 57.52 | 0.39 | G3 V | 2 |
| 08 58 43.93 | −16 07 57.8 | 044075 | 076932 | 5.80 | 0.521 | 46.90 | 0.97 | 47.54 | 0.31 | G2 V | 2 |
| 09 08 51.07 | +33 52 56.0 | 044897 | 078366 | 5.95 | 0.585 | 52.25 | 0.87 | 52.11 | 0.33 | G0 IV-V | 3 |
| 09 12 17.55 | +14 59 45.7 | 045170 | 079096 | 6.49 | 0.731 | 48.83 | 0.92 | 49.11 | 0.54 | G9 V | 3 |
| 09 14 20.54 | +61 25 23.9 | 045333 | 079028 | 5.18 | 0.605 | 51.12 | 0.72 | 51.10 | 0.32 | G0 IV-V | 3 |
| 09 17 53.46 | +28 33 37.9 | 045617 | 079969 | 7.20 | 0.992 | 57.05 | 1.08 | 57.92 | 0.76 | K3 V | 3 |
| 09 22 25.95 | +40 12 03.8 | 045963 | 080715 | 7.69 | 0.987 | 41.19 | 1.08 | 40.10 | 0.64 | K2.5 V | 3 |
| 09 30 28.09 | −32 06 12.2 | 046626 | 082342 | 8.31 | 0.985 | 51.71 | 0.91 | 51.79 | 0.85 | K3.5 V | 2 |
| 09 32 25.57 | −11 11 04.7 | 046816 | 082558 | 7.82 | 0.933 | 54.52 | 0.99 | 53.70 | 0.84 | K0 | 1 |
| 09 32 43.76 | +26 59 18.7 | 046843 | 082443 | 7.05 | 0.779 | 56.35 | 0.89 | 56.20 | 0.60 | G9 V | 3 |
| 09 35 39.50 | +35 48 36.5 | 047080 | 082885 | 5.40 | 0.770 | 89.45 | 0.78 | 87.96 | 0.32 | G8+ V | 3 |
| 09 42 14.42 | −23 54 56.1 | 047592 | 084117 | 4.93 | 0.534 | 67.19 | 0.73 | 66.61 | 0.21 | F8 V | 2 |
| 09 48 35.37 | +46 01 15.6 | 048113 | 084737 | 5.08 | 0.619 | 54.26 | 0.74 | 54.44 | 0.28 | G0 IV-V | 3 |
| 10 01 00.66 | +31 55 25.2 | 049081 | 086728 | 5.37 | 0.676 | 67.14 | 0.83 | 66.46 | 0.32 | G4 V | 3 |
| 10 04 37.66 | −11 43 46.9 | 049366 | 087424 | 8.15 | 0.891 | 43.14 | 1.11 | 41.61 | 0.95 | K2 V | 2 |
| 10 08 43.14 | +34 14 32.1 | 049699 | 087883 | 7.56 | 0.965 | 55.37 | 0.94 | 54.93 | 0.54 | K2.5 V | 3 |
| 10 13 24.73 | −33 01 54.2 | 050075 | 088742 | 6.38 | 0.592 | 43.98 | 0.72 | 43.77 | 0.41 | G0 V | 2 |
| 10 17 14.54 | +23 06 22.4 | 050384 | 089125 | 5.81 | 0.500 | 44.01 | 0.75 | 43.85 | 0.36 | F6 V | 3 |
| 10 18 51.95 | +44 02 54.0 | 050505 | 089269 | 6.66 | 0.653 | 48.45 | 0.85 | 49.41 | 0.50 | G4 V | 3 |
| 10 23 55.27 | −29 38 43.9 | 050921 | 090156 | 6.92 | 0.659 | 45.26 | 0.75 | 44.74 | 0.49 | G5 V | 2 |
| 10 28 03.88 | +48 47 05.6 | 051248 | 090508 | 6.42 | 0.610 | 42.45 | 0.77 | 43.65 | 0.43 | G0 V | 3 |
| 10 30 37.58 | +55 58 49.9 | 051459 | 090839 | 4.82 | 0.541 | 77.82 | 0.65 | 78.25 | 0.28 | F8 V | 3 |
| 10 31 21.82 | −53 42 55.7 | 051523 | 091324 | 4.89 | 0.500 | 45.72 | 0.51 | 45.85 | 0.19 | F9 V | 2 |
| 10 35 11.27 | +84 23 57.6 | 051819 | 090343 | 7.29 | 0.819 | 47.55 | 0.60 | 48.24 | 0.49 | K0 | 1 |
| 10 36 32.38 | −12 13 48.4 | 051933 | 091889 | 5.71 | 0.528 | 40.67 | 0.68 | 39.88 | 0.37 | F8 V | 2 |
| 10 42 13.32 | −13 47 15.8 | 052369 | 092719 | 6.79 | 0.622 | 42.73 | 0.82 | 41.97 | 0.47 | G1.5 V | 2 |
| 10 43 28.27 | −29 03 51.4 | 052462 | 092945 | 7.72 | 0.873 | 46.36 | 0.84 | 46.73 | 0.69 | K1.5 V | 2 |
| 10 56 30.80 | +07 23 18.5 | 053486 | 094765 | 7.37 | 0.920 | 56.98 | 1.03 | 57.79 | 0.87 | K2.5 V | 3 |
| 10 59 27.97 | +40 25 48.9 | 053721 | 095128 | 5.03 | 0.624 | 71.04 | 0.66 | 71.11 | 0.25 | G0 V | 1 |
| 11 04 41.47 | −04 13 15.9 | 054155 | 096064 | 7.64 | 0.770 | 40.57 | 1.40 | 38.06 | 0.99 | G8+ V | 3 |
| 11 08 14.01 | +38 25 35.9 | 054426 | 096612 | 8.35 | 0.942 | 43.91 | 1.03 | 44.29 | 0.84 | K3- V | 3 |
| 11 12 01.19 | −26 08 12.0 | 054704 | 097343 | 7.05 | 0.760 | 46.22 | 0.84 | 45.16 | 0.50 | G8.5 V | 2 |
| 11 12 32.35 | +35 48 50.7 | 054745 | 097334 | 6.41 | 0.600 | 46.04 | 0.90 | 45.61 | 0.44 | G1 V | 3 |
| 11 14 33.16 | +25 42 37.4 | 054906 | 097658 | 7.76 | 0.845 | 46.95 | 0.97 | 47.36 | 0.75 | K1 V | 3 |

Table 1—Continued

| R.A. (J2000.0) (1) | Decl. (J2000.0) (2) | HIP Name (3) | HD Name (4) | *Hipparcos* V (5) | B − V (6) | *Hipparcos* π (mas) (7) | σ_π (mas) (8) | van Leeuwen (2007b) π (mas) (9) | σ_π (mas) (10) | Spec Type (11) | Ref (12) |
|---|---|---|---|---|---|---|---|---|---|---|---|
| 11 18 10.95 | +31 31 45.7 | 055203 | 098230 | 3.79 | 0.606 | 119.70[a] | 0.80[a] | . . . | . . . | G0 V | 1 |
| 11 18 22.01 | −05 04 02.3 | 055210 | 098281 | 7.29 | 0.732 | 45.48 | 1.00 | 46.36 | 0.64 | G8 V | 1 |
| 11 26 45.32 | +03 00 47.2 | 055846 | 099491 | 6.49 | 0.778 | 56.59 | 1.40 | 56.35 | 0.75 | K0 IV | 1 |
| 11 31 44.95 | +14 21 52.2 | 056242 | 100180 | 6.27 | 0.570 | 43.42 | 1.10 | 42.87 | 1.22 | F9.5 V | 3 |
| 11 34 29.49 | −32 49 52.8 | 056452 | 100623 | 5.96 | 0.811 | 104.84 | 0.81 | 104.61 | 0.37 | K0- V | 2 |
| 11 38 44.90 | +45 06 30.3 | 056809 | 101177 | 6.29 | 0.566 | 42.94 | 0.95 | 43.01 | 0.73 | F9.5 V | 3 |
| 11 38 59.72 | +42 19 43.7 | 056829 | 101206 | 8.22 | 0.980 | 50.61 | 1.15 | 50.19 | 1.03 | K5 V | 1 |
| 11 41 03.02 | +34 12 05.9 | 056997 | 101501 | 5.31 | 0.723 | 104.81 | 0.72 | 104.04 | 0.26 | G8 V | 3 |
| 11 46 31.07 | −40 30 01.3 | 057443 | 102365 | 4.89 | 0.664 | 108.23 | 0.70 | 108.45 | 0.22 | G2 V | 2 |
| 11 47 15.81 | −30 17 11.4 | 057507 | 102438 | 6.48 | 0.681 | 56.26 | 0.77 | 57.23 | 0.41 | G6 V | 2 |
| 11 50 41.72 | +01 45 53.0 | 057757 | 102870 | 3.59 | 0.518 | 91.74 | 0.77 | 91.50 | 0.22 | F8 V | 1 |
| 11 52 58.77 | +37 43 07.2 | 057939 | 103095 | 6.42 | 0.754 | 109.21 | 0.78 | 109.99 | 0.41 | K1 V | 3 |
| 11 59 10.01 | −20 21 13.6 | 058451 | 104067 | 7.92 | 0.974 | 48.04 | 1.03 | 47.47 | 0.90 | K3- V | 2 |
| 12 00 44.45 | −10 26 45.6 | 058576 | 104304 | 5.54 | 0.760 | 77.48 | 0.80 | 78.35 | 0.31 | G8 IV-V | 3 |
| 12 09 37.26 | +40 15 07.4 | 059280 | 105631 | 7.46 | 0.794 | 41.07 | 0.98 | 40.77 | 0.66 | G9 V | 3 |
| 12 30 50.14 | +53 04 35.8 | 061053 | 108954 | 6.20 | 0.568 | 45.58 | 0.62 | 45.92 | 0.35 | F9 V | 3 |
| 12 33 31.38 | −68 45 20.9 | 061291 | 109200 | 7.13 | 0.836 | 61.83 | 0.63 | 61.82 | 0.48 | K1 V | 2 |
| 12 33 44.54 | +41 21 26.9 | 061317 | 109358 | 4.24 | 0.588 | 119.46 | 0.83 | 118.49 | 0.20 | G0 V | 3 |
| 12 41 44.52 | +55 43 28.8 | 061946 | 110463 | 8.27 | 0.955 | 43.06 | 0.82 | 42.78 | 0.81 | K3 V | 1 |
| 12 44 14.55 | +51 45 33.5 | 062145 | 110833 | 7.01 | 0.936 | 66.40 | 0.78 | 67.20 | 0.66 | K3 V | 1 |
| 12 44 59.41 | +39 16 44.1 | 062207 | 110897 | 5.95 | 0.557 | 57.57 | 0.64 | 57.55 | 0.32 | F9 V | 3 |
| 12 45 14.41 | −57 21 28.8 | 062229 | 110810 | 7.82 | 0.937 | 49.71 | 0.95 | 48.79 | 0.88 | K2+ V | 2 |
| 12 48 32.31 | −15 43 10.1 | 062505 | 111312 | 7.93 | 0.946 | 47.19 | 1.93 | 41.96 | 3.00 | K2.5 V | 2 |
| 12 48 47.05 | +24 50 24.8 | 062523 | 111395 | 6.29 | 0.703 | 58.23 | 0.99 | 59.06 | 0.45 | G7 V | 3 |
| 12 59 01.56 | −09 50 02.7 | 063366 | 112758 | 7.54 | 0.769 | 47.60 | 0.86 | 47.87 | 0.90 | K2 V | 3 |
| 12 59 32.78 | +41 59 12.4 | 063406 | 112914 | 8.60 | 0.940 | 41.36 | 1.48 | 39.71 | 1.00 | K3- V | 1 |
| 13 03 49.66 | −05 09 42.5 | 063742 | 113449 | 7.69 | 0.847 | 45.20 | 1.27 | 46.10 | 0.81 | K1 V | 3 |
| 13 11 52.39 | +27 52 41.5 | 064394 | 114710 | 4.23 | 0.572 | 109.23 | 0.72 | 109.54 | 0.17 | G0 V | 1 |
| 13 12 03.18 | −37 48 10.9 | 064408 | 114613 | 4.85 | 0.693 | 48.83 | 0.79 | 48.38 | 0.29 | G4 IV | 2 |
| 13 12 43.79 | −02 15 54.1 | 064457 | 114783 | 7.56 | 0.930 | 48.95 | 1.06 | 48.78 | 0.59 | K1 V | 3 |
| 13 13 52.23 | −45 11 08.9 | 064550 | 114853 | 6.93 | 0.643 | 40.87 | 0.84 | 40.95 | 0.56 | G1.5 V | 2 |
| 13 15 26.45 | −87 33 38.5 | 064690 | 113283 | 7.11 | 0.710 | 40.52 | 0.56 | 40.70 | 0.38 | G5 V | 2 |
| 13 16 46.52 | +09 25 27.0 | 064792 | 115383 | 5.19 | 0.585 | 55.71 | 0.85 | 56.95 | 0.26 | G0 Vs | 1 |
| 13 16 51.05 | +17 01 01.9 | 064797 | 115404 | 6.49 | 0.926 | 89.07 | 0.99 | 90.32 | 0.74 | K2.5 V | 3 |
| 13 18 24.31 | −18 18 40.3 | 064924 | 115617 | 4.74 | 0.709 | 117.30 | 0.71 | 116.89 | 0.22 | G7 V | 2 |
| 13 23 39.15 | +02 43 24.0 | 065352 | 116442 | 7.06 | 0.780 | 62.41 | 1.41 | 64.73 | 1.33 | G9 V | 3 |
| 13 25 45.53 | +56 58 13.8 | 065515 | 116956 | 7.29 | 0.804 | 45.76 | 0.72 | 46.31 | 0.51 | G9 V | 3 |
| 13 25 59.86 | +63 15 40.6 | 065530 | 117043 | 6.50 | 0.739 | 46.86 | 0.55 | 47.24 | 0.31 | G6 V | 1 |
| 13 28 25.81 | +13 46 43.6 | 065721 | 117176 | 4.97 | 0.714 | 55.22 | 0.73 | 55.60 | 0.24 | G5 V | 1 |
| 13 41 04.17 | −34 27 51.0 | 066765 | 118972 | 6.92 | 0.855 | 64.08 | 0.81 | 63.88 | 0.49 | K0 V | 2 |
| 13 41 13.40 | +56 43 37.8 | 066781 | 119332 | 7.77 | 0.830 | 42.12 | 0.76 | 40.59 | 0.53 | K0 IV-V | 1 |
| 13 47 15.74 | +17 27 24.9 | 067275 | 120136 | 4.50 | 0.508 | 64.12 | 0.70 | 64.03 | 0.19 | F7 V | 1 |

Table 1—Continued

| R.A. (J2000.0) (1) | Decl. (J2000.0) (2) | HIP Name (3) | HD Name (4) | *Hipparcos* | | *Hipparcos* | | van Leeuwen (2007b) | | Spec Type (11) | Ref (12) |
|---|---|---|---|---|---|---|---|---|---|---|---|
| | | | | $V$ (5) | $B-V$ (6) | $\pi$ (mas) (7) | $\sigma_\pi$ (mas) (8) | $\pi$ (mas) (9) | $\sigma_\pi$ (mas) (10) | | |
| 13 51 20.33 | −24 23 25.3 | 067620 | 120690 | 6.43 | 0.703 | 50.20 | 0.85 | 51.35 | 0.45 | G5+ V | 2 |
| 13 51 40.40 | −57 26 08.4 | 067655 | 120559 | 7.97 | 0.663 | 40.02 | 1.00 | 39.42 | 0.97 | G7 V | 2 |
| 13 52 35.87 | −50 55 18.3 | 067742 | 120780 | 7.37 | 0.891 | 60.86 | 0.95 | 58.55 | 0.68 | K2 V | 2 |
| 13 54 41.08 | +18 23 51.8 | 067927 | 121370 | 2.68 | 0.580 | 88.17 | 0.75 | 87.75 | 1.24 | G0 IV | 1 |
| 13 55 49.99 | +14 03 23.4 | 068030 | 121560 | 6.16 | 0.518 | 41.28 | 0.79 | 40.22 | 0.37 | F6 V | 1 |
| 14 03 32.35 | +10 47 12.4 | 068682 | 122742 | 6.27 | 0.733 | 60.24 | 0.78 | 58.88 | 0.62 | G6 V | 3 |
| 14 11 46.17 | −12 36 42.4 | 069357 | 124106 | 7.93 | 0.865 | 43.35 | 1.40 | 42.76 | 1.22 | K1 V | 2 |
| 14 12 45.24 | −03 19 12.3 | 069414 | 124292 | 7.05 | 0.733 | 44.89 | 1.01 | 45.35 | 0.54 | G8+ V | 3 |
| 14 15 38.68 | −45 00 02.7 | 069671 | 124580 | 6.31 | 0.596 | 47.51 | 0.78 | 47.13 | 0.42 | G0 V | 2 |
| 14 16 00.87 | −06 00 02.0 | 069701 | 124850 | 4.07 | 0.511 | 46.74 | 0.87 | 44.97 | 0.19 | F7 V | 1 |
| 14 19 00.90 | −25 48 55.5 | 069965 | 125276 | 5.87 | 0.518 | 56.23 | 0.94 | 55.45 | 0.82 | F9 V | 2 |
| 14 19 34.86 | −05 09 04.3 | 070016 | 125455 | 7.58 | 0.867 | 48.12 | 1.11 | 47.89 | 0.81 | K1 V | 1 |
| 14 23 15.28 | +01 14 29.6 | 070319 | 126053 | 6.25 | 0.639 | 56.82 | 1.04 | 58.17 | 0.53 | G1.5 V | 3 |
| 14 29 22.30 | +80 48 35.5 | 070857 | 128642 | 6.88 | 0.774 | 51.04 | 0.58 | 50.27 | 0.48 | G5 | 1 |
| 14 29 36.81 | +41 47 45.3 | 070873 | 127334 | 6.36 | 0.702 | 42.43 | 0.59 | 42.12 | 0.38 | G5 V | 3 |
| 14 33 28.87 | +52 54 31.6 | 071181 | 128165 | 7.24 | 0.997 | 74.50 | 0.69 | 75.65 | 0.42 | K3 V | 1 |
| 14 36 00.56 | +09 44 47.5 | 071395 | 128311 | 7.48 | 0.973 | 60.35 | 0.99 | 60.60 | 0.83 | K3- V | 3 |
| 14 39 36.50 | −60 50 02.3 | 071683 | 128620 | −0.01 | 0.710 | 742.12 | 1.40 | 754.81 | 4.11 | G2 V | 2 |
| 14 40 31.11 | −16 12 33.4 | 071743 | 128987 | 7.24 | 0.710 | 42.43 | 0.97 | 42.23 | 0.54 | G8 V | 2 |
| 14 41 52.46 | −75 08 22.1 | 071855 | 128400 | 6.73 | 0.707 | 49.15 | 0.64 | 50.01 | 0.43 | G5 V | 2 |
| 14 45 24.18 | +13 50 46.7 | 072146 | 130004 | 7.87 | 0.931 | 51.20 | 0.98 | 52.92 | 0.82 | K2.5 V | 3 |
| 14 47 16.10 | +02 42 11.6 | 072312 | 130307 | 7.76 | 0.893 | 50.84 | 1.04 | 51.62 | 0.79 | K2.5 V | 3 |
| 14 49 23.72 | −67 14 09.5 | 072493 | 130042 | 7.26 | 0.836 | 41.69 | 1.24 | 40.11 | 0.89 | K1 V | 2 |
| 14 50 15.81 | +23 54 42.6 | 072567 | 130948 | 5.86 | 0.576 | 55.73 | 0.80 | 55.03 | 0.34 | G2 V | 1 |
| 14 51 23.38 | +19 06 01.7 | 072659 | 131156 | 4.54 | 0.720 | 149.26 | 0.76 | 148.98 | 0.48 | G7 V | 3 |
| 14 53 23.77 | +19 09 10.1 | 072848 | 131511 | 6.00 | 0.841 | 86.69 | 0.81 | 86.88 | 0.46 | K0 V | 3 |
| 14 53 41.57 | +23 20 42.6 | 072875 | 131582 | 8.65 | 0.934 | 43.66 | 1.20 | 42.47 | 1.12 | K3 V | 1 |
| 14 55 11.04 | +53 40 49.2 | 073005 | 132142 | 7.77 | 0.785 | 41.83 | 0.63 | 42.76 | 0.45 | K1 V | 1 |
| 14 56 23.04 | +49 37 42.4 | 073100 | 132254 | 5.63 | 0.533 | 40.25 | 0.54 | 39.83 | 0.26 | F8- V | 3 |
| 14 58 08.80 | −48 51 46.8 | 073241 | 131923 | 6.34 | 0.708 | 40.79 | 0.86 | 41.93 | 0.83 | G4 V | 2 |
| 15 03 47.30 | +47 39 14.6 | 073695 | 133640 | 4.83 | 0.647 | 78.39 | 1.03 | 79.95 | 1.56 | G2 V | 1 |
| 15 10 44.74 | −61 25 20.3 | 074273 | 134060 | 6.29 | 0.623 | 41.41 | 0.77 | 41.32 | 0.45 | G0 V | 2 |
| 15 13 50.89 | −01 21 05.0 | 074537 | 135204 | 6.58 | 0.763 | 57.80 | 0.85 | 56.59 | 0.49 | G9 V | 3 |
| 15 15 59.17 | +00 47 46.9 | 074702 | 135599 | 6.92 | 0.830 | 64.19 | 0.97 | 63.11 | 0.70 | K0 V | 3 |
| 15 19 18.80 | +01 45 55.5 | 074975 | 136202 | 5.04 | 0.540 | 40.46 | 0.81 | 39.40 | 0.29 | F8 III-IV | 1 |
| 15 21 48.15 | −48 19 03.5 | 075181 | 136352 | 5.65 | 0.639 | 68.70 | 0.79 | 67.51 | 0.39 | G2- V | 2 |
| 15 22 36.69 | −10 39 40.0 | 075253 | 136713 | 7.97 | 0.970 | 45.83 | 1.41 | 45.22 | 1.12 | K3 IV-V | 3 |
| 15 22 46.83 | +18 55 08.3 | 075277 | 136923 | 7.16 | 0.804 | 49.67 | 0.92 | 51.01 | 0.64 | G9 V | 3 |
| 15 23 12.31 | +30 17 16.1 | 075312 | 137107 | 4.99 | 0.577 | 53.70 | 1.24 | 55.98 | 0.78 | G2 V | 1 |
| 15 28 09.61 | −09 20 53.1 | 075718 | 137763 | 6.89 | 0.788 | 50.34 | 1.11 | 48.58 | 1.33 | G9 V | 3 |
| 15 29 11.18 | +80 26 55.0 | 075809 | 139777 | 6.57 | 0.665 | 45.32 | 0.57 | 45.77 | 0.37 | G1.5 V(n) | 3 |
| 15 36 02.22 | +39 48 08.9 | 076382 | 139341 | 6.78 | 0.906 | 45.85 | 0.79 | 44.83 | 0.60 | K1 V | 3 |

Table 1—Continued

| R.A.<br>(J2000.0)<br>(1) | Decl.<br>(J2000.0)<br>(2) | HIP<br>Name<br>(3) | HD<br>Name<br>(4) | *Hipparcos* | | *Hipparcos* | | van Leeuwen (2007b) | | Spec<br>Type<br>(11) | Ref<br>(12) |
|---|---|---|---|---|---|---|---|---|---|---|---|
| | | | | $V$<br>(5) | $B-V$<br>(6) | $\pi$<br>(mas)<br>(7) | $\sigma_\pi$<br>(mas)<br>(8) | $\pi$<br>(mas)<br>(9) | $\sigma_\pi$<br>(mas)<br>(10) | | |
| 15 44 01.82 | +02 30 54.6 | 077052 | 140538 | 5.86 | 0.684 | 68.16 | 0.87 | 68.22 | 0.66 | G5 V | 1 |
| 15 46 26.61 | +07 21 11.1 | 077257 | 141004 | 4.42 | 0.604 | 85.08 | 0.80 | 82.48 | 0.32 | G0 IV-V | 3 |
| 15 47 29.10 | −37 54 58.7 | 077358 | 140901 | 6.01 | 0.715 | 65.60 | 0.77 | 65.13 | 0.40 | G7 IV-V | 2 |
| 15 48 09.46 | +01 34 18.3 | 077408 | 141272 | 7.44 | 0.801 | 46.84 | 1.05 | 46.97 | 0.80 | G9 V | 3 |
| 15 52 40.54 | +42 27 05.5 | 077760 | 142373 | 4.60 | 0.563 | 63.08 | 0.54 | 62.92 | 0.21 | G0 V | 3 |
| 15 53 12.10 | +13 11 47.8 | 077801 | 142267 | 6.07 | 0.598 | 57.27 | 0.88 | 57.64 | 0.54 | G0 IV | 1 |
| 16 01 02.66 | +33 18 12.6 | 078459 | 143761 | 5.39 | 0.612 | 57.38 | 0.71 | 58.02 | 0.28 | G0 V | 3 |
| 16 01 53.35 | +58 33 54.9 | 078527 | 144284 | 4.01 | 0.528 | 47.79 | 0.54 | 47.54 | 0.12 | F8 IV-V | 1 |
| 16 04 03.71 | +25 15 17.4 | 078709 | 144287 | 7.10 | 0.771 | 46.56 | 0.89 | 45.01 | 0.79 | G8+ V | 3 |
| 16 04 56.79 | +39 09 23.4 | 078775 | 144579 | 6.66 | 0.734 | 69.61 | 0.57 | 68.87 | 0.33 | K0 V | 3 |
| 16 06 29.60 | +38 37 56.1 | 078913 | 144872 | 8.58 | 0.963 | 42.57 | 0.86 | 42.55 | 0.77 | K3 V | 3 |
| 16 09 42.79 | −56 26 42.5 | 079190 | 144628 | 7.11 | 0.856 | 69.66 | 0.90 | 68.17 | 0.64 | K1 V | 2 |
| 16 10 24.31 | +43 49 03.5 | 079248 | 145675 | 6.61 | 0.877 | 55.11 | 0.59 | 56.91 | 0.34 | K0 IV-V | 3 |
| 16 13 18.45 | +13 31 36.9 | 079492 | 145958 | 6.68 | 0.764 | 41.05 | 1.58 | 42.40 | 1.12 | G9 V | 3 |
| 16 13 48.56 | −57 34 13.8 | 079537 | 145417 | 7.53 | 0.815 | 72.75 | 0.82 | 72.01 | 0.68 | K3 V | 2 |
| 16 14 11.93 | −31 39 49.1 | 079578 | 145825 | 6.55 | 0.646 | 45.73 | 0.95 | 46.40 | 0.62 | G2 V | 2 |
| 16 14 40.85 | +33 51 31.0 | 079607 | 146361 | 5.23 | 0.599 | 46.11 | 0.98 | 47.44 | 1.22 | G1 IV-V | 3 |
| 16 15 37.27 | −08 22 10.0 | 079672 | 146233 | 5.49 | 0.652 | 71.30 | 0.89 | 71.94 | 0.37 | G2 V | 3 |
| 16 24 01.29 | −39 11 34.7 | 080337 | 147513 | 5.37 | 0.625 | 77.69 | 0.86 | 78.26 | 0.37 | G1 V | 2 |
| 16 24 19.81 | −13 38 30.0 | 080366 | 147776 | 8.40 | 0.950 | 46.44 | 1.20 | 46.46 | 1.06 | K3- V | 2 |
| 16 28 28.14 | −70 05 03.8 | 080686 | 147584 | 4.90 | 0.555 | 82.61 | 0.57 | 82.53 | 0.52 | F9 V | 2 |
| 16 28 52.67 | +18 24 50.6 | 080725 | 148653 | 6.98 | 0.848 | 51.20 | 1.49 | 50.87 | 0.80 | K2 V | 3 |
| 16 31 30.03 | −39 00 44.2 | 080925 | 148704 | 7.24 | 0.858 | 40.60 | 1.75 | 40.77 | 2.01 | K1 V | 2 |
| 16 36 21.45 | −02 19 28.5 | 081300 | 149661 | 5.77 | 0.827 | 102.27 | 0.85 | 102.55 | 0.40 | K0 V | 3 |
| 16 37 08.43 | +00 15 15.6 | 081375 | 149806 | 7.09 | 0.828 | 49.63 | 0.92 | 49.18 | 0.62 | K0 V | 3 |
| 16 39 04.14 | −58 15 29.5 | 081520 | 149612 | 7.01 | 0.616 | 46.13 | 0.91 | 44.54 | 0.54 | G5 V | 2 |
| 16 42 38.58 | +68 06 07.8 | 081813 | 151541 | 7.56 | 0.769 | 41.15 | 0.57 | 39.97 | 0.45 | K1 V | 1 |
| 16 52 58.80 | −00 01 35.1 | 082588 | 152391 | 6.65 | 0.749 | 59.04 | 0.87 | 57.97 | 0.66 | G8+ V | 3 |
| 16 57 53.18 | +47 22 00.1 | 083020 | 153557 | 7.76 | 0.980 | 55.71 | 1.21 | 54.63 | 0.61 | K3 V | 3 |
| 17 02 36.40 | +47 04 54.8 | 083389 | 154345 | 6.76 | 0.728 | 55.37 | 0.55 | 53.80 | 0.32 | G8 V | 1 |
| 17 04 27.84 | −28 34 57.6 | 083541 | 154088 | 6.59 | 0.814 | 55.31 | 0.89 | 56.06 | 0.50 | K0 IV-V | 2 |
| 17 05 16.82 | +00 42 09.2 | 083601 | 154417 | 6.00 | 0.578 | 49.06 | 0.89 | 48.39 | 0.40 | F9 V | 3 |
| 17 10 10.35 | −60 43 43.6 | 083990 | 154577 | 7.38 | 0.889 | 73.07 | 0.91 | 73.41 | 0.70 | K2.5 V | 2 |
| 17 12 37.62 | +18 21 04.3 | 084195 | 155712 | 7.95 | 0.941 | 48.69 | 1.03 | 47.70 | 0.93 | K2.5 V | 3 |
| 17 15 20.98 | −26 36 10.2 | 084405 | 155885 | 4.33 | 0.855 | 167.08 | 1.07 | 168.54 | 0.54 | K1.5 V | 2 |
| 17 19 03.83 | −46 38 10.4 | 084720 | 156274 | 5.47 | 0.764 | 113.81 | 1.36 | 113.61 | 0.69 | M0 V | 1 |
| 17 20 39.57 | +32 28 03.9 | 084862 | 157214 | 5.38 | 0.619 | 69.48 | 0.56 | 69.80 | 0.25 | G0 V | 1 |
| 17 22 51.29 | −02 23 17.4 | 085042 | 157347 | 6.28 | 0.680 | 51.39 | 0.85 | 51.22 | 0.40 | G3 V | 3 |
| 17 25 00.10 | +67 18 24.1 | 085235 | 158633 | 6.44 | 0.759 | 78.14 | 0.51 | 78.11 | 0.30 | K0 V | 1 |
| 17 30 16.43 | +47 24 07.9 | 085653 | 159062 | 7.22 | 0.737 | 44.77 | 0.59 | 44.91 | 0.50 | G9 V | 3 |
| 17 30 23.80 | −01 03 46.5 | 085667 | 158614 | 5.31 | 0.715 | 60.80 | 1.42 | 61.19 | 0.68 | G8 IV-V | 1 |
| 17 32 00.99 | +34 16 16.1 | 085810 | 159222 | 6.52 | 0.639 | 42.20 | 0.56 | 41.81 | 0.35 | G1 V | 3 |

Table 1—Continued

| R.A. (J2000.0) (1) | Decl. (J2000.0) (2) | HIP Name (3) | HD Name (4) | *Hipparcos* $V$ (5) | $B-V$ (6) | *Hipparcos* $\pi$ (mas) (7) | $\sigma_\pi$ (mas) (8) | van Leeuwen (2007b) $\pi$ (mas) (9) | $\sigma_\pi$ (mas) (10) | Spec Type (11) | Ref (12) |
|---|---|---|---|---|---|---|---|---|---|---|---|
| 17 34 59.59 | +61 52 28.4 | 086036 | 160269 | 5.23 | 0.602 | 70.98 | 0.55 | 70.47 | 0.37 | G0 V | 1 |
| 17 39 16.92 | +03 33 18.9 | 086400 | 160346 | 6.53 | 0.959 | 93.36 | 1.25 | 90.91 | 0.67 | K2.5 V | 3 |
| 17 41 58.10 | +72 09 24.9 | 086620 | 162004 | 5.81 | 0.530 | 44.80 | 1.94 | 43.36 | 0.51 | G0 V | 1 |
| 17 43 15.64 | +21 36 33.1 | 086722 | 161198 | 7.51 | 0.752 | 42.45 | 0.98 | 44.15 | 0.89 | G9 V | 3 |
| 17 44 08.70 | −51 50 02.6 | 086796 | 160691 | 5.12 | 0.694 | 65.46 | 0.80 | 64.47 | 0.31 | G3 IV-V | 2 |
| 17 46 27.53 | +27 43 14.4 | 086974 | 161797 | 3.42 | 0.750 | 119.05 | 0.62 | 120.33 | 0.16 | G5 IV | 1 |
| 17 53 29.94 | +21 19 31.1 | 087579 | ... | 8.50 | 0.940 | 40.22 | 1.04 | 41.06 | 1.04 | K2.5 V | 3 |
| 18 02 30.86 | +26 18 46.8 | 088348 | 164922 | 7.01 | 0.799 | 45.61 | 0.71 | 45.21 | 0.54 | G9 V | 3 |
| 18 05 27.29 | +02 30 00.4 | 088601 | 165341 | 4.03 | 0.860 | 196.62 | 1.38 | 196.72 | 0.83 | K0- V | 3 |
| 18 05 37.46 | +04 39 25.8 | 088622 | 165401 | 6.80 | 0.610 | 41.00 | 0.88 | 41.82 | 0.59 | G0 V | 3 |
| 18 06 23.72 | −36 01 11.2 | 088694 | 165185 | 5.94 | 0.615 | 57.58 | 0.77 | 56.97 | 0.48 | G0 V | 2 |
| 18 07 01.54 | +30 33 43.7 | 088745 | 165908 | 5.05 | 0.528 | 63.88 | 0.55 | 63.93 | 0.34 | F7 V | 1 |
| 18 09 37.42 | +38 27 28.0 | 088972 | 166620 | 6.38 | 0.876 | 90.11 | 0.54 | 90.71 | 0.30 | K2 V | 3 |
| 18 10 26.16 | −62 00 07.9 | 089042 | 165499 | 5.47 | 0.592 | 56.32 | 0.68 | 56.78 | 0.52 | G0 V | 2 |
| 18 15 32.46 | +45 12 33.5 | 089474 | 168009 | 6.30 | 0.641 | 44.08 | 0.51 | 43.82 | 0.29 | G1 V | 3 |
| 18 19 40.13 | −63 53 11.6 | 089805 | 167425 | 6.17 | 0.584 | 43.64 | 0.72 | 43.39 | 0.39 | F9.5 V | 2 |
| 18 31 18.96 | −18 54 31.7 | 090790 | 170657 | 6.81 | 0.861 | 75.71 | 0.89 | 75.46 | 0.70 | K2 V | 2 |
| 18 38 53.40 | −21 03 06.7 | 091438 | 172051 | 5.85 | 0.673 | 77.02 | 0.85 | 76.43 | 0.47 | G6 V | 2 |
| 18 40 54.88 | +31 31 59.1 | 091605 | ... | 8.54 | 0.865 | 41.88 | 1.59 | 42.48 | 1.11 | K2.5 V | 3 |
| 18 55 18.80 | −37 29 54.1 | 092858 | 175073 | 7.98 | 0.857 | 41.84 | 1.19 | 41.31 | 0.98 | K1 V | 2 |
| 18 55 53.22 | +23 33 23.9 | 092919 | 175742 | 8.16 | 0.910 | 46.64 | 1.03 | 46.74 | 0.85 | K0 V | 1 |
| 18 57 01.61 | +32 54 04.6 | 093017 | 176051 | 5.20 | 0.594 | 66.76 | 0.54 | 67.24 | 0.37 | G0 V | 1 |
| 18 58 51.00 | +30 10 50.3 | 093185 | 176377 | 6.80 | 0.606 | 42.68 | 0.64 | 41.94 | 0.47 | G1 V | 3 |
| 19 06 25.11 | −37 03 48.4 | 093825 | 177474 | 4.23 | 0.523 | 55.89 | 1.94 | 57.79 | 0.75 | F8 V | 2 |
| 19 06 52.46 | −37 48 38.4 | 093858 | 177565 | 6.15 | 0.705 | 58.24 | 0.91 | 58.98 | 0.47 | G6 V | 2 |
| 19 07 57.32 | +16 51 12.2 | 093966 | 178428 | 6.08 | 0.705 | 47.72 | 0.77 | 46.66 | 0.48 | G5 IV-V | 3 |
| 19 12 05.03 | +49 51 20.7 | 094336 | 179957 | 5.85 | 0.666 | 40.16 | 0.83 | 40.90 | 0.58 | G3 V | 3 |
| 19 12 11.36 | +57 40 19.1 | 094346 | 180161 | 7.04 | 0.804 | 50.00 | 0.54 | 49.96 | 0.32 | G8 V | 1 |
| 19 21 29.76 | −34 59 00.6 | 095149 | 181321 | 6.48 | 0.628 | 47.95 | 1.28 | 53.10 | 1.41 | G1 V | 2 |
| 19 23 34.01 | +33 13 19.1 | 095319 | 182488 | 6.37 | 0.804 | 64.54 | 0.60 | 63.45 | 0.35 | G9+ V | 3 |
| 19 24 58.20 | +11 56 39.9 | 095447 | 182572 | 5.17 | 0.761 | 66.01 | 0.77 | 65.89 | 0.26 | G8 IVvar | 1 |
| 19 31 07.97 | +58 35 09.6 | 095995 | 184467 | 6.60 | 0.859 | 59.84 | 0.64 | 58.96 | 0.65 | K2 V | 3 |
| 19 32 06.70 | −11 16 29.8 | 096085 | 183870 | 7.53 | 0.922 | 55.50 | 0.90 | 56.73 | 0.72 | K2.5 V | 2 |
| 19 32 21.59 | +69 39 40.2 | 096100 | 185144 | 4.67 | 0.786 | 173.41 | 0.46 | 173.77 | 0.18 | G9 V | 3 |
| 19 33 25.55 | +21 50 25.2 | 096183 | 184385 | 6.89 | 0.745 | 49.61 | 0.94 | 48.64 | 0.63 | G8 V | 3 |
| 19 35 55.61 | +56 59 02.0 | 096395 | 185414 | 6.73 | 0.636 | 41.24 | 0.49 | 41.48 | 0.30 | G0 | 1 |
| 19 41 48.95 | +50 31 30.2 | 096895 | 186408 | 5.99 | 0.643 | 46.25 | 0.50 | 47.44 | 0.27 | G1.5 V | 3 |
| 19 45 33.53 | +33 36 07.2 | 097222 | 186858 | 7.68 | 1.000 | 49.09 | 1.43 | 47.34 | 0.82 | K3+ V | 3 |
| 19 51 01.64 | +10 24 56.6 | 097675 | 187691 | 5.12 | 0.563 | 51.57 | 0.77 | 52.11 | 0.29 | F8 V | 1 |
| 19 59 47.34 | −09 57 29.7 | 098416 | 189340 | 5.87 | 0.598 | 40.75 | 1.35 | 45.04 | 0.99 | F9 V | 3 |
| 20 00 43.71 | +22 42 39.1 | 098505 | 189733 | 7.67 | 0.932 | 51.94 | 0.87 | 51.41 | 0.69 | K2 V | 3 |
| 20 02 34.16 | +15 35 31.5 | 098677 | 190067 | 7.15 | 0.714 | 51.71 | 0.83 | 52.71 | 0.65 | K0 V | 3 |

Table 1—Continued

| R.A. (J2000.0) (1) | Decl. (J2000.0) (2) | HIP Name (3) | HD Name (4) | *Hipparcos* | | *Hipparcos* | | van Leeuwen (2007b) | | Spec Type (11) | Ref (12) |
|---|---|---|---|---|---|---|---|---|---|---|---|
| | | | | $V$ (5) | $B-V$ (6) | $\pi$ (mas) (7) | $\sigma_\pi$ (mas) (8) | $\pi$ (mas) (9) | $\sigma_\pi$ (mas) (10) | | |
| 20 03 37.41 | +29 53 48.5 | 098767 | 190360 | 5.73 | 0.749 | 62.92 | 0.62 | 63.06 | 0.34 | G7 IV-V | 3 |
| 20 03 52.13 | +23 20 26.5 | 098792 | 190404 | 7.28 | 0.815 | 64.17 | 0.85 | 63.43 | 0.57 | K1 V | 3 |
| 20 04 06.22 | +17 04 12.6 | 098819 | 190406 | 5.80 | 0.600 | 56.60 | 0.76 | 56.28 | 0.35 | G0 V | 3 |
| 20 04 10.05 | +25 47 24.8 | 098828 | 190470 | 7.82 | 0.924 | 46.28 | 0.91 | 45.56 | 0.77 | K2.5 V | 3 |
| 20 05 09.78 | +38 28 42.4 | 098921 | 190771 | 6.18 | 0.654 | 52.99 | 0.55 | 53.22 | 0.36 | G2 V | 3 |
| 20 05 32.76 | −67 19 15.2 | 098959 | 189567 | 6.07 | 0.648 | 56.45 | 0.74 | 56.41 | 0.44 | G2 V | 2 |
| 20 07 35.09 | −55 00 57.6 | 099137 | 190422 | 6.26 | 0.530 | 43.08 | 0.79 | 42.68 | 0.45 | F9 V | 2 |
| 20 08 43.61 | −66 10 55.4 | 099240 | 190248 | 3.55 | 0.751 | 163.73 | 0.65 | 163.71 | 0.17 | G8 IV | 2 |
| 20 09 34.30 | +16 48 20.8 | 099316 | 191499 | 7.56 | 0.810 | 41.07 | 1.18 | 42.26 | 0.99 | G9 V | 3 |
| 20 11 06.07 | +16 11 16.8 | 099452 | 191785 | 7.34 | 0.830 | 48.83 | 0.91 | 49.04 | 0.65 | K0 V | 3 |
| 20 11 11.94 | −36 06 04.4 | 099461 | 191408 | 5.32 | 0.868 | 165.24 | 0.90 | 166.25 | 0.27 | K2.5 V | 2 |
| 20 13 59.85 | −00 52 00.8 | 099711 | 192263 | 7.79 | 0.938 | 50.27 | 1.13 | 51.77 | 0.78 | K2.5 V | 3 |
| 20 15 17.39 | −27 01 58.7 | 099825 | 192310 | 5.73 | 0.878 | 113.33 | 0.89 | 112.22 | 0.30 | K2+ V | 2 |
| 20 17 31.33 | +66 51 13.3 | 100017 | 193664 | 5.91 | 0.602 | 56.92 | 0.52 | 56.92 | 0.24 | G0 V | 3 |
| 20 27 44.24 | −30 52 04.2 | 100925 | 194640 | 6.61 | 0.724 | 51.50 | 0.82 | 51.22 | 0.54 | G8 V | 2 |
| 20 32 23.70 | −09 51 12.2 | 101345 | 195564 | 5.66 | 0.689 | 41.26 | 0.87 | 40.98 | 0.33 | G2 V | 3 |
| 20 32 51.64 | +41 53 54.5 | 101382 | 195987 | 7.08 | 0.796 | 44.99 | 0.64 | 45.35 | 0.43 | G9 V | 3 |
| 20 40 02.64 | −60 32 56.0 | 101983 | 196378 | 5.11 | 0.544 | 41.33 | 0.73 | 40.55 | 0.27 | G0 V | 2 |
| 20 40 11.76 | −23 46 25.9 | 101997 | 196761 | 6.36 | 0.719 | 68.28 | 0.82 | 69.53 | 0.40 | G8 V | 2 |
| 20 40 45.14 | +19 56 07.9 | 102040 | 197076 | 6.43 | 0.611 | 47.65 | 0.76 | 47.74 | 0.48 | G1 V | 3 |
| 20 43 16.00 | −29 25 26.1 | 102264 | 197214 | 6.95 | 0.671 | 44.57 | 0.87 | 44.83 | 0.91 | G6 V | 2 |
| 20 49 16.23 | +32 17 05.2 | 102766 | 198425 | 8.25 | 0.939 | 42.23 | 0.98 | 41.58 | 0.83 | K2.5 V | 3 |
| 20 56 47.33 | −26 17 47.0 | 103389 | 199260 | 5.70 | 0.507 | 47.61 | 0.95 | 45.52 | 0.38 | F6 V | 2 |
| 20 57 40.07 | −44 07 45.7 | 103458 | 199288 | 6.52 | 0.587 | 46.26 | 0.81 | 45.17 | 0.46 | G2 V | 2 |
| 21 02 40.76 | +45 53 05.2 | 103859 | 200560 | 7.69 | 0.970 | 51.65 | 0.72 | 51.36 | 0.63 | K2.5 V | 3 |
| 21 07 10.38 | −13 55 22.6 | 104239 | 200968 | 7.12 | 0.901 | 56.67 | 1.18 | 56.90 | 0.60 | G9.5 V | 2 |
| 21 09 20.74 | −82 01 38.1 | 104436 | 199509 | 6.98 | 0.619 | 41.28 | 0.58 | 41.95 | 0.37 | G1 V | 2 |
| 21 09 22.45 | −73 10 22.7 | 104440 | 200525 | 5.67 | 0.590 | 53.38 | 2.18 | 50.59 | 1.52 | F9.5 V | 2 |
| 21 14 28.82 | +10 00 25.1 | 104858 | 202275 | 4.47 | 0.529 | 54.11 | 0.85 | 54.09 | 0.66 | F7 V | 3 |
| 21 18 02.97 | +00 09 41.7 | 105152 | 202751 | 8.15 | 0.990 | 52.03 | 1.23 | 50.46 | 1.03 | K3 V | 3 |
| 21 18 27.27 | −43 20 04.7 | 105184 | 202628 | 6.75 | 0.637 | 42.04 | 0.90 | 40.95 | 0.46 | G1.5 V | 2 |
| 21 19 45.62 | −26 21 10.4 | 105312 | 202940 | 6.56 | 0.737 | 53.40 | 1.09 | 55.65 | 0.62 | G7 V | 2 |
| 21 24 40.64 | −68 13 40.2 | 105712 | 203244 | 6.98 | 0.723 | 48.86 | 0.81 | 48.97 | 0.68 | G8 V | 2 |
| 21 26 58.45 | −56 07 30.9 | 105905 | 203850 | 8.65 | 0.924 | 43.12 | 1.17 | 43.47 | 1.01 | K2.5 V | 2 |
| 21 27 01.33 | −44 48 30.9 | 105911 | 203985 | 7.49 | 0.876 | 42.52 | 1.29 | 42.54 | 1.32 | K2 III-IV | 2 |
| 21 36 41.24 | −50 50 43.4 | 106696 | 205390 | 7.14 | 0.879 | 67.85 | 0.92 | 68.40 | 0.58 | K1.5 V | 3 |
| 21 40 29.77 | −74 04 27.4 | 107022 | 205536 | 7.07 | 0.755 | 45.17 | 0.67 | 45.41 | 0.52 | G9 V | 2 |
| 21 44 08.58 | +28 44 33.5 | 107310 | 206826 | 4.49 | 0.512 | 44.64 | 0.69 | 44.97 | 0.43 | F6 V | 3 |
| 21 44 31.33 | +14 46 19.0 | 107350 | 206860 | 5.96 | 0.587 | 54.37 | 0.85 | 55.91 | 0.45 | G0 IV-V | 3 |
| 21 48 00.05 | −40 15 21.9 | 107625 | 207144 | 8.62 | 0.960 | 42.12 | 1.06 | 42.20 | 0.93 | K3 V | 2 |
| 21 48 15.75 | −47 18 13.0 | 107649 | 207129 | 5.57 | 0.601 | 63.95 | 0.78 | 62.52 | 0.35 | G0 V | 2 |
| 21 53 05.35 | +20 55 49.9 | 108028 | 208038 | 8.18 | 0.937 | 41.71 | 0.98 | 43.40 | 0.75 | K2.5 V | 3 |

Table 1—Continued

| R.A. (J2000.0) (1) | Decl. (J2000.0) (2) | HIP Name (3) | HD Name (4) | *Hipparcos* V (5) | *Hipparcos* B − V (6) | *Hipparcos* π (mas) (7) | *Hipparcos* σ_π (mas) (8) | van Leeuwen (2007b) π (mas) (9) | van Leeuwen (2007b) σ_π (mas) (10) | Spec Type (11) | Ref (12) |
|---|---|---|---|---|---|---|---|---|---|---|---|
| 21 54 45.04 | +32 19 42.9 | 108156 | 208313 | 7.73 | 0.911 | 49.21 | 0.93 | 50.11 | 0.80 | K2 V | 3 |
| 22 09 29.87 | −07 32 55.1 | 109378 | 210277 | 6.54 | 0.773 | 46.97 | 0.79 | 46.38 | 0.48 | G8 V | 3 |
| 22 11 11.91 | +36 15 22.8 | 109527 | 210667 | 7.23 | 0.812 | 44.57 | 0.79 | 43.67 | 0.53 | G9 V | 3 |
| 22 14 38.65 | −41 22 54.0 | 109821 | 210918 | 6.23 | 0.648 | 45.19 | 0.71 | 45.35 | 0.37 | G2 V | 2 |
| 22 15 54.14 | +54 40 22.4 | 109926 | 211472 | 7.50 | 0.810 | 46.62 | 0.67 | 46.43 | 0.50 | K0 V | 3 |
| 22 18 15.62 | −53 37 37.5 | 110109 | 211415 | 5.36 | 0.614 | 73.47 | 0.70 | 72.54 | 0.36 | G0 V | 2 |
| 22 24 56.39 | −57 47 50.7 | 110649 | 212330 | 5.31 | 0.665 | 48.81 | 0.61 | 48.63 | 0.34 | G2 IV-V | 2 |
| 22 25 51.16 | −75 00 56.5 | 110712 | 212168 | 6.12 | 0.599 | 43.39 | 0.96 | 43.39 | 0.50 | G0 V | 2 |
| 22 39 50.77 | +04 06 58.0 | 111888 | 214683 | 8.48 | 0.938 | 44.10 | 1.12 | 41.49 | 0.76 | K3 V | 3 |
| 22 42 36.88 | −47 12 38.9 | 112117 | 214953 | 5.99 | 0.584 | 42.47 | 0.72 | 42.31 | 0.40 | F9.5 V | 2 |
| 22 43 21.30 | −06 24 03.0 | 112190 | 215152 | 8.11 | 0.966 | 46.46 | 1.31 | 46.47 | 0.90 | K3 V | 3 |
| 22 46 41.58 | +12 10 22.4 | 112447 | 215648 | 4.20 | 0.502 | 61.54 | 0.77 | 61.36 | 0.19 | F7 V | 1 |
| 22 47 31.87 | +83 41 49.3 | 112527 | 216520 | 7.53 | 0.867 | 50.15 | 0.64 | 50.83 | 0.44 | K0 V | 3 |
| 22 51 26.36 | +13 58 11.9 | 112870 | 216259 | 8.29 | 0.849 | 47.56 | 1.18 | 46.99 | 1.01 | K2.5 V | 3 |
| 22 57 27.98 | +20 46 07.8 | 113357 | 217014 | 5.45 | 0.666 | 65.10 | 0.76 | 64.07 | 0.38 | G3 V | 3 |
| 22 58 15.54 | −02 23 43.4 | 113421 | 217107 | 6.17 | 0.744 | 50.71 | 0.75 | 50.36 | 0.38 | G8 IV-V | 3 |
| 23 03 04.98 | +20 55 06.9 | 113829 | 217813 | 6.65 | 0.620 | 41.19 | 0.87 | 40.46 | 0.57 | G1 V | 3 |
| 23 10 50.08 | +45 30 44.2 | 114456 | 218868 | 6.98 | 0.750 | 42.65 | 0.74 | 41.15 | 0.54 | G8 V | 3 |
| 23 13 16.98 | +57 10 06.1 | 114622 | 219134 | 5.57 | 1.000 | 153.24 | 0.65 | 152.76 | 0.29 | K3 V | 3 |
| 23 16 18.16 | +30 40 12.8 | 114886 | 219538 | 8.07 | 0.871 | 41.33 | 0.97 | 41.63 | 0.72 | K2 V | 3 |
| 23 16 42.30 | +53 12 48.5 | 114924 | 219623 | 5.58 | 0.556 | 49.31 | 0.58 | 48.77 | 0.26 | F7 V | 1 |
| 23 16 57.69 | −62 00 04.3 | 114948 | 219482 | 5.64 | 0.521 | 48.60 | 0.60 | 48.69 | 0.33 | F6 V | 2 |
| 23 19 26.63 | +79 00 12.7 | 115147 | 220140 | 7.53 | 0.893 | 50.65 | 0.64 | 52.07 | 0.47 | K2 V | 3 |
| 23 21 36.51 | +44 05 52.4 | 115331 | 220182 | 7.36 | 0.801 | 45.63 | 0.83 | 46.46 | 0.53 | G9 V | 3 |
| 23 23 04.89 | −10 45 51.3 | 115445 | 220339 | 7.80 | 0.881 | 51.37 | 1.25 | 52.29 | 0.86 | K2.5 V | 3 |
| 23 31 22.21 | +59 09 55.9 | 116085 | 221354 | 6.76 | 0.839 | 59.31 | 0.67 | 59.06 | 0.45 | K0 V | 3 |
| 23 35 25.61 | +31 09 40.7 | 116416 | 221851 | 7.90 | 0.845 | 42.63 | 0.93 | 42.00 | 0.72 | K1 V | 3 |
| 23 37 58.49 | +46 11 58.0 | 116613 | 222143 | 6.58 | 0.665 | 43.26 | 0.80 | 42.86 | 0.42 | G3 V | 3 |
| 23 39 37.39 | −72 43 19.8 | 116745 | 222237 | 7.09 | 0.989 | 87.72 | 0.64 | 87.56 | 0.51 | K3+ V | 2 |
| 23 39 51.31 | −32 44 36.3 | 116763 | 222335 | 7.18 | 0.802 | 53.52 | 0.86 | 53.85 | 0.63 | G9.5 V | 2 |
| 23 39 57.04 | +05 37 34.6 | 116771 | 222368 | 4.13 | 0.507 | 72.51 | 0.88 | 72.92 | 0.15 | F7 V | 1 |
| 23 52 25.32 | +75 32 40.5 | 117712 | 223778 | 6.36 | 0.977 | 92.68 | 0.55 | 91.82 | 0.30 | K3 V | 3 |
| 23 56 10.67 | −39 03 08.4 | 118008 | 224228 | 8.24 | 0.973 | 45.28 | 1.11 | 45.52 | 0.93 | K2.5 V | 2 |
| 23 58 06.82 | +50 26 51.6 | 118162 | 224465 | 6.72 | 0.694 | 41.35 | 0.76 | 40.77 | 0.49 | G4 V | 3 |

Note. — Column 12 reference codes: (1) The *Hipparcos* catalog; (2) Gray et al. (2006); (3) Gray et al. (2003).

[a]The parallax is from Söderhjelm (1999).

Table 2.   Stars Excluded Due to Large Offset from the Main Sequence

| R.A. (J2000.0) (1) | Decl. (J2000.0) (2) | HIP Name (3) | HD Name (4) | $V$ (5) | $B-V$ (6) | *Hipparcos* $\pi$ (mas) (7) | *Hipparcos* $\sigma_\pi$ (mas) (8) | $M_V$ (9) | MS $M_V$[a] (10) | Spec Type (11) | Ref (12) |
|---|---|---|---|---|---|---|---|---|---|---|---|
| 00 49 09.90 | +05 23 19.0 | 003829 | . . . | 12.37 | 0.554 | 226.95 | 5.35 | 14.15 | 4.21 | DG | 1 |
| 01 55 57.47 | −51 36 32.0 | 009007 | 011937 | 3.69 | 0.844 | 57.19 | 0.62 | 2.48 | 5.72 | G9 IV | 2 |
| 03 43 14.90 | −09 45 48.2 | 017378 | 023249 | 3.52 | 0.915 | 110.58 | 0.88 | 3.74 | 6.08 | K1 III-IV | 2 |
| 05 16 41.36 | +45 59 52.8 | 024608 | 034029 | 0.08 | 0.795 | 77.29 | 0.89 | −0.48 | 5.48 | Late G + Late F III | 3 |
| 07 34 27.43 | +62 56 29.4 | 036834 | . . . | 10.40 | 0.942 | 87.01 | 2.17 | 10.10 | 6.22 | M0 | 4 |
| 07 42 57.10 | −45 10 23.2 | 037606 | 062644 | 5.04 | 0.765 | 41.43 | 0.81 | 3.13 | 5.33 | G8 IV-V | 2 |
| 07 45 18.95 | +28 01 34.3 | 037826 | 062509 | 1.16 | 0.991 | 96.74 | 0.87 | 1.09 | 6.47 | K0 III | 5 |
| 16 41 17.16 | +31 36 09.8 | 081693 | 150680 | 2.81 | 0.650 | 92.63 | 0.60 | 2.64 | 4.73 | F9 IV | 1 |
| 18 21 18.60 | −02 53 55.8 | 089962 | 168723 | 3.23 | 0.941 | 52.81 | 0.75 | 1.84 | 6.22 | K0 III-IV | 5 |
| 19 55 18.79 | +06 24 24.4 | 098036 | 188512 | 3.71 | 0.855 | 72.95 | 0.83 | 3.03 | 5.78 | G9.5 IV | 2 |
| 19 55 50.36 | −26 17 58.2 | 098066 | 188376 | 4.70 | 0.748 | 42.03 | 0.94 | 2.82 | 5.24 | G5 IV | 2 |
| 20 06 21.77 | +35 58 20.9 | 099031 | 191026 | 5.38 | 0.850 | 41.34 | 0.54 | 3.46 | 5.75 | K0 IV | 5 |
| 20 45 17.38 | +61 50 19.6 | 102422 | 198149 | 3.41 | 0.912 | 69.73 | 0.49 | 2.63 | 6.07 | K0 IV | 5 |
| 21 58 24.52 | +75 35 20.6 | 108467 | . . . | 10.56 | 0.742 | 47.95 | 1.08 | 8.96 | 5.21 | M0 | 1 |
| 23 19 06.67 | −13 27 30.8 | 115126 | 219834 | 5.20 | 0.787 | 40.28[b] | 1.51[b] | 3.23 | 5.44 | G8.5 IV | 2 |

Note. — Column 12 reference codes: (1) The *Hipparcos* catalog; (2) Gray et al. (2006); (3) Griffin & Griffin (1986); (4) Endl et al. (2006); (5) Gray et al. (2003).

[a] This column lists the absolute magnitude of the Cox (2000) main sequence corresponding to the $B-V$ color of the star.

[b] The parallax is from Söderhjelm (1999).

Table 3. Stars Excluded Due to Large Parallax Errors ($\sigma_\pi/\pi \geq 0.05$)

| | | | | *Hipparcos* | | *Hipparcos* | | van Leeuwen (2007b) | | | | Alternate | | |
| R.A. (J2000.0) (1) | Decl. (J2000.0) (2) | HIP Name (3) | HD Name (4) | $V$ (5) | $B-V$ (6) | $\pi$ (mas) (7) | $\sigma_\pi$ (mas) (8) | $\pi$ (mas) (9) | $\sigma_\pi$ (mas) (10) | Spec Type (11) | Ref (12) | $\pi$ (mas) (13) | $\sigma_\pi$ (mas) (14) | Method (15) |
|---|---|---|---|---|---|---|---|---|---|---|---|---|---|---|
| 00 22 23.61 | −27 01 57.3 | 001768 | 001815 | 8.30 | 0.888 | 44.57 | 7.12 | 34.63 | 4.58 | K2 V | 1 | 40 | ... | 01 |
| 01 49 23.36 | −10 42 12.8 | 008486 | 011131 | 6.72 | 0.654 | 43.47 | 4.48 | 44.32 | 3.02 | G1 V | 2 | 43 | ... | 01 |
| 02 15 42.55 | +67 40 20.2 | 010531 | 013579 | 7.13 | 0.920 | 42.46 | 2.51 | 53.82 | 1.74 | K2 V | 3 | 71 | ... | 01 |
| 02 57 14.71 | −24 58 10.2 | 013772 | 018455 | 7.33 | 0.863 | 44.49 | 2.55 | 44.51 | 2.09 | K2 V | 1 | 38.87 | 1.50 | 02 |
| 03 47 02.12 | +41 25 38.2 | 017666 | 023439 | 7.67 | 0.796 | 40.83 | 2.24 | 45.65 | 2.63 | K3 V | 2 | 43 | ... | 01 |
| 04 29 44.87 | −29 01 37.4 | 020968 | ... | 11.42 | 0.646 | 120.70 | 56.47 | 25.66 | 10.02 | F9.5 V | 1 | 6 | ... | 01 |
| 04 30 12.58 | +05 17 55.8 | 021000 | ... | 9.83 | 0.600 | 84.76 | 4.74 | 93.67 | 7.62 | F8+... | 3 | 9 | ... | 01 |
| 05 44 56.79 | +09 14 31.5 | 027111 | 247168 | 11.35 | 0.699 | 44.67 | 14.98 | 61.21 | 15.04 | F8: | 3 | 6 | ... | 01 |
| 07 03 58.92 | −43 36 40.8 | 034069 | 053706 | 6.83 | 0.779 | 66.29 | 6.81 | 47.99 | 9.89 | K0.5 V | 1 | 61.54 | 1.05 | 03 |
| 08 35 51.27 | +06 37 22.0 | 042173 | 072946 | 7.25 | 0.710 | 42.71 | 4.61 | 38.11 | 0.85 | G8 V | 2 | 37.68 | 1.41 | 04 |
| 10 04 50.59 | −31 05 28.0 | 049376 | ... | 11.99 | 0.938 | 41.59 | 3.13 | 43.49 | 3.44 | M2+ V | 1 | 8 | ... | 01 |
| 12 29 55.04 | +36 26 42.1 | 060970 | ... | 11.91 | 0.800 | 43.42 | 3.62 | 37.95 | 4.14 | ... | ... | 6 | ... | 01 |
| 12 31 18.92 | +55 07 07.7 | 061100 | 109011 | 8.08 | 0.941 | 42.13 | 3.11 | 39.84 | 1.07 | K2 V | 3 | 48 | ... | 01 |
| 12 59 18.98 | +06 30 33.7 | 063383 | ... | 10.69 | 0.515 | 45.19 | 44.19 | 3.18 | 18.34 | G | 3 | 5 | ... | 01 |
| 13 33 18.71 | −77 34 24.6 | 066125 | ... | 9.31 | 0.914 | 52.09 | 39.91 | 55.78 | 10.02 | K2.5 V | 1 | 28.45 | 1.13 | 05 |
| 15 31 54.05 | +09 39 26.9 | 076051 | ... | 9.80 | 0.787 | 44.59 | 20.41 | 12.33 | 11.19 | G2 V | 2 | 16 | ... | 01 |
| 15 38 39.95 | −08 47 41.0 | 076602 | 139460 | 6.56 | 0.520 | 44.21 | 4.72 | 40.99 | 3.72 | F7 V | 2 | 30 | ... | 01 |
| 15 38 40.08 | −08 47 29.4 | 076603 | 139461 | 6.45 | 0.505 | 40.19 | 3.62 | 37.56 | 4.29 | F6.5 V | 2 | 31 | ... | 01 |
| 16 19 31.52 | −30 54 06.7 | 079979 | 146835 | 7.29 | 0.585 | 56.82 | 23.38 | −4.71 | 13.69 | F9- V | 1 | 20.72 | 1.02 | 06 |
| 22 06 11.82 | +10 05 28.7 | 109119 | ... | 10.20 | 0.668 | 93.81 | 66.11 | 0.40 | 12.19 | F2 IV | 2 | 9 | ... | 01 |
| 22 12 59.72 | −47 23 11.1 | 109670 | ... | 11.48 | 0.660 | 44.20 | 16.43 | 3.51 | 3.07 | G5 V | 1 | 5 | ... | 01 |
| 22 26 34.28 | −16 44 31.7 | 110778 | 212697 | 5.55 | 0.618 | 49.80 | 2.54 | 49.50 | 1.23 | G5 V | 1 | 67 | ... | 01 |
| 23 01 51.54 | −03 50 55.4 | 113718 | 217580 | 7.48 | 0.943 | 59.04 | 3.40 | 58.70 | 0.92 | K2.5 V | 2 | 67 | ... | 01 |
| 23 44 07.36 | −27 11 45.8 | 117081 | 222834 | 9.01 | 0.535 | 58.45 | 47.16 | 63.56 | 21.02 | G1 V | 1 | 10 | ... | 01 |

Note. — Column 12 reference codes: (1) Gray et al. (2006); (2) Gray et al. (2003); (3) The *Hipparcos* catalog. Column 15 notes: (01) Distance estimated by fitting the *Hipparcos* $V$ magnitude and $B-V$ color to the Cox (2000) main sequence. (02) *Hipparcos* parallax of the companion HIP 13769. (03) *Hipparcos* parallax of the companion HIP 34065, which is included in the sample. (04) *Hipparcos* parallax of the companion HIP 42172. (05) *Hipparcos* parallax of the companion HIP 66121. (06) *Hipparcos* parallax of the companion HIP 79980.

Table 4.   CPM Companions Identified

| WDS ID | Disc Desig | Pair ID | Primary Name | CPM Candidate Companion | | | | Reason Code |
|---|---|---|---|---|---|---|---|---|
| | | | | $\rho$ ('') | $\theta$ (°) | Epoch | Name | |

— Physically Associated CPM Companions —

| WDS ID | Disc Desig | Pair ID | Primary Name | $\rho$ ('') | $\theta$ (°) | Epoch | Name | Reason Code |
|---|---|---|---|---|---|---|---|---|
| . . . | . . . | . . . | HD 004391[a] | 49 | 240 | 1993.70 | . . . | 1 |
| 01158−6853 | HJ 3423 | A-CD | HD 007693 | 318 | 315 | 1989.73 | HD 7788 | 2 |
| 01368+4124 | LWR 1 | AD | HD 009826 | 55 | 150 | 1989.77 | $v$ And B | 3 |
| 01398−5612 | DUN 5 | AB | HD 010360 | 13 | 185 | 1997.61 | HD 10361 | 4 |
| 02361+0653 | PLQ 32 | AB | HD 016160 | 162 | 110 | 1990.73 | NLTT 8455 | 1 |
| 02482+2704 | LDS1138 | AB | HD 017382 | 21 | 27 | 1989.98 | NLTT 8996 | 1 |
| 02556+2652 | LDS 883 | AC | HD 018143 | 45 | 267 | 1989.98 | NLTT 9303 | 2 |
| 03042+6142 | LDS9142 | AC | HD 018757 | 262 | 68 | 1993.94 | NLTT 9726 | 1 |
| 03182−6230 | ALB 1 | AB | HD 020807 | 308 | 220 | 1996.87 | HD 20766 | 2 |
| 04076+3804 | ALC 1 | AE | HD 025998 | 746 | 100 | 1989.76 | HD 25893 | 2 |
| 04153−0739 | STF 518 | AC | HD 026965 | 81 | 99 | 1985.96 | LHS 25 | 2 |
| 04153−0739 | STF 518 | AB | HD 026965 | 85 | 104 | 1985.96 | HD 26976 | 4 |
| 04155+0611 | H 4 98 | AB | HD 026923 | 64 | 315 | 1986.77 | HD 26913 | 2 |
| 05023−5605 | LDS 135 | AB | HD 032778 | 81 | 145 | 1989.98 | NLTT 14447 | 1 |
| 05244+1723 | WNO 52 | AC | HD 035296 | 707 | 252 | 1989.91 | HD 35171 | 2 |
| 05369−4758 | HDS 751 | AB | HD 037572 | 18 | 285 | 1997.02 | HIP 26369 | 2 |
| 05413+5329 | ENG 22 | AB | HD 037394 | 99 | 70 | 1991.11 | HD 233153 | 2 |
| . . . | . . . | . . . | HD 043162[a] | 164 | 171 | 1996.04 | . . . | 1 |
| 06173+0506 | LEP 24 | AE | HD 043587 | 102 | 306 | 1990.82 | NLTT 16333 | 1 |
| 06461+3233 | LDS6201 | AB | HD 263175 | 30 | 101 | 1989.84 | LHS 1867 | 1 |
| 07040−4337 | DUN 38 | AB | HD 053705 | 24 | 128 | 1994.93 | HD 53706 | 2 |
| 07040−4337 | DUN 38 | AC | HD 053705 | 184 | 336 | 1994.93 | HD 53680 | 2 |
| 07400−0336 | BGH 3 | AB | HD 061606 | 57 | 112 | 1985.96 | NLTT 18260 | 1 |
| 07291+3147 | ALC 3 | AE | HIP 036357 | 756 | 355 | 1988.93 | HD 58946 | 2 |
| . . . | . . . | . . . | HD 063077 | 914 | 6 | 1991.88 | NLTT 18414 | 1 |
| 07578−6018 | LDS 198 | AB | HD 065907 | 61 | 74 | 1991.21 | LHS 1960 | 1 |
| 08122+1739 | STF1196 | AC | HD 068257 | 6 | 90 | 1997.18 | HD 68256 | 4 |
| 08421−4256 | LDS 230 | AB | HD 074385 | 46 | 184 | 1994.18 | NLTT 20102 | 1 |
| 08526+2820 | LDS6219 | AB | HD 075732 | 84 | 128 | 1998.29 | LHS 2063 | 1 |
| 09305−3206 | LDS5704 | AB | HD 082342 | 11 | 204 | 1992.19 | . . . | 1 |
| 09327+2659 | LDS3903 | AB | HD 082443 | 65 | 68 | 1998.31 | NLTT 22015 | 1 |
| . . . | . . . | . . . | HD 086728 | 133 | 278 | 1989.93 | GJ 376 B | 5 |

Table 4—Continued

| WDS ID | Disc Desig | Pair ID | Primary Name | CPM Candidate Companion | | | | Reason Code |
|--------|-----------|---------|--------------|----------|----------|-------|------|-------------|
| | | | | $\rho$ (″) | $\theta$ (°) | Epoch | Name | |
| 10306+5559 | LDS2863 | AB | HD 090839 | 124 | 303 | 1998.30 | HD 237903 | 2 |
| 11047−0413 | STF1506 | A-BC | HD 096064 | 13 | 218 | 1985.06 | NLTT 26194 | 1 |
| 11268+0301 | STF1540 | AB | HD 099491 | 29 | 149 | 1996.27 | HD 99492 | 2 |
| 11317+1422 | STF1547 | AB | HD 100180 | 17 | 332 | 1993.08 | NLTT 27656 | 1 |
| 11387+4507 | STF1561 | AB | HD 101177 | 8 | 270 | 1997.28 | LHS 2436 | 4 |
| 13237+0243 | STF1740 | AB | HD 116442 | 26 | 76 | 1997.18 | HD 116443 | 2 |
| 14196−0509 | KUI 67 | AB | HD 125455 | 12 | 104 | 1994.19 | LHS 2895 | 6 |
| 14396−6050 | RHD 1 | AB | HD 128620 | 7 | 213 | 1997.19 | HD 128621 | 4 |
| 14396−6050 | LDS 494 | AC | HD 128620 | 7867 | 225 | 1997.19 | HIP 70890 | 2 |
| 15282−0921 | SHJ 202 | AB | HD 137763 | 52 | 133 | 1991.21 | HD 137778 | 2 |
| 15292+8027 | STF1972 | AC | HD 139777 | 33 | 70 | 1994.44 | HD 139813 | 2 |
| 15360+3948 | STT 298 | AB-C | HD 139341 | 123 | 328 | 1993.29 | HD 139323 | 2 |
| 15475−3755 | SEE 249 | AB | HD 140901 | 14 | 138 | 1997.29 | NLTT 41169 | 6 |
| 16048+3910 | WNO 47 | AB | HD 144579 | 70 | 280 | 1991.27 | LHS 3150 | 1 |
| 16133+1332 | STF2021 | AB | HD 145958 | 3 | 340 | 1989.27 | NLTT 42272 | 4 |
| 16147+3352 | STF2032 | AB | HD 146361 | 6 | 248 | 1991.28 | HD 146362 | 4 |
| 16147+3352 | STF2032 | AE | HD 146361 | 634 | 241 | 1991.28 | HIP 79551 | 2 |
| ⋯ | ⋯ | ⋯ | HD 147513 | 345 | 248 | 1993.25 | HIP 80300 | 2 |
| 16579+4722 | STFA 32 | AC | HD 153557 | 113 | 260 | 1991.30 | HD 153525 | 2 |
| 17153−2636 | SHJ 243 | AB | HD 155885 | 4 | 160 | 1997.33 | ⋯ | 4 |
| 17153−2636 | SHJ 243 | AC | HD 155885 | 732 | 74 | 1997.33 | HD 156026 | 2 |
| 17191−4638 | BSO 13 | AB | HD 156274 | 13 | 245 | 1992.43 | NLTT 44525 | 4 |
| ⋯ | ⋯ | ⋯ | HD 157347[a] | 49 | 147 | 1992.35 | ⋯ | 1 |
| 17350+6153 | LDS2736 | AB-C | HD 160269 | 738 | 161 | 1992.36 | HIP 86087 | 2 |
| 17419+7209 | STF2241 | AB | HD 162004 | 31 | 10 | 1993.63 | HD 162003 | 2 |
| 18409+3132 | HJ 1337 | AB | HIP 091605 | 9 | 150 | 1989.34 | LHS 3402 | 7 |
| 19121+4951 | STF2486 | AB | HD 179957 | 7 | 195 | 1992.65 | LHS 3441 | 4 |
| 19418+5032 | STFA 46 | AB | HD 186408 | 40 | 132 | 1991.52 | HD 186427 | 2 |
| 19456+3337 | LEP 93 | AF | HD 186858 | 793 | 55 | 1992.67 | HD 187013 | 2 |
| 19456+3337 | STF2576 | BF | HD 186858 | 810 | 58 | 1992.67 | HD 225732 | 1 |
| 19510+1025 | J 124 | AC | HD 187691 | 23 | 221 | 1991.53 | ⋯ | 1 |
| 20036+2954 | LDS6339 | AB | HD 190360 | 178 | 232 | 1992.49 | LHS 3509 | 2 |
| 20111+1611 | GIC 163 | AE | HD 191785 | 102 | 95 | 1992.74 | LHS 3530 | 1 |
| 20408+1956 | LDS1045 | AC | HD 197076 | 125 | 184 | 1992.67 | NLTT 49681 | 1 |
| 20493+3217 | LDS2931 | AB | HD 198425 | 33 | 247 | 1992.56 | NLTT 49961 | 1 |

Table 4—Continued

| WDS ID | Disc Desig | Pair ID | Primary Name | CPM Candidate Companion | | | | Reason Code |
|---|---|---|---|---|---|---|---|---|
| | | | | $\rho$ ('') | $\theta$ (°) | Epoch | Name | |
| 21270−4449 | LDS6352 | AB | HD 203985 | 88 | 259 | 1993.61 | LTT 8515 | 1 |
| 22159+5440 | GIC 177 | AT | HD 211472 | 76 | 105 | 1991.60 | GJ 4269 | 1 |
| 22259−7501 | DUN 238 | AB | HD 212168 | 20 | 93 | 1996.78 | HIP 110719 | 1 |
| . . . | . . . | . . . | HD 218868[a] | 50 | 90 | 1989.68 | . . . | 1 |

— Refuted CPM Candidates —

| | | | | | | | | |
|---|---|---|---|---|---|---|---|---|
| . . . | . . . | . . . | HD 000166 | 311 | 147 | 1989.83 | SAO 73748 | 8 |
| . . . | . . . | . . . | HD 001581 | 87 | 343 | 1993.64 | . . . | 9 |
| 01333−2411 | LDS2209 | AB | HD 009540 | 336 | 220 | 1996.72 | NLTT 5160 | 10 |
| . . . | . . . | . . . | HD 012846 | 304 | 336 | 1990.87 | . . . | 10 |
| . . . | . . . | . . . | HD 025665 | 287 | 134 | 1993.72 | NLTT 12588 | 8 |
| 04053+2201 | STT 559 | AB | HD 025680 | 174 | 0 | 1989.97 | HIP 19075 | 8 |
| 04155+0611 | STU 18 | AE | HD 026923 | 223 | 63 | 1986.77 | BD+05 617 | 10 |
| . . . | . . . | . . . | HD 068257 | 372 | 107 | 1997.18 | . . . | 8 |
| . . . | . . . | . . . | HD 073667 | 335 | 207 | 1997.10 | NLTT 19982 | 10 |
| . . . | . . . | . . . | HD 075767 | 385 | 41 | 1997.10 | NLTT 20430 | 10 |
| . . . | . . . | . . . | HD 082885 | 328 | 333 | 1998.29 | NLTT 22106 | 8 |
| . . . | . . . | . . . | HD 084117 | 722 | 331 | 1995.09 | NLTT 22384 | 8 |
| . . . | . . . | . . . | HD 096064 | 296 | 142 | 1985.06 | . . . | 10 |
| . . . | . . . | . . . | HD 097658 | 346 | 139 | 1992.32 | . . . | 10 |
| . . . | . . . | . . . | HD 114783 | 240 | 46 | 1996.23 | . . . | 10 |
| . . . | . . . | . . . | HD 141004 | 235 | 200 | 1993.25 | . . . | 10 |
| . . . | . . . | . . . | HD 206860 | 591 | 16 | 1990.80 | . . . | 10 |

Note. — Reason Code values: (1) Photometric distance to the CPM candidate matches the *Hipparcos* distance to the primary (see Table 5). (2) Published parallax and proper motion for the CPM candidate matches the corresponding primary's values from *Hipparcos*. (3) Spectroscopic distance to the CPM candidate matches the *Hipparcos* distance to the primary. (4) Known companion with a published orbit (each of these were seen as comoving diffraction spikes). (5) Published evidence, see §4.3. (6) Companionship implied by proximity to the primary and a matching, large proper motion. (7) Measurements of the pair in the WDS confirm orbital motion. (8) While the proper motion is similar enough to enable selection as a possible CPM companion, the numerical value for the candidate in catalogs (Høg et al. 1998; Salim & Gould 2003; Zacharias et al. 2004b; Lépine

& Shara 2005 *Hipparcos*) are significantly different from the primary's *Hipparcos* values, ruling out a physical association. (9) The candidate companion is a non-stellar artifact such as a plate defect. (10) Photometric distance to the CPM candidate is significantly different than the primary's *Hipparcos* distance (see Table 5).

[a]New companion discovered by this effort.

Table 5.   Spectral Type, Proper Motion, and Photometry of CPM Candidates

| Primary Name (1) | Comp ρ ('') (2) | Spec Type (3) | ref (4) | $\mu_\alpha$ (mas yr$^{-1}$) (5) | $\mu_\delta$ (6) | ref (7) | ====== CCD Magnitudes ====== V (8) | ref (9) | R (10) | ref (11) | I (12) | ref (13) | Nbr Obs (14) | Infrared Magnitudes J (15) | H (16) | K (17) | D (pc) (18) | Error (pc) (19) |
|---|---|---|---|---|---|---|---|---|---|---|---|---|---|---|---|---|---|---|
| — Physically Associated CPM Companions — | | | | | | | | | | | | | | | | | | |
| HD 004391 | 49 | ... | ... | ... | ... | ... | 14.36 | 1 | 13.14 | 1 | 11.46 | 1 | 3 | 9.88 | 9.34 | 9.03 | 19.3 | 3.1 |
| HD 016160 | 162 | M3.5V | 2 | 1813 | 1447 | 3 | 11.68 | 1 | 10.47 | 1 | 8.87 | 1 | 1 | 7.33 | 6.79 | 6.57 | 6.9 | 1.1 |
| HD 017382 | 21 | ... | ... | 275 | −123 | 3 | 16.5 | 4 | 13.89 | 5 | ... | ... | ... | 10.73 | 10.17 | 9.87 | 22.1 | 9.0 |
| HD 018757 | 262 | ... | ... | 717 | −697 | 3 | 12.65 | 6 | ... | ... | 10.07 | 7 | ... | 8.88 | 8.33 | 8.10 | 22.6 | 5.2 |
| HD 032778 | 81 | ... | ... | 656 | −70 | 8 | 10.49 | 1 | 9.64 | 1 | 8.85 | 1 | 1 | 7.86 | 7.32 | 7.06 | 23.9 | 3.9 |
| HD 043162 | 164 | ... | ... | −32 | 106 | 7 | 12.96 | 1 | 11.78 | 1 | 10.21 | 1 | 1 | 8.72 | 8.16 | 7.87 | 13.2 | 2.0 |
| HD 043587 | 102 | ... | ... | −212 | 166 | 7 | 13.29 | 1 | 12.11 | 1 | 10.58 | 1 | 1 | 9.09 | 8.56 | 8.27 | 16.5 | 2.5 |
| HD 061606 | 57 | ... | ... | 67 | −286 | 9 | 8.93 | 10 | 8.09 | 10 | 7.34 | 10 | ... | 6.38 | 5.70 | 5.57 | 12.5 | 2.1 |
| HD 063077 | 914 | DC | 11 | −246 | 1654 | 7 | 16.60 | 1 | 15.97 | 1 | 15.39 | 1 | 1 | 14.78 | 14.55 | 14.40 | 15.3 | 1.1 |
| HD 065907 | 61 | ... | ... | 521 | 103 | 12 | 9.33 | 12 | ... | ... | 7.47 | 7 | ... | 6.91 | 6.28 | 6.06 | 17.5 | 4.8 |
| HD 074385 | 46 | ... | ... | −294 | −84 | 13 | 12.68 | 14 | ... | ... | ... | ... | ... | 9.00 | 8.42 | 8.17 | 22.0 | 3.4 |
| HD 075732 | 84 | ... | ... | −482 | −238 | 8 | 13.26 | 1 | 11.91 | 1 | 10.24 | 1 | 2 | 8.56 | 7.93 | 7.67 | 8.7 | 1.4 |
| HD 082342 | 11 | M3.5V | 15 | −75 | 386 | 13 | 13.14 | 1 | 12.03 | 1 | 10.58 | 1 | 1 | 9.27 | 8.77 | 8.50 | 23.7 | 4.2 |
| HD 082443 | 65 | M5.5V | 15 | −134 | −242 | 3 | 16.8 | 4 | 14.39 | 5 | 12.07 | 7 | ... | 10.36 | 9.86 | 9.47 | 12.6 | 5.1 |
| HD 096064 | 13 | ... | ... | −189 | −113 | 16 | 10.0 | 4 | ... | ... | ... | ... | ... | 7.27 | 6.62 | 6.42 | 17.0$^a$ | 2.6 |
| HD 100180 | 17 | ... | ... | −328 | −189 | 3 | 9.24 | 17 | 7.75 | 7 | ... | ... | ... | 7.04 | 6.52 | 6.37 | 23.5 | 4.3 |
| HD 144579 | 70 | ... | ... | −547 | 55 | 3 | 14.23 | 18 | 12.75 | 18 | 11.47 | 18 | ... | 9.90 | 9.45 | 9.16 | 25.4$^b$ | 7.3 |
| HD 157347 | 49 | M3.0V | 1 | ... | ... | ... | 12.18 | 1 | 11.06 | 1 | 9.64 | 1 | 3 | 8.26 | 7.68 | 7.44 | 13.4 | 2.1 |
| HD 186858 | 810 | F5.5IV-V | 19 | 19 | −447 | 3 | 9.25 | 20 | 8.38 | 21 | ... | ... | ... | 6.64 | 6.12 | 6.00 | 15.4 | 2.9 |
| HD 187691 | 23 | ... | ... | ... | ... | ... | 12.67 | 1 | 11.55 | 1 | 10.04 | 1 | 1 | 8.89 | 8.30 | 8.01 | 20.0 | 4.8 |
| HD 191785 | 102 | M3.5V | 1 | −432 | 392 | 3 | 13.93 | 1 | 12.73 | 1 | 11.14 | 1 | 1 | 9.63 | 9.11 | 8.88 | 20.9 | 3.4 |
| HD 197076 | 125 | M2.5V | 10 | 108 | 312 | 3 | 11.88 | 1 | 10.80 | 1 | 9.47 | 1 | 1 | 8.16 | 7.65 | 7.42 | 15.8 | 2.5 |
| HD 198425 | 33 | ... | ... | −158 | −278 | 3 | 18.6 | 4 | 15.47 | 7 | 13.68 | 7 | ... | 11.78 | 11.18 | 10.86 | 25.7 | 7.3 |
| HD 203985 | 88 | M3.5V | 1 | ... | ... | ... | 13.49 | 1 | 12.29 | 1 | 10.71 | 1 | 1 | 9.22 | 8.62 | 8.35 | 15.9 | 2.4 |
| HD 211472 | 76 | ... | ... | 205 | 66 | 3 | 13.93 | 22 | ... | ... | ... | ... | ... | 9.72 | 9.19 | 8.93 | 22.3 | 3.5 |
| HD 212168 | 20 | G0V | 19 | 60 | 35 | 23 | 8.80 | 1 | 8.09 | 1 | 7.43 | 1 | 1 | 6.56 | 5.94 | 5.81 | 16.9$^c$ | 2.8 |
| HD 218868 | 50 | ... | ... | −50 | −322 | 7 | 15.32 | 22 | ... | ... | 12.57 | 7 | ... | 10.84 | 10.23 | 9.90 | 25.9 | 6.0 |
| HD 263175 | 30 | M0.5V | 10 | −454 | 99 | 3 | 12.15 | 10 | 11.18 | 10 | 10.11 | 10 | ... | 8.99 | 8.43 | 8.18 | 30.6 | 4.8 |
| — Refuted CPM Candidates — | | | | | | | | | | | | | | | | | | |
| HD 009540 | 336 | ... | ... | 288 | −153 | 13 | 12.75 | 1 | 11.94 | 1 | 11.12 | 1 | 1 | 10.10 | 9.47 | 9.35 | 69.1 | 11.7 |
| HD 012846 | 304 | ... | ... | −32 | −102 | 7 | 11.48 | 21 | 10.62 | 21 | 9.84 | 7 | ... | 10.18 | 9.80 | 9.66 | 81.7 | 13.2 |
| HD 026923 | 223 | ... | ... | −15 | −131 | ... | 9.99 | 1 | 9.50 | 1 | 9.06 | 1 | 1 | 8.43 | 8.04 | 7.88 | 55.6 | 8.9 |
| HD 073667 | 385 | ... | ... | 58 | 185 | 3 | 15.35 | 1 | 14.27 | 1 | 12.96 | 1 | 1 | 11.70 | 11.18 | 10.92 | 82.4 | 13.4 |
| HD 075767 | 385 | ... | ... | 58 | −178 | 3 | 13.77 | 1 | 12.74 | 1 | 11.51 | 1 | 1 | 10.22 | 9.57 | 9.30 | 38.1$^b$ | 6.8 |
| HD 096064 | 296 | ... | ... | ... | ... | ... | 16.20 | 1 | 15.36 | 1 | 14.60 | 1 | 1 | 13.63 | 13.06 | 12.85 | 355.8 | 58.7 |
| HD 097658 | 346 | ... | ... | ... | ... | ... | 15.94 | 1 | 14.95 | 1 | 13.61 | 1 | 1 | 12.32 | 11.74 | 11.56 | 109.4 | 18.3 |
| HD 114783 | 240 | ... | ... | ... | ... | ... | 9.78 | 1 | 9.31 | 1 | 8.90 | 1 | 2 | 8.32 | 7.90 | 7.79 | 54.0 | 9.3 |
| HD 141004 | 235 | ... | ... | ... | ... | ... | 18.35 | 1 | 16.92 | 1 | 15.11 | 1 | 2 | 13.40 | 12.85 | 12.59 | 76.9 | 13.2 |
| HD 206860 | 591 | ... | ... | 111 | −89 | 24 | 15.04 | 21 | 14.07 | 7 | 13.63 | 7 | ... | 12.77 | 12.15 | 12.06 | 300.0 | 83.9 |

Note. — Reference codes for columns 4, 7, 9, 11, and 13: (1) CTIO observations obtained for this work; (2) Henry et al. (2002); (3) The LSPM North catalog (Lépine & Shara 2005); (4) Visual Double Stars in *Hipparcos* (Dommanget & Nys 2000); (5) The Guide Star Catalog, Version 2.2.01 (I/271); (6) Catalog of stars with high proper motions (I/306A); (7) The USNO B 1.0 Catalog (Monet et al. 2003); (8) Revised Luyten Half-Second Catalog (Bakos et al. 2002); (9) The Revised NLTT Catalog (Salim & Gould 2003); (10) Reid et al. (2004); (11) Kunkel et al. (1984); (12) All-sky Compiled Catalog (Kharchenko 2001); (13) NLTT Catalog (Luyten 1979); (14) The Catalog of Nearby Stars (Gliese & Jahreiß 1991); (15) Hawley et al. (1996); (16) Yale Trigonometric Parallaxes, Fourth Edition (van Altena et al. 1995); (17) An Astrometric Catalog (Rapaport et al. 2001); (18) Weis (1996); (19) Gray et al. (2003, 2006); (20) The Tycho-2 Catalog (Høg et al. 2000); (21) The NOMAD Catalog (Zacharias et al. 2004a); (22) The Guide Star Catalog, Version 2.3.2 (Lasker et al. 2008); (23) Yale Zones Catalog Integrated (I/141); (24) The DENIS Consortium (The 2005).

[a]While this distance is too low compared to the primary's *Hipparcos* distance of 24.6 pc, the companion is a roughly equal-brightness binary. Adjusting the *Hipparcos* and 2MASS magnitudes accordingly changes the distance estimate to $24.0 \pm 3.7$, a much better match with the distance to the primary.

[b]See § 4.3 for a discussion of these photometric distance estimates and the status of these companions.

[c]Even though the companion is HIP 110719, its *Hipparcos* astrometry has large errors, necessitating the photometric distance check.

Table 6.   Optical WDS Entries

| WDS ID | Disc Desig | Pair ID | Primary Name | Nbr Obs | $\theta$ (deg) | $\rho$ (") | Epoch |
|---|---|---|---|---|---|---|---|
| 00022+2705 | BU 733 | AC | HD 224930 | 64 | 325 | 161.7 | 2000 |
| 00022+2705 | BU 733 | AD | HD 224930 | 15 | 296 | 109.9 | 1921 |
| 00022+2705 | HSW 1 | AE | HD 224930 | 2 | 309 | 100.9 | 1998 |
| 00066+2901 | ENG 1 | A-CD | HD 000166 | 9 | 196 | 142.4 | 1991 |
| 00066+2901 | STT 549 | AB | HD 000166 | 15 | 259 | 186.7 | 2002 |
| 00066+2901 | BU 1338 | CD | HD 000166 | 8 | 210 | 3.2 | 1999 |
| 00200+3814 | S 384 | AB | HD 001562 | 43 | 22 | 100.7 | 2003 |
| 00200+3814 | S 384 | AC | HD 001562 | 4 | 260 | 22.6 | 1998 |
| 00229−1213 | BUP 6 | … | HD 001835 | 5 | 294 | 213.9 | 1998 |
| 00352−0336 | BU 490 | AB-C | HD 003196 | 15 | 324 | 24.4 | 1998 |
| 00394+2115 | STT 550 | AB | HD 003651 | 9 | 80 | 167.6 | 1997 |
| 00484+0517 | BUP 10 | AC | HD 004628 | 4 | 241 | 158.5 | 2000 |
| 00490+1656 | BUP 12 | AB | HD 004676 | 4 | 330 | 82.9 | 1998 |
| 00490+1656 | BUP 12 | AC | HD 004676 | 4 | 162 | 64.2 | 1998 |
| 00491+5749 | STF 60 | AC | HD 004614 | 4 | 258 | 216.0 | 1991 |
| 00491+5749 | STF 60 | AD | HD 004614 | 5 | 1 | 177.0 | 1991 |
| 00491+5749 | STF 60 | AE | HD 004614 | 12 | 127 | 91.8 | 2002 |
| 00491+5749 | STF 60 | AF | HD 004614 | 2 | 275 | 369.3 | 1991 |
| 00491+5749 | STF 60 | AG | HD 004614 | 15 | 256 | 409.8 | 2002 |
| 00491+5749 | STF 60 | AH | HD 004614 | 2 | 355 | 689.2 | 1991 |
| 00491+5749 | STF 60 | BC | HD 004614 | 3 | 237 | 152.0 | 1921 |
| 00491+5749 | STF 60 | BE | HD 004614 | 3 | 121 | 220.8 | 1991 |
| 00491+5749 | STF 60 | BH | HD 004614 | 3 | 356 | 679.9 | 1991 |
| 00491+5749 | STF 60 | FG | HD 004614 | 5 | 188 | 142.3 | 2000 |
| 00498+7027 | ENG 2 | … | HD 004635 | 9 | 277 | 90.9 | 1999 |
| 00531+6107 | BU 497 | AB | HD 005015 | 22 | 172 | 145.0 | 2003 |
| 00531+6107 | BU 497 | AD | HD 005015 | 3 | 145 | 105.6 | 1991 |
| 00531+6107 | BU 497 | AE | HD 005015 | 2 | 42 | 127.4 | 1959 |
| 00531+6107 | BU 497 | BC | HD 005015 | 7 | 162 | 0.9 | 1946 |
| 01083+5455 | STT 551 | AB | HD 006582 | 14 | 270 | 428.9 | 1998 |
| 01083+5455 | BUP 14 | AC | HD 006582 | 2 | 258 | 175.6 | 1991 |
| 01083+5455 | BUP 14 | AE | HD 006582 | 1 | 145 | 87.7 | 1907 |
| 01083+5455 | STT 551 | AF | HD 006582 | 2 | 328 | 53.2 | 1854 |
| 01083+5455 | BUP 14 | CD | HD 006582 | 2 | 115 | 4.3 | 1998 |
| 01291+2143 | HO 9 | AB | HD 008997 | 24 | 47 | 55.7 | 2003 |

Table 6—Continued

| WDS ID | Disc Desig | Pair ID | Primary Name | Nbr Obs | $\theta$ (deg) | $\rho$ (″) | Epoch |
|---|---|---|---|---|---|---|---|
| 01291+2143 | HO 9 | AD | HD 008997 | 11 | 217 | 81.8 | 1999 |
| 01291+2143 | HO 9 | BC | HD 008997 | 23 | 91 | 2.7 | 2001 |
| 01350−2955 | BU 1000 | AB-D | HD 009770 | 12 | 19 | 132.2 | 1998 |
| 01368+4124 | BUP 23 | AB | HD 009826 | 1 | 128 | 114.0 | 1909 |
| 01368+4124 | STF 554 | AC | HD 009826 | 9 | 291 | 271.6 | 2006 |
| 01425+2016 | HJ 2071 | AB | HD 010476 | 6 | 10 | 53.2 | 1998 |
| 01425+2016 | HJ 2071 | AC | HD 010476 | 13 | 4 | 154.0 | 1998 |
| 01441−1556 | BUP 25 | ... | HD 010700 | 5 | 157 | 137.0 | 2000 |
| 01477+6351 | ENG 7 | ... | HD 010780 | 10 | 176 | 45.9 | 2003 |
| 01591+3313 | ENG 9 | AB | HD 012051 | 11 | 137 | 92.8 | 2002 |
| 01591+3313 | BUP 28 | AC | HD 012051 | 3 | 26 | 91.2 | 1934 |
| 02171+3413 | DOR 66 | AB | HD 013974 | 2 | 337 | 65.4 | 1907 |
| 02442+4914 | STF 296 | AC | HD 016895 | 18 | 229 | 77.2 | 1924 |
| 02442+4914 | STF 296 | BC | HD 016895 | 13 | 215 | 73.5 | 1924 |
| 03042+6142 | KUI 11 | AB | HD 018757 | 1 | 132 | 12.7 | 1931 |
| 03091+4937 | BUP 38 | ... | HD 019373 | 1 | 132 | 146.2 | 1911 |
| 03194+0322 | STT 557 | AB | HD 020630 | 16 | 166 | 266.6 | 2002 |
| 03194+0322 | BUP 42 | BC | HD 020630 | 6 | 272 | 214.2 | 2000 |
| 03329−0927 | MBA 1 | AB | HD 022049 | 1 | 326 | 17.1 | 2001 |
| 03329−0927 | MBA 1 | AC | HD 022049 | 1 | 15 | 17.6 | 2001 |
| 03329−0927 | MBA 1 | AD | HD 022049 | 1 | 355 | 44.3 | 2001 |
| 03329−0927 | MBA 1 | AE | HD 022049 | 1 | 69 | 28.7 | 2001 |
| 03329−0927 | MBA 1 | AF | HD 022049 | 1 | 70 | 41.3 | 2001 |
| 03329−0927 | MBA 1 | AG | HD 022049 | 1 | 119 | 41.1 | 2001 |
| 03329−0927 | MBA 1 | AH | HD 022049 | 1 | 321 | 21.0 | 2001 |
| 03329−0927 | MBA 1 | AI | HD 022049 | 1 | 294 | 34.0 | 2001 |
| 03329−0927 | MBA 1 | AJ | HD 022049 | 1 | 145 | 27.9 | 2001 |
| 03329−0927 | MBA 1 | AK | HD 022049 | 1 | 208 | 38.5 | 2001 |
| 03562+5939 | ENG 16 | AB | HD 024409 | 10 | 7 | 134.7 | 1999 |
| 03562+5939 | ENG 16 | AC | HD 024409 | 7 | 50 | 193.4 | 1999 |
| 03562+5939 | BUP 48 | AF | HD 024409 | 2 | 75 | 37.6 | 1925 |
| 04033+3516 | OSO 16 | ... | HD 025329 | 3 | 240 | 16.0 | 1994 |
| 04053+2201 | STT 559 | AC | HD 025680 | 5 | 17 | 149.2 | 1997 |
| 04053+2201 | STT 559 | BC | HD 025680 | 8 | 124 | 58.2 | 1997 |
| 04076+3804 | STT 531 | AC | HD 025998 | 15 | 218 | 225.2 | 2002 |

Table 6—Continued

| WDS ID | Disc Desig | Pair ID | Primary Name | Nbr Obs | $\theta$ (deg) | $\rho$ ($''$) | Epoch |
|---|---|---|---|---|---|---|---|
| 04076+3804 | BU 545 | CD | HD 025998 | 15 | 305 | 1.3 | 1991 |
| 04153−0739 | STF 518 | AD | HD 026965 | 18 | 97 | 77.9 | 1992 |
| 04153−0739 | STF 518 | AE | HD 026965 | 10 | 8 | 211.0 | 1907 |
| 04153−0739 | STF 518 | BD | HD 026965 | 1 | 196 | 147.0 | 1922 |
| 04153−0739 | STF 518 | BE | HD 026965 | 1 | 356 | 279.5 | 1922 |
| 04155+0611 | H 4 98 | AC | HD 026923 | 8 | 48 | 230.1 | 2003 |
| 04155+0611 | H 4 98 | CD | HD 026923 | 7 | 316 | 54.3 | 1987 |
| 05188−1808 | SEE 50 | AB | HD 034721 | 4 | 234 | 45.5 | 1951 |
| 05188−1808 | SEE 50 | BC | HD 034721 | 2 | 101 | 15.8 | 1951 |
| 05191+4006 | STFB 3 | AB | HD 034411 | 1 | 274 | 29.1 | 1900 |
| 05191+4006 | STFB 3 | AC | HD 034411 | 6 | 268 | 41.7 | 1934 |
| 05191+4006 | STFB 3 | AD | HD 034411 | 29 | 349 | 203.4 | 2003 |
| 05191+4006 | DOB 4 | AE | HD 034411 | 5 | 34 | 174.8 | 2003 |
| 05191+4006 | KUI 20 | CB | HD 034411 | 2 | 351 | 27.2 | 1934 |
| 05191+4006 | DOB 4 | DE | HD 034411 | 10 | 112 | 147.7 | 2003 |
| 05226+7914 | STF 634 | AB | HD 033564 | 86 | 135 | 25.8 | 2001 |
| 05226+7914 | STF 634 | AC | HD 033564 | 3 | 333 | 173.1 | 1999 |
| 05244+1723 | S 478 | AB | HD 035296 | 35 | 271 | 102.7 | 2002 |
| 05413+5329 | BUP 82 | AC | HD 037394 | 3 | 307 | 87.1 | 1925 |
| 05413+5329 | BUP 82 | AD | HD 037394 | 2 | 262 | 688.0 | 1950 |
| 05413+5329 | BUP 82 | BE | HD 037394 | 2 | 159 | 129.7 | 1910 |
| 05460+3717 | BLL 16 | ... | HD 038230 | 3 | 104 | 104.0 | 1954 |
| 05584−0439 | A 322 | AC | HD 040397 | 13 | 304 | 195.5 | 2002 |
| 06173+0506 | ENG 26 | AB | HD 043587 | 8 | 241 | 179.7 | 2002 |
| 06173+0506 | BUP 87 | AC | HD 043587 | 1 | 265 | 58.5 | 1911 |
| 06173+0506 | BUP 87 | AD | HD 043587 | 1 | 231 | 69.3 | 1911 |
| 06467+4335 | SHJ 75 | AB | HD 048682 | 47 | 40 | 30.1 | 2007 |
| 06467+4335 | WAL 47 | AC | HD 048682 | 1 | 330 | 80.0 | 1944 |
| 07040−4337 | WRH 38 | AD | HD 053705 | 2 | 268 | 33.3 | 1999 |
| 07096+2544 | HO 519 | AB | HD 054371 | 6 | 103 | 22.2 | 1927 |
| 07096+2544 | STTA 83 | AC | HD 054371 | 32 | 80 | 120.5 | 2002 |
| 07291+3147 | A 2124 | AC | HIP 036357 | 14 | 285 | 221.7 | 1998 |
| 07291+3147 | A 2124 | CD | HIP 036357 | 3 | 270 | 102.3 | 1998 |
| 07549+1914 | ENG 33 | AB | HD 064468 | 12 | 285 | 96.8 | 2007 |
| 07549+1914 | ENG 33 | AC | HD 064468 | 10 | 65 | 125.4 | 2007 |

Table 6—Continued

| WDS ID | Disc Desig | Pair ID | Primary Name | Nbr Obs | $\theta$ (deg) | $\rho$ ($''$) | Epoch |
|---|---|---|---|---|---|---|---|
| 07549+1914 | BUP 109 | AD | HD 064468 | 3 | 28 | 43.1 | 1925 |
| 08116+3227 | STT 564 | AB | HD 068017 | 6 | 327 | 55.0 | 1915 |
| 08116+3227 | STT 564 | AC | HD 068017 | 10 | 66 | 289.3 | 2007 |
| 08122+1739 | STF1196 | AB-D | HD 068257 | 12 | 107 | 275.2 | 2007 |
| 08122+1739 | ENH 1 | AB-E | HD 068257 | 3 | 26 | 557.7 | 1991 |
| 08122+1739 | ENH 1 | AB-F | HD 068257 | 2 | 47 | 629.2 | 1894 |
| 08122+1739 | ENH 1 | AB-G | HD 068257 | 3 | 332 | 664.4 | 1991 |
| 08122+1739 | STF1196 | CD | HD 068257 | 5 | 107 | 274.8 | 1991 |
| 08122+1739 | ENH 1 | EF | HD 068257 | 6 | 106 | 218.7 | 1997 |
| 08379−0648 | HJ 99 | AB | HD 073350 | 15 | 181 | 60.2 | 2002 |
| 08379−0648 | HJ 99 | BC | HD 073350 | 3 | 215 | 9.8 | 1999 |
| 08391−2240 | BU 208 | AC | HD 073752 | 6 | 186 | 113.7 | 1999 |
| 08398+1131 | ENG 36 | AB | HD 073667 | 9 | 337 | 142.3 | 2002 |
| 08398+1131 | BUP 119 | BC | HD 073667 | 10 | 13 | 30.6 | 2002 |
| 09123+1500 | BUP 125 | AB | HD 079096 | 3 | 117 | 88.5 | 1997 |
| 09123+1500 | STT 569 | AC | HD 079096 | 11 | 217 | 204.4 | 2007 |
| 09179+2834 | ABT 6 | AB-C | HD 079969 | 1 | 56 | 152.4 | 1921 |
| 09179+2834 | ABT 6 | AB-D | HD 079969 | 3 | 133 | 166.4 | 1999 |
| 10189+4403 | ENG 43 | ... | HD 089269 | 14 | 97 | 145.2 | 2002 |
| 10306+5559 | ARN 4 | AC | HD 090839 | 3 | 294 | 233.1 | 2002 |
| 10314−5343 | HJ 4329 | ... | HD 091324 | 28 | 103 | 73.9 | 2000 |
| 10365−1214 | KUI 51 | ... | HD 091889 | 6 | 0 | 37.5 | 2007 |
| 11125+3549 | STTA108 | AB | HD 097334 | 23 | 67 | 156.5 | 2004 |
| 11125+3549 | STTA108 | AC | HD 097334 | 4 | 144 | 86.5 | 1998 |
| 11125+3549 | STTA108 | BD | HD 097334 | 2 | 88 | 34.9 | 1910 |
| 11182+3132 | POP1219 | AC | HD 098230 | 3 | 324 | 56.4 | 2007 |
| 11268+0301 | STF1540 | AC | HD 099491 | 3 | 187 | 90.5 | 1937 |
| 11387+4507 | STF1561 | AC | HD 101177 | 22 | 90 | 164.9 | 2006 |
| 11387+4507 | STF1561 | AD | HD 101177 | 2 | 76 | 704.2 | 1991 |
| 11387+4507 | STF1561 | AE | HD 101177 | 6 | 331 | 64.5 | 2001 |
| 11387+4507 | STF1561 | BC | HD 101177 | 17 | 89 | 173.3 | 2002 |
| 11387+4507 | STF1561 | BE | HD 101177 | 5 | 340 | 64.0 | 2000 |
| 11387+4507 | STF1561 | CD | HD 101177 | 3 | 72 | 552.9 | 1991 |
| 11411+3412 | STT 574 | ... | HD 101501 | 11 | 88 | 158.4 | 1998 |
| 11507+0146 | STT 576 | AB | HD 102870 | 11 | 285 | 305.3 | 1984 |

Table 6—Continued

| WDS ID | Disc Desig | Pair ID | Primary Name | Nbr Obs | $\theta$ (deg) | $\rho$ ('') | Epoch |
|---|---|---|---|---|---|---|---|
| 11507+0146 | STT 576 | AC | HD 102870 | 4 | 80 | 421.7 | 2002 |
| 12337+4121 | BU 1433 | AB | HD 109358 | 6 | 200 | 264.7 | 2007 |
| 13119+2753 | STT 578 | ... | HD 114710 | 11 | 238 | 85.8 | 1924 |
| 13168+0925 | KUI 62 | ... | HD 115383 | 1 | 89 | 34.3 | 1958 |
| 13169+1701 | BU 800 | AC | HD 115404 | 7 | 339 | 120.6 | 1999 |
| 13169+1701 | BU 800 | AD | HD 115404 | 1 | 88 | 50.8 | 1990 |
| 13184−1819 | H 6 90 | ... | HD 115617 | 13 | 37 | 376.2 | 2007 |
| 13284+1347 | STT 579 | AB | HD 117176 | 14 | 127 | 268.6 | 2002 |
| 13284+1347 | DIC 3 | AC | HD 117176 | 1 | 263 | 325.5 | 1923 |
| 13547+1824 | SHJ 169 | ... | HD 121370 | 27 | 87 | 113.3 | 2007 |
| 14190−2549 | BU 1246 | AC | HD 125276 | 9 | 117 | 82.6 | 1999 |
| 14514+1906 | STF1888 | AC | HD 131156 | 8 | 342 | 68.6 | 2000 |
| 14514+1906 | STF1888 | AD | HD 131156 | 7 | 286 | 159.6 | 2007 |
| 14514+1906 | ARN 11 | AE | HD 131156 | 6 | 99 | 269.2 | 2007 |
| 14514+1906 | ARN 12 | AF | HD 131156 | 5 | 38 | 333.7 | 2007 |
| 14514+1906 | STF1888 | BC | HD 131156 | 6 | 347 | 60.3 | 1953 |
| 14514+1906 | STF1888 | BE | HD 131156 | 10 | 99 | 274.5 | 2007 |
| 14537+2321 | COU 101 | ... | HD 131582 | 3 | 68 | 51.0 | 2000 |
| 15193+0146 | STF1930 | AC | HD 136202 | 5 | 40 | 127.2 | 1924 |
| 15193+0146 | STF1930 | AD | HD 136202 | 4 | 268 | 674.7 | 1911 |
| 15232+3017 | STF1937 | AB-C | HD 137107 | 8 | 0 | 69.2 | 1984 |
| 15232+3017 | STF1937 | AB-D | HD 137107 | 6 | 41 | 217.5 | 2000 |
| 15282−0921 | SHJ 202 | BC | HD 137763 | 4 | 99 | 152.8 | 1999 |
| 15292+8027 | STF1972 | AC | HD 139777 | 3 | 102 | 153.8 | 1983 |
| 15360+3948 | STT 298 | AB-D | HD 139341 | 4 | 232 | 189.3 | 1934 |
| 15360+3948 | STT 298 | AB-E | HD 139341 | 4 | 335 | 456.4 | 1999 |
| 15360+3948 | STT 298 | CE | HD 139341 | 4 | 337 | 335.8 | 1999 |
| 15440+0231 | A 2230 | AC | HD 140538 | 9 | 208 | 195.5 | 2002 |
| 15440+0231 | A 2230 | AD | HD 140538 | 17 | 285 | 172.8 | 2002 |
| 15440+0231 | A 2230 | CE | HD 140538 | 14 | 235 | 171.0 | 2002 |
| 15475−3755 | SEE 249 | AC | HD 140901 | 6 | 125 | 8.4 | 1956 |
| 15532+1312 | STT 583 | ... | HD 142267 | 9 | 86 | 102.4 | 1998 |
| 16010+3318 | S 676 | ... | HD 143761 | 23 | 49 | 135.3 | 2002 |
| 16133+1332 | STF2021 | AC | HD 145958 | 12 | 118 | 206.7 | 1998 |
| 16147+3352 | STF2032 | AC | HD 146361 | 11 | 93 | 24.2 | 2007 |

Table 6—Continued

| WDS ID | Disc Desig | Pair ID | Primary Name | Nbr Obs | $\theta$ (deg) | $\rho$ (″) | Epoch |
|---|---|---|---|---|---|---|---|
| 16147+3352 | STF2032 | AD | HD 146361 | 107 | 82 | 90.5 | 2006 |
| 16147+3352 | STF2032 | BD | HD 146361 | 66 | 81 | 95.0 | 1998 |
| 16156−0822 | BUP 165 | … | HD 146233 | 2 | 280 | 25.8 | 1958 |
| 16243−1338 | BUP 169 | AB | HD 147776 | 3 | 281 | 103.0 | 1909 |
| 16289+1825 | STF2052 | AC | HD 148653 | 2 | 29 | 143.3 | 1925 |
| 16364−0219 | BUP 171 | … | HD 149661 | 1 | 245 | 100.3 | 1910 |
| 17153−2636 | SHJ 243 | AD | HD 155885 | 11 | 338 | 276.6 | 1998 |
| 17153−2636 | SHJ 243 | AE | HD 155885 | 3 | 312 | 38.4 | 1998 |
| 17153−2636 | SHJ 243 | BD | HD 155885 | 6 | 339 | 284.9 | 1987 |
| 17191−4638 | BSO 13 | AC | HD 156274 | 2 | 279 | 41.8 | 1900 |
| 17191−4638 | BSO 13 | AD | HD 156274 | 1 | 30 | 47.0 | 1900 |
| 17207+3228 | DOR 1 | AB | HD 157214 | 12 | 340 | 308.1 | 2002 |
| 17207+3228 | ARN 14 | AD | HD 157214 | 1 | 59 | 395.0 | 2002 |
| 17207+3228 | ARN 14 | AE | HD 157214 | 3 | 51 | 302.2 | 2002 |
| 17207+3228 | ARN 14 | AF | HD 157214 | 3 | 104 | 383.2 | 2002 |
| 17207+3228 | DOR 1 | BC | HD 157214 | 2 | 216 | 8.8 | 1911 |
| 17350+6153 | SDR 1 | AB-D | HD 160269 | 2 | 245 | 23.9 | 1999 |
| 17419+7209 | STF2241 | AC | HD 162004 | 12 | 108 | 79.2 | 1999 |
| 17419+7209 | STF2241 | AD | HD 162004 | 1 | 84 | 100.5 | 1905 |
| 17419+7209 | STF2241 | CD | HD 162004 | 1 | 19 | 67.6 | 1908 |
| 17465+2743 | ABT 14 | AD | HD 161797 | 1 | 0 | 256.1 | 1921 |
| 17465+2743 | ABT 14 | BC-D | HD 161797 | 1 | 7 | 272.6 | 1921 |
| 18025+2619 | HO 564 | AB | HD 164922 | 6 | 326 | 96.1 | 1999 |
| 18025+2619 | HO 564 | AC | HD 164922 | 2 | 57 | 80.3 | 1924 |
| 18055+0230 | STF2272 | AC | HD 165341 | 53 | 282 | 34.9 | 1947 |
| 18055+0230 | STF2272 | AD | HD 165341 | 18 | 324 | 88.1 | 2007 |
| 18055+0230 | STF2272 | AR | HD 165341 | 24 | 29 | 155.0 | 2000 |
| 18055+0230 | STF2272 | AS | HD 165341 | 35 | 11 | 193.2 | 2000 |
| 18055+0230 | STF2272 | AT | HD 165341 | 23 | 48 | 125.9 | 2007 |
| 18055+0230 | STF2272 | AU | HD 165341 | 15 | 334 | 184.5 | 1945 |
| 18055+0230 | STF2272 | AV | HD 165341 | 37 | 246 | 138.6 | 1946 |
| 18055+0230 | STF2272 | AY | HD 165341 | 2 | 354 | 231.9 | 2000 |
| 18055+0230 | STF2272 | BC | HD 165341 | 4 | 252 | 32.4 | 1932 |
| 18055+0230 | STF2272 | BD | HD 165341 | 1 | 247 | 69.3 | 1900 |
| 18055+0230 | STF2272 | BR | HD 165341 | 22 | 50 | 121.1 | 2000 |

Table 6—Continued

| WDS ID | Disc Desig | Pair ID | Primary Name | Nbr Obs | $\theta$ (deg) | $\rho$ ($''$) | Epoch |
|---|---|---|---|---|---|---|---|
| 18055+0230 | STF2272 | BZ | HD 165341 | 2 | 163 | 68.3 | 2000 |
| 18055+0230 | STF2272 | VT | HD 165341 | 21 | 72 | 247.3 | 1946 |
| 18055+0230 | STF2272 | VW | HD 165341 | 4 | 270 | 180.9 | 1910 |
| 18055+0230 | STF2272 | VX | HD 165341 | 2 | 254 | 17.4 | 2002 |
| 18070+3034 | AC 15 | AC | HD 165908 | 6 | 59 | 96.2 | 1998 |
| 18070+3034 | AC 15 | AD | HD 165908 | 4 | 103 | 140.0 | 1998 |
| 18070+3034 | AC 15 | AE | HD 165908 | 4 | 78 | 168.8 | 1998 |
| 18070+3034 | AC 15 | AF | HD 165908 | 2 | 165 | 162.5 | 1998 |
| 18070+3034 | AC 15 | AG | HD 165908 | 5 | 360 | 177.0 | 1998 |
| 18570+3254 | BU 648 | AB-C | HD 176051 | 7 | 289 | 64.7 | 1960 |
| 18570+3254 | BU 648 | AB-D | HD 176051 | 4 | 198 | 85.0 | 1998 |
| 18570+3254 | BU 648 | AE | HD 176051 | 3 | 320 | 105.3 | 1934 |
| 18570+3254 | BU 648 | AF | HD 176051 | 3 | 87 | 100.3 | 1934 |
| 19080+1651 | ENG 66 | AB | HD 178428 | 10 | 288 | 132.3 | 2007 |
| 19121+4951 | STF2486 | AC | HD 179957 | 4 | 101 | 26.8 | 2005 |
| 19121+4951 | STF2486 | AD | HD 179957 | 6 | 102 | 192.9 | 2000 |
| 19250+1157 | STT 588 | AB | HD 182572 | 29 | 286 | 101.7 | 2006 |
| 19250+1157 | STT 588 | AC | HD 182572 | 7 | 282 | 140.6 | 2000 |
| 19250+1157 | COM 7 | AD | HD 182572 | 3 | 140 | 78.8 | 1914 |
| 19250+1157 | STT 588 | BC | HD 182572 | 24 | 266 | 44.2 | 2006 |
| 19324+6940 | STT 590 | . . . | HD 185144 | 6 | 339 | 492.8 | 1999 |
| 19464+3344 | STF2580 | AC | HD 186858 | 32 | 125 | 110.5 | 2005 |
| 19464+3344 | KPR 4 | AD | HD 186858 | 4 | 63 | 105.6 | 2005 |
| 19464+3344 | KPR 4 | AE | HD 186858 | 1 | 61 | 40.9 | 1960 |
| 19464+3344 | STF2580 | BC | HD 186858 | 21 | 139 | 101.3 | 1999 |
| 19464+3344 | STF2576 | FH | HD 186858 | 3 | 348 | 51.8 | 1998 |
| 19464+3344 | STF2576 | FI | HD 186858 | 7 | 333 | 35.1 | 1998 |
| 19464+3344 | TKA 1 | FJ | HD 186858 | 2 | 249 | 25.9 | 2006 |
| 19464+3344 | STF2576 | GI | HD 186858 | 2 | 257 | 15.9 | 1919 |
| 19510+1025 | J 124 | AB | HD 187691 | 4 | 203 | 14.4 | 1958 |
| 19510+1025 | POP1228 | AD | HD 187691 | 3 | 121 | 53.5 | 2002 |
| 19510+1025 | POP1228 | AE | HD 187691 | 3 | 147 | 84.6 | 2002 |
| 20041+1704 | STT 592 | AB | HD 190406 | 20 | 289 | 166.9 | 2002 |
| 20041+1704 | STT 592 | AC | HD 190406 | 22 | 333 | 213.9 | 2001 |
| 20041+1704 | BUP 202 | AD | HD 190406 | 6 | 360 | 83.7 | 2002 |

Table 6—Continued

| WDS ID | Disc Desig | Pair ID | Primary Name | Nbr Obs | $\theta$ (deg) | $\rho$ ($''$) | Epoch |
|---|---|---|---|---|---|---|---|
| 20041+1704 | BUP 202 | AE | HD 190406 | 3 | 51 | 169.0 | 2000 |
| 20041+1704 | BUP 202 | AF | HD 190406 | 2 | 312 | 142.3 | 2000 |
| 20041+1704 | ENG 69 | BC | HD 190406 | 19 | 204 | 151.5 | 2000 |
| 20041+1704 | STTA202 | BG | HD 190406 | 15 | 232 | 182.7 | 2002 |
| 20041+1704 | BUP 202 | CH | HD 190406 | 3 | 185 | 95.4 | 2000 |
| 20052+3829 | BU 1481 | AB | HD 190771 | 1 | 230 | 12.4 | 1906 |
| 20052+3829 | WAL 126 | AC | HD 190771 | 3 | 180 | 40.0 | 1944 |
| 20096+1648 | STF2634 | AC | HD 191499 | 2 | 312 | 74.8 | 1924 |
| 20111+1611 | ENG 71 | AB | HD 191785 | 9 | 148 | 205.4 | 2002 |
| 20111+1611 | HZG 15 | AD | HD 191785 | 7 | 267 | 40.8 | 1998 |
| 20111+1611 | BUP 205 | BC | HD 191785 | 4 | 273 | 61.7 | 2002 |
| 20140−0052 | BU 1485 | A-BC | HD 192263 | 19 | 102 | 73.1 | 2003 |
| 20140−0052 | ABT 15 | AD | HD 192263 | 1 | 244 | 71.3 | 1921 |
| 20140−0052 | J 551 | BC | HD 192263 | 4 | 265 | 0.2 | 1949 |
| 20140−0052 | BU 1485 | BC-D | HD 192263 | 8 | 65 | 23.5 | 1998 |
| 20324−0951 | BU 668 | AC | HD 195564 | 6 | 200 | 103.2 | 1921 |
| 20408+1956 | BUP 215 | AB | HD 197076 | 2 | 25 | 93.7 | 1924 |
| 21028+4551 | BU 1138 | AB[a] | HD 200560 | 54 | 170 | 0.1 | 1985 |
| 21028+4551 | BU 1138 | CA | HD 200560 | 19 | 150 | 153.1 | 2002 |
| 21028+4551 | BU 1138 | CE | HD 200560 | 6 | 250 | 5.6 | 1962 |
| 21072−1355 | BU 157 | AC | HD 200968 | 15 | 287 | 26.2 | 1999 |
| 21072−1355 | KPR 5 | AD | HD 200968 | 1 | 78 | 276.2 | 1950 |
| 21145+1000 | STF2777 | AB-C | HD 202275 | 80 | 6 | 72.5 | 2005 |
| 21180+0010 | ENG 82 | AB | HD 202751 | 10 | 44 | 157.0 | 2003 |
| 21180+0010 | TOB 317 | AD | HD 202751 | 2 | 56 | 129.1 | 2000 |
| 21180+0010 | BUP 228 | BC | HD 202751 | 3 | 117 | 12.3 | 2000 |
| 21180+0010 | LYS 44 | BD | HD 202751 | 2 | 182 | 41.8 | 2000 |
| 21198−2621 | BU 271 | AC | HD 202940 | 3 | 72 | 81.7 | 1909 |
| 21198−2621 | BU 271 | AD | HD 202940 | 1 | 62 | 247.2 | 1917 |
| 21198−2621 | BU 271 | AE | HD 202940 | 2 | 32 | 180.6 | 1999 |
| 21441+2845 | STF2822 | AC | HD 206826 | 14 | 290 | 72.6 | 1999 |
| 21441+2845 | STF2822 | AD | HD 206826 | 46 | 45 | 197.5 | 2001 |
| 21441+2845 | STF2822 | BD | HD 206826 | 23 | 46 | 198.7 | 1991 |
| 21441+2845 | ES 521 | DE | HD 206826 | 3 | 284 | 16.9 | 1999 |
| 21483−4718 | BSO 15 | . . . | HD 207129 | 29 | 351 | 75.2 | 1999 |

Table 6—Continued

| WDS ID | Disc Desig | Pair ID | Primary Name | Nbr Obs | $\theta$ (deg) | $\rho$ ($''$) | Epoch |
|---|---|---|---|---|---|---|---|
| 22159+5440 | BU 377 | AB | HD 211472 | 17 | 62 | 38.0 | 2006 |
| 22159+5440 | BU 377 | AC | HD 211472 | 6 | 52 | 35.2 | 2006 |
| 22159+5440 | BU 377 | AD | HD 211472 | 5 | 158 | 22.2 | 2006 |
| 22159+5440 | BU 377 | AQ | HD 211472 | 3 | 259 | 56.6 | 1999 |
| 22159+5440 | BU 377 | AR | HD 211472 | 2 | 256 | 62.7 | 1999 |
| 22159+5440 | BU 377 | AS | HD 211472 | 3 | 336 | 55.4 | 2006 |
| 22159+5440 | BU 377 | BC | HD 211472 | 11 | 303 | 6.7 | 2007 |
| 22159+5440 | BU 377 | QR | HD 211472 | 2 | 233 | 6.6 | 1999 |
| 22249−5748 | I 383 | . . . | HD 212330 | 2 | 237 | 81.2 | 1914 |
| 22467+1210 | HJ 301 | AC | HD 215648 | 5 | 15 | 145.0 | 1924 |
| 22514+1358 | STT 597 | AB | HD 216259 | 12 | 329 | 201.2 | 1998 |
| 22514+1358 | BUP 233 | BC | HD 216259 | 3 | 194 | 114.1 | 1998 |
| 23108+4531 | HJ 1853 | . . . | HD 218868 | 2 | 281 | 31.4 | 1905 |
| 23133+5710 | STT 599 | . . . | HD 219134 | 12 | 244 | 271.6 | 2002 |
| 23167+5313 | BUP 235 | . . . | HD 219623 | 2 | 320 | 129.3 | 1930 |
| 23399+0538 | BUP 240 | AB | HD 222368 | 4 | 305 | 119.2 | 2002 |
| 23399+0538 | BUP 240 | AC | HD 222368 | 3 | 21 | 307.3 | 2002 |
| 23524+7533 | BU 996 | AC | HD 223778 | 11 | 141 | 145.7 | 2000 |

[a]While the AB pair has an orbital solution in ORB6 and is likely physically bound, it is listed here to identify that component B is only optically associated with the sample star, which is component C. As noted in the next line, the blinking of archival images also helps identify AC as an optical pair. Thus, the physical pair AB is not associated with sample star C.

Table 7.   *Hipparcos* Component Solutions

| HD Name (1) | HIP Name (2) | Companion ID (3) | Solution Quality (4) | Companion Status (5) | Reason (6) |
|---|---|---|---|---|---|
| 000123 | 000518 | B | A | YES | 1 |
| 003196 | 002762 | B | A | YES | 1 |
| 003443 | 002941 | B | A | YES | 1 |
| 004614 | 003821 | B | A | YES | 1 |
| 007693 | 005842 | D | A | YES | 1 |
| 009770 | 007372 | B | C | YES | 1 |
| 010360 | 007751 | A | D | YES | 1 |
| 016765 | 012530 | B | B | YES | 2 |
| 018143 | 013642 | B | A | YES | 1 |
| 020010 | 014879 | B | A | YES | 1 |
| 024409 | 018413 | D | A | YES | 1 |
| 025893 | 019255 | B | A | YES | 1 |
| 035112 | 025119 | B | A | YES | 1 |
| 037572 | 026373 | B | B | YES | 3 |
| 039855 | 027922 | B | A | YES | 4 |
| 048189 | 031711 | B | A | YES | 2 |
| 053705 | 034065 | B | A | YES | 3 |
| 057095 | 035296 | B | A | YES | 1 |
| 064096 | 038382 | B | A | YES | 1 |
| 064606 | 038625 | B | C | NO | 5 |
| 068255 | 040167 | B | B | YES | 1 |
| 068255 | 040167 | C | B | YES | 1 |
| 073752 | 042430 | B | A | YES | 1 |
| 096064 | 054155 | B | C | YES | 4 |
| 096064 | 054155 | C | C | YES | 1 |
| 099491 | 055846 | B | D | YES | 3 |
| 100180 | 056242 | B | A | YES | 4 |
| 101177 | 056809 | B | A | YES | 1 |
| 111312 | 062505 | B | C | MAY | 5 |
| 115404 | 064797 | B | A | YES | 1 |
| 116442 | 065352 | B | A | YES | 3 |
| 128620 | 071683 | B | D | YES | 1 |
| 130042 | 072493 | B | A | YES | 2 |
| 131156 | 072659 | B | A | YES | 1 |

Table 7—Continued

| HD Name (1) | HIP Name (2) | Companion ID (3) | Solution Quality (4) | Companion Status (5) | Reason (6) |
|---|---|---|---|---|---|
| 133640 | 073695 | B | B | YES | 1 |
| 137107 | 075312 | B | A | YES | 1 |
| 139341 | 076382 | B | A | YES | 1 |
| 145958 | 079492 | B | A | YES | 1 |
| 146361 | 079607 | B | B | YES | 1 |
| 148653 | 080725 | B | A | YES | 1 |
| 148704 | 080925 | C | A | NO | 5 |
| 153557 | 083020 | B | B | YES | 6 |
| 155885 | 084405 | A | A | YES | 1 |
| 156274 | 084720 | B | A | YES | 1 |
| 158614 | 085667 | B | A | YES | 1 |
| 160269 | 086036 | B | A | YES | 1 |
| 162004 | 086620 | B | D | YES | 3 |
| 165341 | 088601 | B | D | YES | 1 |
| 165908 | 088745 | B | A | YES | 1 |
| . . . | 091605 | B | B | YES | 2 |
| 176051 | 093017 | B | A | YES | 1 |
| 177474 | 093825 | B | A | YES | 1 |
| 179957 | 094336 | B | A | YES | 1 |
| 184467 | 095995 | S | B | YES | 1 |
| 186858 | 097222 | B | A | YES | 1 |
| 189340 | 098416 | B | A | YES | 1 |
| 191499 | 099316 | B | A | YES | 5 |
| 200968 | 104239 | B | A | YES | 2 |
| 202275 | 104858 | B | A | YES | 1 |
| 202940 | 105312 | B | A | YES | 1 |
| 206826 | 107310 | B | A | YES | 1 |
| 212168 | 110712 | B | A | YES | 4 |

Note. — Column 5 values: 'YES' identifies physically associated companions, 'NO' marks unrelated field stars, and 'MAY' denotes unconfirmed candidates retained for further investigations. Column 6 notes: (1) Visual binary and/or spectroscopic binary with an or-

bital solution. (2) Measurements of the pair in the WDS confirm orbital motion. (3) Companion has matching independent parallax and proper motion measurements with the primary. (4) Companion has matching proper motion and photometric distance with the primary (see Table 4). (5) See individual system notes in §4.3. (6) Companionship confirmed based on matching large proper motion and proximity to the primary.

Table 8. Accelerating Proper Motion Solutions

| HD Name (1) | HIP Name (2) | H59 (3) | ====== Hipparcos ====== | | ====== Tycho-2 ====== | | $\mu$ difference | MK05[a] (9) | F07[b] (10) | Companion | Reason (12) |
|---|---|---|---|---|---|---|---|---|---|---|---|
| | | | $\mu_\alpha$ (mas yr$^{-1}$) (4) | $\mu_\delta$ (mas yr$^{-1}$) (5) | $\mu_\alpha$ (mas yr$^{-1}$) (6) | $\mu_\delta$ (mas yr$^{-1}$) (7) | ($\sigma$) (8) | | | Status (11) | |
| 000123 | 000518 | C | $247.36 \pm 0.81$ | $17.77 \pm 0.70$ | $270.6 \pm 1.6$ | $30.1 \pm 1.7$ | 16.2 | ... | ... | YES | 1 |
| 003196 | 002762 | C | $407.68 \pm 1.31$ | $-36.47 \pm 0.61$ | $415.9 \pm 0.5$ | $-23.2 \pm 0.5$ | 22.6 | ... | ... | YES | 1 |
| 003443 | 002941 | C | $1422.09 \pm 2.84$ | $-17.15 \pm 1.23$ | $1391.0 \pm 2.3$ | $-13.0 \pm 2.3$ | 11.1 | ... | ... | YES | 1 |
| 004747 | 003850 | ... | $516.74 \pm 1.04$ | $119.52 \pm 0.72$ | $518.8 \pm 1.4$ | $124.7 \pm 1.4$ | 4.0 | ... | ... | YES | 1 |
| 007788[c] | 005896 | C | $411.11 \pm 0.50$ | $127.43 \pm 0.48$ | $404.9 \pm 3.3$ | $108.3 \pm 3.0$ | 6.6 | ... | ... | YES | 1 |
| 010307 | 007918 | G | $791.35 \pm 0.65$ | $-180.16 \pm 0.47$ | $806.6 \pm 1.0$ | $-152.2 \pm 1.0$ | 31.8 | Y | Y | YES | 1 |
| 010360 | 007751 | C | $286.10 \pm 1.01$ | $16.66 \pm 1.41$ | $302.6 \pm 1.4$ | $-14.1 \pm 1.3$ | 24.8 | ... | ... | YES | 1 |
| 013445 | 010138 | ... | $2092.84 \pm 0.50$ | $654.32 \pm 0.55$ | $2150.3 \pm 2.5$ | $673.2 \pm 2.4$ | 24.3 | Y | Y | YES | 2 |
| 014802 | 011072 | G | $197.34 \pm 0.77$ | $-4.39 \pm 0.51$ | ... | ... | ... | ... | ... | YES | 1 |
| 017382 | 013081 | G | $264.17 \pm 1.24$ | $-127.75 \pm 0.81$ | $274.5 \pm 1.1$ | $-122.6 \pm 1.1$ | 9.6 | Y | Y | YES | 1 |
| 018143 | 013642 | C | $262.72 \pm 1.86$ | $-191.44 \pm 1.15$ | $274.0 \pm 1.7$ | $-185.4 \pm 1.6$ | 7.1 | ... | ... | YES | 1 |
| 024409 | 018413 | C | $-284.06 \pm 0.90$ | $159.32 \pm 0.96$ | $-265.1 \pm 0.9$ | $169.7 \pm 1.0$ | 23.5 | ... | ... | YES | 1 |
| 025998 | 019335 | G | $163.93 \pm 0.65$ | $-203.52 \pm 0.55$ | $166.8 \pm 1.2$ | $-203.1 \pm 1.3$ | 2.4 | ... | Y | MAY | 3 |
| 026491 | 019233 | ... | $185.91 \pm 0.46$ | $336.76 \pm 0.51$ | $196.7 \pm 1.1$ | $333.7 \pm 1.2$ | 10.1 | Y | Y | YES | 4 |
| 035112 | 025119 | C | $54.20 \pm 1.44$ | $-139.41 \pm 0.94$ | $69.7 \pm 1.3$ | $-152.1 \pm 1.3$ | 14.5 | ... | ... | YES | 1 |
| 036705 | 025647 | G | $32.14 \pm 0.53$ | $150.97 \pm 0.73$ | $48.9 \pm 1.3$ | $137.6 \pm 1.2$ | 17.0 | Y | Y | YES | 2 |
| 039587 | 027913 | ... | $-163.17 \pm 1.06$ | $-98.92 \pm 0.60$ | $-174.6 \pm 0.7$ | $-89.9 \pm 0.7$ | 16.8 | Y | Y | YES | 1 |
| 040397 | 028267 | ... | $71.23 \pm 0.93$ | $-203.34 \pm 0.66$ | $73.6 \pm 1.0$ | $-206.5 \pm 1.1$ | 3.7 | ... | ... | YES | 2 |
| 043587 | 029860 | ... | $-189.37 \pm 0.71$ | $171.18 \pm 0.50$ | $-195.4 \pm 1.0$ | $164.6 \pm 1.0$ | 8.9 | Y | Y | YES | 1 |
| 045088 | 030630 | ... | $-119.32 \pm 1.06$ | $-164.06 \pm 0.76$ | $-115.3 \pm 0.8$ | $-167.8 \pm 0.8$ | 6.0 | ... | Y | YES | 1 |
| 048189 | 031711 | C | $-50.08 \pm 0.73$ | $72.69 \pm 0.66$ | $-26.0 \pm 3.8$ | $72.4 \pm 3.4$ | 6.3 | ... | ... | YES | 2 |
| 052698 | 033817 | G | $206.58 \pm 0.48$ | $40.89 \pm 0.72$ | $203.3 \pm 1.4$ | $37.5 \pm 1.3$ | 3.5 | ... | Y | YES | 5 |
| 053680[d] | 034052 | G | $-75.43 \pm 0.75$ | $401.32 \pm 2.10$ | $-93.0 \pm 1.2$ | $395.3 \pm 1.1$ | 14.9 | Y | Y | YES | 5 |
| ...... | 036357 | G | $160.40 \pm 1.56$ | $174.68 \pm 1.02$ | $159.7 \pm 1.0$ | $175.8 \pm 1.1$ | 1.1 | ... | Y | MAY | 3 |
| 058946[e] | 036366 | ... | $159.33 \pm 1.26$ | $193.82 \pm 0.56$ | $157.2 \pm 0.6$ | $186.9 \pm 0.6$ | 11.7 | Y | Y | YES | 6 |
| 063077 | 037853 | G | $-220.83 \pm 0.46$ | $1722.89 \pm 0.55$ | $-274.1 \pm 1.1$ | $1687.0 \pm 1.1$ | 58.4 | Y | Y | YES | 7 |
| 064096 | 038382 | C | $-68.46 \pm 1.11$ | $-344.83 \pm 1.03$ | $-60.0 \pm 0.7$ | $-338.9 \pm 0.7$ | 9.6 | ... | ... | YES | 1 |
| 064606 | 038625 | C | $-251.57 \pm 2.07$ | $-62.07 \pm 1.48$ | $-259.1 \pm 1.5$ | $-47.7 \pm 1.4$ | 10.4 | ... | ... | YES | 1 |
| 065430 | 039064 | ... | $180.46 \pm 0.91$ | $-544.36 \pm 0.50$ | $180.1 \pm 1.1$ | $-550.8 \pm 1.0$ | 6.4 | Y | Y | YES | 1 |
| 067199 | 039342 | ... | $-157.34 \pm 0.47$ | $-130.52 \pm 0.66$ | $-155.9 \pm 1.4$ | $-126.7 \pm 1.3$ | 3.1 | ... | ... | MAY | 8 |
| 068017 | 040118 | ... | $-460.69 \pm 1.17$ | $-644.64 \pm 0.61$ | $-464.7 \pm 0.8$ | $-646.5 \pm 0.8$ | 4.1 | ... | ... | MAY | 8 |
| 068255 | 040167 | C | $28.29 \pm 2.00$ | $-150.94 \pm 1.15$ | $79.8 \pm 1.7$ | $-129.4 \pm 1.7$ | 28.7 | ... | ... | YES | 2 |

| HD Name (1) | HIP Name (2) | H59 (3) | ====== Hipparcos ====== | | ====== Tycho-2 ====== | | $\mu$ difference | MK05[a] (9) | F07[b] (10) | Companion | Reason (12) |
| | | | $\mu_\alpha$ (mas yr$^{-1}$) (4) | $\mu_\delta$ (mas yr$^{-1}$) (5) | $\mu_\alpha$ (mas yr$^{-1}$) (6) | $\mu_\delta$ (mas yr$^{-1}$) (7) | $(\sigma)$ (8) | | | Status (11) | |
|---|---|---|---|---|---|---|---|---|---|---|---|
| 072760 | 042074 | … | $-194.28 \pm 1.05$ | $23.42 \pm 0.82$ | $-197.5 \pm 1.0$ | $18.9 \pm 0.9$ | 5.9 | Y | Y | YES | 2 |
| 079969 | 045617 | … | $49.78 \pm 1.15$ | $-507.62 \pm 0.51$ | $52.8 \pm 1.2$ | $-510.5 \pm 1.1$ | 3.6 | … | … | YES | 1 |
| 082885 | 047080 | … | $-730.05 \pm 0.71$ | $-260.62 \pm 0.46$ | $-723.0 \pm 1.3$ | $-247.8 \pm 1.4$ | 10.6 | Y | Y | YES | 1 |
| 101177 | 056809 | C | $-593.87 \pm 0.68$ | $14.80 \pm 0.52$ | $-577.3 \pm 1.0$ | $1.5 \pm 1.0$ | 21.2 | … | … | YES | 1 |
| 110833 | 062145 | O | $-378.76 \pm 0.59$ | $-183.86 \pm 0.60$ | $-389.8 \pm 1.1$ | $-176.9 \pm 1.2$ | 11.6 | … | … | YES | 1 |
| 111312 | 062505 | C | $79.70 \pm 1.90$ | $44.66 \pm 1.46$ | $59.9 \pm 1.9$ | $39.2 \pm 1.7$ | 10.9 | … | … | YES | 1 |
| 113283 | 064690 | … | $-221.35 \pm 0.55$ | $-155.49 \pm 0.57$ | $-226.1 \pm 2.2$ | $-160.4 \pm 2.0$ | 3.3 | … | … | MAY | 8 |
| 113449 | 063742 | O | $-189.79 \pm 1.16$ | $-219.55 \pm 1.10$ | $-189.6 \pm 0.7$ | $-223.2 \pm 0.8$ | 3.3 | … | … | YES | 2 |
| 115404 | 064797 | C | $631.21 \pm 0.90$ | $-260.84 \pm 0.63$ | $623.9 \pm 1.2$ | $-259.0 \pm 1.1$ | 6.3 | … | … | YES | 1 |
| 120136 | 067275 | … | $-480.33 \pm 0.61$ | $54.18 \pm 0.47$ | $-480.8 \pm 0.4$ | $50.4 \pm 0.4$ | 8.1 | Y | Y | YES | 1 |
| 120690 | 067620 | … | $-580.93 \pm 0.95$ | $-244.87 \pm 0.71$ | $-584.2 \pm 3.0$ | $-290.9 \pm 3.0$ | 15.4 | Y | Y | YES | 1 |
| 120780 | 067742 | G | $-583.27 \pm 0.91$ | $-60.27 \pm 0.59$ | $-596.4 \pm 1.1$ | $-45.9 \pm 1.0$ | 18.7 | Y | Y | YES | 5 |
| 122742 | 068682 | O | $85.26 \pm 0.64$ | $-304.04 \pm 0.47$ | $91.1 \pm 1.2$ | $-307.5 \pm 1.2$ | 5.7 | … | … | YES | 1 |
| 125276 | 069965 | … | $-356.39 \pm 0.81$ | $366.76 \pm 0.78$ | $-351.9 \pm 1.1$ | $365.3 \pm 1.1$ | 4.3 | … | … | MAY | 4 |
| 128642 | 070857 | O | $-73.73 \pm 0.65$ | $-132.93 \pm 0.59$ | $-72.0 \pm 0.8$ | $-129.6 \pm 0.9$ | 4.3 | … | … | YES | 1 |
| 130042 | 072493 | C | $-107.28 \pm 0.70$ | $-320.91 \pm 0.95$ | $-109.4 \pm 1.7$ | $-330.8 \pm 1.7$ | 5.9 | … | … | YES | 6 |
| 131582 | 072875 | G | $-824.15 \pm 1.27$ | $2.29 \pm 1.07$ | $-824.4 \pm 0.8$ | $11.0 \pm 0.8$ | 8.1 | Y | Y | YES | 5 |
| 131923 | 073241 | G | $-15.47 \pm 0.89$ | $-337.07 \pm 0.88$ | $-15.5 \pm 1.0$ | $-327.8 \pm 1.2$ | 7.7 | Y | Y | YES | 1 |
| 133640 | 073695 | C | $-436.24 \pm 1.20$ | $18.94 \pm 1.17$ | $-443.7 \pm 1.2$ | $9.9 \pm 1.2$ | 9.8 | … | Y | YES | 1 |
| 137763 | 075718 | G | $72.69 \pm 1.09$ | $-363.37 \pm 0.79$ | $76.4 \pm 1.0$ | $-359.0 \pm 1.0$ | 5.5 | … | Y | YES | 1 |
| 139341 | 076382 | C | $-482.47 \pm 2.25$ | $27.52 \pm 1.47$ | $-455.2 \pm 0.8$ | $51.0 \pm 0.9$ | 20.1 | … | … | YES | 1 |
| 140538 | 077052 | … | $-44.96 \pm 0.95$ | $-144.73 \pm 0.78$ | $-48.0 \pm 0.8$ | $-147.2 \pm 0.8$ | 4.4 | … | … | YES | 6 |
| 144287 | 078709 | G | $-488.79 \pm 0.58$ | $696.64 \pm 0.73$ | $-532.9 \pm 1.0$ | $683.0 \pm 1.1$ | 45.8 | Y | Y | YES | 1 |
| 145825 | 079578 | … | $-81.41 \pm 0.89$ | $-252.61 \pm 1.04$ | $-78.4 \pm 1.2$ | $-256.9 \pm 1.2$ | 4.4 | … | … | YES | 1 |
| 146361 | 079607 | C | $-266.47 \pm 0.86$ | $-86.88 \pm 1.12$ | $-289.0 \pm 3.0$ | $-85.1 \pm 2.8$ | 7.5 | … | … | YES | 1 |
| 147584 | 080686 | … | $199.89 \pm 0.31$ | $110.77 \pm 0.51$ | $197.8 \pm 0.7$ | $111.5 \pm 0.7$ | 3.2 | … | … | YES | 1 |
| 148653 | 080725 | C | $-345.93 \pm 1.56$ | $385.98 \pm 1.36$ | $-339.9 \pm 1.2$ | $383.3 \pm 1.3$ | 4.3 | … | … | YES | 1 |
| 148704 | 080925 | C | $-428.05 \pm 1.47$ | $-333.41 \pm 1.43$ | $-427.6 \pm 1.1$ | $-326.2 \pm 1.2$ | 5.1 | … | … | YES | 1 |
| 153557 | 083020 | C | $-146.90 \pm 1.15$ | $272.21 \pm 1.35$ | $-153.9 \pm 1.3$ | $267.8 \pm 1.4$ | 6.2 | … | … | YES | 2 |
| 156274 | 084720 | C | $1035.25 \pm 1.38$ | $109.22 \pm 0.65$ | $1053.5 \pm 2.1$ | $144.0 \pm 2.0$ | 19.4 | … | … | YES | 1 |
| 158614 | 085667 | C | $-126.64 \pm 1.72$ | $-172.00 \pm 0.91$ | $-126.7 \pm 1.1$ | $-179.6 \pm 1.1$ | 6.9 | … | … | YES | 1 |
| 160269 | 086036 | C | $277.38 \pm 0.54$ | $-525.62 \pm 0.60$ | $265.4 \pm 3.3$ | $-520.9 \pm 3.2$ | 3.9 | … | … | YES | 1 |

| HD Name (1) | HIP Name (2) | H59 (3) | ====== Hipparcos ====== | | ====== Tycho-2 ====== | | $\mu$ difference ($\sigma$) (8) | MK05[a] (9) | F07[b] (10) | Companion Status (11) | Reason (12) |
|---|---|---|---|---|---|---|---|---|---|---|---|
| | | | $\mu_\alpha$ (mas yr$^{-1}$) (4) | $\mu_\delta$ (mas yr$^{-1}$) (5) | $\mu_\alpha$ (mas yr$^{-1}$) (6) | $\mu_\delta$ (mas yr$^{-1}$) (7) | | | | | |
| 161198 | 086722 | G | $-123.15 \pm 1.00$ | $-619.84 \pm 0.88$ | $-123.1 \pm 1.1$ | $-628.0 \pm 1.0$ | 8.2 | Y | Y | YES | 1 |
| 161797 | 086974 | ... | $-291.42 \pm 0.49$ | $-750.00 \pm 0.53$ | $-310.3 \pm 0.4$ | $-750.3 \pm 0.4$ | 38.5 | Y | Y | YES | 1 |
| 165341 | 088601 | C | $124.56 \pm 1.15$ | $-962.66 \pm 0.91$ | $276.3 \pm 2.3$ | $-1091.8 \pm 2.3$ | 86.6 | ... | ... | YES | 1 |
| 165401 | 088622 | ... | $-30.66 \pm 0.88$ | $-322.06 \pm 0.75$ | $-26.0 \pm 1.1$ | $-316.7 \pm 1.1$ | 6.5 | Y | Y | YES | 8 |
| 165499 | 089042 | G | $-77.60 \pm 0.59$ | $234.68 \pm 0.44$ | $-81.5 \pm 0.9$ | $221.2 \pm 0.9$ | 15.6 | Y | Y | YES | 5 |
| 167425 | 089805 | ... | $38.89 \pm 0.57$ | $-276.16 \pm 0.51$ | $37.9 \pm 1.4$ | $-280.4 \pm 1.4$ | 3.1 | ... | ... | YES | 9 |
| 177474 | 093825 | C | $96.93 \pm 2.44$ | $-279.67 \pm 1.34$ | $87.7 \pm 1.2$ | $-283.9 \pm 1.2$ | 4.9 | ... | ... | YES | 1 |
| 179957 | 094336 | C | $-205.02 \pm 0.97$ | $624.33 \pm 0.89$ | $-209.5 \pm 1.3$ | $622.2 \pm 1.4$ | 3.8 | ... | ... | YES | 1 |
| 181321 | 095149 | G | $78.88 \pm 4.08$ | $-108.93 \pm 2.50$ | $87.6 \pm 1.2$ | $-86.4 \pm 1.3$ | 9.3 | Y | Y | YES | 5 |
| 186858 | 097222 | C | $13.30 \pm 1.07$ | $-440.57 \pm 1.35$ | $18.9 \pm 2.6$ | $-445.8 \pm 2.4$ | 3.1 | ... | ... | YES | 1 |
| 189340 | 098416 | C | $-246.73 \pm 2.31$ | $-392.36 \pm 1.63$ | $-282.0 \pm 1.0$ | $-399.7 \pm 1.0$ | 15.9 | ... | ... | YES | 1 |
| 190771 | 098921 | ... | $263.35 \pm 0.46$ | $111.57 \pm 0.46$ | $259.2 \pm 1.1$ | $115.7 \pm 1.1$ | 5.3 | Y | Y | YES | 8 |
| 191408 | 099461 | ... | $456.89 \pm 0.89$ | $-1574.91 \pm 0.61$ | $458.4 \pm 1.1$ | $-1569.3 \pm 1.1$ | 5.3 | Y | Y | YES | 6 |
| 191499 | 099316 | C | $3.79 \pm 1.00$ | $175.79 \pm 1.02$ | $3.9 \pm 1.4$ | $167.9 \pm 1.4$ | 5.6 | ... | ... | YES | 2 |
| 193664 | 100017 | ... | $468.52 \pm 0.55$ | $296.81 \pm 0.41$ | $472.5 \pm 1.1$ | $296.1 \pm 1.2$ | 3.7 | ... | ... | MAY | 8 |
| 195564 | 101345 | ... | $307.59 \pm 0.88$ | $106.07 \pm 0.66$ | $307.2 \pm 0.7$ | $103.7 \pm 0.7$ | 3.4 | ... | ... | YES | 2 |
| 195987 | 101382 | O | $-156.89 \pm 0.53$ | $452.80 \pm 0.47$ | $-154.3 \pm 0.9$ | $454.5 \pm 0.9$ | 3.4 | ... | ... | YES | 1 |
| 200560 | 103859 | ... | $402.30 \pm 0.66$ | $141.72 \pm 0.57$ | $396.1 \pm 0.8$ | $141.0 \pm 0.8$ | 7.8 | Y | Y | YES | 2 |
| 200968 | 104239 | C | $382.32 \pm 1.31$ | $-46.55 \pm 0.60$ | $382.3 \pm 0.9$ | $-39.9 \pm 0.8$ | 8.3 | ... | ... | YES | 6 |
| 202940 | 105312 | C | $-582.35 \pm 1.11$ | $-357.67 \pm 0.62$ | $-568.3 \pm 1.5$ | $-353.7 \pm 1.5$ | 9.7 | ... | ... | YES | 1 |
| 203985 | 105911 | G | $264.07 \pm 1.28$ | $184.71 \pm 0.97$ | $253.2 \pm 1.1$ | $179.6 \pm 1.2$ | 9.5 | Y | Y | YES | 5 |
| 206826 | 107310 | C | $260.33 \pm 0.61$ | $-242.73 \pm 0.57$ | $277.4 \pm 2.7$ | $-251.1 \pm 2.6$ | 7.1 | ... | ... | YES | 1 |
| 211415 | 110109 | ... | $439.86 \pm 0.53$ | $-632.60 \pm 0.43$ | $436.8 \pm 0.9$ | $-632.8 \pm 0.9$ | 3.4 | ... | ... | YES | 6 |
| 212330 | 110649 | ... | $180.71 \pm 0.41$ | $-331.27 \pm 0.38$ | $150.6 \pm 1.1$ | $-344.9 \pm 1.1$ | 30.0 | Y | Y | YES | 8 |
| 214953 | 112117 | ... | $6.15 \pm 0.63$ | $-331.43 \pm 0.51$ | $3.1 \pm 1.1$ | $-326.9 \pm 1.0$ | 5.3 | Y | Y | YES | 9 |
| 223778 | 117712 | ... | $341.82 \pm 0.53$ | $41.88 \pm 0.47$ | $325.8 \pm 1.0$ | $45.6 \pm 1.1$ | 16.4 | Y | Y | YES | 1 |
| 224930 | 000171 | X | $778.59 \pm 2.81$ | $-918.72 \pm 1.81$ | $829.9 \pm 1.2$ | $-989.4 \pm 1.1$ | 43.1 | Y | Y | YES | 1 |

Note. — Column 11 values: 'YES' identifies physically associated companions and 'MAY' denotes unconfirmed candidates retained for further investigations. Column 12 notes: (1) Visual and/or spectroscopic binary with an orbital solution. (2) Nearby companion is likely responsible for the proper motion acceleration (see §4.3). (3) *Hipparcos* G flag and the $\chi^2$ test in Frankowski et al. (2007) suggest an unseen companion, but because the *Hipparcos* and Tycho-2 proper

motions differ by less than $3\sigma$, this is retained as a candidate for further investigations. (4) See individual system notes in §4.3. (5) *Hipparcos* 'G' flag and a greater than $3\sigma$ difference in proper motion indicate an unseen companion. (6) Nearby companion with evidence of orbital motion based on WDS measurements is likely responsible for the proper motion acceleration. (7) Radial velocity variations indicate a spectroscopic binary, but not enough observations exist to derive an orbit. (8) Greater than $3\sigma$ difference in proper motion is the only evidence of a companion. The companion is considered physical if it also passed the $\chi^2$ test in Frankowski et al. (2007), otherwise is retained as a candidate. (9) Nearby companion with matching photometric distance is likely likely responsible for the proper motion acceleration (see Table 10).

[a]'Y' in this column indicates that this was identified as a proper-motion binary by Makarov & Kaplan (2005)

[b]'Y' in this column indicates that this was identified as a proper-motion binary by Frankowski et al. (2007)

[c]Wide companion, $319''$ away from HD 7693.

[d]Wide companion, $185''$ away from HD 53705.

[e]Wide companion, $756''$ away from HIP 36357.

Table 9.   *Hipparcos* Orbital Solutions

| HD Name (1) | HIP Name (2) | Period (days) (3) | Companion Status (4) | Reason Code (5) |
|---|---|---|---|---|
| 001273 | 001349 | 411.4 | YES | 1 |
| 006582 | 005336 | 7816.0 | YES | 1 |
| 010476 | 007981 | 207.3 | NO | 2 |
| 013974 | 010644 | 10.0 | YES | 1 |
| 014214 | 010723 | 93.5 | YES | 1 |
| 016739 | 012623 | 331.0 | YES | 1 |
| 032850 | 023786 | 204.4 | YES | 1 |
| 110833 | 062145 | 270.2 | YES | 1 |
| 112914 | 063406 | 736.8 | YES | 1 |
| 113449 | 063742 | 231.2 | YES | 3 |
| 121370 | 067927 | 494.2 | YES | 1 |
| 122742 | 068682 | 3614.9 | YES | 1 |
| 128642 | 070857 | 179.7 | YES | 1 |
| 131511 | 072848 | 125.4 | YES | 1 |
| 142373 | 077760 | 51.3 | NO | 2 |
| 143761 | 078459 | 78.0 | MAY | 3 |
| 160346 | 086400 | 83.9 | YES | 1 |
| 195987 | 101382 | 57.3 | YES | 1 |
| 203244 | 105712 | 1060.6 | NO | 2 |

Note. — Column 4 values: 'YES' identifies physically associated companions, 'NO' marks unrelated field stars, and 'MAY' denotes unconfirmed candidates retained for further investigations. Column 5 notes: (1) Visual binary and/or spectroscopic binary with an orbital solution. (2) The pho-

tocentric orbital solution with a fairly high incli-
nation is refuted by constant radial velocities mea-
sured over several years (Nidever et al. 2002; Abt &
Biggs 1972; Gontcharov 2006; Latham et al. 2010).
*Hipparcos* and Tycho-2 proper motions match to
within $3\sigma$, providing further evidence against a
companion for the moderately inclined orbits. (3)
See individual system notes in § 4.3.

Table 10.   WDS Companions Confirmed by Photometry

| WDS ID (1) | Pair ID (2) | HD Name (3) | Sep (″) (4) | PA (deg) (5) | Epoch (year) (6) | N (7) | ΔT (years) (8) | $\mu_\alpha$ (9) | $\mu_\delta$ (mas yr$^{-1}$) (10) | ref (11) | V (12) | ref (13) | R (14) | ref (15) | I (16) | ref (17) | J (18) | H (19) | K (20) | D (pc) (21) | Error (pc) (22) |
|---|---|---|---|---|---|---|---|---|---|---|---|---|---|---|---|---|---|---|---|---|---|
| 00458−4733 | AB | 004391 | 16.6 | 307 | 1993 | 3 | 98 | ... | ... | | 12.7 | 1 | ... | | ... | | 8.44 | 7.95 | 7.64 | 12.0 | 1.9 |
| 03562+5939 | AE | 024409 | 8.9 | 228 | 1999 | 3 | 92 | ... | ... | | 12.9 | 2 | ... | | ... | | 9.22 | 8.66 | 8.50 | 26.3 | 4.1 |
| 05287−6527 | AB | 036705 | 9.2 | 345 | 1998 | 3 | 69 | ... | ... | | 13.0 | 2 | ... | | ... | | 8.17 | 7.66 | 7.34 | 7.2[a] | 1.1 |
| 05545−1942 | AB | 039855 | 10.6 | 20 | 2003 | 12 | 127 | ... | ... | | ... | | ... | | 9.13 | 3 | 7.99 | 7.40 | 7.22 | 19.0 | 2.9 |
| 05584−0439 | AD | 040397 | 89.3 | 313 | 2000 | 2 | 40 | 78 | −216 | 3 | ... | | 15.15 | 4 | 13.28 | 3 | 11.11 | 10.64 | 10.31 | 19.0 | 3.6 |
| 10172+2306 | AB | 089125 | 7.7 | 299 | 2005 | 30 | 154 | ... | ... | | 11.4 | 2 | ... | | ... | | 8.36 | 7.79 | 7.59 | 24.9 | 3.9 |
| 15193+0146 | AB | 136202 | 11.4 | 35 | 2000 | 45 | 175 | 376 | −517 | 5 | 10.2 | 2 | ... | | ... | | 7.49 | 6.91 | 6.75 | 20.6 | 3.5 |
| 16371+0015 | AB | 149806 | 6.3 | 19 | 2000 | 3 | 54 | ... | ... | | 13.0 | 2 | ... | | ... | | 8.09 | 7.57 | 7.28 | 6.7[a] | 1.0 |
| 17465+2743 | A-BC | 161797 | 34.9 | 248 | 2007 | 108 | 226 | −333 | −753 | 5 | 10.4 | 2 | ... | | ... | | 6.52[b] | 5.92[b] | 5.70[b] | 6.2 | 0.9 |
| 18197−6353 | AB | 167425 | 7.8 | 352 | 2000 | 6 | 103 | ... | ... | | 10.8 | 2 | ... | | ... | | 8.00 | 7.42 | 7.19 | 23.7 | 3.7 |
| 20096+1648 | AB | 191499 | 4.0 | 14 | 2003 | 51 | 221 | ... | ... | | 9.4 | 2 | ... | | ... | | 6.21 | 5.88 | 5.95 | 12.9[a] | 3.5 |
| 22426−4713 | AB | 214953 | 7.5 | 125 | 1998 | 15 | 104 | 6 | −320 | 6 | 10.3 | 2 | ... | | ... | | 7.44 | 6.86 | 6.63 | 17.8 | 2.8 |

Note. — Columns 1, 2, and 4–8 are from the WDS, and the astrometry listed in Columns 4–6 is for the most recent observation in the catalog as of 2008 July. Reference codes for columns 11, 13, 15, 17 are as follows: (1) Differential photometry with CTIO 0.9m $V$-band image; (2) Visual Double Stars in *Hipparcos* (Dommanget & Nys 2000); (3) The USNO B 1.0 Catalog (Monet et al. 2003); (4) The NOMAD Catalog (Zacharias et al. 2004a); (5) The LSPM North catalog (Lépine & Shara 2005); (6) The NLTT Catalog (Luyten 1979).

[a] See §4.3 for a discussion of these photometric distance estimates and the status of these companions.

[b] The companion is a visual binary. The *Hipparcos* input catalog lists individual $V$ magnitudes of 10.2 and 10.7 for the components. 2MASS lists combined magnitudes for the two stars. The 2MASS magnitudes have been accordingly adjusted for one component.

Table 11.   Visual Orbit Solutions

| WDS Coord (1) | HD Name (2) | HIP Name (3) | Discoverer ID (4) | N (5) | ΔT (years) (6) | P (7) | G (8) | Ref (9) | SB (10) | Comp Stat (11) |
|---|---|---|---|---|---|---|---|---|---|---|
| | | | | | | | | | | |

Resolved-pair Visual Orbits

| WDS Coord (1) | HD Name (2) | HIP Name (3) | Discoverer ID (4) | N (5) | ΔT (years) (6) | P (7) | G (8) | Ref (9) | SB (10) | Comp Stat (11) |
|---|---|---|---|---|---|---|---|---|---|---|
| 00022+2705 | 224930 | 000171 | BU 733AB | 178 | 127 | 26.28 y | 2 | Sod1999 | 1 | YES |
| 00063+5826 | 000123 | 000518 | STF3062 | 582 | 183 | 106.7 y | 2 | Sod1999 | ... | YES |
| 00352−0336 | 003196 | 002762 | HO 212AB | 201 | 121 | 6.89 y | 1 | Msn2005 | 2 | YES |
| 00373−2446 | 003443 | 002941 | BU 395 | 163 | 132 | 25.09 y | 1 | Pbx2000b | 2 | YES |
| 00490+1656 | 004676 | 003810 | 64 PscAa,Ab | ... | ... | 13.82 d | 8 | Bod1999b | 2 | YES |
| 00491+5749 | 004614 | 003821 | STF 60AB | 1029 | 228 | 480 y | 3 | Str1969a | ... | YES |
| 01083+5455 | 006582 | 005336 | WCK 1Aa,Ab | 14 | 31 | 21.75 y | 4 | Dru1995 | 1 | YES |
| 01158−6853 | 007788 | 005896 | HJ 3423AB | 70 | 165 | 857.0 y | 5 | Sca2005b | ... | YES |
| 01158−6853 | 007693 | 005842 | I 27CD | 69 | 106 | 85.2 y | 3 | Sod1999 | ... | YES |
| 01350−2955 | 009770 | 007372 | DAW 31AB | 88 | 79 | 4.56 y | 1 | Msn1999c | ... | YES |
| 01350−2955 | 009770 | 007372 | BU 1000AB-C | 44 | 120 | 111.8 y | 4 | Nwb1969a | ... | YES |
| 01398−5612 | 010361 | 007751 | DUN 5 | 159 | 177 | 483.66 y | 5 | vAb1957 | ... | YES |
| 01418+4237 | 010307 | 007918 | MCY 2 | 6 | 13 | 19.5 y | 4 | Sod1999 | 1 | YES |
| 02104−5049 | 013445 | 010138 | ESG 1 | 4 | 5 | 69.7 y | 5 | Lgr2006 | ... | YES |
| 02171+3413 | 013974 | 010644 | MKT 5Aa,Ab | 21 | 2 | 10.02 d | 1 | MkT1995 | 2 | YES |
| 02422+4012 | 016739 | 012623 | MCA 8 | 41 | 23 | 330.98 d | 8 | Bgn2006 | 2 | YES |
| 02442+4914 | 016895 | 012777 | STF 296AB | 71 | 224 | 2720 y | 5 | Hop1958 | ... | YES |
| 02556+2652 | 018143 | 013642 | STF 326AB | 76 | 175 | ... | 5 | Hop1967 | ... | YES |
| 03121−2859 | 020010 | 014879 | HJ 3555 | 94 | 167 | 269 y | 4 | Sod1999 | ... | YES |
| 03128−0112 | 019994 | 014954 | HJ 663 | 15 | 154 | 1420 y | 5 | Hle1994 | ... | YES |
| 03236−4005 | 021175 | 015799 | I 468 | 3 | 56 | 111 y | 5 | Sod1999 | ... | YES |
| 04076+3804 | 025893 | 019255 | STT 531AB | 171 | 153 | 590 y | 5 | Hei1986b | ... | YES |

| WDS Coord (1) | HD Name (2) | HIP Name (3) | Discoverer ID (4) | $N$ (5) | $\Delta T$ (years) (6) | $P$ (7) | G (8) | Ref (9) | SB (10) | Comp Stat (11) |
|---|---|---|---|---|---|---|---|---|---|---|
| 04153−0739 | 026976 | 000000 | STF 518BC | 173 | 149 | 252.1 y | 4 | Hei1974c | ... | YES |
| 05074+1839 | 032923 | 023835 | A 3010 | 19 | 76 | 1.19 y | 3 | Egg1956 | ... | NO |
| 05226+0236 | 035112 | 025119 | A 2641 | 21 | 88 | 93 y | 4 | Sod1999 | ... | YES |
| 06173+0506 | 043587 | 029860 | CAT 1Aa,Ab | 4 | 3 | 28.8 y | 5 | Cat2006 | 1 | YES |
| 06262+1845 | 045088 | 030630 | BU 1191 | 14 | 106 | 600 y | 5 | Hle1994 | ... | YES |
| 07175−4659 | 057095 | 035296 | I 7 | 60 | 107 | 94.0 y | 4 | Hei1995 | ... | YES |
| 07518−1354 | 064096 | 038382 | BU 101 | 229 | 132 | 22.70 y | 2 | Pbx2000b | 2 | YES |
| 08122+1739 | 068257 | 040167 | STF1196AB | 1133 | 182 | 59.58 y | 1 | WSI2006b | ... | YES |
| 08122+1739 | 068257 | 040167 | STF1196AB-C | 515 | 207 | 1115 y | 4 | Hei1996b | ... | YES |
| 08122+1739 | 068256 | 040167 | HUT 1Ca,Cb | 5 | 19 | 17.32 y | 5 | Hei1996b | 1 | YES |
| 08391−2240 | 073752 | 042430 | BU 208AB | 126 | 124 | 123.0 y | 3 | Hei1990c | ... | YES |
| 09123+1500 | 079096 | 045170 | FIN 347Aa,Ab | 99 | 40 | 2.71 y | 1 | Msn1996a | 2 | YES |
| 09179+2834 | 079969 | 045617 | STF3121AB | 388 | 173 | 34.17 y | 1 | Sod1999 | ... | YES |
| 09357+3549 | 082885 | 047080 | HU 1128 | 19 | 101 | 201 y | 5 | Hei1988d | ... | YES |
| 10281+4847 | 090508 | 051248 | KUI 50 | 11 | 67 | 765 y | 5 | Hle1994 | ... | YES |
| 11047−0413 | 096064 | 054155 | A 676BC | 53 | 95 | 23.23 y | 2 | Doc2001e | ... | YES |
| 11182+3132 | 098231 | 055203 | STF1523AB | 1560 | 227 | 59.88 y | 1 | Msn1995 | ... | YES |
| 11268+0301 | 099491 | 055846 | STF1540AB | 120 | 227 | 32000 y | 5 | Hop1960a | ... | YES |
| 11387+4507 | 101177 | 056809 | STF1561AB | 110 | 224 | 2050 y | 5 | Hle1994 | ... | YES |
| 13169+1701 | 115404 | 064797 | BU 800AB | 194 | 123 | 770.0 y | 4 | Hle1994 | ... | YES |
| 13473+1727 | 120136 | 067275 | STT 270 | 56 | 169 | 2000 y | 5 | Hle1994 | ... | YES |
| 14396−6050 | 128620 | 071683 | RHD 1AB | 438 | 255 | 79.91 y | 2 | Pbx2002 | 2 | YES |
| 14514+1906 | 131156 | 072659 | STF1888AB | 1358 | 227 | 151.6 y | 2 | Sod1999 | ... | YES |
| 15038+4739 | 133640 | 073695 | STF1909 | 769 | 226 | 206 y | 2 | Sod1999 | ... | YES |
| 15232+3017 | 137107 | 075312 | STF1937AB | 1002 | 225 | 41.56 y | 1 | WSI2006b | 2 | YES |

| WDS Coord (1) | HD Name (2) | HIP Name (3) | Discoverer ID (4) | N (5) | $\Delta T$ (years) (6) | P (7) | G (8) | Ref (9) | SB (10) | Comp Stat (11) |
|---|---|---|---|---|---|---|---|---|---|---|
| 15360+3948 | 139341 | 076382 | STT 298AB | 515 | 164 | 55.6 y | 1 | Sod1999 | ... | YES |
| 16133+1332 | 145958 | 079492 | STF2021AB | 395 | 224 | 1354 y | 4 | Hop1964b | ... | YES |
| 16147+3352 | 146361 | 079607 | STF2032AB | 1041 | 226 | 726 y | 4 | Rag2009 | ... | YES |
| 16289+1825 | 148653 | 080725 | STF2052AB | 528 | 185 | 224 y | 2 | Sod1999 | ... | YES |
| 17153−2636 | 155885 | 084405 | SHJ 243AB | 264 | 230 | 470.9 y | 4 | Irw1996 | ... | YES |
| 17191−4638 | 156274 | 084720 | BSO 13AB | 127 | 127 | 693.24 y | 5 | Wie1957 | ... | YES |
| 17304−0104 | 158614 | 085667 | STF2173 | 692 | 176 | 46.40 y | 1 | Hei1994a | 2 | YES |
| 17350+6153 | 160269 | 086036 | BU 962AB | 129 | 125 | 76.1 y | 3 | Sod1999 | 1 | YES |
| 17419+7209 | 162003 | 086614 | STF2241AB | 141 | 207 | 12500 y | 5 | Rmn1994 | ... | YES |
| 17433+2137 | 161198 | 086722 | DUQ 1 | 6 | 5 | 2558.4 d | 5 | Duq1996 | 1 | YES |
| 17465+2743 | 161797 | 086794 | AC 7BC | 343 | 153 | 43.20 y | 2 | Cou1960b | ... | YES |
| 18055+0230 | 165341 | 088601 | STF2272AB | 1678 | 230 | 88.38 y | 1 | Pbx2000b | 2 | YES |
| 18070+3034 | 165908 | 088745 | AC 15AB | 205 | 146 | 56.4 y | 2 | Sod1999 | ... | YES |
| 18570+3254 | 176051 | 093017 | BU 648AB | 332 | 127 | 61.18 y | 2 | Doc2008f | 1 | YES |
| 19064−3704 | 177474 | 093825 | HJ 5084 | 260 | 165 | 121.76 y | 2 | Hei1986b | ... | YES |
| 19121+4951 | 179958 | 094336 | STF2486AB | 258 | 188 | 3100 y | 4 | Hle1994 | ... | YES |
| 19311+5835 | 184467 | 095995 | MCA 56 | 28 | 21 | 1.35 y | 1 | Pbx2000b | 2 | YES |
| 19418+5032 | 186408 | 096895 | STFA 46AB | 534 | 206 | 18212 y | 4 | Mrc1999 | ... | YES |
| 19456+3337 | 186858 | 097222 | STF2576FG | 426 | 179 | 232 y | 2 | Sod1999 | ... | YES |
| 19598−0957 | 189340 | 098416 | HO 276 | 36 | 39 | 4.90 y | 2 | Pbx2000b | 2 | YES |
| 20329+4154 | 195987 | 101382 | BLA 8 | 1 | 0 | 57.32 d | 8 | Trr2002 | 2 | YES |
| 21145+1000 | 202275 | 104858 | STT 535AB | 486 | 155 | 2084.03 d | 1 | Mut2008 | 2 | YES |
| 21198−2621 | 202940 | 105312 | BU 271AB | 49 | 117 | 261.62 y | 5 | Jas1997 | ... | YES |
| 21441+2845 | 206826 | 107310 | STF2822AB | 693 | 229 | 789 y | 4 | Hei1995 | ... | YES |
| 23524+7533 | 223778 | 117712 | BU 996AB | 15 | 103 | 290.0 y | 5 | Hle1994 | ... | YES |

| WDS Coord (1) | HD Name (2) | HIP Name (3) | Discoverer ID (4) | N (5) | ΔT (years) (6) | P (7) | G (8) | Ref (9) | SB (10) | Comp Stat (11) |
|---|---|---|---|---|---|---|---|---|---|---|
| | | | | | | | | | | |
| colspan=11 | Photocentric-motion Visual Orbits | | | | | | | | | |
| 00169−5239 | 001273 | 001349 | GJ 13 | . . . | . . . | 411.4 d | 9 | Jnc2005 | 1 | YES |
| 01425+2016 | 010476 | 007981 | 107 Psc | . . . | . . . | 207.27 d | 9 | HIP1997d | . . . | NO |
| 02180+0145 | 014214 | 010723 | GC 2770 | . . . | . . . | 93.29 d | 9 | Fek2007 | 1 | YES |
| 02225−2349 | 014802 | 011072 | κ For | . . . | . . . | 26.5 y | 9 | Gon2000 | . . . | YES |
| 02361+0653 | 016160 | 012114 | PLQ 32Aa,P | 6 | 5 | 61 y | 9 | Hei1994b | . . . | YES |
| 02481+2704 | 017382 | 013081 | GC 3359 | . . . | . . . | 15.5 y | 9 | Hei1990d | 1 | YES |
| 05067+1427 | 032850 | 023786 | GJ 3330 | . . . | . . . | 204.38 d | 9 | HIP1997d | 1 | YES |
| 05544+2017 | 039587 | 027913 | χ¹ Ori | . . . | . . . | 5156.29 d | 9 | HaI2002 | 1 | YES |
| 11182+3132 | 098231 | 055203 | ξ UMa Aa,Ab | . . . | . . . | 1.83 y | 9 | Hei1996b | 1 | YES |
| 12442+5146 | 110833 | 062145 | GC 17326 | . . . | . . . | 270.21 d | 9 | HIP1997d | 1 | YES |
| 12595+4159 | 112914 | 063406 | LTT 13738 | . . . | . . . | 710.6 d | 9 | Jnc2005 | 1 | YES |
| 13038−0510 | 113449 | 063742 | GC 17714 | . . . | . . . | 231.23 d | 9 | HIP1997d | . . . | YES |
| 13546+1825 | 121370 | 067927 | η Boo | . . . | . . . | 494.2 d | 9 | Jnc2005 | 1 | YES |
| 14035+1047 | 122742 | 068682 | GJ 538 | . . . | . . . | 3614.89 d | 9 | HIP1997d | 1 | YES |
| 14160−0600 | 124850 | 069701 | ι Vir | . . . | . . . | 55 y | 9 | Gon2003 | . . . | MAY |
| 14294+8049 | 128642 | 070857 | GC 19630 | . . . | . . . | 179.73 d | 9 | HIP1997d | 1 | YES |
| 14534+1909 | 131511 | 072848 | DE Boo | . . . | . . . | 125.4 d | 9 | Jnc2005 | 1 | YES |
| 15282−0921 | 137763 | 075718 | BAG 25Aa,Ab | 1 | 0 | 889.6 d | 9 | Jnc2005 | 2 | YES |
| 15527+4227 | 142373 | 077760 | χ Her | . . . | . . . | 51.29 d | 9 | HIP1997d | . . . | NO |
| 16010+3318 | 143761 | 078459 | ρ CrB | . . . | . . . | 0.11 y | 9 | Gat2001a | 1 | MAY |
| 16147+3352 | 146361 | 079607 | σ CrB C | . . . | . . . | 52 y | 9 | Hei1990d | . . . | YES |
| 16285−7005 | 147584 | 080686 | ζ TrA | . . . | . . . | 12.9 d | 9 | Jnc2005 | 1 | YES |

| WDS Coord (1) | HD Name (2) | HIP Name (3) | Discoverer ID (4) | $N$ (5) | $\Delta T$ (years) (6) | $P$ (7) | G (8) | Ref (9) | SB (10) | Comp Stat (11) |
|---|---|---|---|---|---|---|---|---|---|---|
| 17393+0333 | 160346 | 086400 | GJ 688 | ... | ... | 83.70 d | 9 | Jnc2005 | 1 | YES |
| 17465+2743 | 161797 | 086974 | TRN 2Aa,Ab | 2 | 0 | 65 y | 9 | Hei1994a | ... | YES |
| 21094−7310 | 200525 | 104440 | I 379A | 4 | 34 | 2145 d | 9 | Gln2006 | ... | YES |
| 21247−6814 | 203244 | 105712 | GC 29928 | ... | ... | 1060.61 d | 9 | HIP1997d | ... | NO |

Note. — Column 11 values: 'YES' identifies physically associated companions, 'NO' marks unrelated field stars, and 'MAY' denotes unconfirmed candidates retained for further investigations.

Table 12.　CNS Candidates Refuted by Constant RV Measures

| HD Name (1) | Comp Claim (2) | Ref (3) | RV Constant References (4) |
|---|---|---|---|
| 004813 | SB? | 1 | 2,3 |
| 005133 | SB | 1,4,5 | 6,7 |
| 017925 | RV-Var | 1 | 6,8 |
| 020630 | SB? | 1 | 6,8 |
| 020794 | SB | 1 | 5,7,9 |
| 022484 | SB? | 1 | 6,8 |
| 025457 | RV-Var | 1 | 5,7,9 |
| 034721 | SB | 1 | 6,8 |
| 042807 | SB? | 1 | 8 |
| 046588 | SB? | 1,10 | 11,12 |
| 073752 | SB | 1 | 5,7,13 |
| 090839 | SB? | 1 | 6,8 |
| 102870 | SB? | 1 | 6 |
| 109358 | SB | 1,14 | 6,8,15 |
| 114613 | RV-Var | 1 | 7,16 |
| 130948 | SB | 1 | 6,8 |
| 141004 | SB1 | 1 | 1,6,8 |
| 155885 | RV-Var | 1 | 6,8 |
| 158633 | SB | 1 | 8 |
| 182572 | SB?,RV-Var? | 1 | 1 |
| 184385 | RV-Var | 1 | 6,8 |
| 192310 | SB? | 1 | 5,7,9 |
| 222368 | SB? | 1 | 6 |

Note. — Reference codes in Columns 3 & 4:
(1) Gliese (1969); Gliese & Jahreiß (1979, 1991);
(2) DM91; (3) Abt & Willmarth (2006); (4) Joy
(1947); (5) Abt & Biggs (1972); (6) Nidever et al.
(2002); (7) Gontcharov (2006); (8) Latham et al.
(2010); (9) Duflot et al. (1995); (10) Batten et al.
(1989); (11) Gomez & Abt (1982); (12) Pourbaix
et al. (2004); (13) Heintz (1968); (14) Abt & Levy
(1976); (15) Morbey & Griffin (1987); (16) Mur-

doch et al. (1993).

Table 13.   Survey Stars and their Stellar, Brown Dwarf, and Planetary Companions

| R.A. (J2000.0) (1) | Decl. (J2000.0) (2) | HD Name (3) | Other Name (4) | N (5) | Comp ID (6) | Period (7) | Ang Sep ($''$) (8) | Lin Sep (AU) (9) | Sts (10) | VB (11) | SB (12) | CP (13) | OT (14) | CH (15) |
|---|---|---|---|---|---|---|---|---|---|---|---|---|---|---|
| ... | ... | ... | Sun | ... | A | ... | ... | ... | ... | ... | ... | ... | ... | ... |
| ... | ... | ... | Mercury | ... | ... | 0.24 y | ... | 0.39 | Y | ... | ... | ... | ... | ... |
| ... | ... | ... | Venus | ... | ... | 0.62 y | ... | 0.72 | Y | ... | ... | ... | ... | ... |
| ... | ... | ... | Earth | ... | ... | 1.00 y | ... | 1.00 | Y | ... | ... | ... | ... | ... |
| ... | ... | ... | Mars | ... | ... | 1.88 y | ... | 1.52 | Y | ... | ... | ... | ... | ... |
| ... | ... | ... | Jupiter | ... | ... | 11.86 y | ... | 5.20 | Y | ... | ... | ... | ... | ... |
| ... | ... | ... | Saturn | ... | ... | 29.42 y | ... | 9.54 | Y | ... | ... | ... | ... | ... |
| ... | ... | ... | Uranus | ... | ... | 84.01 y | ... | 19.19 | Y | ... | ... | ... | ... | ... |
| ... | ... | ... | Neptune | ... | ... | 164.8 y | ... | 30.06 | Y | ... | ... | ... | ... | ... |
| 00 02 10.16 | +27 04 56.1 | 224930 | 85 Peg | ... | A | ... | ... | ... | ... | ... | ... | ... | ... | ... |
| ... | ... | ... | ... | ... | AB | 26.28 y | 0.83 | 10.1 | Y | O | 1 | ... | M | ... |
| 00 06 15.81 | +58 26 12.2 | 000123 | HIP 518 | N | A | ... | ... | ... | ... | ... | ... | ... | ... | ... |
| ... | ... | ... | ... | ... | AB | 106.7 y | 1.44 | 30.9 | Y | O | ... | ... | M | ... |
| ... | ... | ... | ... | ... | Ba,Bb | 47.69 d | ... | ... | Y | ... | 1 | ... | ... | ... |
| 00 06 36.78 | +29 01 17.4 | 000166 | HIP 544 | ... | ... | ... | ... | ... | ... | ... | ... | ... | ... | ... |
| 00 12 50.25 | −57 54 45.4 | 000870 | HIP 1031 | ... | ... | ... | ... | ... | ... | ... | ... | ... | ... | ... |
| 00 16 12.68 | −79 51 04.3 | 001237 | HIP 1292 | N | A | ... | ... | ... | ... | ... | ... | ... | ... | ... |
| ... | ... | ... | GJ 3021 b | ... | ... | 133.71 d | ... | 0.50 | Y | ... | 1 | ... | ... | ... |
| ... | ... | ... | HD 1237B | ... | AB | ... | 3.87 | 67.7 | Y | ... | ... | S | ... | ... |
| 00 16 53.89 | −52 39 04.1 | 001273 | HIP 1349 | ... | A | ... | ... | ... | ... | ... | ... | ... | ... | ... |
| ... | ... | ... | ... | ... | Aa,Ab | 1.13 y | ... | ... | Y | U | 1 | ... | ... | ... |
| 00 18 41.87 | −08 03 10.8 | 001461 | HIP 1499 | ... | ... | ... | ... | ... | ... | ... | ... | ... | ... | ... |
| 00 20 00.41 | +38 13 38.6 | 001562 | HIP 1598 | ... | ... | ... | ... | ... | ... | ... | ... | ... | ... | ... |
| 00 20 04.26 | −64 52 29.2 | 001581 | ζ Tuc | ... | ... | ... | ... | ... | ... | ... | ... | ... | ... | ... |
| 00 22 51.79 | −12 12 34.0 | 001835 | HIP 1803 | ... | ... | ... | ... | ... | ... | ... | ... | ... | ... | ... |
| 00 24 25.93 | −27 01 36.4 | 002025 | HIP 1936 | ... | ... | ... | ... | ... | ... | ... | ... | ... | ... | ... |
| 00 25 45.07 | −77 15 15.3 | 002151 | β Hyi | ... | ... | ... | ... | ... | ... | ... | ... | ... | ... | ... |
| 00 35 14.88 | −03 35 34.2 | 003196 | 13 Cet | ... | A | ... | ... | ... | ... | ... | ... | ... | ... | ... |
| ... | ... | ... | ... | ... | Aa,Ab | 2.08 d | ... | ... | Y | ... | 1 | ... | ... | ... |
| ... | ... | ... | ... | ... | AB | 6.89 y | 0.24 | 5.10 | Y | O | 2 | ... | M | S |
| 00 37 20.70 | −24 46 02.2 | 003443 | HIP 2941 | ... | A | ... | ... | ... | ... | ... | ... | ... | ... | ... |

| R.A. (J2000.0) (1) | Decl. (J2000.0) (2) | HD Name (3) | Other Name (4) | N (5) | Comp ID (6) | Period (7) | Ang Sep (″) (8) | Lin Sep (AU) (9) | Sts (10) | VB (11) | SB (12) | CP (13) | OT (14) | CH (15) |
|---|---|---|---|---|---|---|---|---|---|---|---|---|---|---|
| ... | ... | ... | GJ 25 B | ... | AB | 25.09 y | 0.67 | 10.3 | Y | O | 2 | ... | M | ... |
| 00 39 21.81 | +21 15 01.7 | 003651 | 54 Psc | N | A | ... | ... | ... | ... | ... | ... | ... | ... | ... |
| ... | ... | ... | HD 3651 b | ... | ... | 62.24 d | ... | 0.30 | Y | ... | 1 | ... | ... | ... |
| ... | ... | ... | HD 3651B | ... | AB | ... | 43.07 | 476 | Y | ... | ... | S | ... | ... |
| 00 40 49.27 | +40 11 13.8 | 003765 | HIP 3206 | ... | ... | ... | ... | ... | ... | ... | ... | ... | ... | ... |
| 00 44 39.27 | −65 38 58.3 | 004308 | HIP 3497 | ... | A | ... | ... | ... | ... | ... | ... | ... | ... | ... |
| ... | ... | ... | HD 4308 b | ... | ... | 15.56 d | ... | 0.12 | Y | ... | 1 | ... | ... | ... |
| 00 45 04.89 | +01 47 07.9 | 004256 | HIP 3535 | ... | ... | ... | ... | ... | ... | ... | ... | ... | ... | ... |
| 00 45 45.59 | −47 33 07.2 | 004391 | HIP 3583 | N | A | ... | ... | ... | ... | ... | ... | ... | ... | ... |
| ... | ... | ... | ... | ... | AB | ... | 16.6 | 251 | Y | ... | ... | P | ... | ... |
| ... | ... | ... | ... | ... | AC | ... | 49.0 | 742 | Y | ... | ... | P | ... | ... |
| 00 48 22.98 | +05 16 50.2 | 004628 | HIP 3765 | N | A | ... | ... | ... | ... | ... | ... | ... | ... | ... |
| ... | ... | ... | ... | ... | Aa,Ab | ... | 2.7 | 20.1 | M | ... | ... | R | ... | ... |
| 00 48 58.71 | +16 56 26.3 | 004676 | 64 Psc | N | A | ... | ... | ... | ... | ... | ... | ... | ... | ... |
| ... | ... | ... | ... | ... | Aa,Ab | 13.82 d | 0.01 | 0.23 | Y | O | 2 | ... | ... | S |
| 00 49 06.29 | +57 48 54.7 | 004614 | η Cas | ... | A | ... | ... | ... | ... | ... | ... | ... | ... | ... |
| ... | ... | ... | LHS 122 | ... | AB | 480 y | 11.99 | 71.4 | Y | O | ... | ... | ... | ... |
| 00 49 26.77 | −23 12 44.9 | 004747 | HIP 3850 | ... | A | ... | ... | ... | ... | ... | ... | ... | ... | ... |
| ... | ... | ... | ... | ... | Aa,Ab | 18.7 y | ... | ... | Y | ... | 1 | ... | M | ... |
| 00 49 46.48 | +70 26 58.1 | 004635 | HIP 3876 | ... | ... | ... | ... | ... | ... | ... | ... | ... | ... | ... |
| 00 50 07.59 | −10 38 39.6 | 004813 | HIP 3909 | ... | ... | ... | ... | ... | ... | ... | ... | ... | ... | ... |
| 00 51 10.85 | −05 02 21.4 | 004915 | HIP 3979 | ... | ... | ... | ... | ... | ... | ... | ... | ... | ... | ... |
| 00 53 01.13 | −30 21 24.9 | 005133 | HIP 4148 | ... | ... | ... | ... | ... | ... | ... | ... | ... | ... | ... |
| 00 53 04.20 | +61 07 26.3 | 005015 | HIP 4151 | ... | ... | ... | ... | ... | ... | ... | ... | ... | ... | ... |
| 01 08 16.39 | +54 55 13.2 | 006582 | μ Cas | ... | A | ... | ... | ... | ... | ... | ... | ... | ... | ... |
| ... | ... | ... | μ Cas B | ... | Aa,Ab | 21.75 y | 1.01 | 7.63 | Y | P | 1 | ... | ... | ... |
| 01 15 00.99 | −68 49 08.1 | 007693 | HIP 5842 | ... | C[a] | ... | ... | ... | ... | ... | ... | ... | ... | ... |
| ... | ... | ... | GJ 55.1B | ... | CD | 85.2 y | 1.14 | 24.7 | Y | O | ... | ... | M | ... |
| ... | ... | ... | HD 7788 | ... | A,CD | ... | 318 | 6883 | Y | ... | ... | T | ... | ... |
| ... | ... | ... | GJ 55.3B | ... | AB | 857 y | 5.96 | 129 | Y | P | ... | ... | ... | ... |
| 01 15 11.12 | −45 31 54.0 | 007570 | ν Phe | ... | ... | ... | ... | ... | ... | ... | ... | ... | ... | ... |

| R.A. (J2000.0) (1) | Decl. (J2000.0) (2) | HD Name (3) | Other Name (4) | N (5) | Comp ID (6) | Period (7) | Ang Sep (″) (8) | Lin Sep (AU) (9) | Sts (10) | VB (11) | SB (12) | CP (13) | OT (14) | CH (15) |
|---|---|---|---|---|---|---|---|---|---|---|---|---|---|---|
| 01 16 29.25 | +42 56 21.9 | 007590 | HIP 5944 | ... | ... | ... | ... | ... | ... | ... | ... | ... | ... | ... |
| 01 21 59.12 | +76 42 37.0 | 007924 | HIP 6379 | ... | ... | ... | ... | ... | ... | ... | ... | ... | ... | ... |
| ... | ... | ... | HD 7924 b | ... | ... | 5.40 d | ... | 0.06 | Y | ... | 1 | ... | ... | ... |
| 01 29 04.90 | +21 43 23.4 | 008997 | HIP 6917 | ... | A | ... | ... | ... | ... | ... | ... | ... | ... | ... |
| ... | ... | ... | ... | ... | Aa,Ab | 10.98 d | ... | ... | Y | ... | 2 | ... | ... | V |
| 01 33 15.81 | −24 10 40.7 | 009540 | HIP 7235 | ... | ... | ... | ... | ... | ... | ... | ... | ... | ... | ... |
| 01 34 33.26 | +68 56 53.3 | 009407 | HIP 7339 | ... | ... | ... | ... | ... | ... | ... | ... | ... | ... | ... |
| 01 35 01.01 | −29 54 37.2 | 009770 | HIP 7372 | ... | A | ... | ... | ... | ... | ... | ... | ... | ... | ... |
| ... | ... | ... | ... | ... | AB | 4.56 y | 0.18 | 3.89 | Y | O | ... | ... | ... | ... |
| ... | ... | ... | ... | ... | Ba,Bb | 11.50 h | ... | 0.01 | Y | ... | ... | ... | E | ... |
| ... | ... | ... | ... | ... | AB,C | 111.8 y | 1.42 | 30.7 | Y | P | ... | ... | ... | ... |
| 01 36 47.84 | +41 24 19.7 | 009826 | υ And | N | A | ... | ... | ... | ... | ... | ... | ... | ... | ... |
| ... | ... | ... | υ And d | ... | ... | 4.62 d | ... | 0.06 | Y | ... | 1 | ... | ... | ... |
| ... | ... | ... | υ And c | ... | ... | 241.31 d | ... | 0.83 | Y | ... | 1 | ... | ... | ... |
| ... | ... | ... | υ And b | ... | ... | 3.55 y | ... | 2.55 | Y | ... | 1 | ... | ... | ... |
| ... | ... | ... | υ And B | ... | AD | ... | 55 | 742 | Y | ... | ... | S | ... | ... |
| 01 37 35.47 | −06 45 37.5 | 010008 | HIP 7576 | ... | ... | ... | ... | ... | ... | ... | ... | ... | ... | ... |
| 01 39 36.02 | +45 52 40.0 | 010086 | HIP 7734 | ... | ... | ... | ... | ... | ... | ... | ... | ... | ... | ... |
| 01 39 47.54 | −56 11 47.0 | 010360 | HIP 7751 | ... | B[a] | ... | ... | ... | ... | ... | ... | ... | ... | ... |
| ... | ... | ... | HD 10361 | ... | AB | 483.66 y | 7.82 | 61.2 | Y | P | ... | ... | M | ... |
| 01 41 47.14 | +42 36 48.1 | 010307 | HIP 7918 | ... | A | ... | ... | ... | ... | ... | ... | ... | ... | ... |
| ... | ... | ... | ... | ... | Aa,Ab | 19.5 y | 0.58 | 7.39 | Y | P | 1 | ... | M | ... |
| 01 42 29.32 | −53 44 27.0 | 010647 | q¹ Eri | ... | A | ... | ... | ... | ... | ... | ... | ... | ... | ... |
| ... | ... | ... | HD 10647 b | ... | ... | 2.75 y | ... | 2.03 | Y | ... | 1 | ... | ... | ... |
| 01 42 29.76 | +20 16 06.6 | 010476 | 107 Psc | ... | ... | ... | ... | ... | ... | ... | ... | ... | ... | ... |
| 01 44 04.08 | −15 56 14.9 | 010700 | τ Cet | ... | ... | ... | ... | ... | ... | ... | ... | ... | ... | ... |
| 01 47 44.83 | +63 51 09.0 | 010780 | HIP 8362 | ... | ... | ... | ... | ... | ... | ... | ... | ... | ... | ... |
| 01 59 06.63 | +33 12 34.9 | 012051 | HIP 9269 | ... | ... | ... | ... | ... | ... | ... | ... | ... | ... | ... |
| 02 06 30.24 | +24 20 02.4 | 012846 | HIP 9829 | ... | ... | ... | ... | ... | ... | ... | ... | ... | ... | ... |
| 02 10 25.93 | −50 49 25.4 | 013445 | HIP 10138 | N | A | ... | ... | ... | ... | ... | ... | ... | ... | ... |
| ... | ... | ... | Gl 86 b | ... | ... | 15.76 d | ... | 0.11 | Y | P | 1 | ... | ... | ... |

| R.A. (J2000.0) (1) | Decl. (J2000.0) (2) | HD Name (3) | Other Name (4) | N (5) | Comp ID (6) | Period (7) | Ang Sep (″) (8) | Lin Sep (AU) (9) | Sts (10) | VB (11) | SB (12) | CP (13) | OT (14) | CH (15) |
|---|---|---|---|---|---|---|---|---|---|---|---|---|---|---|
| ... | ... | ... | GJ 86 B | ... | AB | 69.7 y | 1.69 | 18.2 | Y | P | V | ... | M | ... |
| 02 17 03.23 | +34 13 27.2 | 013974 | δ Tri | ... | A | ... | ... | ... | ... | ... | ... | ... | ... | ... |
| ... | ... | ... | ... | ... | Aa,Ab | 10.02 d | 0.01 | 0.11 | Y | O | 2 | ... | ... | ... |
| 02 18 01.44 | +01 45 28.1 | 014214 | HIP 10723 | ... | A | ... | ... | ... | ... | ... | ... | ... | ... | ... |
| ... | ... | ... | ... | ... | Aa,Ab | 93.29 d | ... | ... | Y | U | 1 | ... | ... | ... |
| 02 18 58.50 | −25 56 44.5 | 014412 | HIP 10798 | ... | ... | ... | ... | ... | ... | ... | ... | ... | ... | ... |
| 02 22 32.55 | −23 48 58.8 | 014802 | κ For | ... | A | ... | ... | ... | ... | ... | ... | ... | ... | ... |
| ... | ... | ... | ... | ... | AB | 26.5 y | 0.5 | 11.0 | Y | U | V | R | M | ... |
| 02 36 04.89 | +06 53 12.7 | 016160 | HIP 12114 | ... | A | ... | ... | ... | ... | ... | ... | ... | ... | ... |
| ... | ... | ... | ... | ... | Aa,Ab | 61 y | 3.2 | 23.0 | Y | U | ... | R | ... | ... |
| ... | ... | ... | NLTT 8455 | ... | Aa,B | ... | 164 | 1177 | Y | ... | ... | P | ... | ... |
| 02 36 41.76 | −03 09 22.1 | 016287 | HIP 12158 | ... | A | ... | ... | ... | ... | ... | ... | ... | ... | ... |
| ... | ... | ... | ... | ... | Aa,Ab | 14.84 d | 0.01 | 0.24 | Y | ... | 1 | ... | ... | ... |
| 02 40 12.42 | −09 27 10.3 | 016673 | HIP 12444 | ... | A | ... | ... | ... | ... | ... | ... | ... | ... | ... |
| ... | ... | ... | ... | ... | Aa,Ab | ... | ... | ... | Y | ... | V | ... | ... | ... |
| 02 41 14.00 | −00 41 44.4 | 016765 | HIP 12530 | ... | A | ... | ... | ... | ... | ... | ... | ... | ... | ... |
| ... | ... | ... | ... | ... | AB | ... | 3.3 | 74.5 | Y | ... | V | O | ... | ... |
| 02 42 14.92 | +40 11 38.2 | 016739 | 12 Per | ... | A | ... | ... | ... | ... | ... | ... | ... | ... | ... |
| ... | ... | ... | ... | ... | Aa,Ab | 330.98 d | 0.05 | 1.21 | Y | O | 2 | ... | ... | S |
| 02 42 33.47 | −50 48 01.1 | 017051 | ι Hor | ... | A | ... | ... | ... | ... | ... | ... | ... | ... | ... |
| ... | ... | ... | HR 810 b | ... | ... | 302.8 d | ... | 0.93 | Y | ... | 1 | ... | ... | ... |
| 02 44 11.99 | +49 13 42.4 | 016895 | θ Per | ... | A | ... | ... | ... | ... | ... | ... | ... | ... | ... |
| ... | ... | ... | NLTT 8787 | ... | AB | 2720 y | 22.29 | 248 | Y | P | ... | ... | ... | ... |
| 02 48 09.14 | +27 04 07.1 | 017382 | HIP 13081 | ... | A | ... | ... | ... | ... | ... | ... | ... | ... | ... |
| ... | ... | ... | ... | ... | Aa,Ab | 15.27 y | ... | ... | Y | U | 1 | ... | M | ... |
| ... | ... | ... | NLTT 8996 | ... | AB | ... | 20.7 | 509 | Y | ... | ... | P | ... | ... |
| 02 52 32.13 | −12 46 11.0 | 017925 | HIP 13402 | ... | ... | ... | ... | ... | ... | ... | ... | ... | ... | ... |
| 02 55 39.06 | +26 52 23.6 | 018143 | HIP 13642 | ... | A | ... | ... | ... | ... | ... | ... | ... | ... | ... |
| ... | ... | ... | HD 18143 B | ... | AB | ... | 5 | 117 | Y | P | ... | ... | M | ... |
| ... | ... | ... | NLTT 9303 | ... | AC | ... | 44 | 1033 | Y | ... | ... | T | ... | ... |
| 03 00 02.81 | +07 44 59.1 | 018632 | HIP 13976 | ... | ... | ... | ... | ... | ... | ... | ... | ... | ... | ... |

| R.A. (J2000.0) (1) | Decl. (J2000.0) (2) | HD Name (3) | Other Name (4) | N (5) | Comp ID (6) | Period (7) | Ang Sep (″) (8) | Lin Sep (AU) (9) | Sts (10) | VB (11) | SB (12) | CP (13) | OT (14) | CH (15) |
|---|---|---|---|---|---|---|---|---|---|---|---|---|---|---|
| 03 02 26.03 | +26 36 33.3 | 018803 | 51 Ari | ... | ... | ... | ... | ... | ... | ... | ... | ... | ... | ... |
| 03 04 09.64 | +61 42 21.0 | 018757 | HIP 14286 | ... | A | ... | ... | ... | ... | ... | ... | ... | ... | ... |
| ... | ... | ... | NLTT 9726 | ... | AC | ... | 263.2 | 6377 | Y | ... | ... | P | ... | ... |
| 03 09 04.02 | +49 36 47.8 | 019373 | ι Per | ... | ... | ... | ... | ... | ... | ... | ... | ... | ... | ... |
| 03 12 04.53 | −28 59 15.4 | 020010 | 12 Eri | N | A | ... | ... | ... | ... | ... | ... | ... | ... | ... |
| ... | ... | ... | GJ 127 B | ... | AB | 269 y | 4.0 | 56.9 | Y | P | ... | ... | ... | ... |
| ... | ... | ... | ... | ... | Ba,Bb | ... | ... | ... | M | ... | V | ... | ... | ... |
| 03 12 46.44 | −01 11 46.0 | 019994 | 94 Cet | ... | A | ... | ... | ... | ... | ... | ... | ... | ... | ... |
| ... | ... | ... | HD 19994 b | ... | ... | 1.47 y | ... | 1.43 | Y | ... | 1 | ... | ... | ... |
| ... | ... | ... | ... | ... | AB | 1420 y | 6.77 | 152 | Y | P | ... | ... | ... | ... |
| 03 14 47.23 | +08 58 50.9 | 020165 | HIP 15099 | ... | ... | ... | ... | ... | ... | ... | ... | ... | ... | ... |
| 03 15 06.39 | −45 39 53.4 | 020407 | HIP 15131 | ... | ... | ... | ... | ... | ... | ... | ... | ... | ... | ... |
| 03 18 12.82 | −62 30 22.9 | 020807 | ζ² Ret | N | A | ... | ... | ... | ... | ... | ... | ... | ... | ... |
| ... | ... | ... | HD 20766 | ... | AB | ... | 309.2 | 3720 | Y | ... | ... | T | ... | ... |
| 03 19 01.89 | −02 50 35.5 | 020619 | HIP 15442 | ... | ... | ... | ... | ... | ... | ... | ... | ... | ... | ... |
| 03 19 21.70 | +03 22 12.7 | 020630 | κ Cet | ... | ... | ... | ... | ... | ... | ... | ... | ... | ... | ... |
| 03 19 55.65 | −43 04 11.2 | 020794 | HIP 15510 | ... | ... | ... | ... | ... | ... | ... | ... | ... | ... | ... |
| 03 21 54.76 | +52 19 53.4 | 232781 | HIP 15673 | ... | ... | ... | ... | ... | ... | ... | ... | ... | ... | ... |
| 03 23 35.26 | −40 04 35.0 | 021175 | HIP 15799 | N | A | ... | ... | ... | ... | ... | ... | ... | ... | ... |
| ... | ... | ... | ... | ... | AB | 111 y | 1.8 | 31.4 | Y | P | ... | ... | ... | ... |
| 03 32 55.84 | −09 27 29.7 | 022049 | ε Eri | N | A | ... | ... | ... | ... | ... | ... | ... | ... | ... |
| ... | ... | ... | ε Eri b | ... | ... | 7.2 y | ... | 3.5 | Y | ... | 1 | ... | ... | ... |
| 03 36 52.38 | +00 24 06.0 | 022484 | 10 Tau | ... | ... | ... | ... | ... | ... | ... | ... | ... | ... | ... |
| 03 40 22.06 | −03 13 01.1 | 022879 | HIP 17147 | ... | ... | ... | ... | ... | ... | ... | ... | ... | ... | ... |
| 03 43 55.34 | −19 06 39.2 | 023356 | HIP 17420 | ... | ... | ... | ... | ... | ... | ... | ... | ... | ... | ... |
| 03 44 09.17 | −38 16 54.4 | 023484 | HIP 17439 | N | A | ... | ... | ... | ... | ... | ... | ... | ... | ... |
| ... | ... | ... | ... | ... | Aa,Ab | ... | ... | ... | M | ... | V | ... | ... | ... |
| 03 54 28.03 | +16 36 57.8 | 024496 | HIP 18267 | N | A | ... | ... | ... | ... | ... | ... | ... | ... | ... |
| ... | ... | ... | ... | ... | AB | ... | 2.7 | 55.2 | Y | ... | ... | M | ... | ... |
| 03 55 03.84 | +61 10 00.5 | 024238 | HIP 18324 | ... | ... | ... | ... | ... | ... | ... | ... | ... | ... | ... |
| 03 56 11.52 | +59 38 30.8 | 024409 | HIP 18413 | ... | A | ... | ... | ... | ... | ... | ... | ... | ... | ... |

| R.A. (J2000.0) (1) | Decl. (J2000.0) (2) | HD Name (3) | Other Name (4) | N (5) | Comp ID (6) | Period (7) | Ang Sep ('') (8) | Lin Sep (AU) (9) | Sts (10) | VB (11) | SB (12) | CP (13) | OT (14) | CH (15) |
|---|---|---|---|---|---|---|---|---|---|---|---|---|---|---|
| ... | ... | ... | ... | ... | AD | 34.57 y | 0.4 | 8.79 | Y | ... | 1 | ... | M | ... |
| ... | ... | ... | ... | ... | AE | ... | 8.9 | 195 | Y | ... | ... | P | ... | ... |
| 04 02 36.74 | −00 16 08.1 | 025457 | HIP 18859 | ... | ... | ... | ... | ... | ... | ... | ... | ... | ... | ... |
| 04 03 15.00 | +35 16 23.8 | 025329 | HIP 18915 | ... | ... | ... | ... | ... | ... | ... | ... | ... | ... | ... |
| 04 05 20.26 | +22 00 32.1 | 025680 | 39 Tau | N | A | ... | ... | ... | ... | ... | ... | ... | ... | ... |
| ... | ... | ... | ... | ... | Aa,Ab | ... | 0.4 | 6.78 | Y | ... | ... | M | ... | ... |
| 04 07 21.54 | −64 13 20.2 | 026491 | HIP 19233 | N | A | ... | ... | ... | ... | ... | ... | ... | ... | ... |
| ... | ... | ... | ... | ... | Aa,Ab | 19.19 y | ... | ... | Y | ... | 1 | ... | M | ... |
| 04 08 36.62 | +38 02 23.0 | 025998 | HIP 19335 | ... | E[b] | ... | ... | ... | ... | ... | ... | ... | ... | ... |
| ... | ... | ... | ... | ... | Ea,Eb | ... | ... | ... | M | ... | ... | ... | M | ... |
| ... | ... | ... | HD 25893 | ... | AE | ... | 746 | 15662 | Y | ... | ... | T | ... | ... |
| ... | ... | ... | ... | ... | AB | 590 y | 3.87 | 81.3 | Y | P | V | ... | ... | ... |
| 04 09 35.04 | +69 32 29.0 | 025665 | HIP 19422 | ... | ... | ... | ... | ... | ... | ... | ... | ... | ... | ... |
| 04 15 16.32 | −07 39 10.3 | 026965 | HIP 19849 | ... | A | ... | ... | ... | ... | ... | ... | ... | ... | ... |
| ... | ... | ... | LHS 25 | ... | A,BC | ... | 83 | 413 | Y | ... | ... | T | ... | ... |
| ... | ... | ... | HD 26976 | ... | BC | 252.1 y | 6.94 | 34.6 | Y | P | ... | ... | ... | ... |
| 04 15 28.80 | +06 11 12.7 | 026923 | HIP 19859 | ... | A | ... | ... | ... | ... | ... | ... | ... | ... | ... |
| ... | ... | ... | HD 26913 | ... | AB | ... | 64.5 | 1375 | Y | ... | ... | T | ... | ... |
| 04 43 35.44 | +27 41 14.6 | 029883 | HIP 21988 | ... | ... | ... | ... | ... | ... | ... | ... | ... | ... | ... |
| 04 45 38.58 | −50 04 27.2 | 030501 | HIP 22122 | ... | ... | ... | ... | ... | ... | ... | ... | ... | ... | ... |
| 04 47 36.29 | −16 56 04.0 | 030495 | HIP 22263 | ... | ... | ... | ... | ... | ... | ... | ... | ... | ... | ... |
| 04 49 52.33 | −35 06 27.5 | 030876 | HIP 22451 | ... | ... | ... | ... | ... | ... | ... | ... | ... | ... | ... |
| 05 02 17.06 | −56 04 49.9 | 032778 | HIP 23437 | ... | A | ... | ... | ... | ... | ... | ... | ... | ... | ... |
| ... | ... | ... | NLTT 14447 | ... | AB | ... | 79.1 | 1778 | Y | ... | ... | P | ... | ... |
| 05 05 30.66 | −57 28 21.7 | 033262 | ζ Dor | ... | ... | ... | ... | ... | ... | ... | ... | ... | ... | ... |
| 05 06 42.22 | +14 26 46.4 | 032850 | HIP 23786 | ... | A | ... | ... | ... | ... | ... | ... | ... | ... | ... |
| ... | ... | ... | ... | ... | Aa,Ab | 205.68 d | ... | ... | Y | U | 1 | ... | ... | ... |
| 05 07 27.01 | +18 38 42.2 | 032923 | 104 Tau | N | ... | ... | ... | ... | ... | ... | ... | ... | ... | ... |
| 05 18 50.47 | −18 07 48.2 | 034721 | HIP 24786 | ... | ... | ... | ... | ... | ... | ... | ... | ... | ... | ... |
| 05 19 08.47 | +40 05 56.6 | 034411 | HIP 24813 | ... | ... | ... | ... | ... | ... | ... | ... | ... | ... | ... |
| 05 22 33.53 | +79 13 52.1 | 033564 | HIP 25110 | ... | A | ... | ... | ... | ... | ... | ... | ... | ... | ... |

| R.A. (J2000.0) (1) | Decl. (J2000.0) (2) | HD Name (3) | Other Name (4) | N (5) | Comp ID (6) | Period (7) | Ang Sep (″) (8) | Lin Sep (AU) (9) | Sts (10) | VB (11) | SB (12) | CP (13) | OT (14) | CH (15) |
|---|---|---|---|---|---|---|---|---|---|---|---|---|---|---|
| ... | ... | ... | HD 33564 b | ... | ... | 1.06 y | ... | 1.12 | Y | ... | 1 | ... | ... | ... |
| 05 22 37.49 | +02 36 11.5 | 035112 | HIP 25119 | ... | A | ... | ... | ... | ... | ... | ... | ... | ... | ... |
| ... | ... | ... | ... | ... | AB | 93 y | 1.1 | 22.2 | Y | P | ... | ... | M | ... |
| 05 24 25.46 | +17 23 00.7 | 035296 | HIP 25278 | N | A | ... | ... | ... | ... | ... | ... | ... | ... | ... |
| ... | ... | ... | HD 35171 | ... | AC | ... | 707.2 | 10174 | Y | ... | ... | T | ... | ... |
| 05 26 14.74 | −32 30 17.2 | 035854 | HIP 25421 | ... | ... | ... | ... | ... | ... | ... | ... | ... | ... | ... |
| 05 27 39.35 | −60 24 57.6 | 036435 | HIP 25544 | ... | ... | ... | ... | ... | ... | ... | ... | ... | ... | ... |
| 05 28 44.83 | −65 26 54.9 | 036705 | AB Dor | N | A | ... | ... | ... | ... | ... | ... | ... | ... | ... |
| ... | ... | ... | ... | ... | Aa,Ab | ... | 0.16 | 2.43 | Y | ... | ... | S | M | ... |
| ... | ... | ... | ... | ... | AB | ... | 9.2 | 139 | Y | ... | ... | P | ... | ... |
| ... | ... | ... | ... | ... | Ba,Bb | ... | 0.07 | 1.06 | Y | ... | ... | R | ... | ... |
| 05 36 56.85 | −47 57 52.9 | 037572 | UY Pic | ... | A | ... | ... | ... | ... | ... | ... | ... | ... | ... |
| ... | ... | ... | HIP 26369 | ... | AB | ... | 18.3 | 459 | Y | ... | ... | T | ... | ... |
| 05 37 09.89 | −80 28 08.8 | 039091 | π Men | ... | A | ... | ... | ... | ... | ... | ... | ... | ... | ... |
| ... | ... | ... | HD 39091 b | ... | ... | 5.89 y | ... | 3.38 | Y | ... | 1 | ... | ... | ... |
| 05 38 11.86 | +51 26 44.7 | 037008 | HIP 26505 | ... | ... | ... | ... | ... | ... | ... | ... | ... | ... | ... |
| 05 41 20.34 | +53 28 51.8 | 037394 | HIP 26779 | ... | A | ... | ... | ... | ... | ... | ... | ... | ... | ... |
| ... | ... | ... | HD 233153 | ... | AB | ... | 98.8 | 1213 | Y | ... | ... | T | ... | ... |
| 05 46 01.89 | +37 17 04.7 | 038230 | HIP 27207 | ... | ... | ... | ... | ... | ... | ... | ... | ... | ... | ... |
| 05 48 34.94 | −04 05 40.7 | 038858 | HIP 27435 | ... | ... | ... | ... | ... | ... | ... | ... | ... | ... | ... |
| 05 54 04.24 | −60 01 24.5 | 040307 | HIP 27887 | ... | A | ... | ... | ... | ... | ... | ... | ... | ... | ... |
| ... | ... | ... | HD 40307 b | ... | ... | 4.31 d | ... | 0.05 | Y | ... | 1 | ... | ... | ... |
| ... | ... | ... | HD 40307 c | ... | ... | 9.62 d | ... | 0.08 | Y | ... | 1 | ... | ... | ... |
| ... | ... | ... | HD 40307 d | ... | ... | 20.46 d | ... | 0.13 | Y | ... | 1 | ... | ... | ... |
| 05 54 22.98 | +20 16 34.2 | 039587 | χ¹ Ori | ... | A | ... | ... | ... | ... | ... | ... | ... | ... | ... |
| ... | ... | ... | ... | ... | Aa,Ab | 14.06 y | 0.5 | 4.33 | Y | U | 1 | R | M | ... |
| 05 54 30.16 | −19 42 15.7 | 039855 | HIP 27922 | ... | A | ... | ... | ... | ... | ... | ... | ... | ... | ... |
| ... | ... | ... | BD-19 1297B | ... | AB | ... | 10.6 | 249 | Y | ... | ... | P | ... | ... |
| 05 58 21.54 | −04 39 02.4 | 040397 | HIP 28267 | N | A | ... | ... | ... | ... | ... | ... | ... | ... | ... |
| ... | ... | ... | ... | ... | AB | ... | 3.9 | 91.9 | Y | ... | ... | M | M | ... |
| ... | ... | ... | NLTT 15867 | ... | AD | ... | 89.3 | 2103 | Y | ... | ... | P | ... | ... |

| R.A. (J2000.0) (1) | Decl. (J2000.0) (2) | HD Name (3) | Other Name (4) | N (5) | Comp ID (6) | Period (7) | Ang Sep (″) (8) | Lin Sep (AU) (9) | Sts (10) | VB (11) | SB (12) | CP (13) | OT (14) | CH (15) |
|---|---|---|---|---|---|---|---|---|---|---|---|---|---|---|
| 06 06 40.48 | +15 32 31.6 | 041593 | HIP 28954 | ... | ... | ... | ... | ... | ... | ... | ... | ... | ... | ... |
| 06 10 14.47 | −74 45 11.0 | 043834 | α Men | N | A | ... | ... | ... | ... | ... | ... | ... | ... | ... |
| ... | ... | ... | ... | ... | Aa,Ab | ... | 3.05 | 31.1 | Y | ... | V | M | ... | ... |
| 06 12 00.57 | +06 46 59.1 | 042618 | HIP 29432 | ... | ... | ... | ... | ... | ... | ... | ... | ... | ... | ... |
| 06 13 12.50 | +10 37 37.7 | 042807 | HIP 29525 | ... | ... | ... | ... | ... | ... | ... | ... | ... | ... | ... |
| 06 13 45.30 | −23 51 43.0 | 043162 | HIP 29568 | ... | A | ... | ... | ... | ... | ... | ... | ... | ... | ... |
| ... | ... | ... | ... | ... | AB | ... | 164 | 2742 | Y | ... | ... | P | ... | ... |
| 06 17 16.14 | +05 06 00.4 | 043587 | HIP 29860 | ... | A | ... | ... | ... | ... | ... | ... | ... | ... | ... |
| ... | ... | ... | ... | ... | Aa,Ab | 28.8 y | 0.62 | 11.9 | Y | P | 1 | ... | M | ... |
| ... | ... | ... | NLTT 16333 | ... | AE | ... | 103.1 | 1984 | Y | ... | ... | P | ... | ... |
| 06 22 30.94 | −60 13 07.2 | 045270 | HIP 30314 | N | A | ... | ... | ... | ... | ... | ... | ... | ... | ... |
| ... | ... | ... | ... | ... | AB | ... | 16.2 | 385 | M | ... | ... | R | ... | ... |
| 06 24 43.88 | −28 46 48.4 | 045184 | HIP 30503 | ... | ... | ... | ... | ... | ... | ... | ... | ... | ... | ... |
| 06 26 10.25 | +18 45 24.8 | 045088 | OU Gem | ... | A | ... | ... | ... | ... | ... | ... | ... | ... | ... |
| ... | ... | ... | ... | ... | Aa,Ab | 6.99 d | ... | ... | Y | ... | 2 | ... | ... | V |
| ... | ... | ... | ... | ... | AB | 600 y | 3.23 | 47.6 | Y | P | ... | ... | M | ... |
| 06 38 00.36 | −61 32 00.2 | 048189 | HIP 31711 | N | A | ... | ... | ... | ... | ... | ... | ... | ... | ... |
| ... | ... | ... | ... | ... | AB | ... | 0.3 | 6.39 | Y | ... | ... | O | M | ... |
| 06 46 05.05 | +32 33 20.4 | 263175 | HIP 32423 | ... | A | ... | ... | ... | ... | ... | ... | ... | ... | ... |
| ... | ... | ... | HD 263175B | ... | AB | ... | 30 | 787 | Y | ... | ... | P | ... | ... |
| 06 46 14.15 | +79 33 53.3 | 046588 | HIP 32439 | ... | ... | ... | ... | ... | ... | ... | ... | ... | ... | ... |
| 06 46 44.34 | +43 34 38.7 | 048682 | $\psi^5$ Aur | ... | ... | ... | ... | ... | ... | ... | ... | ... | ... | ... |
| 06 55 18.67 | +25 22 32.5 | 050692 | HIP 33277 | ... | ... | ... | ... | ... | ... | ... | ... | ... | ... | ... |
| 06 58 11.75 | +22 28 33.2 | 051419 | HIP 33537 | ... | ... | ... | ... | ... | ... | ... | ... | ... | ... | ... |
| 06 59 59.66 | −61 20 10.3 | 053143 | HIP 33690 | ... | ... | ... | ... | ... | ... | ... | ... | ... | ... | ... |
| 07 01 13.74 | −25 56 55.4 | 052698 | HIP 33817 | ... | A | ... | ... | ... | ... | ... | ... | ... | ... | ... |
| ... | ... | ... | ... | ... | Aa,Ab | ... | ... | ... | Y | ... | V | ... | M | ... |
| 07 01 38.59 | +48 22 43.2 | 051866 | HIP 33852 | ... | ... | ... | ... | ... | ... | ... | ... | ... | ... | ... |
| 07 03 30.46 | +29 20 13.5 | 052711 | HIP 34017 | ... | ... | ... | ... | ... | ... | ... | ... | ... | ... | ... |
| 07 03 57.32 | −43 36 28.9 | 053705 | HIP 34065 | ... | A | ... | ... | ... | ... | ... | ... | ... | ... | ... |
| ... | ... | ... | HD 53706 | ... | AB | ... | 20.9 | 345 | Y | ... | ... | T | ... | ... |

| R.A. (J2000.0) (1) | Decl. (J2000.0) (2) | HD Name (3) | Other Name (4) | N (5) | Comp ID (6) | Period (7) | Ang Sep (") (8) | Lin Sep (AU) (9) | Sts (10) | VB (11) | SB (12) | CP (13) | OT (14) | CH (15) |
|---|---|---|---|---|---|---|---|---|---|---|---|---|---|---|
| ... | ... | ... | HD 53680 | ... | AC | ... | 184.9 | 3053 | Y | ... | ... | T | ... | ... |
| ... | ... | ... | ... | ... | Ca,Cb | ... | ... | ... | Y | ... | ... | ... | M | ... |
| 07 08 04.24 | +29 50 04.2 | 053927 | HIP 34414 | ... | ... | ... | ... | ... | ... | ... | ... | ... | ... | ... |
| 07 09 35.39 | +25 43 43.1 | 054371 | HIP 34567 | ... | A | ... | ... | ... | ... | ... | ... | ... | ... | ... |
| ... | ... | ... | ... | ... | Aa,Ab | 32.81 d | ... | ... | Y | ... | 1 | ... | ... | ... |
| 07 15 50.14 | +47 14 23.9 | 055575 | HIP 35136 | ... | ... | ... | ... | ... | ... | ... | ... | ... | ... | ... |
| 07 17 29.56 | −46 58 45.3 | 057095 | HIP 35296 | ... | A | ... | ... | ... | ... | ... | ... | ... | ... | ... |
| ... | ... | ... | ... | ... | AB | 94.0 y | 0.75 | 10.9 | Y | P | ... | ... | ... | ... |
| 07 27 25.47 | −51 24 09.4 | 059468 | HIP 36210 | ... | ... | ... | ... | ... | ... | ... | ... | ... | ... | ... |
| 07 29 01.77 | +31 59 37.8 | ... | HIP 36357 | N | E[a] | ... | ... | ... | ... | ... | ... | ... | ... | ... |
| ... | ... | ... | ... | ... | Ea,Eb | ... | ... | ... | M | ... | ... | ... | M | ... |
| ... | ... | ... | HD 58946 | ... | AE | ... | 756.1 | 13351 | Y | ... | ... | T | ... | ... |
| ... | ... | ... | GJ 274B | ... | AB | ... | 3.4 | 60.0 | Y | ... | ... | O | M | ... |
| 07 30 42.51 | −37 20 21.7 | 059967 | HIP 36515 | ... | ... | ... | ... | ... | ... | ... | ... | ... | ... | ... |
| 07 33 00.58 | +37 01 47.4 | 059747 | HIP 36704 | ... | ... | ... | ... | ... | ... | ... | ... | ... | ... | ... |
| 07 34 26.17 | −06 53 48.0 | 060491 | HIP 36827 | ... | ... | ... | ... | ... | ... | ... | ... | ... | ... | ... |
| 07 39 59.33 | −03 35 51.0 | 061606 | HIP 37349 | ... | A | ... | ... | ... | ... | ... | ... | ... | ... | ... |
| ... | ... | ... | NLTT 18260 | ... | AB | ... | 58.3 | 828 | Y | ... | ... | P | ... | ... |
| 07 45 35.02 | −34 10 20.5 | 063077 | HIP 37853 | ... | A | ... | ... | ... | ... | ... | ... | ... | ... | ... |
| ... | ... | ... | ... | ... | Aa,Ab | ... | ... | ... | Y | ... | V | ... | M | ... |
| ... | ... | ... | NLTT 18414 | ... | AB | ... | 914 | 13901 | Y | ... | ... | P | ... | ... |
| 07 49 55.06 | +27 21 47.4 | 063433 | HIP 38228 | ... | ... | ... | ... | ... | ... | ... | ... | ... | ... | ... |
| 07 51 46.30 | −13 53 52.9 | 064096 | 9 Pup | ... | A | ... | ... | ... | ... | ... | ... | ... | ... | ... |
| ... | ... | ... | ... | ... | AB | 22.7 y | 0.60 | 9.90 | Y | O | 2 | ... | M | ... |
| 07 54 34.18 | −01 24 44.1 | 064606 | HIP 38625 | N | A | ... | ... | ... | ... | ... | ... | ... | ... | ... |
| ... | ... | ... | ... | ... | Aa,Ab | 1.23 y | ... | ... | Y | ... | 1 | ... | M | ... |
| 07 54 54.07 | +19 14 10.8 | 064468 | HIP 38657 | ... | A | ... | ... | ... | ... | ... | ... | ... | ... | ... |
| ... | ... | ... | ... | ... | Aa,Ab | 161.2 d | ... | ... | Y | ... | 1 | ... | ... | ... |
| 07 56 17.23 | +80 15 55.9 | 062613 | HIP 38784 | ... | ... | ... | ... | ... | ... | ... | ... | ... | ... | ... |
| 07 57 46.91 | −60 18 11.1 | 065907 | HIP 38908 | N | A | ... | ... | ... | ... | ... | ... | ... | ... | ... |
| ... | ... | ... | LHS 1960 | ... | AB | ... | 60.3 | 977 | Y | ... | ... | P | ... | ... |

| R.A.<br>(J2000.0)<br>(1) | Decl.<br>(J2000.0)<br>(2) | HD<br>Name<br>(3) | Other<br>Name<br>(4) | N<br>(5) | Comp<br>ID<br>(6) | Period<br>(7) | Ang<br>Sep<br>(″)<br>(8) | Lin<br>Sep<br>(AU)<br>(9) | Sts<br>(10) | VB<br>(11) | SB<br>(12) | CP<br>(13) | OT<br>(14) | CH<br>(15) |
|---|---|---|---|---|---|---|---|---|---|---|---|---|---|---|
| . . . | . . . | . . . | . . . | . . . | BC | . . . | 3 | 48.6 | Y | . . . | . . . | M | . . . | . . . |
| 07 59 33.93 | +20 50 38.0 | 065430 | HIP 39064 | . . . | A | . . . | . . . | . . . | . . . | . . . | . . . | . . . | M | . . . |
| . . . | . . . | . . . | . . . | . . . | Aa,Ab | 8.59 y | . . . | . . . | Y | . . . | 1 | . . . | M | . . . |
| 08 00 32.13 | +29 12 44.5 | 065583 | HIP 39157 | . . . | . . . | . . . | . . . | . . . | . . . | . . . | . . . | . . . | . . . | . . . |
| 08 02 31.19 | −66 01 15.4 | 067199 | HIP 39342 | . . . | A | . . . | . . . | . . . | . . . | . . . | . . . | . . . | . . . | . . . |
| . . . | . . . | . . . | . . . | . . . | Aa,Ab | . . . | . . . | . . . | M | . . . | . . . | . . . | M | . . . |
| 08 07 45.86 | +21 34 54.5 | 067228 | μ Cnc | . . . | . . . | . . . | . . . | . . . | . . . | . . . | . . . | . . . | . . . | . . . |
| 08 11 38.64 | +32 27 25.7 | 068017 | HIP 40118 | . . . | A | . . . | . . . | . . . | . . . | . . . | . . . | . . . | M | . . . |
| . . . | . . . | . . . | . . . | . . . | Aa,Ab | . . . | . . . | . . . | M | . . . | . . . | . . . | M | . . . |
| 08 12 12.73 | +17 38 52.0 | 068257 | ζ Cnc C | N | A | . . . | . . . | . . . | . . . | . . . | . . . | . . . | M | . . . |
| . . . | . . . | . . . | HD 68255 | . . . | AB | 59.58 y | 0.86 | 21.6 | Y | O | . . . | . . . | M | . . . |
| . . . | . . . | . . . | HD 68256 | . . . | AB,C | 1115 y | 7.70 | 193 | Y | P | . . . | . . . | M | . . . |
| . . . | . . . | . . . | . . . | . . . | Ca,Cb | 17.32 y | 0.18 | 4.51 | Y | P | 1 | . . . | . . . | . . . |
| . . . | . . . | . . . | . . . | . . . | Cb1,Cb2 | . . . | . . . | . . . | Y | . . . | . . . | . . . | L | . . . |
| . . . | . . . | . . . | . . . | . . . | Cb1,Cb3 | . . . | . . . | . . . | M | . . . | . . . | R | . . . | . . . |
| 08 18 23.95 | −12 37 55.8 | 069830 | HIP 40693 | . . . | A | . . . | . . . | . . . | . . . | . . . | . . . | . . . | . . . | . . . |
| . . . | . . . | . . . | HD 69830 b | . . . | . . . | 8.67 d | . . . | 0.08 | Y | . . . | 1 | . . . | . . . | . . . |
| . . . | . . . | . . . | HD 69830 c | . . . | . . . | 31.56 d | . . . | 0.19 | Y | . . . | 1 | . . . | . . . | . . . |
| . . . | . . . | . . . | HD 69830 d | . . . | . . . | 197 d | . . . | 0.63 | Y | . . . | 1 | . . . | . . . | . . . |
| 08 19 19.05 | +01 20 19.9 | . . . | HIP 40774 | . . . | . . . | . . . | . . . | . . . | . . . | . . . | . . . | . . . | . . . | . . . |
| 08 27 36.79 | +45 39 10.8 | 071148 | HIP 41484 | . . . | . . . | . . . | . . . | . . . | . . . | . . . | . . . | . . . | . . . | . . . |
| 08 32 51.50 | −31 30 03.1 | 072673 | HIP 41926 | . . . | A | . . . | . . . | . . . | . . . | . . . | . . . | . . . | . . . | . . . |
| 08 34 31.65 | −00 43 33.8 | 072760 | HIP 42074 | N | A | . . . | . . . | . . . | . . . | . . . | . . . | . . . | . . . | . . . |
| . . . | . . . | . . . | . . . | . . . | Aa,Ab | . . . | 0.96 | 20.3 | Y | . . . | . . . | R | M | . . . |
| 08 37 50.29 | −06 48 24.8 | 073350 | HIP 42333 | N | . . . | . . . | . . . | . . . | . . . | . . . | . . . | . . . | . . . | . . . |
| 08 39 07.90 | −22 39 42.8 | 073752 | HIP 42430 | N | A | . . . | . . . | . . . | . . . | . . . | . . . | . . . | . . . | . . . |
| . . . | . . . | . . . | . . . | . . . | AB | 123.0 y | 1.71 | 33.2 | Y | O | . . . | . . . | . . . | . . . |
| 08 39 11.70 | +65 01 15.3 | 072905 | π¹ UMa | . . . | . . . | . . . | . . . | . . . | . . . | . . . | . . . | . . . | . . . | . . . |
| 08 39 50.79 | +11 31 21.6 | 073667 | HIP 42499 | . . . | . . . | . . . | . . . | . . . | . . . | . . . | . . . | . . . | . . . | . . . |
| 08 42 07.52 | −42 55 46.0 | 074385 | HIP 42697 | . . . | A | . . . | . . . | . . . | . . . | . . . | . . . | . . . | . . . | . . . |
| . . . | . . . | . . . | NLTT 20102 | . . . | AB | . . . | 45 | 1029 | Y | . . . | . . . | P | . . . | . . . |

| R.A. (J2000.0) (1) | Decl. (J2000.0) (2) | HD Name (3) | Other Name (4) | N (5) | Comp ID (6) | Period (7) | Ang Sep ($''$) (8) | Lin Sep (AU) (9) | Sts (10) | VB (11) | SB (12) | CP (13) | OT (14) | CH (15) |
|---|---|---|---|---|---|---|---|---|---|---|---|---|---|---|
| 08 43 18.03 | −38 52 56.6 | 074576 | HIP 42808 | ... | A | ... | ... | ... | ... | ... | ... | ... | ... | ... |
| 08 52 16.39 | +08 03 46.5 | 075767 | HIP 43557 | N | A | ... | ... | ... | ... | ... | ... | ... | ... | ... |
| ... | ... | ... | ... | ... | Aa,Ab | 10.25 d | ... | ... | Y | ... | 1 | ... | ... | ... |
| ... | ... | ... | ... | ... | AB | ... | 3.4 | 81.7 | Y | ... | ... | M | ... | ... |
| ... | ... | ... | ... | ... | Ba,Bb | ... | ... | ... | Y | ... | 2 | ... | ... | ... |
| 08 52 35.81 | +28 19 50.9 | 075732 | 55 Cnc | ... | A | ... | ... | ... | ... | ... | ... | ... | ... | ... |
| ... | ... | ... | 55 Cnc e | ... | ... | 2.80 d | ... | 0.04 | Y | ... | 1 | ... | ... | ... |
| ... | ... | ... | 55 Cnc b | ... | ... | 14.65 d | ... | 0.11 | Y | ... | 1 | ... | ... | ... |
| ... | ... | ... | 55 Cnc c | ... | ... | 44.38 d | ... | 0.24 | Y | ... | 1 | ... | ... | ... |
| ... | ... | ... | 55 Cnc f | ... | ... | 260.7 d | ... | 0.78 | Y | ... | 1 | ... | ... | ... |
| ... | ... | ... | 55 Cnc d | ... | ... | 14.70 y | ... | 5.84 | Y | ... | 1 | ... | ... | ... |
| ... | ... | ... | LHS 2063 | ... | AB | ... | 84.7 | 1045 | Y | ... | ... | P | ... | ... |
| 08 54 17.95 | −05 26 04.1 | 076151 | HIP 43726 | ... | ... | ... | ... | ... | ... | ... | ... | ... | ... | ... |
| 08 58 43.93 | −16 07 57.8 | 076932 | HIP 44075 | ... | ... | ... | ... | ... | ... | ... | ... | ... | ... | ... |
| 09 08 51.07 | +33 52 56.0 | 078366 | HIP 44897 | ... | ... | ... | ... | ... | ... | ... | ... | ... | ... | ... |
| 09 12 17.55 | +14 59 45.7 | 079096 | 81 Cnc | N | A | ... | ... | ... | ... | ... | ... | ... | ... | ... |
| ... | ... | ... | ... | ... | Aa,Ab | 2.71 y | 0.12 | 2.44 | Y | O | 2 | ... | ... | S |
| ... | ... | ... | Gl 337C | ... | AE | ... | 43 | 875 | Y | ... | ... | P | ... | ... |
| ... | ... | ... | ... | ... | Ea,Eb | ... | 0.53 | 10.8 | Y | ... | ... | M | ... | ... |
| 09 14 20.54 | +61 25 23.9 | 079028 | HIP 45333 | ... | A | ... | ... | ... | ... | ... | ... | ... | ... | ... |
| ... | ... | ... | ... | ... | Aa,Ab | 16.24 d | ... | ... | Y | ... | 1 | ... | ... | ... |
| 09 17 53.46 | +28 33 37.9 | 079969 | HIP 45617 | ... | A | ... | ... | ... | ... | ... | ... | ... | ... | ... |
| ... | ... | ... | ... | ... | AB | 34.17 y | 0.68 | 11.7 | Y | O | ... | ... | M | ... |
| 09 22 25.95 | +40 12 03.8 | 080715 | HIP 45963 | ... | A | ... | ... | ... | ... | ... | ... | ... | ... | ... |
| ... | ... | ... | ... | ... | Aa,Ab | 3.8 d | ... | ... | Y | ... | 2 | ... | ... | V |
| 09 30 28.09 | −32 06 12.2 | 082342 | HIP 46626 | ... | A | ... | ... | ... | ... | ... | ... | ... | ... | ... |
| ... | ... | ... | ... | ... | AB | ... | 12 | 231 | Y | ... | ... | P | ... | ... |
| 09 32 25.57 | −11 11 04.7 | 082558 | HIP 46816 | ... | ... | ... | ... | ... | ... | ... | ... | ... | ... | ... |
| 09 32 43.76 | +26 59 18.7 | 082443 | HIP 46843 | ... | A | ... | ... | ... | ... | ... | ... | ... | ... | ... |
| ... | ... | ... | NLTT 22015 | ... | AB | ... | 65.2 | 1160 | Y | ... | ... | P | ... | ... |
| 09 35 39.50 | +35 48 36.5 | 082885 | HIP 47080 | ... | A | ... | ... | ... | ... | ... | ... | ... | ... | ... |

| R.A. (J2000.0) (1) | Decl. (J2000.0) (2) | HD Name (3) | Other Name (4) | N (5) | Comp ID (6) | Period (7) | Ang Sep (″) (8) | Lin Sep (AU) (9) | Sts (10) | VB (11) | SB (12) | CP (13) | OT (14) | CH (15) |
|---|---|---|---|---|---|---|---|---|---|---|---|---|---|---|
| ... | ... | ... | ... | ... | AB | 201 y | 3.84 | 43.7 | Y | P | ... | ... | M | ... |
| 09 42 14.42 | −23 54 56.1 | 084117 | HIP 47592 | ... | ... | ... | ... | ... | ... | ... | ... | ... | ... | ... |
| 09 48 35.37 | +46 01 15.6 | 084737 | HIP 48113 | ... | ... | ... | ... | ... | ... | ... | ... | ... | ... | ... |
| 10 01 00.66 | +31 55 25.2 | 086728 | HIP 49081 | N | A | ... | ... | ... | ... | ... | ... | ... | ... | ... |
| ... | ... | ... | GJ 376 B | ... | AB | ... | 133 | 2001 | Y | ... | ... | P | ... | ... |
| ... | ... | ... | ... | ... | Ba,Bb | ... | ... | ... | M | ... | ... | ... | L | ... |
| 10 04 37.66 | −11 43 46.9 | 087424 | HIP 49366 | ... | ... | ... | ... | ... | ... | ... | ... | ... | ... | ... |
| 10 08 43.14 | +34 14 32.1 | 087883 | HIP 49699 | ... | ... | ... | ... | ... | ... | ... | ... | ... | ... | ... |
| ... | ... | ... | HD 87883 b | ... | ... | 7.54 y | ... | 3.6 | Y | ... | 1 | ... | ... | ... |
| 10 13 24.73 | −33 01 54.2 | 088742 | HIP 50075 | ... | ... | ... | ... | ... | ... | ... | ... | ... | ... | ... |
| 10 17 14.54 | +23 06 22.4 | 089125 | HIP 50384 | ... | A | ... | ... | ... | ... | ... | ... | ... | ... | ... |
| ... | ... | ... | GJ 387 B | ... | AB | ... | 7.7 | 175 | Y | ... | ... | P | ... | ... |
| 10 18 51.95 | +44 02 54.0 | 089269 | HIP 50505 | ... | ... | ... | ... | ... | ... | ... | ... | ... | ... | ... |
| 10 23 55.27 | −29 38 43.9 | 090156 | HIP 50921 | ... | ... | ... | ... | ... | ... | ... | ... | ... | ... | ... |
| 10 28 03.88 | +48 47 05.6 | 090508 | HIP 51248 | ... | A | ... | ... | ... | ... | ... | ... | ... | ... | ... |
| ... | ... | ... | LHS 2266 | ... | AB | 765 y | 7.08 | 162 | Y | P | ... | ... | ... | ... |
| 10 30 37.58 | +55 58 49.9 | 090839 | HIP 51459 | N | A | ... | ... | ... | ... | ... | ... | ... | ... | ... |
| ... | ... | ... | HD 237903 | ... | AB | ... | 122.5 | 1565 | Y | ... | ... | T | ... | ... |
| ... | ... | ... | ... | ... | Ba,Bb | ... | ... | ... | M | ... | V | ... | ... | ... |
| 10 31 21.82 | −53 42 55.7 | 091324 | HIP 51523 | ... | ... | ... | ... | ... | ... | ... | ... | ... | ... | ... |
| 10 35 11.27 | +84 23 57.6 | 090343 | HIP 51819 | ... | ... | ... | ... | ... | ... | ... | ... | ... | ... | ... |
| 10 36 32.38 | −12 13 48.4 | 091889 | HIP 51933 | ... | ... | ... | ... | ... | ... | ... | ... | ... | ... | ... |
| 10 42 13.32 | −13 47 15.8 | 092719 | HIP 52369 | ... | ... | ... | ... | ... | ... | ... | ... | ... | ... | ... |
| 10 43 28.27 | −29 03 51.4 | 092945 | HIP 52462 | ... | ... | ... | ... | ... | ... | ... | ... | ... | ... | ... |
| 10 56 30.80 | +07 23 18.5 | 094765 | HIP 53486 | ... | ... | ... | ... | ... | ... | ... | ... | ... | ... | ... |
| 10 59 27.97 | +40 25 48.9 | 095128 | 47 UMa | ... | A | ... | ... | ... | ... | ... | ... | ... | ... | ... |
| ... | ... | ... | 47 UMa b | ... | ... | 3.00 y | ... | 2.14 | Y | ... | 1 | ... | ... | ... |
| ... | ... | ... | 47 UMa c | ... | ... | 6.00 y | ... | 3.39 | Y | ... | 1 | ... | ... | ... |
| 11 04 41.47 | −04 13 15.9 | 096064 | HIP 54155 | ... | A | ... | ... | ... | ... | ... | ... | ... | ... | ... |
| ... | ... | ... | NLTT 26194 | ... | A,BC | ... | 11.8 | 310 | Y | ... | ... | P | ... | ... |
| ... | ... | ... | BD-033040C | ... | BC | 23.23 y | 0.34 | 8.93 | Y | O | V | ... | ... | ... |

| R.A. (J2000.0) (1) | Decl. (J2000.0) (2) | HD Name (3) | Other Name (4) | N (5) | Comp ID (6) | Period (7) | Ang Sep ('') (8) | Lin Sep (AU) (9) | Sts (10) | VB (11) | SB (12) | CP (13) | OT (14) | CH (15) |
|---|---|---|---|---|---|---|---|---|---|---|---|---|---|---|
| 11 08 14.01 | +38 25 35.9 | 096612 | HIP 54426 | ... | ... | ... | ... | ... | ... | ... | ... | ... | ... | ... |
| 11 12 01.19 | −26 08 12.0 | 097343 | HIP 54704 | ... | ... | ... | ... | ... | ... | ... | ... | ... | ... | ... |
| 11 12 32.35 | +35 48 50.7 | 097334 | HIP 54745 | N | A | ... | ... | ... | ... | ... | ... | ... | ... | ... |
| ... | ... | ... | Gl 417B | ... | AE | ... | 89.7 | 1966 | Y | ... | ... | R | ... | ... |
| ... | ... | ... | ... | ... | Ea,Eb | ... | 0.1 | 2.19 | Y | ... | ... | M | ... | ... |
| 11 14 33.16 | +25 42 37.4 | 097658 | HIP 54906 | ... | ... | ... | ... | ... | ... | ... | ... | ... | ... | ... |
| 11 18 10.95 | +31 31 45.7 | 098230 | ξ UMa B | N | Bᵃ | ... | ... | ... | ... | ... | ... | ... | ... | ... |
| ... | ... | ... | ... | ... | Ba,Bb | 3.98 d | ... | ... | Y | ... | 1 | ... | ... | ... |
| ... | ... | ... | HD 98231 | ... | AB | 59.88 y | 2.54 | 21.2 | Y | O | ... | ... | ... | S |
| ... | ... | ... | ... | ... | Aa,Ab | 1.84 y | ... | ... | Y | U | 1 | ... | ... | ... |
| 11 18 22.01 | −05 04 02.3 | 098281 | HIP 55210 | ... | ... | ... | ... | ... | ... | ... | ... | ... | ... | ... |
| 11 26 45.32 | +03 00 47.2 | 099491 | 83 Leo | ... | A | ... | ... | ... | ... | ... | ... | ... | ... | ... |
| ... | ... | ... | HD 99492 | ... | AB | 32000 y | 40.76 | 723 | Y | P | ... | T | ... | ... |
| ... | ... | ... | HD 99492 b | ... | ... | 17.04 d | ... | 0.12 | Y | ... | 1 | ... | ... | ... |
| 11 31 44.95 | +14 21 52.2 | 100180 | 88 Leo | N | A | ... | ... | ... | ... | ... | ... | ... | ... | ... |
| ... | ... | ... | ... | ... | Aa,Ab | ... | 0.1 | 2.33 | M | ... | ... | ... | ... | ... |
| ... | ... | ... | NLTT 27656 | ... | AB | ... | 15.3 | 356 | Y | ... | ... | P | ... | ... |
| 11 34 29.49 | −32 49 52.8 | 100623 | HIP 56452 | N | A | ... | ... | ... | ... | ... | ... | ... | ... | ... |
| ... | ... | ... | LHS 309 | ... | AB | ... | 17 | 162 | Y | ... | ... | M | ... | ... |
| 11 38 44.90 | +45 06 30.3 | 101177 | HIP 56809 | ... | A | ... | ... | ... | ... | ... | ... | ... | ... | ... |
| ... | ... | ... | LHS 2436 | ... | AB | 2050 y | 10.33 | 240 | Y | P | ... | ... | M | ... |
| ... | ... | ... | ... | ... | Ba,Bb | 23.54 d | ... | ... | Y | ... | 2 | ... | ... | ... |
| 11 38 59.72 | +42 19 43.7 | 101206 | HIP 56829 | ... | A | ... | ... | ... | ... | ... | ... | ... | ... | ... |
| ... | ... | ... | ... | ... | Aa,Ab | 12.92 d | ... | ... | Y | ... | 1 | ... | ... | ... |
| 11 41 03.02 | +34 12 05.9 | 101501 | 61 UMa | ... | ... | ... | ... | ... | ... | ... | ... | ... | ... | ... |
| 11 46 31.07 | −40 30 01.3 | 102365 | HIP 57443 | N | A | ... | ... | ... | ... | ... | ... | ... | ... | ... |
| ... | ... | ... | LHS 313 | ... | AB | ... | 22.9 | 211 | Y | ... | ... | M | ... | ... |
| 11 47 15.81 | −30 17 11.4 | 102438 | HIP 57507 | ... | ... | ... | ... | ... | ... | ... | ... | ... | ... | ... |
| ... | ... | ... | ... | ... | Aa,Ab | ... | ... | ... | M | ... | V | ... | ... | ... |
| 11 50 41.72 | +01 45 53.0 | 102870 | β Vir | ... | ... | ... | ... | ... | ... | ... | ... | ... | ... | ... |
| 11 52 58.77 | +37 43 07.2 | 103095 | HIP 57939 | N | ... | ... | ... | ... | ... | ... | ... | ... | ... | ... |

| R.A. (J2000.0) (1) | Decl. (J2000.0) (2) | HD Name (3) | Other Name (4) | N (5) | Comp ID (6) | Period (7) | Ang Sep ($''$) (8) | Lin Sep (AU) (9) | Sts (10) | VB (11) | SB (12) | CP (13) | OT (14) | CH (15) |
|---|---|---|---|---|---|---|---|---|---|---|---|---|---|---|
| 11 59 10.01 | −20 21 13.6 | 104067 | HIP 58451 | … | … | … | … | … | … | … | … | … | … | … |
| 12 00 44.45 | −10 26 45.6 | 104304 | HIP 58576 | … | … | … | … | … | … | … | … | … | … | … |
| 12 09 37.26 | +40 15 07.4 | 105631 | HIP 59280 | … | … | … | … | … | … | … | … | … | … | … |
| 12 30 50.14 | +53 04 35.8 | 108954 | HIP 61053 | … | … | … | … | … | … | … | … | … | … | … |
| 12 33 31.38 | −68 45 20.9 | 109200 | HIP 61291 | … | … | … | … | … | … | … | … | … | … | … |
| 12 33 44.54 | +41 21 26.9 | 109358 | HIP 61317 | N | … | … | … | … | … | … | … | … | … | … |
| 12 41 44.52 | +55 43 28.8 | 110463 | HIP 61946 | … | A | … | … | … | … | … | … | … | … | … |
| … | … | … | … | … | Aa,Ab | … | … | … | M | … | V | … | … | … |
| 12 44 14.55 | +51 45 33.5 | 110833 | HIP 62145 | … | A | … | … | … | … | … | … | … | … | … |
| … | … | … | … | … | Aa,Ab | 271.17 d | … | … | Y | U | 1 | … | M | … |
| 12 44 59.41 | +39 16 44.1 | 110897 | HIP 62207 | … | … | … | … | … | … | … | … | … | … | … |
| 12 45 14.41 | −57 21 28.8 | 110810 | HIP 62229 | … | … | … | … | … | … | … | … | … | … | … |
| 12 48 32.31 | −15 43 10.1 | 111312 | HIP 62505 | N | A | … | … | … | … | … | … | … | … | … |
| … | … | … | … | … | Aa,Ab | 2.68 y | 0.1 | 2.38 | Y | … | 2 | … | M | … |
| … | … | … | … | … | Aa,B | … | 2.7 | 64.3 | M | … | … | R | M | … |
| 12 48 47.05 | +24 50 24.8 | 111395 | HIP 62523 | … | … | … | … | … | … | … | … | … | … | … |
| 12 59 01.56 | −09 50 02.7 | 112758 | HIP 63366 | N | A | … | … | … | … | … | … | … | … | … |
| … | … | … | … | … | Aa,Ab | 103.17 d | … | … | Y | … | … | … | … | … |
| … | … | … | … | … | AB | … | 0.8 | 16.7 | Y | … | … | M | … | … |
| 12 59 32.78 | +41 59 12.4 | 112914 | HIP 63406 | … | A | … | … | … | … | … | … | … | … | … |
| … | … | … | … | … | Aa,Ab | 1.95 y | … | … | Y | U | 1 | … | … | … |
| 13 03 49.66 | −05 09 42.5 | 113449 | HIP 63742 | N | A | … | … | … | … | … | … | … | … | … |
| … | … | … | … | … | Aa,Ab | 231.23 d | … | … | Y | U | V | R | M | … |
| 13 11 52.39 | +27 52 41.5 | 114710 | $\beta$ Com | … | … | … | … | … | … | … | … | … | … | … |
| 13 12 03.18 | −37 48 10.9 | 114613 | HIP 64408 | … | … | … | … | … | … | … | … | … | … | … |
| 13 12 43.79 | −02 15 54.1 | 114783 | HIP 64457 | … | A | … | … | … | … | … | … | … | … | … |
| … | … | … | HD 114783 b | … | … | 1.35 y | … | 1.17 | Y | … | 1 | … | … | … |
| 13 13 52.23 | −45 11 08.9 | 114853 | HIP 64550 | … | … | … | … | … | … | … | … | … | … | … |
| 13 15 26.45 | −87 33 38.5 | 113283 | HIP 64690 | … | A | … | … | … | … | … | … | … | … | … |
| … | … | … | … | … | Aa,Ab | … | … | … | M | … | … | … | M | … |
| 13 16 46.52 | +09 25 27.0 | 115383 | 59 Vir | … | … | … | … | … | … | … | … | … | … | … |

| R.A. (J2000.0) (1) | Decl. (J2000.0) (2) | HD Name (3) | Other Name (4) | N (5) | Comp ID (6) | Period (7) | Ang Sep (") (8) | Lin Sep (AU) (9) | Sts (10) | VB (11) | SB (12) | CP (13) | OT (14) | CH (15) |
|---|---|---|---|---|---|---|---|---|---|---|---|---|---|---|
| 13 16 51.05 | +17 01 01.9 | 115404 | HIP 64797 | ... | A | ... | ... | ... | ... | ... | ... | ... | ... | ... |
| ... | ... | ... | LHS 2714 | ... | AB | 770 y | 8.06 | 89.2 | Y | P | ... | ... | M | ... |
| 13 18 24.31 | −18 18 40.3 | 115617 | 61 Vir | ... | ... | ... | ... | ... | ... | ... | ... | ... | ... | ... |
| 13 23 39.15 | +02 43 24.0 | 116442 | HIP 65352 | ... | A | ... | ... | ... | ... | ... | ... | ... | ... | ... |
| ... | ... | ... | HD 116443 | ... | AB | ... | 26.2 | 404 | Y | ... | ... | T | ... | ... |
| 13 25 45.53 | +56 58 13.8 | 116956 | HIP 65515 | ... | ... | ... | ... | ... | ... | ... | ... | ... | ... | ... |
| 13 25 59.86 | +63 15 40.6 | 117043 | HIP 65530 | ... | ... | ... | ... | ... | ... | ... | ... | ... | ... | ... |
| 13 28 25.81 | +13 46 43.6 | 117176 | 70 Vir | ... | A | ... | ... | ... | ... | ... | ... | ... | ... | ... |
| ... | ... | ... | 70 Vir b | ... | ... | 116.69 d | ... | 0.48 | Y | ... | 1 | ... | ... | ... |
| 13 41 04.17 | −34 27 51.0 | 118972 | HIP 66765 | ... | ... | ... | ... | ... | ... | ... | ... | ... | ... | ... |
| 13 41 13.40 | +56 43 37.8 | 119332 | HIP 66781 | ... | ... | ... | ... | ... | ... | ... | ... | ... | ... | ... |
| 13 47 15.74 | +17 27 24.9 | 120136 | τ Boo | N | A | ... | ... | ... | ... | ... | ... | ... | ... | ... |
| ... | ... | ... | τ Boo b | ... | ... | 3.31 d | ... | 0.05 | Y | ... | 1 | ... | ... | ... |
| ... | ... | ... | HD 120136B | ... | AB | 2000 y | 14.39 | 224 | Y | P | ... | ... | M | ... |
| 13 51 20.33 | −24 23 25.3 | 120690 | HIP 67620 | ... | A | ... | ... | ... | ... | ... | ... | ... | ... | ... |
| 13 51 40.40 | ... | ... | ... | ... | Aa,Ab | 10.3 y | ... | ... | Y | ... | 1 | ... | M | ... |
| 13 51 40.40 | −57 26 08.4 | 120559 | HIP 67655 | ... | ... | ... | ... | ... | ... | ... | ... | ... | ... | ... |
| 13 52 35.87 | −50 55 18.3 | 120780 | HIP 67742 | N | A | ... | ... | ... | ... | ... | ... | ... | ... | ... |
| ... | ... | ... | ... | ... | Aa,Ab | ... | ... | ... | Y | ... | ... | ... | M | ... |
| ... | ... | ... | ... | ... | AB | ... | 5.8 | 99.1 | Y | ... | ... | M | ... | ... |
| 13 54 41.08 | +18 23 51.8 | 121370 | η Boo | ... | A | ... | ... | ... | ... | ... | ... | ... | ... | ... |
| ... | ... | ... | ... | ... | Aa,Ab | 1.34 y | ... | ... | Y | U | 1 | ... | ... | ... |
| 13 55 49.99 | +14 03 23.4 | 121560 | HIP 68030 | ... | ... | ... | ... | ... | ... | ... | ... | ... | ... | ... |
| 14 03 32.35 | +10 47 12.4 | 122742 | HIP 68682 | ... | A | ... | ... | ... | ... | ... | ... | ... | ... | ... |
| ... | ... | ... | ... | ... | Aa,Ab | 9.9 y | ... | ... | Y | U | 1 | ... | M | ... |
| 14 11 46.17 | −12 36 42.4 | 124106 | HIP 69357 | ... | ... | ... | ... | ... | ... | ... | ... | ... | ... | ... |
| 14 12 45.24 | −03 19 12.3 | 124292 | HIP 69414 | ... | ... | ... | ... | ... | ... | ... | ... | ... | ... | ... |
| 14 15 38.68 | −45 00 02.7 | 124580 | HIP 69671 | N | ... | ... | ... | ... | ... | ... | ... | ... | ... | ... |
| 14 16 00.87 | −06 00 02.0 | 124850 | ι Vir | ... | A | ... | ... | ... | ... | ... | ... | ... | ... | ... |
| ... | ... | ... | ... | ... | Aa,Ab | 55 y | ... | ... | M | U | ... | ... | ... | ... |
| 14 19 00.90 | −25 48 55.5 | 125276 | HIP 69965 | N | A | ... | ... | ... | ... | ... | ... | ... | ... | ... |

| R.A. (J2000.0) (1) | Decl. (J2000.0) (2) | HD Name (3) | Other Name (4) | N (5) | Comp ID (6) | Period (7) | Ang Sep ('') (8) | Lin Sep (AU) (9) | Sts (10) | VB (11) | SB (12) | CP (13) | OT (14) | CH (15) |
|---|---|---|---|---|---|---|---|---|---|---|---|---|---|---|
| ... | ... | ... | ... | ... | Aa,Ab | ... | 8.0 | 144 | M | ... | ... | R | M | ... |
| 14 19 34.86 | −05 09 04.3 | 125455 | HIP 70016 | N | A | ... | ... | ... | ... | ... | ... | ... | ... | ... |
| ... | ... | ... | LHS 2895 | ... | AB | ... | 8.1 | 169 | Y | ... | ... | M | ... | ... |
| 14 23 15.28 | +01 14 29.6 | 126053 | HIP 70319 | ... | ... | ... | ... | ... | ... | ... | ... | ... | ... | ... |
| 14 29 22.30 | +80 48 35.5 | 128642 | HIP 70857 | ... | A | ... | ... | ... | ... | ... | ... | ... | ... | ... |
| ... | ... | ... | ... | ... | Aa,Ab | 178.78 d | ... | ... | Y | U | 1 | ... | M | ... |
| 14 29 36.81 | +41 47 45.3 | 127334 | HIP 70873 | ... | ... | ... | ... | ... | ... | ... | ... | ... | ... | ... |
| 14 33 28.87 | +52 54 31.6 | 128165 | HIP 71181 | ... | ... | ... | ... | ... | ... | ... | ... | ... | ... | ... |
| 14 36 00.56 | +09 44 47.5 | 128311 | HIP 71395 | ... | A | ... | ... | ... | ... | ... | ... | ... | ... | ... |
| ... | ... | ... | HD 128311 b | ... | ... | 1.25 y | ... | 1.10 | Y | ... | 1 | ... | ... | ... |
| ... | ... | ... | HD 128311 c | ... | ... | 2.50 y | ... | 1.74 | Y | ... | 1 | ... | ... | ... |
| 14 39 36.50 | −60 50 02.3 | 128620 | α Cen | N | A | ... | ... | ... | ... | ... | ... | ... | ... | ... |
| ... | ... | ... | HD 128621 | ... | AB | 79.91 y | 17.57 | 23.3 | Y | O | 2 | ... | ... | ... |
| ... | ... | ... | Proxima Cen | ... | AC | ... | 7867 | 10422 | Y | ... | ... | T | ... | ... |
| 14 40 31.11 | −16 12 33.4 | 128987 | HIP 71743 | ... | ... | ... | ... | ... | ... | ... | ... | ... | ... | ... |
| 14 41 52.46 | −75 08 22.1 | 128400 | HIP 71855 | ... | ... | ... | ... | ... | ... | ... | ... | ... | ... | ... |
| 14 45 24.18 | +13 50 46.7 | 130004 | HIP 72146 | ... | ... | ... | ... | ... | ... | ... | ... | ... | ... | ... |
| 14 47 16.10 | +02 42 11.6 | 130307 | HIP 72312 | ... | ... | ... | ... | ... | ... | ... | ... | ... | ... | ... |
| 14 49 23.72 | −67 14 09.5 | 130042 | HIP 72493 | ... | A | ... | ... | ... | ... | ... | ... | ... | ... | ... |
| ... | ... | ... | ... | ... | AB | ... | 1.5 | 37.4 | Y | ... | ... | O | M | ... |
| 14 50 15.81 | +23 54 42.6 | 130948 | HIP 72567 | N | A | ... | ... | ... | ... | ... | ... | ... | ... | ... |
| ... | ... | ... | HD 130948 B | ... | AB | ... | 2.6 | 47.2 | Y | ... | ... | M | ... | ... |
| ... | ... | ... | HD 130948 C | ... | BC | ... | 0.1 | 1.82 | Y | ... | ... | M | ... | ... |
| 14 51 23.38 | +19 06 01.7 | 131156 | ξ Boo | ... | A | ... | ... | ... | ... | ... | ... | ... | ... | ... |
| ... | ... | ... | HD 131156B | ... | AB | 151.6 y | 4.94 | 33.2 | Y | O | ... | ... | ... | ... |
| 14 53 23.77 | +19 09 10.1 | 131511 | HIP 72848 | ... | A | ... | ... | ... | ... | ... | ... | ... | ... | ... |
| ... | ... | ... | ... | ... | Aa,Ab | 125.4 d | ... | ... | Y | U | 1 | ... | ... | S |
| 14 53 41.57 | +23 20 42.6 | 131582 | HIP 72875 | ... | A | ... | ... | ... | ... | ... | ... | ... | ... | ... |
| ... | ... | ... | ... | ... | Aa,Ab | ... | ... | ... | Y | ... | ... | ... | M | ... |
| 14 55 11.04 | +53 40 49.2 | 132142 | HIP 73005 | ... | ... | ... | ... | ... | ... | ... | ... | ... | ... | ... |
| 14 56 23.04 | +49 37 42.4 | 132254 | HIP 73100 | ... | ... | ... | ... | ... | ... | ... | ... | ... | ... | ... |

| R.A. (J2000.0) (1) | Decl. (J2000.0) (2) | HD Name (3) | Other Name (4) | N (5) | Comp ID (6) | Period (7) | Ang Sep (″) (8) | Lin Sep (AU) (9) | Sts (10) | VB (11) | SB (12) | CP (13) | OT (14) | CH (15) |
|---|---|---|---|---|---|---|---|---|---|---|---|---|---|---|
| 14 58 08.80 | −48 51 46.8 | 131923 | HIP 73241 | ... | A | ... | ... | ... | ... | ... | ... | ... | ... | ... |
| ... | ... | ... | ... | ... | Aa,Ab | 14.87 y | ... | ... | Y | ... | 1 | ... | M | ... |
| 15 03 47.30 | +47 39 14.6 | 133640 | 44 Boo | ... | A | ... | ... | ... | ... | ... | ... | ... | ... | ... |
| ... | ... | ... | NLTT 39210 | ... | AB | 206 y | 3.8 | 47.5 | Y | O | V | ... | M | ... |
| ... | ... | ... | ... | ... | Ba,Bb | 6.43 h | ... | ... | Y | ... | 2 | ... | E | ... |
| 15 10 44.74 | −61 25 20.3 | 134060 | HIP 74273 | ... | ... | ... | ... | ... | ... | ... | ... | ... | ... | ... |
| 15 13 50.89 | −01 21 05.0 | 135204 | HIP 74537 | N | A | ... | ... | ... | ... | ... | ... | ... | ... | ... |
| ... | ... | ... | ... | ... | AB | ... | 0.1 | 1.77 | Y | ... | ... | O | ... | ... |
| 15 15 59.17 | +00 47 46.9 | 135599 | HIP 74702 | ... | ... | ... | ... | ... | ... | ... | ... | ... | ... | ... |
| 15 19 18.80 | +01 45 55.5 | 136202 | HIP 74975 | ... | A | ... | ... | ... | ... | ... | ... | ... | ... | ... |
| ... | ... | ... | LHS 3060 | ... | AB | ... | 11.4 | 289 | Y | ... | ... | P | ... | ... |
| 15 21 48.15 | −48 19 03.5 | 136352 | HIP 75181 | ... | ... | ... | ... | ... | ... | ... | ... | ... | ... | ... |
| 15 22 36.69 | −10 39 40.0 | 136713 | HIP 75253 | ... | ... | ... | ... | ... | ... | ... | ... | ... | ... | ... |
| 15 22 46.83 | +18 55 08.3 | 136923 | HIP 75277 | ... | ... | ... | ... | ... | ... | ... | ... | ... | ... | ... |
| 15 23 12.31 | +30 17 16.1 | 137107 | η CrB | N | A | ... | ... | ... | ... | ... | ... | ... | ... | ... |
| ... | ... | ... | HD 137108 | ... | AB | 41.56 y | 0.87 | 15.5 | Y | O | 2 | ... | ... | ... |
| ... | ... | ... | GJ 584 C | ... | AB,E | ... | 193.5 | 3456 | Y | ... | ... | S | ... | ... |
| 15 28 09.61 | −09 20 53.1 | 137763 | HIP 75718 | N | A | ... | ... | ... | ... | ... | ... | ... | ... | ... |
| ... | ... | ... | ... | ... | Aa,Ab | 2.44 y | 0.1 | 2.06 | Y | U | 2 | R | M | S |
| ... | ... | ... | HD 137778 | ... | AB | ... | 52.3 | 1076 | Y | ... | ... | T | ... | ... |
| ... | ... | ... | GJ 586C | ... | AC | ... | 1212 | 24948 | Y | ... | ... | T | ... | ... |
| 15 29 11.18 | +80 26 55.0 | 139777 | HIP 75809 | ... | A | ... | ... | ... | ... | ... | ... | ... | ... | ... |
| ... | ... | ... | HD 139813 | ... | AB | ... | 31.3 | 683 | Y | ... | ... | T | ... | ... |
| 15 36 02.22 | +39 48 08.9 | 139341 | HIP 76382 | ... | A | ... | ... | ... | ... | ... | ... | ... | ... | ... |
| ... | ... | ... | ... | ... | AB | 55.6 y | 0.79 | 17.6 | Y | O | ... | ... | M | ... |
| ... | ... | ... | HD 139323 | ... | AB,C | ... | 121.5 | 2710 | Y | ... | ... | T | ... | ... |
| 15 44 01.82 | +02 30 54.6 | 140538 | ψ Ser | ... | A | ... | ... | ... | ... | ... | ... | ... | ... | ... |
| ... | ... | ... | ... | ... | AB | ... | 4.4 | 64.5 | Y | ... | ... | O | M | ... |
| 15 46 26.61 | +07 21 11.1 | 141004 | HIP 77257 | ... | ... | ... | ... | ... | ... | ... | ... | ... | ... | ... |
| 15 47 29.10 | −37 54 58.7 | 140901 | HIP 77358 | N | A | ... | ... | ... | ... | ... | ... | ... | ... | ... |
| ... | ... | ... | NLTT 41169 | ... | AB | ... | 15 | 230 | Y | ... | ... | M | ... | ... |

| R.A.<br>(J2000.0)<br>(1) | Decl.<br>(J2000.0)<br>(2) | HD<br>Name<br>(3) | Other<br>Name<br>(4) | N<br>(5) | Comp<br>ID<br>(6) | Period<br>(7) | Ang<br>Sep<br>(″)<br>(8) | Lin<br>Sep<br>(AU)<br>(9) | Sts<br>(10) | VB<br>(11) | SB<br>(12) | CP<br>(13) | OT<br>(14) | CH<br>(15) |
|---|---|---|---|---|---|---|---|---|---|---|---|---|---|---|
| 15 48 09.46 | +01 34 18.3 | 141272 | HIP 77408 | N | A | ... | ... | ... | ... | ... | ... | ... | ... | ... |
| ... | ... | ... | ... | ... | AB | ... | 17.9 | 381 | Y | ... | ... | S | ... | ... |
| 15 52 40.54 | +42 27 05.5 | 142373 | χ Her | ... | ... | ... | ... | ... | ... | ... | ... | ... | ... | ... |
| 15 53 12.10 | +13 11 47.8 | 142267 | 39 Ser | ... | A | ... | ... | ... | ... | ... | ... | ... | ... | ... |
| ... | ... | ... | ... | ... | Aa,Ab | 138.56 d | ... | ... | Y | ... | 1 | ... | ... | ... |
| 16 01 02.66 | +33 18 12.6 | 143761 | ρ CrB | N | A | ... | ... | ... | ... | ... | ... | ... | ... | ... |
| ... | ... | ... | ρ CrB b | ... | ... | 39.84 d | ... | 0.23 | M | ... | 1 | ... | ... | ... |
| ... | ... | ... | ... | ... | Aa,Ab | 40.18 d | ... | ... | M | U | ... | ... | ... | ... |
| 16 01 53.35 | +58 33 54.9 | 144284 | θ Dra | N | A | ... | ... | ... | ... | ... | ... | ... | ... | ... |
| ... | ... | ... | ... | ... | Aa,Ab | 3.07 d | ... | ... | Y | ... | 2 | ... | ... | V |
| 16 04 03.71 | +25 15 17.4 | 144287 | HIP 78709 | ... | A | ... | ... | ... | ... | ... | ... | ... | ... | ... |
| ... | ... | ... | ... | ... | Aa,Ab | 12.19 y | 0.2 | 4.44 | Y | ... | 1 | ... | M | ... |
| 16 04 56.79 | +39 09 23.4 | 144579 | HIP 78775 | N | A | ... | ... | ... | ... | ... | ... | ... | ... | ... |
| ... | ... | ... | LHS 3150 | ... | AB | ... | 70.3 | 1020 | Y | ... | ... | P | ... | ... |
| 16 06 29.60 | +38 37 56.1 | 144872 | HIP 78913 | ... | ... | ... | ... | ... | ... | ... | ... | ... | ... | ... |
| 16 09 42.79 | −56 26 42.5 | 144628 | HIP 79190 | ... | ... | ... | ... | ... | ... | ... | ... | ... | ... | ... |
| 16 10 24.31 | +43 49 03.5 | 145675 | 14 Her | ... | A | ... | ... | ... | ... | ... | ... | ... | ... | ... |
| ... | ... | ... | 14 Her b | ... | ... | 4.81 y | ... | 2.85 | Y | ... | 1 | ... | ... | ... |
| 16 13 18.45 | +13 31 36.9 | 145958 | HIP 79492 | N | A | ... | ... | ... | ... | ... | ... | ... | ... | ... |
| ... | ... | ... | NLTT 42272 | ... | AB | 1354 y | 5.09 | 120 | Y | P | ... | ... | ... | ... |
| ... | ... | ... | ... | ... | AD | ... | 1623 | 38278 | M | ... | ... | S | ... | ... |
| 16 13 48.56 | −57 34 13.8 | 145417 | HIP 79537 | ... | ... | ... | ... | ... | ... | ... | ... | ... | ... | ... |
| 16 14 11.93 | −31 39 49.1 | 145825 | HIP 79578 | ... | A | ... | ... | ... | ... | ... | ... | ... | ... | ... |
| ... | ... | ... | ... | ... | Aa,Ab | 7.14 y | ... | ... | Y | ... | 1 | ... | M | ... |
| 16 14 40.85 | +33 51 31.0 | 146361 | σ² CrB | N | A | ... | ... | ... | ... | ... | ... | ... | ... | ... |
| ... | ... | ... | ... | ... | Aa,Ab | 1.14 d | ... | ... | Y | O | 2 | ... | ... | V |
| ... | ... | ... | HD 146362 | ... | AB | 726 y | 5.26 | 110 | Y | P | ... | ... | M | ... |
| ... | ... | ... | HIP 79551 | ... | AE | ... | 633.7 | 13357 | Y | ... | ... | T | ... | ... |
| ... | ... | ... | σ CrB D | ... | Ea,Eb | 52 y | ... | ... | Y | U | ... | ... | ... | ... |
| 16 15 37.27 | −08 22 10.0 | 146233 | 18 Sco | ... | ... | ... | ... | ... | ... | ... | ... | ... | ... | ... |
| 16 24 01.29 | −39 11 34.7 | 147513 | HIP 80337 | ... | A | ... | ... | ... | ... | ... | ... | ... | ... | ... |

| R.A. (J2000.0) (1) | Decl. (J2000.0) (2) | HD Name (3) | Other Name (4) | N (5) | Comp ID (6) | Period (7) | Ang Sep (″) (8) | Lin Sep (AU) (9) | Sts (10) | VB (11) | SB (12) | CP (13) | OT (14) | CH (15) |
|---|---|---|---|---|---|---|---|---|---|---|---|---|---|---|
| ... | ... | ... | HD 147513 b | ... | ... | 1.45 y | ... | 1.31 | Y | ... | 1 | ... | ... | ... |
| ... | ... | ... | ... | ... | AB | ... | 345 | 4408 | Y | ... | ... | T | ... | ... |
| 16 24 19.81 | −13 38 30.0 | 147776 | HIP 80366 | N | A | ... | ... | ... | ... | ... | ... | ... | ... | ... |
| ... | ... | ... | ... | ... | AC | ... | 6.4 | 137 | M | ... | ... | R | ... | ... |
| ... | ... | ... | ... | ... | AD | ... | 9.7 | 208 | Y | ... | ... | M | ... | ... |
| 16 28 28.14 | −70 05 03.8 | 147584 | ζ TrA | ... | A | ... | ... | ... | ... | ... | ... | ... | ... | ... |
| ... | ... | ... | ... | ... | Aa,Ab | 12.98 d | ... | ... | Y | U | 1 | ... | M | ... |
| 16 28 52.67 | +18 24 50.6 | 148653 | HIP 80725 | ... | A | ... | ... | ... | ... | ... | ... | ... | ... | ... |
| ... | ... | ... | LHS 3204 | ... | AB | 224 y | 2.21 | 43.4 | Y | O | ... | ... | M | ... |
| 16 31 30.03 | −39 00 44.2 | 148704 | HIP 80925 | N | A | ... | ... | ... | ... | ... | ... | ... | ... | ... |
| ... | ... | ... | ... | ... | Aa,Ab | 31.86 d | ... | ... | Y | ... | 2 | ... | M | ... |
| 16 36 21.45 | −02 19 28.5 | 149661 | HIP 81300 | ... | ... | ... | ... | ... | ... | ... | ... | ... | ... | ... |
| 16 37 08.43 | +00 15 15.6 | 149806 | HIP 81375 | N | A | ... | ... | ... | ... | ... | ... | ... | ... | ... |
| ... | ... | ... | ... | ... | AB | ... | 6.3 | 128 | Y | ... | ... | P | ... | ... |
| 16 39 04.14 | −58 15 29.5 | 149612 | HIP 81520 | ... | ... | ... | ... | ... | ... | ... | ... | ... | ... | ... |
| 16 42 38.58 | +68 06 07.8 | 151541 | HIP 81813 | ... | ... | ... | ... | ... | ... | ... | ... | ... | ... | ... |
| 16 52 58.80 | −00 01 35.1 | 152391 | HIP 82588 | ... | ... | ... | ... | ... | ... | ... | ... | ... | ... | ... |
| 16 57 53.18 | +47 22 00.1 | 153557 | HIP 83020 | N | A | ... | ... | ... | ... | ... | ... | ... | ... | ... |
| ... | ... | ... | ... | ... | AB | ... | 4.9 | 89.7 | Y | ... | ... | M | M | ... |
| ... | ... | ... | HD 153525 | ... | AC | ... | 112.1 | 2051 | Y | ... | ... | T | ... | ... |
| 17 02 36.40 | +47 04 54.8 | 154345 | HIP 83389 | ... | A | ... | ... | ... | ... | ... | ... | ... | ... | ... |
| ... | ... | ... | HD 154345 b | ... | ... | 9.10 y | ... | 4.18 | Y | ... | 1 | ... | ... | ... |
| 17 04 27.84 | −28 34 57.6 | 154088 | HIP 83541 | ... | ... | ... | ... | ... | ... | ... | ... | ... | ... | ... |
| 17 05 16.82 | +00 42 09.2 | 154417 | HIP 83601 | ... | ... | ... | ... | ... | ... | ... | ... | ... | ... | ... |
| 17 10 10.35 | −60 43 43.6 | 154577 | HIP 83990 | ... | ... | ... | ... | ... | ... | ... | ... | ... | ... | ... |
| 17 12 37.62 | +18 21 04.3 | 155712 | HIP 84195 | ... | ... | ... | ... | ... | ... | ... | ... | ... | ... | ... |
| 17 15 20.98 | −26 36 10.2 | 155885 | 36 Oph | ... | A | ... | ... | ... | ... | ... | ... | ... | ... | ... |
| ... | ... | ... | ... | ... | AB | 470.9 y | 13.0 | 77.1 | Y | P | ... | ... | ... | ... |
| ... | ... | ... | HD 156026 | ... | AC | ... | 731.6 | 4340 | Y | ... | ... | T | ... | ... |
| 17 19 03.83 | −46 38 10.4 | 156274 | 41 Ara | ... | A | ... | ... | ... | ... | ... | ... | ... | ... | ... |
| ... | ... | ... | ... | ... | Aa,Ab | 88.03 d | ... | ... | Y | ... | 1 | ... | M | ... |

| R.A. (J2000.0) (1) | Decl. (J2000.0) (2) | HD Name (3) | Other Name (4) | N (5) | Comp ID (6) | Period (7) | Ang Sep ('') (8) | Lin Sep (AU) (9) | Sts (10) | VB (11) | SB (12) | CP (13) | OT (14) | CH (15) |
|---|---|---|---|---|---|---|---|---|---|---|---|---|---|---|
| ... | ... | ... | NLTT 44525 | ... | AB | 693.24 y | 10.42 | 91.7 | Y | P | ... | ... | ... | ... |
| 17 20 39.57 | +32 28 03.9 | 157214 | 72 Her | ... | ... | ... | ... | ... | ... | ... | ... | ... | ... | ... |
| 17 22 51.29 | −02 23 17.4 | 157347 | HIP 85042 | ... | A | ... | ... | ... | ... | ... | ... | ... | ... | ... |
| ... | ... | ... | HR 6465 | ... | AB | ... | 49 | 956 | Y | ... | ... | P | ... | ... |
| 17 25 00.10 | +67 18 24.1 | 158633 | HIP 85235 | ... | ... | ... | ... | ... | ... | ... | ... | ... | ... | ... |
| 17 30 16.43 | +47 24 07.9 | 159062 | HIP 85653 | ... | ... | ... | ... | ... | ... | ... | ... | ... | ... | ... |
| 17 30 23.80 | −01 03 46.5 | 158614 | HIP 85667 | ... | A | ... | ... | ... | ... | ... | ... | ... | ... | ... |
| ... | ... | ... | ... | ... | AB | 46.40 y | 0.98 | 16.0 | Y | O | 2 | ... | M | ... |
| 17 32 00.99 | +34 16 16.1 | 159222 | HIP 85810 | ... | ... | ... | ... | ... | ... | ... | ... | ... | ... | ... |
| 17 34 59.59 | +61 52 28.4 | 160269 | 26 Dra | ... | A | ... | ... | ... | ... | ... | ... | ... | ... | ... |
| ... | ... | ... | ... | ... | AB | 76.1 y | 1.53 | 21.7 | Y | O | 1 | ... | M | ... |
| ... | ... | ... | HIP 86087 | ... | AB,C | ... | 737.6 | 10466 | Y | ... | ... | ... | T | ... |
| 17 39 16.92 | +03 33 18.9 | 160346 | HIP 86400 | ... | A | ... | ... | ... | ... | ... | ... | ... | ... | ... |
| ... | ... | ... | ... | ... | Aa,Ab | 83.73 d | ... | ... | Y | U | 1 | ... | ... | ... |
| 17 41 58.10 | +72 09 24.9 | 162004 | 31 Dra B | ... | B[a] | ... | ... | ... | ... | ... | ... | ... | ... | ... |
| ... | ... | ... | HD 162003 | ... | AB | 12500 y | 55.2 | 1273 | Y | P | ... | T | ... | ... |
| 17 43 15.64 | +21 36 33.1 | 161198 | HIP 86722 | ... | A | ... | ... | ... | ... | ... | ... | ... | ... | ... |
| ... | ... | ... | ... | ... | Aa,Ab | 7.0 y | 0.17 | 3.85 | Y | P | 1 | ... | M | ... |
| 17 44 08.70 | −51 50 02.6 | 160691 | μ Ara | ... | A | ... | ... | ... | ... | ... | ... | ... | ... | ... |
| ... | ... | ... | HD 160691 d | ... | ... | 9.55 d | ... | 0.09 | Y | ... | 1 | ... | ... | ... |
| ... | ... | ... | HD 160691 e | ... | ... | 310.55 d | ... | 0.94 | Y | ... | 1 | ... | ... | ... |
| ... | ... | ... | HD 160691 b | ... | ... | 1.72 y | ... | 1.51 | Y | ... | 1 | ... | ... | ... |
| ... | ... | ... | HD 160691 c | ... | ... | 6.82 y | ... | 3.78 | Y | ... | 1 | ... | ... | ... |
| 17 46 27.53 | +27 43 14.4 | 161797 | μ Her A | ... | A | ... | ... | ... | ... | ... | ... | ... | ... | ... |
| ... | ... | ... | ... | ... | Aa,Ab | 65.0 y | 1.43 | 11.9 | Y | U | V | R | M | ... |
| ... | ... | ... | NLTT 45430 | ... | Aa,BC | ... | 34 | 282 | Y | ... | ... | P | ... | ... |
| ... | ... | ... | ... | ... | BC | 43.2 y | 1.36 | 11.3 | Y | O | ... | ... | ... | ... |
| 17 53 29.94 | +21 19 31.1 | ... | HIP 87579 | ... | ... | ... | ... | ... | ... | ... | ... | ... | ... | ... |
| 18 02 30.86 | +26 18 46.8 | 164922 | HIP 88348 | ... | A | ... | ... | ... | ... | ... | ... | ... | ... | ... |
| ... | ... | ... | HD 164922 b | ... | ... | 3.16 y | ... | 2.11 | Y | ... | 1 | ... | ... | ... |
| 18 05 27.29 | +02 30 00.4 | 165341 | 70 Oph | N | A | ... | ... | ... | ... | ... | ... | ... | ... | ... |

| R.A. (J2000.0) (1) | Decl. (J2000.0) (2) | HD Name (3) | Other Name (4) | N (5) | Comp ID (6) | Period (7) | Ang Sep (″) (8) | Lin Sep (AU) (9) | Sts (10) | VB (11) | SB (12) | CP (13) | OT (14) | CH (15) |
|---|---|---|---|---|---|---|---|---|---|---|---|---|---|---|
| ... | ... | ... | NLTT 45900 | ... | AB | 88.38 y | 4.55 | 23.1 | Y | O | 2 | ... | M | ... |
| 18 05 37.46 | +04 39 25.8 | 165401 | HIP 88622 | ... | A | ... | ... | ... | ... | ... | ... | ... | ... | ... |
| ... | ... | ... | ... | ... | Aa,Ab | ... | ... | ... | Y | ... | V | ... | M | ... |
| 18 06 23.72 | −36 01 11.2 | 165185 | HIP 88694 | ... | ... | ... | ... | ... | ... | ... | ... | ... | ... | ... |
| 18 07 01.54 | +30 33 43.7 | 165908 | HIP 88745 | N | A | ... | ... | ... | ... | ... | ... | ... | ... | ... |
| ... | ... | ... | ... | ... | Aa,Ab | ... | 0.2 | 3.13 | M | ... | V | R | ... | ... |
| ... | ... | ... | ... | ... | AB | 56.4 y | 1.0 | 15.6 | Y | O | ... | ... | ... | ... |
| 18 09 37.42 | +38 27 28.0 | 166620 | HIP 88972 | ... | ... | ... | ... | ... | ... | ... | ... | ... | ... | ... |
| 18 10 26.16 | −62 00 07.9 | 165499 | ι Pav | ... | A | ... | ... | ... | ... | ... | ... | ... | ... | ... |
| ... | ... | ... | ... | ... | Aa,Ab | ... | ... | ... | Y | ... | ... | ... | M | ... |
| 18 15 32.46 | +45 12 33.5 | 168009 | HIP 89474 | ... | ... | ... | ... | ... | ... | ... | ... | ... | ... | ... |
| 18 19 40.13 | −63 53 11.6 | 167425 | HIP 89805 | ... | A | ... | ... | ... | ... | ... | ... | ... | ... | ... |
| ... | ... | ... | ... | ... | AB | ... | 7.8 | 179 | Y | ... | ... | P | M | ... |
| 18 31 18.96 | −18 54 31.7 | 170657 | HIP 90790 | ... | ... | ... | ... | ... | ... | ... | ... | ... | ... | ... |
| 18 38 53.40 | −21 03 06.7 | 172051 | HIP 91438 | ... | ... | ... | ... | ... | ... | ... | ... | ... | ... | ... |
| 18 40 54.88 | +31 31 59.1 | ... | HIP 91605 | ... | A | ... | ... | ... | ... | ... | ... | ... | ... | ... |
| ... | ... | ... | LHS 3402 | ... | AB | ... | 9.3 | 218 | Y | ... | ... | O | ... | ... |
| 18 55 18.80 | −37 29 54.1 | 175073 | HIP 92858 | ... | ... | ... | ... | ... | ... | ... | ... | ... | ... | ... |
| 18 55 53.22 | +23 33 23.9 | 175742 | HIP 92919 | ... | A | ... | ... | ... | ... | ... | ... | ... | ... | ... |
| ... | ... | ... | ... | ... | Aa,Ab | 2.88 d | ... | ... | Y | ... | 1 | ... | ... | ... |
| 18 57 01.61 | +32 54 04.6 | 176051 | HIP 93017 | ... | A | ... | ... | ... | ... | ... | ... | ... | ... | ... |
| ... | ... | ... | ... | ... | AB | 61.18 y | 1.24 | 18.4 | Y | O | 1 | ... | ... | ... |
| 18 58 51.00 | +30 10 50.3 | 176377 | HIP 93185 | ... | ... | ... | ... | ... | ... | ... | ... | ... | ... | ... |
| 19 06 25.11 | −37 03 48.4 | 177474 | γ CrA A | ... | A | ... | ... | ... | ... | ... | ... | ... | ... | ... |
| ... | ... | ... | ... | ... | AB | 121.76 y | 1.90 | 32.9 | Y | O | ... | ... | M | ... |
| 19 06 52.46 | −37 48 38.4 | 177565 | HIP 93858 | ... | ... | ... | ... | ... | ... | ... | ... | ... | ... | ... |
| 19 07 57.32 | +16 51 12.2 | 178428 | HIP 93966 | N | A | ... | ... | ... | ... | ... | ... | ... | ... | ... |
| ... | ... | ... | ... | ... | Aa,Ab | 21.96 d | ... | ... | Y | ... | 1 | ... | ... | ... |
| 19 12 05.03 | +49 51 20.7 | 179957 | HIP 94336 | ... | A | ... | ... | ... | ... | ... | ... | ... | ... | ... |
| ... | ... | ... | HD 179958 | ... | AB | 3100 y | 12.75 | 311 | Y | P | ... | ... | M | ... |
| 19 12 11.36 | +57 40 19.1 | 180161 | HIP 94346 | ... | ... | ... | ... | ... | ... | ... | ... | ... | ... | ... |

| R.A. (J2000.0) (1) | Decl. (J2000.0) (2) | HD Name (3) | Other Name (4) | N (5) | Comp ID (6) | Period (7) | Ang Sep (") (8) | Lin Sep (AU) (9) | Sts (10) | VB (11) | SB (12) | CP (13) | OT (14) | CH (15) |
|---|---|---|---|---|---|---|---|---|---|---|---|---|---|---|
| 19 21 29.76 | −34 59 00.6 | 181321 | HIP 95149 | ... | A | ... | ... | ... | ... | ... | ... | ... | ... | ... |
| ... | ... | ... | ... | ... | Aa,Ab | ... | ... | ... | Y | ... | ... | ... | M | ... |
| 19 23 34.01 | +33 13 19.1 | 182488 | HIP 95319 | ... | ... | ... | ... | ... | ... | ... | ... | ... | ... | ... |
| 19 24 58.20 | +11 56 39.9 | 182572 | 31 Aql | ... | ... | ... | ... | ... | ... | ... | ... | ... | ... | ... |
| 19 31 07.97 | +58 35 09.6 | 184467 | HIP 95995 | ... | A | ... | ... | ... | ... | ... | ... | ... | ... | ... |
| ... | ... | ... | ... | ... | AB | 1.35 y | 0.09 | 1.53 | Y | O | 2 | ... | ... | ... |
| 19 32 06.70 | −11 16 29.8 | 183870 | HIP 96085 | ... | ... | ... | ... | ... | ... | ... | ... | ... | ... | ... |
| 19 32 21.59 | +69 39 40.2 | 185144 | σ Dra | ... | ... | ... | ... | ... | ... | ... | ... | ... | ... | ... |
| 19 33 25.55 | +21 50 25.2 | 184385 | HIP 96183 | ... | ... | ... | ... | ... | ... | ... | ... | ... | ... | ... |
| 19 35 55.61 | +56 59 02.0 | 185414 | HIP 96395 | ... | A | ... | ... | ... | ... | ... | ... | ... | ... | ... |
| ... | ... | ... | ... | ... | Aa,Ab | 13.08 y | ... | ... | Y | ... | 1 | ... | ... | ... |
| 19 41 48.95 | +50 31 30.2 | 186408 | 16 Cyg A | N | A | ... | ... | ... | ... | ... | ... | ... | ... | ... |
| ... | ... | ... | ... | ... | Aa,Ab | ... | 3.4 | 71.7 | Y | ... | V | R | ... | ... |
| ... | ... | ... | HD 186427 | ... | Aa,B | 18212 y | 40.79 | 859 | Y | P | ... | T | ... | ... |
| ... | ... | ... | 16 Cyg B b | ... | ... | 2.19 y | ... | 1.68 | Y | ... | 1 | ... | ... | ... |
| 19 45 33.53 | +33 36 07.2 | 186858 | HIP 97222 | ... | F[a] | ... | ... | ... | ... | ... | ... | ... | ... | ... |
| ... | ... | ... | ... | ... | FG | 232 y | 2.07 | 43.7 | Y | O | ... | ... | M | ... |
| ... | ... | ... | HD 187013 | ... | AF | ... | 780 | 16476 | Y | ... | ... | T | ... | ... |
| ... | ... | ... | ... | ... | Aa,Ab | ... | ... | ... | M | ... | V | ... | ... | ... |
| ... | ... | ... | HD 225732 | ... | AB | ... | 26 | 549 | Y | ... | ... | P | ... | ... |
| 19 51 01.64 | +10 24 56.6 | 187691 | HIP 97675 | ... | A | ... | ... | ... | ... | ... | ... | ... | ... | ... |
| ... | ... | ... | ... | ... | AC | ... | 22.5 | 431 | Y | ... | ... | P | ... | ... |
| 19 59 47.34 | −09 57 29.7 | 189340 | HIP 98416 | ... | A | ... | ... | ... | ... | ... | ... | ... | ... | ... |
| ... | ... | ... | ... | ... | AB | 4.90 y | 0.15 | 3.33 | Y | O | 2 | ... | M | ... |
| 20 00 43.71 | +22 42 39.1 | 189733 | HIP 98505 | ... | A | ... | ... | ... | ... | ... | ... | ... | ... | ... |
| ... | ... | ... | HD 189733 b | ... | ... | 2.22 d | ... | 0.03 | Y | ... | 1 | ... | E | ... |
| 20 02 34.16 | +15 35 31.5 | 190067 | HIP 98677 | N | A | ... | ... | ... | ... | ... | ... | ... | ... | ... |
| ... | ... | ... | ... | ... | AB | ... | 2.9 | 55.0 | Y | ... | ... | M | ... | ... |
| 20 03 37.41 | +29 53 48.5 | 190360 | HIP 98767 | ... | A | ... | ... | ... | ... | ... | ... | ... | ... | ... |
| ... | ... | ... | HD 190360 c | ... | ... | 17.11 d | ... | 0.13 | Y | ... | 1 | ... | ... | ... |
| ... | ... | ... | HD 190360 b | ... | ... | 8.01 y | ... | 4.02 | Y | ... | 1 | ... | ... | ... |

| R.A. (J2000.0) (1) | Decl. (J2000.0) (2) | HD Name (3) | Other Name (4) | N (5) | Comp ID (6) | Period (7) | Ang Sep ('') (8) | Lin Sep (AU) (9) | Sts (10) | VB (11) | SB (12) | CP (13) | OT (14) | CH (15) |
|---|---|---|---|---|---|---|---|---|---|---|---|---|---|---|
| ... | ... | ... | LHS 3509 | ... | AB | ... | 178.2 | 2825 | Y | ... | ... | T | ... | ... |
| 20 03 52.13 | +23 20 26.5 | 190404 | HIP 98792 | ... | ... | ... | ... | ... | ... | ... | ... | ... | ... | ... |
| 20 04 06.22 | +17 04 12.6 | 190406 | HIP 98819 | N | A | ... | ... | ... | ... | ... | ... | ... | ... | ... |
| ... | ... | ... | HD 354613 | ... | Aa,Ab | ... | 0.79 | 14.0 | Y | ... | ... | M | ... | ... |
| 20 04 10.05 | +25 47 24.8 | 190470 | HIP 98828 | ... | ... | ... | ... | ... | ... | ... | ... | ... | ... | ... |
| 20 05 09.78 | +38 28 42.4 | 190771 | HIP 98921 | ... | A | ... | ... | ... | ... | ... | ... | ... | ... | ... |
| ... | ... | ... | ... | ... | Aa,Ab | ... | ... | ... | Y | ... | V | ... | M | ... |
| 20 05 32.76 | −67 19 15.2 | 189567 | HIP 98959 | ... | ... | ... | ... | ... | ... | ... | ... | ... | ... | ... |
| 20 07 35.09 | −55 00 57.6 | 190422 | HIP 99137 | ... | ... | ... | ... | ... | ... | ... | ... | ... | ... | ... |
| 20 08 43.61 | −66 10 55.4 | 190248 | δ Pav | ... | ... | ... | ... | ... | ... | ... | ... | ... | ... | ... |
| 20 09 34.30 | +16 48 20.8 | 191499 | HIP 99316 | N | A | ... | ... | ... | ... | ... | ... | ... | ... | ... |
| ... | ... | ... | ADS 13434B | ... | AB | ... | 4.45 | 105 | Y | ... | ... | P | M | ... |
| 20 11 06.07 | +16 11 16.8 | 191785 | HIP 99452 | ... | A | ... | ... | ... | ... | ... | ... | ... | ... | ... |
| ... | ... | ... | ... | ... | AE | ... | 103.8 | 2116 | Y | ... | ... | P | ... | ... |
| 20 11 11.94 | −36 06 04.4 | 191408 | HIP 99461 | ... | A | ... | ... | ... | ... | ... | ... | ... | ... | ... |
| ... | ... | ... | LHS 487 | ... | AB | ... | 7.1 | 42.7 | Y | ... | ... | O | M | ... |
| 20 13 59.85 | −00 52 00.8 | 192263 | HIP 99711 | ... | A | ... | ... | ... | ... | ... | ... | ... | ... | ... |
| ... | ... | ... | HD 192263 b | ... | ... | 24.36 d | ... | 0.15 | Y | ... | 1 | ... | ... | ... |
| 20 15 17.39 | −27 01 58.7 | 192310 | HIP 99825 | ... | ... | ... | ... | ... | ... | ... | ... | ... | ... | ... |
| 20 17 31.33 | +66 51 13.3 | 193664 | HIP 100017 | ... | A | ... | ... | ... | ... | ... | ... | ... | ... | ... |
| ... | ... | ... | ... | ... | Aa,Ab | ... | ... | ... | M | ... | ... | ... | M | ... |
| 20 27 44.24 | −30 52 04.2 | 194640 | HIP 100925 | ... | ... | ... | ... | ... | ... | ... | ... | ... | ... | ... |
| 20 32 23.70 | −09 51 12.2 | 195564 | HIP 101345 | N | A | ... | ... | ... | ... | ... | ... | ... | ... | ... |
| ... | ... | ... | LTT 8128 | ... | AB | ... | 4.4 | 107 | Y | ... | ... | M | M | ... |
| 20 32 51.64 | +41 53 54.5 | 195987 | HIP 101382 | ... | A | ... | ... | ... | ... | ... | ... | ... | ... | ... |
| ... | ... | ... | ... | ... | Aa,Ab | 57.32 d | 0.02 | 0.44 | Y | O | 2 | ... | M | ... |
| 20 40 02.64 | −60 32 56.0 | 196378 | φ² Pav | ... | ... | ... | ... | ... | ... | ... | ... | ... | ... | ... |
| 20 40 11.76 | −23 46 25.9 | 196761 | HIP 101997 | ... | ... | ... | ... | ... | ... | ... | ... | ... | ... | ... |
| 20 40 45.14 | +19 56 07.9 | 197076 | HIP 102040 | ... | A | ... | ... | ... | ... | ... | ... | ... | ... | ... |
| ... | ... | ... | NLTT 49681 | ... | AC | ... | 125.1 | 2620 | Y | ... | ... | P | ... | ... |
| 20 43 16.00 | −29 25 26.1 | 197214 | HIP 102264 | ... | ... | ... | ... | ... | ... | ... | ... | ... | ... | ... |

| R.A. (J2000.0) (1) | Decl. (J2000.0) (2) | HD Name (3) | Other Name (4) | N (5) | Comp ID (6) | Period (7) | Ang Sep (″) (8) | Lin Sep (AU) (9) | Sts (10) | VB (11) | SB (12) | CP (13) | OT (14) | CH (15) |
|---|---|---|---|---|---|---|---|---|---|---|---|---|---|---|
| ... | ... | ... | ... | ... | Aa,Ab | ... | ... | ... | Y | ... | V | ... | ... | ... |
| 20 49 16.23 | +32 17 05.2 | 198425 | HIP 102766 | ... | A | ... | ... | ... | ... | ... | ... | ... | ... | ... |
| ... | ... | ... | ... | ... | Aa,Ab | ... | ... | ... | M | ... | V | ... | ... | ... |
| ... | ... | ... | NLTT 49961 | ... | AB | ... | 33.0 | 793 | Y | ... | ... | P | ... | ... |
| 20 56 47.33 | −26 17 47.0 | 199260 | HIP 103389 | ... | ... | ... | ... | ... | ... | ... | ... | ... | ... | ... |
| 20 57 40.07 | −44 07 45.7 | 199288 | HIP 103458 | ... | ... | ... | ... | ... | ... | ... | ... | ... | ... | ... |
| 21 02 40.76 | +45 53 05.2 | 200560 | HIP 103859 | N | C | ... | ... | ... | ... | ... | ... | ... | ... | ... |
| ... | ... | ... | GJ 816.1B | ... | CD^c | ... | 3.3 | 64.3 | Y | ... | ... | M | M | ... |
| 21 07 10.38 | −13 55 22.6 | 200968 | HIP 104239 | ... | A | ... | ... | ... | ... | ... | ... | ... | ... | ... |
| ... | ... | ... | GJ 819B | ... | AB | ... | 4.31 | 75.7 | Y | ... | ... | O | M | ... |
| 21 09 20.74 | −82 01 38.1 | 199509 | HIP 104436 | ... | ... | ... | ... | ... | ... | ... | ... | ... | ... | ... |
| 21 09 22.45 | −73 10 22.7 | 200525 | HIP 104440 | N | A | ... | ... | ... | ... | ... | ... | ... | ... | ... |
| ... | ... | ... | ... | ... | AB | 5.87 y | 0.2 | 3.95 | Y | U | ... | R | ... | ... |
| ... | ... | ... | NLTT 50542 | ... | AB,C | ... | 7.2 | 142 | Y | ... | ... | M | ... | ... |
| 21 14 28.82 | +10 00 25.1 | 202275 | δ Equ | N | A | ... | ... | ... | ... | ... | ... | ... | ... | ... |
| ... | ... | ... | ... | ... | AB | 5.71 y | 0.23 | 4.25 | Y | O | 2 | ... | ... | S |
| 21 18 02.97 | +00 09 41.7 | 202751 | HIP 105152 | ... | ... | ... | ... | ... | ... | ... | ... | ... | ... | ... |
| 21 18 27.27 | −43 20 04.7 | 202628 | HIP 105184 | ... | ... | ... | ... | ... | ... | ... | ... | ... | ... | ... |
| 21 19 45.62 | −26 21 10.4 | 202940 | HIP 105312 | ... | A | ... | ... | ... | ... | ... | ... | ... | ... | ... |
| ... | ... | ... | ... | ... | Aa,Ab | 21.35 d | ... | ... | Y | ... | 1 | ... | ... | ... |
| ... | ... | ... | LHS 3656 | ... | AB | 261.62 y | 3.01 | 54.1 | Y | P | ... | ... | M | ... |
| 21 24 40.64 | −68 13 40.2 | 203244 | HIP 105712 | ... | ... | ... | ... | ... | ... | ... | ... | ... | ... | ... |
| 21 26 58.45 | −56 07 30.9 | 203850 | HIP 105905 | ... | ... | ... | ... | ... | ... | ... | ... | ... | ... | ... |
| 21 27 01.33 | −44 48 30.9 | 203985 | HIP 105911 | ... | A | ... | ... | ... | ... | ... | ... | ... | ... | ... |
| ... | ... | ... | ... | ... | Aa,Ab | ... | ... | ... | Y | ... | ... | ... | M | ... |
| ... | ... | ... | LTT 8515 | ... | AB | ... | 88.0 | 2068 | Y | ... | ... | P | ... | ... |
| 21 36 41.24 | −50 50 43.4 | 205390 | HIP 106696 | ... | ... | ... | ... | ... | ... | ... | ... | ... | ... | ... |
| 21 40 29.77 | −74 04 27.4 | 205536 | HIP 107022 | ... | ... | ... | ... | ... | ... | ... | ... | ... | ... | ... |
| 21 44 08.58 | +28 44 33.5 | 206826 | μ Cyg A | ... | A | ... | ... | ... | ... | ... | ... | ... | ... | ... |
| ... | ... | ... | HD 206827 | ... | AB | 789 y | 5.32 | 118 | Y | P | ... | ... | M | ... |
| 21 44 31.33 | +14 46 19.0 | 206860 | HIP 107350 | N | A | ... | ... | ... | ... | ... | ... | ... | ... | ... |

| R.A. (J2000.0) (1) | Decl. (J2000.0) (2) | HD Name (3) | Other Name (4) | N (5) | Comp ID (6) | Period (7) | Ang Sep (″) (8) | Lin Sep (AU) (9) | Sts (10) | VB (11) | SB (12) | CP (13) | OT (14) | CH (15) |
|---|---|---|---|---|---|---|---|---|---|---|---|---|---|---|
| . . . | . . . | . . . | HN Peg B | . . . | AB | . . . | 43.2 | 772 | Y | . . . | . . . | M | . . . | . . . |
| 21 48 00.05 | −40 15 21.9 | 207144 | HIP 107625 | . . . | . . . | . . . | . . . | . . . | . . . | . . . | . . . | . . . | . . . | . . . |
| 21 48 15.75 | −47 18 13.0 | 207129 | HIP 107649 | . . . | . . . | . . . | . . . | . . . | . . . | . . . | . . . | . . . | . . . | . . . |
| 21 53 05.35 | +20 55 49.9 | 208038 | HIP 108028 | . . . | . . . | . . . | . . . | . . . | . . . | . . . | . . . | . . . | . . . | . . . |
| 21 54 45.04 | +32 19 42.9 | 208313 | HIP 108156 | . . . | . . . | . . . | . . . | . . . | . . . | . . . | . . . | . . . | . . . | . . . |
| 22 09 29.87 | −07 32 55.1 | 210277 | HIP 109378 | . . . | A | . . . | . . . | . . . | . . . | . . . | . . . | . . . | . . . | . . . |
| . . . | . . . | . . . | HD 210277 b | . . . | . . . | 1.21 y | . . . | 1.14 | Y | . . . | 1 | . . . | . . . | . . . |
| 22 11 11.91 | +36 15 22.8 | 210667 | HIP 109527 | . . . | . . . | . . . | . . . | . . . | . . . | . . . | . . . | . . . | . . . | . . . |
| 22 14 38.65 | −41 22 54.0 | 210918 | HIP 109821 | . . . | . . . | . . . | . . . | . . . | . . . | . . . | . . . | . . . | . . . | . . . |
| 22 15 54.14 | +54 40 22.4 | 211472 | HIP 109926 | . . . | A | . . . | . . . | . . . | . . . | . . . | . . . | . . . | . . . | . . . |
| . . . | . . . | . . . | GJ 4269 | . . . | AT | . . . | 77.2 | 1662 | Y | . . . | . . . | P | . . . | . . . |
| 22 18 15.62 | −53 37 37.5 | 211415 | HIP 110109 | . . . | A | . . . | . . . | . . . | . . . | . . . | . . . | . . . | . . . | . . . |
| . . . | . . . | . . . | . . . | . . . | AB | . . . | 3.4 | 46.9 | Y | . . . | . . . | O | M | . . . |
| 22 24 56.39 | −57 47 50.7 | 212330 | HIP 110649 | . . . | A | . . . | . . . | . . . | . . . | . . . | . . . | . . . | . . . | . . . |
| . . . | . . . | . . . | . . . | . . . | Aa,Ab | . . . | . . . | . . . | Y | . . . | . . . | . . . | M | . . . |
| 22 25 51.16 | −75 00 56.5 | 212168 | HIP 110712 | . . . | A | . . . | . . . | . . . | . . . | . . . | . . . | . . . | . . . | . . . |
| . . . | . . . | . . . | HIP 110719 | . . . | AB | . . . | 20.8 | 479 | Y | . . . | . . . | P | . . . | . . . |
| 22 39 50.77 | +04 06 58.0 | 214683 | HIP 111888 | . . . | . . . | . . . | . . . | . . . | . . . | . . . | . . . | . . . | . . . | . . . |
| 22 42 36.88 | −47 12 38.9 | 214953 | HIP 112117 | . . . | A | . . . | . . . | . . . | . . . | . . . | . . . | . . . | . . . | . . . |
| . . . | . . . | . . . | NLTT 54607 | . . . | AB | . . . | 7.8 | 184 | Y | . . . | . . . | P | M | . . . |
| 22 43 21.30 | −06 24 03.0 | 215152 | HIP 112190 | . . . | . . . | . . . | . . . | . . . | . . . | . . . | . . . | . . . | . . . | . . . |
| 22 46 41.58 | +12 10 22.4 | 215648 | ξ Peg | N | A | . . . | . . . | . . . | . . . | . . . | . . . | . . . | . . . | . . . |
| . . . | . . . | . . . | . . . | . . . | AB | . . . | 11.1 | 180 | Y | . . . | . . . | O | . . . | . . . |
| 22 47 31.87 | +83 41 49.3 | 216520 | HIP 112527 | . . . | . . . | . . . | . . . | . . . | . . . | . . . | . . . | . . . | . . . | . . . |
| 22 51 26.36 | +13 58 11.9 | 216259 | HIP 112870 | . . . | . . . | . . . | . . . | . . . | . . . | . . . | . . . | . . . | . . . | . . . |
| 22 57 27.98 | +20 46 07.8 | 217014 | 51 Peg | . . . | A | . . . | . . . | . . . | . . . | . . . | . . . | . . . | . . . | . . . |
| . . . | . . . | . . . | 51 Peg b | . . . | . . . | 4.23 d | . . . | 0.05 | Y | . . . | 1 | . . . | . . . | . . . |
| 22 58 15.54 | −02 23 43.4 | 217107 | HIP 113421 | N | A | . . . | . . . | . . . | . . . | . . . | . . . | . . . | . . . | . . . |
| . . . | . . . | . . . | HD 217107 b | . . . | . . . | 7.13 d | . . . | 0.07 | Y | . . . | 1 | . . . | . . . | . . . |
| . . . | . . . | . . . | HD 217107 c | . . . | . . . | 11.14 y | . . . | 5.15 | M | . . . | 1 | . . . | . . . | . . . |
| . . . | . . . | . . . | . . . | . . . | AB | . . . | 0.3 | 5.96 | M | . . . | . . . | R | . . . | . . . |

| R.A.<br>(J2000.0)<br>(1) | Decl.<br>(J2000.0)<br>(2) | HD<br>Name<br>(3) | Other<br>Name<br>(4) | N<br><br>(5) | Comp<br>ID<br>(6) | Period<br><br>(7) | Ang<br>Sep<br>(″)<br>(8) | Lin<br>Sep<br>(AU)<br>(9) | Sts<br><br>(10) | VB<br><br>(11) | SB<br><br>(12) | CP<br><br>(13) | OT<br><br>(14) | CH<br><br>(15) |
|---|---|---|---|---|---|---|---|---|---|---|---|---|---|---|
| 23 03 04.98 | +20 55 06.9 | 217813 | HIP 113829 | ... | ... | ... | ... | ... | ... | ... | ... | ... | ... | ... |
| 23 10 50.08 | +45 30 44.2 | 218868 | HIP 114456 | ... | A | ... | ... | ... | ... | ... | ... | ... | ... | ... |
| ... | ... | ... | ... | ... | AB | ... | 50 | 1215 | Y | ... | ... | P | ... | ... |
| 23 13 16.98 | +57 10 06.1 | 219134 | HIP 114622 | ... | ... | ... | ... | ... | ... | ... | ... | ... | ... | ... |
| 23 16 18.16 | +30 40 12.8 | 219538 | HIP 114886 | ... | ... | ... | ... | ... | ... | ... | ... | ... | ... | ... |
| 23 16 42.30 | +53 12 48.5 | 219623 | HIP 114924 | ... | ... | ... | ... | ... | ... | ... | ... | ... | ... | ... |
| 23 16 57.69 | −62 00 04.3 | 219482 | HIP 114948 | ... | ... | ... | ... | ... | ... | ... | ... | ... | ... | ... |
| 23 19 26.63 | +79 00 12.7 | 220140 | HIP 115147 | N | A | ... | ... | ... | ... | ... | ... | ... | ... | ... |
| ... | ... | ... | NLTT 56532 | ... | AB | ... | 10.8 | 207 | Y | ... | ... | M | ... | ... |
| ... | ... | ... | ... | ... | AC | ... | 962.6 | 18486 | Y | ... | ... | T | ... | ... |
| 23 21 36.51 | +44 05 52.4 | 220182 | HIP 115331 | ... | ... | ... | ... | ... | ... | ... | ... | ... | ... | ... |
| 23 23 04.89 | −10 45 51.3 | 220339 | HIP 115445 | ... | ... | ... | ... | ... | ... | ... | ... | ... | ... | ... |
| 23 31 22.21 | +59 09 55.9 | 221354 | HIP 116085 | ... | ... | ... | ... | ... | ... | ... | ... | ... | ... | ... |
| 23 35 25.61 | +31 09 40.7 | 221851 | HIP 116416 | ... | ... | ... | ... | ... | ... | ... | ... | ... | ... | ... |
| 23 37 58.49 | +46 11 58.0 | 222143 | HIP 116613 | ... | ... | ... | ... | ... | ... | ... | ... | ... | ... | ... |
| 23 39 37.39 | −72 43 19.8 | 222237 | HIP 116745 | ... | ... | ... | ... | ... | ... | ... | ... | ... | ... | ... |
| 23 39 51.31 | −32 44 36.3 | 222335 | HIP 116763 | ... | ... | ... | ... | ... | ... | ... | ... | ... | ... | ... |
| 23 39 57.04 | +05 37 34.6 | 222368 | ι Psc | ... | ... | ... | ... | ... | ... | ... | ... | ... | ... | ... |
| 23 52 25.32 | +75 32 40.5 | 223778 | HIP 117712 | ... | A | ... | ... | ... | ... | ... | ... | ... | ... | ... |
| ... | ... | ... | ... | ... | Aa,Ab | 7.75 d | ... | ... | Y | ... | 2 | ... | ... | V |
| ... | ... | ... | ... | ... | AB | 290 y | 4.14 | 45.1 | Y | P | ... | ... | M | ... |
| 23 56 10.67 | −39 03 08.4 | 224228 | HIP 118008 | ... | ... | ... | ... | ... | ... | ... | ... | ... | ... | ... |
| 23 58 06.82 | +50 26 51.6 | 224465 | HIP 118162 | ... | A | ... | ... | ... | ... | ... | ... | ... | ... | ... |
| ... | ... | ... | ... | ... | Aa,Ab | 52.41 d | ... | ... | Y | ... | 1 | ... | ... | ... |

[a]The sample star is not the system's primary, which is identified as component A below.

[b]The brightest component of the system is HD 25998, but is designated as component E in the WDS. Component A is the wide CPM companion, HD 25893, which is about 2 magnitudes fainter and itself a visual binary. We have retained the component designations of the WDS, so the fainter visual pair is AB and the wide CPM companion is E. WDS components C and D are optical, and E itself might have a close companion, as evidenced by its accelerating proper motion (see Table 8).

[c]WDS lists these entries for HD 200595, a bright binary 153″ away from the sample star HD 200560, but one that is not physically associated with it. HD 200560 is itself is a close CPM pair and listed in the WDS as CD. We have retained the WDS designations, which makes C and D the only physically associated components of this system.

Table 14.   Summary of Confirmed Companions

| Type | Total | Unique | VB | SB | CP | OT |
|------|-------|--------|-----|-----|-----|-----|
| VB | 95 | 21 | ... | 45 | 10 | 46 |
| SB | 88 | 23 | 45 | ... | 7 | 39 |
| CP | 125 | 96 | 10 | 7 | ... | 21 |
| OT | 88 | 9 | 46 | 39 | 21 | ... |

Table 15.   Summary of Systematic Companion Searches

| Search Method (1) | Nbr (%) Searched (2) | Companions Total (3) | Companions New (4) |
|---|---|---|---|
| Latham et al. (2010) radial velocities | 344[a]  (76%) | 59 | 4 |
| CCPS radial velocities | 241[b]  (53%) | 20 | 1 |
| Speckle interferometry | 453 (100%) | 45 | 0 |
| CHARA SFP | 296  (65%) | 8 | 0 |
| CPM: blinking archival plates | 409  (90%) | 70 | 4 |
| The *Hipparcos* survey | 453 (100%) | 83 | ... |
| AO surveys[c] | 82  (18%) | 6 | 2 |

[a]This number includes only primaries.  Latham et al. (2010) also reports the velocities of 38 companions.

[b]This number includes only primaries. We analyzed CCPS velocities for 14 companions as well.

[c]Turner et al. (2001); Luhman & Jayawardhana (2002); Chauvin et al. (2006); Metchev & Hillenbrand (2009)

Table 16.   Multiplicity Statistics

| | | === | Percentages | === | |
|---|---|---|---|---|---|
| Sample | N | Single | Binary | Triple | Quad+ |

### Current Results

| | | | | | |
|---|---|---|---|---|---|
| Observed  .................. | 454 | $56 \pm 2$ | $33 \pm 2$ | $8 \pm 1$ | $3 \pm 1$ |
| Including candidates  ...... | 454 | $54 \pm 2$ | $34 \pm 2$ | $9 \pm 2$ | $3 \pm 1$ |
| Compl analysis adjusted  ... | 454 | $54 \pm 2$ | ... | ... | ... |

### Distance-limited Subsamples

| | | | | | |
|---|---|---|---|---|---|
| Distance $\leq 15$ pc ............ | 103 | $52 \pm 5$ | $37 \pm 5$ | $8 \pm 3$ | $2 \pm 1$ |
| Distance $\leq 20$ pc ............ | 239 | $57 \pm 5$ | $33 \pm 5$ | $9 \pm 3$ | $2 \pm 1$ |
| Distance $\leq 23$ pc ............ | 359 | $57 \pm 3$ | $32 \pm 3$ | $8 \pm 2$ | $3 \pm 1$ |
| Distance $\leq 25$ pc ............ | 454 | $56 \pm 2$ | $33 \pm 2$ | $8 \pm 1$ | $3 \pm 1$ |

### DM91 Results

| | | | | | |
|---|---|---|---|---|---|
| Observed  .................. | 164 | $57 \pm 4$ | $38 \pm 4$ | $4 \pm 1$ | $1 \pm 1$ |
| Incl $P(\chi^2) < 0.01$  ......... | 164 | $50 \pm 4$ | $41 \pm 4$ | $7 \pm 2$ | $2 \pm 1$ |
| Incompl $(q > 0.1)$  ......... | 164 | 43 | ... | ... | ... |
| Incompl $(M_2 < 10$ M$_J)$ ...... | 164 | 33 | ... | ... | ... |

### Current Work & DM91 Comparison

| | | | | | |
|---|---|---|---|---|---|
| DM91, common stars  ...... | 106 | $56 \pm 5$ | $39 \pm 5$ | $4 \pm 2$ | $2 \pm 1$ |
| This work, common stars  ... | 106 | $49 \pm 5$ | $40 \pm 5$ | $9 \pm 3$ | $2 \pm 1$ |
| This work, random subset  ... | 106 | $55 \pm 5$ | $34 \pm 4$ | $8 \pm 3$ | $3 \pm 2$ |
| This work, dec $> -15°$  ...... | 307 | $54 \pm 3$ | $35 \pm 3$ | $8 \pm 2$ | $3 \pm 1$ |
| This work, F7 $\leq$ SpT $\leq$ G9 | 281 | $56 \pm 3$ | $34 \pm 3$ | $7 \pm 2$ | $3 \pm 2$ |

Table 17.   Physical Parameters of the Sample Stars

| HD Name (1) | HIP Name (2) | C (3) | Spec Type (4) | Ref (5) | Mass ( M$_\odot$) (6) | Ref (7) | [Fe/H] (8) | Ref (9) | log($R'_{HK}$) (10) | Ref (11) |
|---|---|---|---|---|---|---|---|---|---|---|
| Sun | . . . | . . . | G2V | . . . | 1.00 | . . . | 0.00 | . . . | . . . | . . . |
| 000123 | 000518 | A | G4V | 1 | . . . | . . . | −0.01 | 2 | −4.644 | 1 |
| 000166 | 000544 | . . . | G8V | 1 | . . . | . . . | 0.12 | 3 | −4.458 | 1 |
| 000870 | 001031 | . . . | K0V | 4 | . . . | . . . | −0.20 | 2 | −4.824 | 4 |
| 001237 | 001292 | A | G8.5V | 4 | . . . | . . . | −0.09 | 2 | −4.496 | 4 |
| 001273 | 001349 | Aa | G5V | 4 | . . . | . . . | −0.65 | 2 | −4.802 | 4 |
| 001461 | 001499 | . . . | G3V | 1 | . . . | . . . | 0.16 | 3 | −5.030 | 5 |
| 001562 | 001598 | . . . | G1V | 1 | . . . | . . . | −0.34 | 2 | −4.979 | 1 |
| 001581 | 001599 | . . . | F9.5V | 4 | . . . | . . . | −0.18 | 3 | −4.855 | 4 |
| 001835 | 001803 | . . . | G5V | 4 | . . . | . . . | 0.22 | 3 | −4.440 | 5 |
| 002025 | 001936 | . . . | K3V | 4 | . . . | . . . | −0.27 | 3 | −4.933 | 4 |
| 002151 | 002021 | . . . | G0V | 4 | . . . | . . . | −0.09 | 3 | −5.006 | 4 |
| 003196 | 002762 | Aa | F8.5V | 1 | . . . | . . . | −0.07 | 2 | −4.461 | 1 |
| 003443 | 002941 | A | G7V | 4 | 0.94 | 6 | −0.14 | 2 | −4.940 | 5 |
| 003651 | 003093 | A | K0V | 1 | . . . | . . . | 0.16 | 3 | −5.020 | 5 |
| 003765 | 003206 | . . . | K2.5V | 1 | . . . | . . . | 0.12 | 3 | −5.101 | 1 |
| 004256 | 003535 | . . . | K3IV-V | 1 | . . . | . . . | 0.22 | 3 | −5.042 | 1 |
| 004308 | 003497 | A | G6V | 4 | . . . | . . . | −0.18 | 3 | −4.853 | 4 |
| 004391 | 003583 | A | G5V | 4 | . . . | . . . | −0.25 | 2 | −4.669 | 4 |
| 004614 | 003821 | A | G0V | 7 | . . . | . . . | −0.17 | 3 | −4.930 | 5 |
| 004628 | 003765 | Aa | K2V | 1 | . . . | . . . | −0.19 | 3 | −5.071 | 1 |
| 004635 | 003876 | . . . | K2.5V+ | 1 | . . . | . . . | . . . | . . . | −4.670 | 5 |
| 004676 | 003810 | Aa | F8V | 1 | 1.22 | 8 | −0.06 | 2 | −4.917 | 1 |
| 004747 | 003850 | Aa | G9V | 4 | . . . | . . . | −0.25 | 3 | −4.720 | 5 |
| 004813 | 003909 | . . . | F7V | 1 | . . . | . . . | −0.15 | 2 | −4.780 | 1 |
| 004915 | 003979 | . . . | G6V | 1 | . . . | . . . | −0.18 | 3 | −4.860 | 5 |
| 005015 | 004151 | . . . | F8V | 7 | . . . | . . . | . . . | . . . | −5.02 | 9 |
| 005133 | 004148 | . . . | K2.5V | 4 | . . . | . . . | −0.13 | 3 | −4.751 | 4 |
| 006582 | 005336 | Aa | K1V | 1 | . . . | . . . | −0.83 | 2 | −5.031 | 1 |
| 007570 | 005862 | . . . | F9V | 4 | . . . | . . . | 0.14 | 3 | −4.861 | 4 |
| 007590 | 005944 | . . . | G0-V | 1 | . . . | . . . | −0.07 | 3 | −4.530 | 5 |
| 007693 | 005842 | C | K2+V | 4 | 0.92 | 10 | 0.05 | 3 | −4.580 | 4 |
| 007924 | 006379 | . . . | K0 | 7 | . . . | . . . | −0.12 | 3 | −4.830 | 5 |
| 008997 | 006917 | Aa | K2.5V | 1 | . . . | . . . | −0.59 | 2 | −4.557 | 1 |
| 009407 | 007339 | . . . | G6.5V | 1 | . . . | . . . | . . . | . . . | −5.010 | 5 |
| 009540 | 007235 | . . . | G8.5V | 4 | . . . | . . . | −0.04 | 3 | −4.600 | 5 |
| 009770 | 007372 | A | K2V | 4 | 0.74 | 10 | −0.65 | 2 | −4.354 | 4 |
| 009826 | 007513 | A | F8V | 7 | 1.2 | 11 | 0.12 | 3 | −5.040 | 5 |
| 010008 | 007576 | . . . | G9V | 1 | . . . | . . . | −0.03 | 1 | −4.530 | 1 |
| 010086 | 007734 | . . . | G5V | 1 | . . . | . . . | 0.09 | 3 | −4.600 | 5 |
| 010307 | 007918 | Aa | G1V | 1 | . . . | . . . | −0.05 | 2 | −5.017 | 1 |
| 010360 | 007751 | B | K2V | 4 | . . . | . . . | −0.20 | 3 | −4.899 | 4 |

Table 17—Continued

| HD Name (1) | HIP Name (2) | C (3) | Spec Type (4) | Ref (5) | Mass ( M$_\odot$ ) (6) | Ref (7) | [Fe/H] (8) | Ref (9) | log($R'_{HK}$) (10) | Ref (11) |
|---|---|---|---|---|---|---|---|---|---|---|
| 010476 | 007981 | ... | K0V | 1 | ... | ... | −0.07 | 3 | −4.950 | 5 |
| 010647 | 007978 | A | F9V | 4 | ... | ... | −0.08 | 3 | −4.675 | 4 |
| 010700 | 008102 | ... | G8.5V | 4 | ... | ... | −0.36 | 3 | −4.980 | 5 |
| 010780 | 008362 | ... | G9V | 1 | ... | ... | −0.06 | 3 | −4.690 | 5 |
| 012051 | 009269 | ... | G9V | 1 | ... | ... | 0.15 | 3 | −5.050 | 5 |
| 012846 | 009829 | ... | G2V- | 1 | ... | ... | −0.20 | 3 | −4.980 | 5 |
| 013445 | 010138 | A | K1V | 4 | ... | ... | −0.20 | 3 | −4.768 | 4 |
| 013974 | 010644 | Aa | G0V | 7 | ... | ... | −0.39 | 2 | −4.710 | 5 |
| 014214 | 010723 | Aa | G0IV- | 1 | ... | ... | 0.06 | 2 | −5.114 | 1 |
| 014412 | 010798 | ... | G8V | 4 | ... | ... | −0.45 | 3 | −4.850 | 5 |
| 014802 | 011072 | A | G0V | 4 | ... | ... | −0.08 | 2 | −5.050 | 5 |
| 016160 | 012114 | Aa | K3V | 1 | ... | ... | 0.08 | 2 | −5.094 | 1 |
| 016287 | 012158 | Aa | K2.5V | 1 | ... | ... | 0.08 | 3 | −4.504 | 1 |
| 016673 | 012444 | Aa | F8V | 1 | ... | ... | −0.11 | 2 | −4.586 | 1 |
| 016739 | 012623 | Aa | F9IV-V | 1 | 1.38 | 12 | 0.14 | 2 | −5.012 | 1 |
| 016765 | 012530 | A | F7V | 1 | ... | ... | −0.24 | 2 | −4.400 | 1 |
| 016895 | 012777 | A | F7V | 7 | ... | ... | 0.02 | 3 | −4.970 | 5 |
| 017051 | 012653 | A | F9V | 4 | ... | ... | 0.09 | 3 | −4.625 | 4 |
| 017382 | 013081 | Aa | K0V | 1 | ... | ... | ... | ... | −4.450 | 5 |
| 017925 | 013402 | ... | K1.5V | 4 | ... | ... | 0.11 | 3 | −4.357 | 4 |
| 018143 | 013642 | A | K2IV | 1 | ... | ... | 0.21 | 3 | −5.119 | 1 |
| 018632 | 013976 | ... | K2.5V | 1 | ... | ... | 0.18 | 3 | −4.418 | 1 |
| 018757 | 014286 | A | G1.5V | 1 | ... | ... | −0.44 | 2 | −4.987 | 1 |
| 018803 | 014150 | A | G6V | 1 | ... | ... | 0.09 | 3 | −4.880 | 5 |
| 019373 | 014632 | ... | F9.5V | 1 | ... | ... | 0.13 | 3 | −5.020 | 5 |
| 019994 | 014954 | A | F8.5V | 1 | ... | ... | 0.17 | 3 | −4.880 | 5 |
| 020010 | 014879 | A | F6V | 4 | ... | ... | −0.16 | 2 | −4.901 | 4 |
| 020165 | 015099 | ... | K1V | 1 | ... | ... | −0.04 | 3 | −4.860 | 5 |
| 020407 | 015131 | ... | G5V | 4 | ... | ... | −0.46 | 2 | −4.734 | 4 |
| 020619 | 015442 | ... | G2V | 1 | ... | ... | −0.18 | 3 | −4.830 | 5 |
| 020630 | 015457 | ... | G5Vvar | 7 | ... | ... | 0.10 | 3 | −4.47 | 9 |
| 020794 | 015510 | ... | G8V | 4 | ... | ... | −0.23 | 3 | −4.998 | 4 |
| 020807 | 015371 | A | G0V | 4 | ... | ... | −0.23 | 3 | −4.827 | 4 |
| 021175 | 015799 | A | K1V | 4 | ... | ... | 0.12 | 2 | −4.773 | 4 |
| 022049 | 016537 | A | K2V | 4 | ... | ... | −0.15 | 4 | −4.510 | 5 |
| 022484 | 016852 | ... | F9V | 7 | ... | ... | −0.03 | 3 | −5.120 | 5 |
| 022879 | 017147 | ... | F9V | 7 | ... | ... | −0.76 | 3 | −4.920 | 5 |
| 023356 | 017420 | ... | K2.5V | 4 | ... | ... | −0.07 | 3 | −4.807 | 4 |
| 023484 | 017439 | Aa | K2V | 4 | ... | ... | 0.04 | 3 | −4.534 | 4 |
| 024238 | 018324 | ... | K2V | 1 | ... | ... | −0.32 | 3 | −4.980 | 5 |
| 024409 | 018413 | A | G3V | 1 | ... | ... | −0.25 | 2 | −4.927 | 1 |
| 024496 | 018267 | A | G7V | 1 | ... | ... | −0.01 | 3 | −4.870 | 5 |
| 025329 | 018915 | ... | K3Vp | 1 | ... | ... | −1.69 | 2 | −4.940 | 5 |

Table 17—Continued

| HD Name (1) | HIP Name (2) | C (3) | Spec Type (4) | Ref (5) | Mass ( M⊙ ) (6) | Ref (7) | [Fe/H] (8) | Ref (9) | $\log(R'_{HK})$ (10) | Ref (11) |
|---|---|---|---|---|---|---|---|---|---|---|
| 025457 | 018859 | ... | F7V | 1 | ... | ... | −0.10 | 2 | −4.390 | 5 |
| 025665 | 019422 | ... | K2.5V | 1 | ... | ... | −0.06 | 3 | −4.847 | 1 |
| 025680 | 019076 | Aa | G1V | 1 | ... | ... | 0.04 | 3 | −4.608 | 1 |
| 025998 | 019335 | Ea | F8V | 1 | ... | ... | 0.02 | 2 | −4.468 | 1 |
| 026491 | 019233 | Aa | G1V | 4 | ... | ... | −0.06 | 3 | −4.889 | 4 |
| 026923 | 019859 | A | G0V | 4 | ... | ... | −0.13 | 2 | −4.521 | 4 |
| 026965 | 019849 | A | K0.5V | 4 | ... | ... | −0.08 | 3 | −4.900 | 5 |
| 029883 | 021988 | ... | K5III | 7 | ... | ... | −0.16 | 3 | −4.87 | 9 |
| 030495 | 022263 | ... | G1.5V | 4 | ... | ... | −0.01 | 3 | −4.600 | 5 |
| 030501 | 022122 | ... | K2V | 4 | ... | ... | −0.09 | 4 | −4.762 | 4 |
| 030876 | 022451 | ... | K2V | 7 | ... | ... | −0.11 | 3 | −4.55 | 9 |
| 032778 | 023437 | A | G7V | 4 | ... | ... | −0.48 | 3 | −4.870 | 4 |
| 032850 | 023786 | Aa | G9V | 1 | ... | ... | −0.19 | 2 | −4.600 | 5 |
| 032923 | 023835 | ... | G1V | 1 | ... | ... | −0.13 | 3 | −5.030 | 5 |
| 033262 | 023693 | ... | F9V | 4 | ... | ... | −0.20 | 2 | −4.373 | 4 |
| 033564 | 025110 | A | F7V | 1 | ... | ... | −0.06 | 2 | −4.949 | 1 |
| 034411 | 024813 | ... | G1V | 1 | ... | ... | 0.09 | 3 | −5.050 | 5 |
| 034721 | 024786 | ... | F9-V | 4 | ... | ... | −0.08 | 3 | −5.030 | 5 |
| 035112 | 025119 | A | K2.5V | 1 | ... | ... | −0.27 | 2 | −4.879 | 1 |
| 035296 | 025278 | A | F8V | 1 | ... | ... | −0.15 | 2 | −4.353 | 1 |
| 035854 | 025421 | ... | K3-V | 4 | ... | ... | −0.04 | 3 | −4.922 | 4 |
| 036435 | 025544 | ... | G9V | 4 | ... | ... | −0.18 | 2 | −4.499 | 4 |
| 036705 | 025647 | Aa | K2V | 4 | 0.87 | 13 | −0.59 | 2 | −3.880 | 4 |
| 037008 | 026505 | ... | K1V | 1 | ... | ... | −0.31 | 2 | −4.960 | 5 |
| 037394 | 026779 | A | K0V | 1 | ... | ... | 0.16 | 3 | −4.553 | 1 |
| 037572 | 026373 | A | K1.5V | 4 | ... | ... | −0.49 | 2 | −4.234 | 4 |
| 038230 | 027207 | ... | K0V | 1 | ... | ... | −0.02 | 3 | −4.990 | 5 |
| 038858 | 027435 | ... | G2V | 1 | ... | ... | −0.21 | 3 | −4.950 | 5 |
| 039091 | 026394 | A | G0V | 4 | ... | ... | 0.04 | 3 | −4.941 | 4 |
| 039587 | 027913 | Aa | G0V | 4 | 1.10 | 14 | −0.16 | 2 | −4.426 | 4 |
| 039855 | 027922 | A | G8V | 4 | ... | ... | −0.49 | 2 | −4.932 | 4 |
| 040307 | 027887 | A | K2.5V | 4 | ... | ... | −0.25 | 3 | −5.037 | 4 |
| 040397 | 028267 | A | G7V | 1 | ... | ... | −0.05 | 3 | −4.980 | 5 |
| 041593 | 028954 | ... | G9V | 1 | ... | ... | ... | ... | −4.390 | 5 |
| 042618 | 029432 | ... | G3V | 1 | ... | ... | −0.09 | 3 | −4.940 | 5 |
| 042807 | 029525 | ... | G5V | 1 | ... | ... | −0.21 | 2 | −4.465 | 1 |
| 043162 | 029568 | A | G6.5V | 4 | ... | ... | −0.10 | 2 | −4.400 | 5 |
| 043587 | 029860 | Aa | G0V | 4 | ... | ... | −0.08 | 3 | −5.000 | 5 |
| 043834 | 029271 | Aa | G7V | 4 | ... | ... | 0.05 | 3 | −4.940 | 4 |
| 045088 | 030630 | Aa | K3V | 1 | 0.83 | 15 | −0.85 | 2 | −4.266 | 1 |
| 045184 | 030503 | ... | G1.5V | 4 | ... | ... | 0.03 | 3 | −4.950 | 5 |
| 045270 | 030314 | A | G0Vp | 4 | ... | ... | −0.18 | 2 | −4.378 | 4 |
| 046588 | 032439 | ... | F8V | 1 | ... | ... | −0.25 | 2 | −4.885 | 1 |

Table 17—Continued

| HD Name (1) | HIP Name (2) | C (3) | Spec Type (4) | Ref (5) | Mass ( M$_\odot$ ) (6) | Ref (7) | [Fe/H] (8) | Ref (9) | log($R'_{HK}$) (10) | Ref (11) |
|---|---|---|---|---|---|---|---|---|---|---|
| 048189 | 031711 | A | G1V | 4 | ... | ... | −0.23 | 2 | −4.268 | 4 |
| 048682 | 032480 | ... | F9V | 1 | ... | ... | 0.09 | 3 | −4.850 | 5 |
| 050692 | 033277 | ... | G0V | 1 | ... | ... | −0.13 | 3 | −4.940 | 5 |
| 051419 | 033537 | ... | G5V | 1 | ... | ... | −0.33 | 3 | −4.870 | 5 |
| 051866 | 033852 | ... | K3V | 1 | ... | ... | ... | ... | −4.889 | 1 |
| 052698 | 033817 | Aa | K1V | 4 | ... | ... | ... | ... | −4.590 | 5 |
| 052711 | 034017 | ... | G0V | 1 | ... | ... | −0.10 | 3 | −4.960 | 5 |
| 053143 | 033690 | ... | K0IV-V | 4 | ... | ... | −0.01 | 2 | −4.547 | 4 |
| 053705 | 034065 | A | G0V | 4 | ... | ... | −0.15 | 3 | −4.981 | 4 |
| 053927 | 034414 | ... | K2.5V | 1 | ... | ... | −0.27 | 2 | −4.984 | 1 |
| 054371 | 034567 | Aa | G6V | 1 | ... | ... | −0.10 | 2 | −4.462 | 4 |
| 055575 | 035136 | ... | F9V | 1 | ... | ... | −0.35 | 2 | −4.950 | 5 |
| 057095 | 035296 | A | K2.5V | 4 | ... | ... | ... | ... | −4.538 | 4 |
| 059468 | 036210 | ... | G6.5V | 4 | ... | ... | 0.01 | 3 | −4.946 | 4 |
| 059747 | 036704 | ... | K1V | 1 | ... | ... | −0.03 | 3 | −4.370 | 5 |
| 059967 | 036515 | ... | G2V | 4 | ... | ... | −0.24 | 2 | −4.372 | 4 |
| 060491 | 036827 | ... | K2.5V | 1 | ... | ... | −0.26 | 3 | −4.430 | 5 |
| 061606 | 037349 | A | K3-V | 1 | ... | ... | −0.05 | 3 | −4.390 | 5 |
| 062613 | 038784 | ... | G8V | 7 | ... | ... | −0.23 | 2 | −4.840 | 5 |
| 063077 | 037853 | Aa | F9V | 4 | ... | ... | −0.75 | 2 | −4.970 | 5 |
| 063433 | 038228 | ... | G5V | 1 | ... | ... | 0.02 | 3 | −4.390 | 5 |
| 064096 | 038382 | A | G0V | 4 | 0.93 | 6 | −0.18 | 2 | −4.883 | 4 |
| 064468 | 038657 | Aa | K2.5V | 1 | ... | ... | 0.09 | 3 | −5.146 | 1 |
| 064606 | 038625 | Aa | K0V | 1 | ... | ... | −0.80 | 2 | −4.940 | 5 |
| 065430 | 039064 | Aa | K0V | 1 | ... | ... | −0.04 | 3 | −5.010 | 5 |
| 065583 | 039157 | ... | K0V | 1 | ... | ... | −0.48 | 3 | −4.950 | 5 |
| 065907 | 038908 | A | F9.5V | 4 | ... | ... | −0.15 | 3 | −4.846 | 4 |
| 067199 | 039342 | Aa | K2V | 4 | ... | ... | 0.01 | 2 | −4.843 | 4 |
| 067228 | 039780 | ... | G2IV | 1 | ... | ... | 0.17 | 3 | −5.120 | 5 |
| 068017 | 040118 | Aa | G3V | 1 | ... | ... | −0.30 | 3 | −4.920 | 5 |
| 068257 | 040167 | A | F8V | 1 | ... | ... | 0.08 | 2 | ... | ... |
| 069830 | 040693 | A | G8+V | 4 | ... | ... | −0.08 | 3 | −4.950 | 5 |
| 071148 | 041484 | ... | G1V | 1 | ... | ... | −0.01 | 3 | −4.950 | 5 |
| 072673 | 041926 | ... | G9V | 4 | ... | ... | −0.33 | 3 | −4.950 | 5 |
| 072760 | 042074 | Aa | K0-V | 1 | ... | ... | 0.03 | 3 | −4.380 | 5 |
| 072905 | 042438 | ... | G1.5Vb | 7 | ... | ... | −0.25 | 2 | −4.400 | 5 |
| 073350 | 042333 | ... | G5V | 1 | ... | ... | 0.04 | 3 | −4.490 | 5 |
| 073667 | 042499 | ... | K2V | 1 | ... | ... | −0.36 | 3 | −4.970 | 5 |
| 073752 | 042430 | A | G5IV | 4 | 1.25 | 10 | 0.31 | 2 | −5.031 | 4 |
| 074385 | 042697 | A | K2+V | 4 | ... | ... | −0.03 | 2 | −4.902 | 4 |
| 074576 | 042808 | ... | K2.5V | 4 | ... | ... | −0.24 | 2 | −4.402 | 4 |
| 075732 | 043587 | A | K0IV-V | 1 | ... | ... | 0.25 | 3 | −5.040 | 5 |
| 075767 | 043557 | Aa | G1.5V | 1 | ... | ... | −0.18 | 2 | −4.638 | 1 |

Table 17—Continued

| HD Name (1) | HIP Name (2) | C (3) | Spec Type (4) | Ref (5) | Mass ( M$_\odot$ ) (6) | Ref (7) | [Fe/H] (8) | Ref (9) | log($R'_{HK}$) (10) | Ref (11) |
|---|---|---|---|---|---|---|---|---|---|---|
| 076151 | 043726 | ... | G3V | 4 | ... | ... | 0.07 | 3 | −4.853 | 4 |
| 076932 | 044075 | ... | G2V | 4 | ... | ... | −0.70 | 2 | −4.781 | 4 |
| 078366 | 044897 | ... | G0IV-V | 1 | ... | ... | 0.03 | 3 | −4.555 | 1 |
| 079028 | 045333 | Aa | G0IV-V | 1 | ... | ... | −0.10 | 2 | −5.073 | 1 |
| 079096 | 045170 | Aa | G9V | 1 | 0.89 | 6 | −0.30 | 2 | −4.846 | 1 |
| 079969 | 045617 | A | K3V | 1 | 0.74 | 10 | −0.22 | 2 | −4.736 | 1 |
| 080715 | 045963 | Aa | K2.5V | 1 | ... | ... | −0.58 | 1 | −4.099 | 1 |
| 082342 | 046626 | A | K3.5V | 4 | ... | ... | ... | ... | −5.131 | 4 |
| 082443 | 046843 | A | K1V | 4 | ... | ... | ... | ... | −4.234 | 4 |
| 082558 | 046816 | ... | K0 | 7 | ... | ... | −0.21 | 3 | −4.09 | 9 |
| 082885 | 047080 | A | G8+V | 1 | ... | ... | ... | ... | −4.68 | 9 |
| 084117 | 047592 | ... | F8V | 4 | ... | ... | −0.08 | 3 | −4.862 | 4 |
| 084737 | 048113 | ... | G0IV-V | 1 | ... | ... | 0.14 | 3 | −5.230 | 5 |
| 086728 | 049081 | A | G4V | 1 | ... | ... | 0.11 | 3 | −5.060 | 5 |
| 087424 | 049366 | ... | K2V | 4 | ... | ... | −0.14 | 3 | −4.440 | 5 |
| 087883 | 049699 | ... | K2.5V | 1 | ... | ... | 0.04 | 3 | −4.999 | 1 |
| 088742 | 050075 | ... | G0V | 4 | ... | ... | −0.04 | 3 | −4.806 | 4 |
| 089125 | 050384 | A | F6V | 1 | ... | ... | −0.44 | 2 | −4.832 | 1 |
| 089269 | 050505 | ... | G4V | 1 | ... | ... | −0.18 | 3 | −4.940 | 5 |
| 090156 | 050921 | ... | G5V | 4 | ... | ... | −0.21 | 3 | −4.950 | 5 |
| 090343 | 051819 | ... | K0 | 7 | ... | ... | ... | ... | −4.58 | 9 |
| 090508 | 051248 | A | G0V | 1 | ... | ... | −0.40 | 2 | −5.005 | 1 |
| 090839 | 051459 | A | F8V | 1 | ... | ... | −0.05 | 3 | −4.860 | 5 |
| 091324 | 051523 | ... | F9V | 4 | ... | ... | −0.28 | 2 | −4.766 | 4 |
| 091889 | 051933 | ... | F8V | 4 | ... | ... | −0.27 | 2 | −4.849 | 4 |
| 092719 | 052369 | ... | G1.5V | 4 | ... | ... | −0.21 | 2 | −4.826 | 4 |
| 092945 | 052462 | ... | K1.5V | 4 | ... | ... | −0.12 | 3 | −4.320 | 5 |
| 094765 | 053486 | ... | K2.5V | 1 | ... | ... | −0.03 | 3 | −4.546 | 1 |
| 095128 | 053721 | A | G0V | 7 | ... | ... | 0.02 | 3 | −5.020 | 5 |
| 096064 | 054155 | A | G8+V | 1 | ... | ... | −0.13 | 2 | −4.373 | 1 |
| 096612 | 054426 | ... | K3-V | 1 | ... | ... | ... | ... | −4.836 | 1 |
| 097334 | 054745 | A | G1V | 1 | ... | ... | 0.08 | 3 | −4.368 | 1 |
| 097343 | 054704 | ... | G8.5V | 4 | ... | ... | −0.05 | 3 | −5.000 | 5 |
| 097658 | 054906 | ... | K1V | 1 | ... | ... | −0.27 | 3 | −4.920 | 5 |
| 098230 | 055203 | Ba | G2V | 16 | ... | ... | −0.30 | 2 | ... | ... |
| 098281 | 055210 | ... | G8V | 7 | ... | ... | −0.17 | 3 | −4.940 | 5 |
| 099491 | 055846 | A | K0IV | 7 | ... | ... | 0.24 | 3 | −4.840 | 5 |
| 100180 | 056242 | Aa | F9.5V | 1 | ... | ... | −0.02 | 3 | −4.940 | 5 |
| 100623 | 056452 | A | K0-V | 4 | ... | ... | −0.32 | 3 | −4.890 | 5 |
| 101177 | 056809 | A | F9.5V | 1 | ... | ... | −0.17 | 3 | −4.930 | 5 |
| 101206 | 056829 | Aa | K5V | 7 | ... | ... | ... | ... | −4.52 | 9 |
| 101501 | 056997 | ... | G8V | 1 | ... | ... | −0.03 | 3 | −4.550 | 5 |
| 102365 | 057443 | A | G2V | 4 | ... | ... | −0.26 | 3 | −4.957 | 4 |

Table 17—Continued

| HD Name (1) | HIP Name (2) | C (3) | Spec Type (4) | Ref (5) | Mass ( M⊙ ) (6) | Ref (7) | [Fe/H] (8) | Ref (9) | $\log(R'_{HK})$ (10) | Ref (11) |
|---|---|---|---|---|---|---|---|---|---|---|
| 102438 | 057507 | ... | G6V | 4 | ... | ... | −0.23 | 3 | −4.924 | 4 |
| 102870 | 057757 | ... | F8V | 7 | ... | ... | 0.16 | 3 | −4.940 | 5 |
| 103095 | 057939 | ... | K1V | 1 | ... | ... | −1.16 | 3 | −4.850 | 5 |
| 104067 | 058451 | ... | K3-V | 4 | ... | ... | 0.04 | 3 | −4.751 | 4 |
| 104304 | 058576 | ... | G8IV | 4 | ... | ... | 0.16 | 3 | −4.920 | 5 |
| 105631 | 059280 | ... | G9V | 1 | ... | ... | 0.14 | 3 | −4.650 | 5 |
| 108954 | 061053 | ... | F9V | 1 | ... | ... | −0.13 | 2 | −4.921 | 1 |
| 109200 | 061291 | ... | K1V | 4 | ... | ... | −0.23 | 3 | −5.124 | 4 |
| 109358 | 061317 | ... | G0V | 1 | ... | ... | −0.10 | 3 | −4.920 | 5 |
| 110463 | 061946 | Aa | K3V | 7 | ... | ... | ... | ... | −4.47 | 9 |
| 110810 | 062229 | ... | K2+V | 4 | ... | ... | −0.03 | 3 | −4.441 | 4 |
| 110833 | 062145 | Aa | K3V | 7 | ... | ... | 0.08 | 2 | −4.70 | 9 |
| 110897 | 062207 | ... | F9V | 1 | ... | ... | −0.59 | 2 | −4.869 | 1 |
| 111312 | 062505 | Aa | K2.5V | 4 | ... | ... | ... | ... | −4.571 | 4 |
| 111395 | 062523 | ... | G7V | 1 | ... | ... | 0.06 | 3 | −4.580 | 5 |
| 112758 | 063366 | Aa | G9V | 4 | ... | ... | −0.38 | 2 | −5.067 | 4 |
| 112914 | 063406 | Aa | K3-V | 1 | ... | ... | −0.26 | 3 | −5.043 | 1 |
| 113283 | 064690 | Aa | G5V | 4 | ... | ... | −0.15 | 2 | −4.720 | 4 |
| 113449 | 063742 | Aa | K1V | 1 | ... | ... | ... | ... | −4.340 | 1 |
| 114613 | 064408 | ... | G4IV | 4 | ... | ... | 0.16 | 3 | −5.118 | 4 |
| 114710 | 064394 | ... | G0V | 7 | ... | ... | 0.04 | 3 | −4.760 | 5 |
| 114783 | 064457 | A | K1V | 1 | ... | ... | 0.10 | 3 | −5.056 | 1 |
| 114853 | 064550 | ... | G1.5V | 4 | ... | ... | −0.24 | 3 | −4.936 | 4 |
| 115383 | 064792 | ... | G0Vs | 7 | ... | ... | 0.21 | 3 | −4.400 | 5 |
| 115404 | 064797 | A | K2.5V | 1 | ... | ... | −0.49 | 1 | −4.640 | 1 |
| 115617 | 064924 | ... | G7V | 4 | ... | ... | 0.09 | 3 | −5.040 | 5 |
| 116442 | 065352 | A | G9V | 1 | ... | ... | −0.30 | 3 | −4.940 | 5 |
| 116956 | 065515 | ... | G9V | 1 | ... | ... | −0.13 | 2 | −4.447 | 1 |
| 117043 | 065530 | ... | G6V | 7 | ... | ... | 0.06 | 2 | −4.96 | 9 |
| 117176 | 065721 | A | G5V | 7 | ... | ... | −0.01 | 3 | −4.990 | 5 |
| 118972 | 066765 | ... | K0V | 4 | ... | ... | −0.09 | 3 | −4.439 | 4 |
| 119332 | 066781 | ... | K0IV-V | 7 | ... | ... | ... | ... | −4.70 | 9 |
| 120136 | 067275 | A | F7V | 7 | ... | ... | 0.25 | 3 | −4.700 | 5 |
| 120559 | 067655 | ... | G7V | 4 | ... | ... | −0.95 | 2 | −5.029 | 4 |
| 120690 | 067620 | Aa | G5+V | 4 | ... | ... | 0.05 | 3 | −4.750 | 5 |
| 120780 | 067742 | Aa | K2V | 4 | ... | ... | −0.26 | 3 | −4.888 | 4 |
| 121370 | 067927 | Aa | G0IV | 7 | ... | ... | 0.27 | 2 | ... | ... |
| 121560 | 068030 | ... | F6V | 7 | ... | ... | −0.41 | 3 | −4.920 | 5 |
| 122742 | 068682 | Aa | G6V | 1 | ... | ... | −0.08 | 2 | −4.955 | 1 |
| 124106 | 069357 | ... | K1V | 4 | ... | ... | −0.13 | 3 | −4.630 | 5 |
| 124292 | 069414 | ... | G8+V | 1 | ... | ... | −0.10 | 3 | −4.970 | 5 |
| 124580 | 069671 | ... | G0V | 4 | ... | ... | −0.28 | 2 | −4.597 | 4 |
| 124850 | 069701 | Aa | F7V | 7 | ... | ... | −0.04 | 2 | ... | ... |

Table 17—Continued

| HD Name (1) | HIP Name (2) | C (3) | Spec Type (4) | Ref (5) | Mass ( M☉ ) (6) | Ref (7) | [Fe/H] (8) | Ref (9) | log($R'_{HK}$) (10) | Ref (11) |
|---|---|---|---|---|---|---|---|---|---|---|
| 125276 | 069965 | Aa | F9V | 4 | … | … | −0.62 | 2 | −4.641 | 4 |
| 125455 | 070016 | A | K1V | 7 | … | … | −0.15 | 3 | −4.930 | 5 |
| 126053 | 070319 | … | G1.5V | 1 | … | … | −0.29 | 3 | −4.940 | 5 |
| 127334 | 070873 | … | G5V | 1 | … | … | 0.22 | 3 | −5.060 | 5 |
| 128165 | 071181 | … | K3V | 7 | … | … | −0.09 | 3 | −4.67 | 9 |
| 128311 | 071395 | A | K3-V | 1 | … | … | 0.01 | 3 | −4.489 | 1 |
| 128400 | 071855 | … | G5V | 4 | … | … | −0.14 | 2 | −4.518 | 4 |
| 128620 | 071683 | A | G2V | 4 | 1.11 | 17 | 0.19 | 3 | −5.059 | 4 |
| 128642 | 070857 | Aa | G5 | 7 | … | … | −0.39 | 2 | −4.910 | 5 |
| 128987 | 071743 | … | G8V | 4 | … | … | 0.02 | 2 | −4.439 | 4 |
| 130004 | 072146 | … | K2.5V | 1 | … | … | … | … | −4.919 | 1 |
| 130042 | 072493 | A | K1V | 4 | … | … | −0.21 | 2 | −4.949 | 4 |
| 130307 | 072312 | … | K2.5V | 1 | … | … | −0.20 | 3 | −4.560 | 5 |
| 130948 | 072567 | A | G2V | 7 | … | … | −0.19 | 2 | −4.500 | 5 |
| 131156 | 072659 | A | G7V | 1 | 0.94 | 10 | −0.07 | 3 | −4.472 | 1 |
| 131511 | 072848 | Aa | K0V | 1 | … | … | 0.11 | 3 | −4.510 | 1 |
| 131582 | 072875 | Aa | K3V | 7 | … | … | −0.54 | 2 | … | … |
| 131923 | 073241 | Aa | G4V | 4 | … | … | 0.09 | 3 | −5.059 | 4 |
| 132142 | 073005 | … | K1V | 7 | … | … | −0.30 | 3 | −5.000 | 5 |
| 132254 | 073100 | … | F8-V | 1 | … | … | 0.02 | 2 | −5.030 | 1 |
| 133640 | 073695 | A | G2V | 7 | … | … | −0.42 | 2 | −4.61 | 9 |
| 134060 | 074273 | … | G0V | 4 | … | … | 0.08 | 3 | −5.042 | 4 |
| 135204 | 074537 | A | G9V | 1 | … | … | −0.06 | 2 | −5.107 | 1 |
| 135599 | 074702 | … | K0V | 1 | … | … | −0.09 | 3 | −4.520 | 5 |
| 136202 | 074975 | A | F8III-IV | 7 | … | … | −0.04 | 2 | … | … |
| 136352 | 075181 | … | G2-V | 4 | … | … | −0.23 | 3 | −5.013 | 4 |
| 136713 | 075253 | … | K3IV-V | 1 | … | … | 0.15 | 3 | −4.878 | 1 |
| 136923 | 075277 | … | G9V | 1 | … | … | −0.07 | 3 | −4.770 | 5 |
| 137107 | 075312 | A | G2V | 7 | 1.19 | 6 | −0.10 | 2 | −4.76 | 9 |
| 137763 | 075718 | Aa | G9V | 1 | … | … | 0.02 | 2 | −4.970 | 5 |
| 139341 | 076382 | A | K1V | 1 | 0.85 | 10 | … | … | −5.102 | 1 |
| 139777 | 075809 | A | G1.5V(n) | 1 | … | … | −0.31 | 2 | −4.405 | 1 |
| 140538 | 077052 | A | G5V | 7 | … | … | 0.06 | 3 | −4.830 | 5 |
| 140901 | 077358 | A | G7IV-V | 4 | … | … | 0.08 | 3 | −4.802 | 4 |
| 141004 | 077257 | … | G0IV-V | 1 | … | … | 0.09 | 3 | −4.970 | 5 |
| 141272 | 077408 | A | G9V | 1 | … | … | −0.05 | 2 | −4.390 | 5 |
| 142267 | 077801 | Aa | G0IV | 7 | … | … | −0.38 | 3 | −4.850 | 5 |
| 142373 | 077760 | … | G0V | 1 | … | … | −0.39 | 3 | −5.110 | 5 |
| 143761 | 078459 | Aa | G0V | 1 | 1.04 | 18 | −0.14 | 3 | −5.080 | 5 |
| 144284 | 078527 | Aa | F8IV-V | 7 | 1.2 | 19 | 0.04 | 2 | … | … |
| 144287 | 078709 | Aa | G8+V | 1 | … | … | −0.13 | 2 | −5.020 | 5 |
| 144579 | 078775 | A | K0V | 1 | … | … | −0.49 | 3 | −4.970 | 5 |
| 144628 | 079190 | … | K1V | 4 | … | … | −0.30 | 3 | −5.128 | 4 |

Table 17—Continued

| HD Name (1) | HIP Name (2) | C (3) | Spec Type (4) | Ref (5) | Mass ( M$_\odot$ ) (6) | Ref (7) | [Fe/H] (8) | Ref (9) | log($R'_{HK}$) (10) | Ref (11) |
|---|---|---|---|---|---|---|---|---|---|---|
| 144872 | 078913 | ... | K3V | 1 | ... | ... | −0.30 | 2 | −4.804 | 1 |
| 145417 | 079537 | ... | K3V | 4 | ... | ... | −1.37 | 2 | −5.205 | 4 |
| 145675 | 079248 | A | K0IV-V | 1 | ... | ... | 0.41 | 3 | −5.060 | 5 |
| 145825 | 079578 | Aa | G2V | 4 | ... | ... | 0.03 | 3 | −4.793 | 4 |
| 145958 | 079492 | A | G8V | 1 | ... | ... | −0.07 | 3 | −4.940 | 5 |
| 146233 | 079672 | ... | G2V | 1 | ... | ... | 0.03 | 3 | −4.950 | 5 |
| 146361 | 079607 | Aa | G1IV-V | 1 | 1.14 | 20 | −0.26 | 2 | −3.827 | 1 |
| 147513 | 080337 | A | G1V | 4 | ... | ... | 0.07 | 3 | −4.656 | 4 |
| 147584 | 080686 | Aa | F9V | 4 | ... | ... | −0.19 | 2 | −4.585 | 4 |
| 147776 | 080366 | A | K3-V | 4 | ... | ... | −0.26 | 3 | −4.810 | 4 |
| 148653 | 080725 | A | K2V | 1 | 0.79 | 10 | ... | ... | −4.655 | 1 |
| 148704 | 080925 | Aa | K1V | 4 | ... | ... | −0.60 | 2 | −5.101 | 4 |
| 149612 | 081520 | ... | G5V | 4 | ... | ... | −0.40 | 3 | −4.954 | 4 |
| 149661 | 081300 | ... | K0V | 4 | ... | ... | 0.05 | 3 | −4.570 | 5 |
| 149806 | 081375 | A | K0V | 1 | ... | ... | 0.17 | 3 | −4.830 | 5 |
| 151541 | 081813 | ... | K1V | 7 | ... | ... | −0.18 | 3 | −4.990 | 5 |
| 152391 | 082588 | ... | G8.5V | 4 | ... | ... | −0.05 | 3 | −4.440 | 5 |
| 153557 | 083020 | A | K3V | 1 | ... | ... | −0.33 | 2 | −4.508 | 1 |
| 154088 | 083541 | ... | K0IV-V | 4 | ... | ... | 0.28 | 3 | −5.020 | 5 |
| 154345 | 083389 | A | G8V | 7 | ... | ... | −0.10 | 3 | −4.910 | 5 |
| 154417 | 083601 | ... | F9V | 1 | ... | ... | 0.03 | 3 | −4.590 | 5 |
| 154577 | 083990 | ... | K2.5V | 4 | ... | ... | −0.56 | 3 | −5.080 | 4 |
| 155712 | 084195 | ... | K2.5V | 1 | ... | ... | ... | ... | −4.988 | 1 |
| 155885 | 084405 | A | K1.5V | 4 | ... | ... | ... | ... | −4.711 | 4 |
| 156274 | 084720 | Aa | M0V | 7 | ... | ... | −0.27 | 3 | ... | ... |
| 157214 | 084862 | ... | G0V | 7 | ... | ... | −0.15 | 3 | −5.040 | 5 |
| 157347 | 085042 | A | G3V | 1 | ... | ... | 0.03 | 3 | −5.040 | 5 |
| 158614 | 085667 | A | G8IV-V | 7 | 0.98 | 6 | −0.01 | 2 | −5.12 | 9 |
| 158633 | 085235 | ... | K0V | 7 | ... | ... | −0.33 | 3 | −4.930 | 5 |
| 159062 | 085653 | ... | G9V | 1 | ... | ... | −0.51 | 2 | −5.030 | 5 |
| 159222 | 085810 | ... | G1V | 1 | ... | ... | 0.09 | 3 | −4.900 | 5 |
| 160269 | 086036 | A | G0V | 7 | 1.08 | 10 | −0.22 | 2 | −4.61 | 9 |
| 160346 | 086400 | Aa | K2.5V | 1 | ... | ... | −0.09 | 2 | −4.956 | 1 |
| 160691 | 086796 | A | G3IV-V | 4 | ... | ... | 0.26 | 3 | −5.101 | 4 |
| 161198 | 086722 | Aa | G9V | 1 | ... | ... | −0.39 | 2 | −4.970 | 5 |
| 161797 | 086974 | Aa | G5IV | 7 | ... | ... | 0.24 | 3 | −5.110 | 5 |
| 162004 | 086620 | B | G0V | 7 | ... | ... | −0.18 | 2 | −4.86 | 9 |
| 164922 | 088348 | A | G9V | 1 | ... | ... | 0.17 | 3 | −5.050 | 5 |
| 165185 | 088694 | ... | G0V | 4 | ... | ... | −0.25 | 2 | −4.545 | 4 |
| 165341 | 088601 | A | K0-V | 1 | 0.90 | 6 | −0.29 | 2 | −4.698 | 1 |
| 165401 | 088622 | A | G0V | 1 | ... | ... | −0.50 | 2 | −4.668 | 1 |
| 165499 | 089042 | Aa | G0V | 4 | ... | ... | −0.17 | 2 | −4.935 | 4 |
| 165908 | 088745 | Aa | F7V | 7 | ... | ... | −0.55 | 2 | −5.020 | 5 |

Table 17—Continued

| HD Name (1) | HIP Name (2) | C (3) | Spec Type (4) | Ref (5) | Mass ( M$_\odot$ ) (6) | Ref (7) | [Fe/H] (8) | Ref (9) | log($R'_{HK}$) (10) | Ref (11) |
|---|---|---|---|---|---|---|---|---|---|---|
| 166620 | 088972 | ... | K2V | 1 | ... | ... | −0.05 | 3 | −4.970 | 5 |
| 167425 | 089805 | A | F9.5V | 4 | ... | ... | 0.02 | 2 | −4.606 | 4 |
| 168009 | 089474 | ... | G1V | 1 | ... | ... | −0.02 | 3 | −5.080 | 5 |
| 170657 | 090790 | ... | K2V | 4 | ... | ... | −0.15 | 3 | −4.650 | 5 |
| 172051 | 091438 | ... | G6V | 4 | ... | ... | −0.24 | 3 | −4.900 | 5 |
| 175073 | 092858 | ... | K1V | 4 | ... | ... | −0.14 | 2 | −4.889 | 4 |
| 175742 | 092919 | Aa | K0V | 7 | ... | ... | ... | ... | ... | ... |
| 176051 | 093017 | A | G0V | 7 | 1.07 | 10 | −0.19 | 2 | ... | ... |
| 176377 | 093185 | ... | G1V | 1 | ... | ... | −0.23 | 3 | −4.870 | 5 |
| 177474 | 093825 | A | F8V | 4 | ... | ... | −0.22 | 2 | −4.890 | 4 |
| 177565 | 093858 | ... | G6V | 4 | ... | ... | 0.07 | 3 | −4.973 | 4 |
| 178428 | 093966 | Aa | G4V | 4 | ... | ... | 0.05 | 2 | −5.110 | 4 |
| 179957 | 094336 | A | G3V | 1 | ... | ... | ... | ... | −5.050 | 5 |
| 180161 | 094346 | ... | G8V | 7 | ... | ... | ... | ... | −4.520 | 5 |
| 181321 | 095149 | Aa | G1V | 4 | ... | ... | −0.26 | 2 | −4.372 | 4 |
| 182488 | 095319 | ... | G9+V | 1 | ... | ... | 0.12 | 3 | −4.940 | 5 |
| 182572 | 095447 | ... | G8IVvar | 7 | ... | ... | 0.36 | 3 | −5.100 | 5 |
| 183870 | 096085 | ... | K2.5V | 4 | ... | ... | −0.05 | 3 | −4.512 | 4 |
| 184385 | 096183 | ... | G8V | 1 | ... | ... | 0.11 | 3 | −4.560 | 5 |
| 184467 | 095995 | A | K2V | 1 | 0.8 | 6 | −0.22 | 2 | −5.047 | 1 |
| 185144 | 096100 | ... | G9V | 1 | ... | ... | −0.16 | 3 | −4.850 | 5 |
| 185414 | 096395 | Aa | G0 | 7 | ... | ... | −0.16 | 2 | −4.88 | 9 |
| 186408 | 096895 | Aa | G1.5V | 1 | ... | ... | 0.08 | 3 | −5.100 | 5 |
| 186858 | 097222 | F | K3+V | 1 | 0.69 | 10 | ... | ... | −4.726 | 1 |
| 187691 | 097675 | A | F8V | 7 | ... | ... | 0.12 | 3 | −5.050 | 5 |
| 189340 | 098416 | A | F9V | 1 | 1.12 | 10 | −0.05 | 2 | −4.951 | 1 |
| 189567 | 098959 | ... | G2V | 4 | ... | ... | −0.18 | 3 | −4.857 | 4 |
| 189733 | 098505 | A | K2V | 1 | ... | ... | −0.12 | 2 | −4.553 | 1 |
| 190067 | 098677 | A | K0V | 1 | ... | ... | −0.30 | 3 | −4.880 | 5 |
| 190248 | 099240 | ... | G8IV | 4 | ... | ... | 0.26 | 3 | −5.092 | 4 |
| 190360 | 098767 | A | G7IV-V | 4 | ... | ... | 0.19 | 3 | −5.090 | 4 |
| 190404 | 098792 | ... | K1V | 1 | ... | ... | −0.44 | 3 | −4.980 | 5 |
| 190406 | 098819 | Aa | G0V | 4 | ... | ... | 0.02 | 3 | −4.770 | 5 |
| 190422 | 099137 | ... | F9V | 4 | ... | ... | −0.21 | 2 | −4.458 | 4 |
| 190470 | 098828 | ... | K2.5V | 1 | ... | ... | −0.16 | 1 | −4.828 | 1 |
| 190771 | 098921 | Aa | G2V | 1 | ... | ... | 0.14 | 3 | −4.430 | 5 |
| 191408 | 099461 | A | K2.5V | 4 | ... | ... | −0.33 | 3 | −5.079 | 4 |
| 191499 | 099316 | A | G9V | 1 | ... | ... | ... | ... | −5.076 | 1 |
| 191785 | 099452 | A | K0V | 1 | ... | ... | −0.09 | 3 | −5.030 | 5 |
| 192263 | 099711 | A | K2.5V | 1 | ... | ... | −0.07 | 3 | −4.676 | 1 |
| 192310 | 099825 | ... | K2+V | 4 | ... | ... | 0.14 | 2 | −5.048 | 4 |
| 193664 | 100017 | Aa | G0V | 1 | ... | ... | −0.11 | 3 | −4.927 | 1 |
| 194640 | 100925 | ... | G8V | 4 | ... | ... | −0.06 | 3 | −4.924 | 4 |

Table 17—Continued

| HD Name (1) | HIP Name (2) | C (3) | Spec Type (4) | Ref (5) | Mass ( M$_\odot$ ) (6) | Ref (7) | [Fe/H] (8) | Ref (9) | log($R'_{HK}$) (10) | Ref (11) |
|---|---|---|---|---|---|---|---|---|---|---|
| 195564 | 101345 | A | G2V | 1 | ... | ... | 0.01 | 3 | −5.130 | 5 |
| 195987 | 101382 | Aa | G9V | 1 | 0.84 | 21 | −0.38 | 2 | −4.970 | 5 |
| 196378 | 101983 | ... | G0V | 4 | ... | ... | −0.32 | 3 | −4.837 | 4 |
| 196761 | 101997 | ... | G8V | 4 | ... | ... | −0.25 | 3 | −4.920 | 5 |
| 197076 | 102040 | A | G1V | 1 | ... | ... | −0.09 | 3 | −4.920 | 5 |
| 197214 | 102264 | ... | G6V | 4 | ... | ... | −0.50 | 2 | −4.920 | 5 |
| 198425 | 102766 | Aa | K2.5V | 1 | ... | ... | ... | ... | −4.726 | 1 |
| 199260 | 103389 | ... | F6V | 4 | ... | ... | −0.17 | 2 | −4.402 | 4 |
| 199288 | 103458 | ... | G2V | 4 | ... | ... | −0.47 | 3 | −4.847 | 4 |
| 199509 | 104436 | ... | G1V | 4 | ... | ... | −0.27 | 3 | −4.925 | 4 |
| 200525 | 104440 | A | F9.5V | 4 | ... | ... | −0.13 | 2 | −4.667 | 4 |
| 200560 | 103859 | C | K2.5V | 1 | ... | ... | 0.01 | 2 | −4.512 | 1 |
| 200968 | 104239 | A | G9.5V | 4 | ... | ... | 0.02 | 3 | −4.650 | 4 |
| 202275 | 104858 | A | F7V | 1 | 1.19 | 6 | −0.07 | 2 | −4.905 | 1 |
| 202628 | 105184 | ... | G1.5V | 4 | ... | ... | −0.01 | 3 | −4.782 | 4 |
| 202751 | 105152 | ... | K3V | 1 | ... | ... | −0.09 | 3 | −5.111 | 1 |
| 202940 | 105312 | Aa | G7V | 4 | ... | ... | −0.34 | 2 | −4.988 | 4 |
| 203244 | 105712 | ... | G8V | 4 | ... | ... | −0.32 | 2 | −4.555 | 4 |
| 203850 | 105905 | ... | K2.5V | 4 | ... | ... | −0.67 | 2 | −5.033 | 4 |
| 203985 | 105911 | Aa | K2III-IV | 4 | ... | ... | ... | ... | ... | ... |
| 205390 | 106696 | ... | K1.5V | 4 | ... | ... | −0.18 | 3 | −4.702 | 4 |
| 205536 | 107022 | ... | G9V | 4 | ... | ... | −0.03 | 3 | −5.084 | 4 |
| 206826 | 107310 | A | F6V | 1 | ... | ... | −0.20 | 2 | −4.783 | 1 |
| 206860 | 107350 | A | G0V | 4 | ... | ... | −0.01 | 2 | −4.400 | 4 |
| 207129 | 107649 | ... | G0V | 4 | ... | ... | −0.04 | 3 | −5.020 | 4 |
| 207144 | 107625 | ... | K3V | 4 | ... | ... | ... | ... | −4.990 | 4 |
| 208038 | 108028 | ... | K2.5V | 1 | ... | ... | ... | ... | −4.569 | 1 |
| 208313 | 108156 | ... | K2V | 1 | ... | ... | −0.08 | 4 | −4.987 | 1 |
| 210277 | 109378 | A | G8V | 1 | ... | ... | 0.20 | 3 | −5.060 | 5 |
| 210667 | 109527 | ... | G9V | 1 | ... | ... | 0.16 | 3 | −4.500 | 5 |
| 210918 | 109821 | ... | G2V | 4 | ... | ... | −0.07 | 3 | −5.121 | 4 |
| 211415 | 110109 | A | G0V | 4 | ... | ... | −0.36 | 2 | −4.918 | 4 |
| 211472 | 109926 | A | K0V | 1 | ... | ... | 0.03 | 2 | −4.538 | 1 |
| 212168 | 110712 | A | G0V | 4 | ... | ... | −0.04 | 3 | −4.981 | 4 |
| 212330 | 110649 | Aa | G2IV-V | 4 | ... | ... | 0.01 | 3 | −5.157 | 4 |
| 214683 | 111888 | ... | K3V | 1 | ... | ... | ... | ... | −4.554 | 1 |
| 214953 | 112117 | A | F9.5V | 4 | ... | ... | 0.03 | 3 | −4.988 | 4 |
| 215152 | 112190 | ... | K3V | 1 | ... | ... | −0.17 | 1 | −4.925 | 1 |
| 215648 | 112447 | A | F7V | 7 | ... | ... | −0.16 | 3 | −5.070 | 5 |
| 216259 | 112870 | ... | K2.5V | 1 | ... | ... | −0.47 | 3 | −4.950 | 5 |
| 216520 | 112527 | ... | K0V | 1 | ... | ... | ... | ... | −4.980 | 5 |
| 217014 | 113357 | A | G2V+ | 4 | ... | ... | 0.15 | 3 | −5.080 | 5 |
| 217107 | 113421 | A | G8IV-V | 1 | ... | ... | 0.27 | 3 | −5.080 | 5 |

Table 17—Continued

| HD Name (1) | HIP Name (2) | C (3) | Spec Type (4) | Ref (5) | Mass ( M$_\odot$) (6) | Ref (7) | [Fe/H] (8) | Ref (9) | log($R'_{HK}$) (10) | Ref (11) |
|---|---|---|---|---|---|---|---|---|---|---|
| 217813 | 113829 | ... | G1V | 1 | ... | ... | 0.02 | 3 | −4.470 | 5 |
| 218868 | 114456 | A | G8V | 1 | ... | ... | 0.19 | 3 | −4.750 | 5 |
| 219134 | 114622 | ... | K3V | 1 | ... | ... | 0.09 | 3 | −5.089 | 1 |
| 219482 | 114948 | ... | F6V | 4 | ... | ... | −0.20 | 2 | −4.434 | 4 |
| 219538 | 114886 | ... | K2V | 1 | ... | ... | −0.05 | 3 | −4.840 | 5 |
| 219623 | 114924 | ... | F7V | 7 | ... | ... | −0.09 | 2 | −4.84 | 9 |
| 220140 | 115147 | A | K2V | 1 | ... | ... | −0.64 | 2 | −4.074 | 1 |
| 220182 | 115331 | ... | G9V | 1 | ... | ... | 0.01 | 2 | −4.370 | 5 |
| 220339 | 115445 | ... | K2.5V | 1 | ... | ... | −0.24 | 3 | −4.896 | 1 |
| 221354 | 116085 | ... | K0V | 1 | ... | ... | −0.01 | 3 | −5.149 | 1 |
| 221851 | 116416 | ... | K1V | 1 | ... | ... | −0.13 | 2 | −4.735 | 1 |
| 222143 | 116613 | ... | G3V | 1 | ... | ... | 0.12 | 3 | −4.555 | 1 |
| 222237 | 116745 | ... | K3+V | 4 | ... | ... | −0.20 | 3 | −4.959 | 4 |
| 222335 | 116763 | ... | G9.5V | 4 | ... | ... | −0.16 | 3 | −4.909 | 4 |
| 222368 | 116771 | ... | F7V | 7 | ... | ... | −0.08 | 3 | −4.76 | 9 |
| 223778 | 117712 | Aa | K3V | 1 | 0.79 | 15 | −0.71 | 2 | −4.518 | 1 |
| 224228 | 118008 | ... | K2.5V | 4 | ... | ... | ... | ... | −4.468 | 4 |
| 224465 | 118162 | Aa | G4V | 1 | ... | ... | −0.01 | 2 | −4.969 | 1 |
| 224930 | 000171 | A | G5V | 1 | 0.91 | 10 | −0.78 | 2 | −4.880 | 1 |
| 232781 | 015673 | ... | K3.5V | 1 | ... | ... | ... | ... | −4.691 | 1 |
| 263175 | 032423 | A | K3V | 1 | ... | ... | −0.59 | 1 | −4.903 | 1 |
| ... | 036357 | Ea | K2.5V | 1 | ... | ... | −0.61 | 1 | −4.526 | 1 |
| ... | 040774 | ... | G5 | 7 | ... | ... | ... | ... | −4.38 | 9 |
| ... | 087579 | ... | K2.5V | 1 | ... | ... | ... | ... | −4.529 | 1 |
| ... | 091605 | A | K2.5V | 1 | ... | ... | −0.68 | 1 | −4.909 | 1 |

Note. — Reference codes for columns 5, 7, 9, and 11: (1) Gray et al. (2003); (2) Nordström et al. (2004); (3) Valenti & Fischer (2005); (4) Gray et al. (2006); (5) Wright et al. (2004); (6) Pourbaix (2000); (7) The *Hipparcos* catalog; (8) Boden et al. (1999); (9) Mason et al. (2010); (10) Söderhjelm (1999); (11) Lowrance et al. (2002); (12) Bagnuolo et al. (2006); (13) Close et al. (2005); (14) Catala et al. (2006); (15) Raghavan et al. (2010b); (16) Ball et al. (2005); (17) Pourbaix et al. (2002); (18) Gatewood et al. (2001); (19) Mazeh et al. (2002); (20) Raghavan et al. (2009); (21) Torres et al. (2002).

Table 18.   Spectral Types and Masses of the Companions

| Comp Name (1) | Alt Name (2) | Spec Type (3) | Ref (4) | Mass ( M$_\odot$) (5) | Ref (6) |
|---|---|---|---|---|---|
| HD 000123 Ba | ... | ... | ... | 0.95 | 1 |
| HD 000123 Bb | ... | ... | ... | 0.22 | 1 |
| HD 001237 B | HD 1237B | M4±1V | 2 | 0.13 | 3 |
| HD 003196 Ab | ... | ... | ... | 0.40[a] | 4 |
| HD 003196 B | ... | G2V | 5 | 1.00 | 4 |
| HD 003443 B | GJ 25 B | ... | ... | 0.70 | 6 |
| HD 003651 B | HD 3651B | T7.5 ± 0.5 | 7 | 0.05 | 7 |
| HD 004391 B | ... | M4V | 8 | ... | ... |
| HD 004391 C | ... | M5V | 8 | ... | ... |
| HD 004614 B | LHS 122 | K7V | 8 | ... | ... |
| HD 004676 Ab | ... | ... | ... | 1.17 | 9 |
| HD 006582 Ab | $\mu$ Cas B | M3V | 5 | ... | ... |
| HD 007693 A | HD 7788 | F5V | 10 | 1.32 | 11 |
| HD 007693 B | GJ 55.3B | K1V | 10 | ... | ... |
| HD 007693 D | GJ 55.1B | ... | ... | 0.86 | 12 |
| HD 008997 Ab | ... | G7V | 13 | ... | ... |
| HD 009770 Ba | ... | ... | ... | 0.73 | 14 |
| HD 009770 Bb | ... | ... | ... | 0.67 | 14 |
| HD 009770 C | ... | M3V | 5 | ... | ... |
| HD 009826 D | $v$ And B | M4.5V | 15 | 0.2 | 15 |
| HD 010307 Ab | ... | ... | ... | 0.2 | 12 |
| HD 010360 A | HD 10361 | K2V | 10 | ... | ... |
| HD 013445 B | GJ 86 B | WD | 16 | 0.5 | 17 |
| HD 013974 Ab | ... | K3V | 13 | ... | ... |
| HD 014802 B | ... | M2V | 5 | ... | ... |
| HD 016160 B | NLTT 8455 | M3.5V | 18 | ... | ... |
| HD 016739 Ab | ... | ... | ... | 1.24 | 19 |
| HD 016765 B | ... | K2V | 5 | ... | ... |
| HD 016895 B | NLTT 8787 | M2V | 8 | ... | ... |
| HD 017382 B | NLTT 8996 | M7V | 8 | ... | ... |
| HD 018143 B | HD 18143 B | K9V | 5 | ... | ... |
| HD 018143 C | NLTT 9303 | M7V | 8 | ... | ... |
| HD 018757 C | NLTT 9726 | M2V | 8 | ... | ... |
| HD 019994 B | ... | M2V | 5 | ... | ... |
| HD 020010 Ba | GJ 127 B | K2V | 5 | ... | ... |
| HD 020807 B | HD 020766 | G2V | 10 | 1.12 | 20 |
| HD 021175 B | ... | M3V | 5 | ... | ... |
| HD 024409 E | ... | M2V | 8 | ... | ... |
| HD 024496 B | ... | M2V | 5 | ... | ... |
| HD 025998 A | HD 25893 | G9V | 21 | ... | ... |
| HD 025998 B | ... | K2V | 5 | ... | ... |
| HD 026923 B | HD 26913 | G6V | 21 | 0.87 | 11 |
| HD 026965 B | LHS 25 | M4.5 | 22 | ... | ... |

Table 18—Continued

| Comp Name (1) | Alt Name (2) | Spec Type (3) | Ref (4) | Mass ( M$_\odot$ ) (5) | Ref (6) |
|---|---|---|---|---|---|
| HD 026965 C | HD 26976 | DA3 | 23 | ... | ... |
| HD 032778 B | NLTT 14447 | M0V | 8 | ... | ... |
| HD 032923 B | ... | G1V | 5 | ... | ... |
| HD 035112 B | ... | M1V | 5 | ... | ... |
| HD 035296 C | HD 35171 | K7V | 8 | ... | ... |
| HD 036705 Ab | ... | M8V | 24 | 0.09 | 25 |
| HD 036705 Ba | ... | M5V | 26 | 0.16 | 26 |
| HD 036705 Bb | ... | M4V ± 1 | 26 | 0.14 | 26 |
| HD 037394 B | HD 233153 | M1V | 8 | ... | ... |
| HD 037572 B | HIP 26369 | K5V | 10 | ... | ... |
| HD 039587 Ab | ... | ... | ... | 0.15 | 27 |
| HD 039855 B | BD-19 1297B | K9V | 8 | ... | ... |
| HD 040397 B | ... | M4V | 5 | ... | ... |
| HD 040397 D | NLTT 15867 | M5V | 8 | ... | ... |
| HD 043162 B | ... | M4V | 8 | ... | ... |
| HD 043587 Ab | ... | ... | ... | 0.54 | 28 |
| HD 043587 E | NLTT 16333 | M4V | 8 | ... | ... |
| HD 043834 Ab | ... | M5V ± 1.5 | 29 | 0.14 | 29 |
| HD 045088 Ab | ... | ... | ... | 0.71 | 30 |
| HD 045088 B | ... | M4V | 5 | ... | ... |
| HD 045270 B | ... | M0V | 5 | ... | ... |
| HD 048189 B | ... | K7V | 5 | ... | ... |
| HD 053705 B | HD 053706 | K0.5V | 10 | 0.78 | 20 |
| HD 053705 Ca | HD 053680 | K6V | 10 | ... | ... |
| HD 057095 B | ... | K6V | 5 | ... | ... |
| HD 061606 B | NLTT 18260 | K9V | 8 | ... | ... |
| HD 063077 B | NLTT 18414 | DC | 31 | 0.56 | 32 |
| HD 064096 B | ... | ... | ... | 0.9 | 6 |
| HD 065907 B | LHS 1960 | M0V | 8 | ... | ... |
| HD 065907 C | ... | M5V | 5 | ... | ... |
| HD 068257 B | HD 068255 | F8V | 21 | 1.52 | 11 |
| HD 068257 Ca | HD 068256 | G0 IV-V | 21 | ... | ... |
| HD 068257 Cb1 | ... | M2V | ... | ... | ... |
| HD 068257 Cb2 | ... | M2V | ... | ... | ... |
| HD 068257 Cb3 | ... | M2V–M4V | ... | ... | ... |
| HD 072760 Ab | ... | ... | ... | 0.13 | 33 |
| HD 073752 B | ... | K2V | 5 | 1.07 | 12 |
| HD 074385 B | NLTT 20102 | M2V | 8 | ... | ... |
| HD 075732 B | LHS 2063 | M6V | 8 | ... | ... |
| HD 075767 Ba | ... | M3V | 34 | ... | ... |
| HD 075767 Bb | ... | M4V | 34 | ... | ... |
| HD 079096 Ab | ... | ... | ... | 0.85 | 6 |
| HD 079096 Ea | Gl 337C | L8 | 35 | ... | ... |

Table 18—Continued

| Comp Name (1) | Alt Name (2) | Spec Type (3) | Ref (4) | Mass ( M$_\odot$) (5) | Ref (6) |
|---|---|---|---|---|---|
| HD 079096 Eb | . . . | L8 | 36 | . . . | . . . |
| HD 079969 B | . . . | K4V | 5 | 0.74 | 12 |
| HD 080715 Ab | . . . | K3V | 13 | . . . | . . . |
| HD 082342 B | . . . | M3.5V | 37 | . . . | . . . |
| HD 082443 B | NLTT 22015 | M5.5V | 37 | . . . | . . . |
| HD 082885 B | . . . | M8V | 5 | . . . | . . . |
| HD 086728 Ba | GJ 376 B | M6.5V | 38 | . . . | . . . |
| HD 086728 Bb | . . . | M6.5V | 38 | . . . | . . . |
| HD 089125 B | GJ 387 B | M1V | 8 | . . . | . . . |
| HD 090508 B | LHS 2266 | M2V | 5 | . . . | . . . |
| HD 090839 Ba | HD 237903 | K5 | 8 | . . . | . . . |
| HD 096064 B | NLTT 26194 | K7 | 5 | . . . | . . . |
| HD 096064 C | BD-033040C | K7 | 5 | . . . | . . . |
| HD 097334 Ea | Gl 417B | L4.5V | 39 | 0.1 | 40 |
| HD 097334 Eb | . . . | L4.5V | 40 | 0.1 | 40 |
| HD 098230 Ab | . . . | M3V | 41 | . . . | . . . |
| HD 098230 Aa | HD 98231 | F9V | 42 | . . . | . . . |
| HD 099491 B | HD 099492 | K2V | 43 | 1.24 | 20 |
| HD 100180 B | NLTT 27656 | K5V | 8 | . . . | . . . |
| HD 100623 B | LHS 309 | DC | 18 | . . . | . . . |
| HD 101177 Ba | LHS 2436 | K3V | 8 | . . . | . . . |
| HD 101177 Bb | . . . | M2V | 13 | . . . | . . . |
| HD 102365 B | LHS 313 | M4V | 37 | . . . | . . . |
| HD 111312 Ab | . . . | M2V | 5 | 0.58 | 13 |
| HD 111312 B | . . . | K8V | 13 | . . . | . . . |
| HD 112758 B | . . . | M2V | 5 | . . . | . . . |
| HD 115404 B | LHS 2714 | M0.5V | 37 | . . . | . . . |
| HD 116442 B | HD 116443 | K2V | 21 | 0.78 | 20 |
| HD 120136 B | HD 120136B | M2V | 5 | . . . | . . . |
| HD 120780 B | . . . | M4V | 5 | . . . | . . . |
| HD 125455 B | LHS 2895 | M6 | 8 | . . . | . . . |
| HD 128620 B | HD 128621 | K2IV | 10 | 0.93 | 44 |
| HD 128620 C | HIP 070890 | . . . | . . . | 0.11 | 45 |
| HD 130042 B | . . . | K8V | 5 | . . . | . . . |
| HD 130948 B | HD 130948 B | L2 ± 2 | 46 | 0.07 | 46 |
| HD 130948 C | HD 130948 C | L2 ± 2 | 46 | 0.07 | 46 |
| HD 131156 B | HD 131156B | . . . | . . . | 0.67 | 12 |
| HD 133640 Ba | NLTT 39210 | K2V | 47 | . . . | . . . |
| HD 133640 Bb | . . . | M2V | 13 | . . . | . . . |
| HD 135204 B | . . . | G9V | 5 | . . . | . . . |
| HD 136202 B | LHS 3060 | K9V | 8 | . . . | . . . |
| HD 137107 B | HD 137108 | . . . | . . . | 1.05 | 6 |
| HD 137107 E | GJ 584 C | L8V | 39 | 0.06 | 39 |

Table 18—Continued

| Comp Name (1) | Alt Name (2) | Spec Type (3) | Ref (4) | Mass ( M$_\odot$ ) (5) | Ref (6) |
|---|---|---|---|---|---|
| HD 137763 Ab | . . . | K9 | 13 | 0.57 | 13 |
| HD 137763 B | HD 137778 | K2V | 21 | 0.87 | 20 |
| HD 137763 C | GJ 586C | M4.5V | 48 | . . . | . . . |
| HD 139341 B | . . . | K1V | 5 | 0.83 | 12 |
| HD 139341 C | HD 139323 | K2IV-V | 21 | 1.13 | 20 |
| HD 139777 B | HD 139813 | K2V | 8 | 1.15 | 20 |
| HD 140538 B | . . . | M5V | 5 | . . . | . . . |
| HD 140901 B | NLTT 41169 | M2 | 5 | . . . | . . . |
| HD 141272 B | . . . | M3V ± 0.5 | 49 | 0.26 | 49 |
| HD 143761 Ab | . . . | . . . | . . . | 0.14 | 50 |
| HD 144284 Ab | . . . | . . . | . . . | 0.46 | 51 |
| HD 144287 Ab | . . . | K4V | 5 | . . . | . . . |
| HD 144579 B | LHS 3150 | M4V | 8 | . . . | . . . |
| HD 145958 B | NLTT 42272 | G9V | 21 | . . . | . . . |
| HD 145958 D | . . . | T6 | 52 | . . . | . . . |
| HD 146361 Ab | . . . | . . . | . . . | 1.09 | 53 |
| HD 146361 B | HD 146362 | . . . | . . . | 1.0 | 53 |
| HD 146361 Ea | HIP 079551 | M2.5V | 48 | . . . | . . . |
| HD 146361 Eb | sig CrB D | . . . | . . . | 0.1 | 54 |
| HD 147513 B | . . . | DA2 | 23 | . . . | . . . |
| HD 148653 B | LHS 3204 | . . . | . . . | 0.77 | 12 |
| HD 148704 Ab | . . . | K1V | 13 | . . . | . . . |
| HD 149806 B | . . . | M6V | 8 | . . . | . . . |
| HD 153557 B | . . . | M2V | 5 | . . . | . . . |
| HD 153557 C | HD 153525 | . . . | . . . | 0.73 | 11 |
| HD 155885 B | . . . | K1.5V | 5 | . . . | . . . |
| HD 155885 C | HD 156026 | K5V | 10 | . . . | . . . |
| HD 156274 B | NLTT 44525 | K7V | 37 | . . . | . . . |
| HD 157347 B | HR 6465 | M3V | 55 | . . . | . . . |
| HD 158614 B | . . . | . . . | . . . | 0.90 | 6 |
| HD 160269 B | . . . | . . . | . . . | 0.65 | 12 |
| HD 160269 C | HIP 86087 | M0.5V | 22 | . . . | . . . |
| HD 161797 Ab | . . . | M8V | 5 | . . . | . . . |
| HD 161797 B | NLTT 45430 | M3V | 5 | . . . | . . . |
| HD 161797 C | . . . | M3V | 5 | . . . | . . . |
| HD 162004 A | HD 162003 | F5V | 8 | 1.38 | 11 |
| HD 165341 B | NLTT 45900 | . . . | . . . | 0.78 | 6 |
| HD 165908 B | . . . | K6V | 5 | . . . | . . . |
| HD 167425 B | . . . | M0V | 8 | . . . | . . . |
| HD 176051 B | . . . | K3V | 5 | 0.71 | 12 |
| HD 177474 B | . . . | F8V | 5 | . . . | . . . |
| HD 179957 B | HD 179958 | G4V | 5 | . . . | . . . |
| HD 184467 B | . . . | . . . | . . . | 0.8 | 6 |

Table 18—Continued

| Comp Name (1) | Alt Name (2) | Spec Type (3) | Ref (4) | Mass ( M⊙ ) (5) | Ref (6) |
|---|---|---|---|---|---|
| HD 186408 Ab | . . . | M0V | 5 | . . . | . . . |
| HD 186408 B | HD 186427 | G3V | 10 | 1.10 | 20 |
| HD 186858 Aa | HD 187013 | F5.5IV-V | 21 | 1.24 | 11 |
| HD 186858 B | . . . | K4V | 5 | . . . | . . . |
| HD 186858 G | HD 225732 | K3V | 5 | 0.68 | 12 |
| HD 187691 C | . . . | M3V | 8 | . . . | . . . |
| HD 189340 B | . . . | . . . | . . . | 0.94 | 12 |
| HD 190067 B | . . . | K7V | 5 | . . . | . . . |
| HD 190360 B | LHS 3509 | M4.5V | 22 | . . . | . . . |
| HD 190406 Ab | HD 354613 | L4.5 ± 1.5 | 56 | 0.06 | 56 |
| HD 191408 B | LHS 487 | M5V | 5 | . . . | . . . |
| HD 191499 B | ADS 13434B | K5V | 5 | . . . | . . . |
| HD 191785 E | . . . | M3.5V | 55 | . . . | . . . |
| HD 195564 B | LTT 8128 | M2V | 5 | . . . | . . . |
| HD 195987 Ab | . . . | . . . | . . . | 0.67 | 57 |
| HD 197076 C | NLTT 49681 | M2.5V | 22 | . . . | . . . |
| HD 198425 B | NLTT 49961 | M6V | 8 | . . . | . . . |
| HD 200525 B | . . . | G0V | 5 | . . . | . . . |
| HD 200525 C | NLTT 50542 | M3V | 5 | . . . | . . . |
| HD 200560 D | GJ 816.1B | M3V | 5 | . . . | . . . |
| HD 200968 B | GJ 819B | K7V | 37 | . . . | . . . |
| HD 202275 B | . . . | . . . | . . . | 1.12 | 6 |
| HD 202940 B | LHS 3656 | K9V | 5 | . . . | . . . |
| HD 203985 B | LTT 8515 | M3.5V | 55 | . . . | . . . |
| HD 206826 B | HD 206827 | G5V | 5 | . . . | . . . |
| HD 206860 B | HN Peg B | T2.5 ± 0.5 | 7 | 0.02 | 7 |
| HD 211415 B | . . . | K9V | 5 | . . . | . . . |
| HD 211472 T | GJ 4269 | M4V | 8 | . . . | . . . |
| HD 212168 B | HIP 110719 | G0V | 10 | 1.13 | 20 |
| HD 214953 B | NLTT 54607 | M0.5V | 37 | . . . | . . . |
| HD 215648 B | . . . | M4V | 5 | . . . | . . . |
| HD 218868 B | . . . | M5V | 8 | . . . | . . . |
| HD 220140 B | NLTT 56532 | M3V | 8 | . . . | . . . |
| HD 220140 C | . . . | M7V | 8 | 0.1 | 58 |
| HD 223778 Ab | . . . | . . . | . . . | 0.75 | 30 |
| HD 223778 B | . . . | M6V | 5 | . . . | . . . |
| HD 224930 B | . . . | K5±1 V | 42 | 0.58 | 12 |
| HD 263175 B | HD 263175B | M0.5V | 22 | . . . | . . . |
| HIP 036357 A | HD 58946 | F1V | 21 | 1.40 | 11 |
| HIP 036357 B | GJ 274B | M7V | 5 | . . . | . . . |
| HIP 091605 B | LHS 3402 | M2V | 8 | . . . | . . . |

Note. — Reference codes for columns 4 and 6 are as follows: (1) Griffin (1999); (2) Chauvin et al. (2007); (3) Chauvin et al. (2006); (4)

DM91; (5) Estimated using primary's spectral type and $\Delta V$ from the WDS or other catalogs; (6) Pourbaix (2000); (7) Luhman et al. (2007); (8) Estimated from optical and infrared magnitudes from catalogs; (9) Boden et al. (1999); (10) Gray et al. (2006); (11) Nordström et al. (2004); (12) Söderhjelm (1999); (13) Estimated using primary's mass or spectral type and mass-ratio from an SB2 orbital solution; (14) Cutispoto et al. (1997); (15) Lowrance et al. (2002); (16) Mugrauer & Neuhäuser (2005); (17) Lagrange et al. (2006); (18) Henry et al. (2002); (19) Bagnulo et al. (2006); (20) Valenti & Fischer (2005); (21) Gray et al. (2003); (22) Reid et al. (2004); (23) Holberg et al. (2002); (24) Boccaletti et al. (2008); (25) Close et al. (2007); (26) Janson et al. (2007); (27) König et al. (2002); (28) Catala et al. (2006); (29) Eggenberger et al. (2007); (30) Raghavan et al. (2010b); (31) Kunkel et al. (1984); (32) Holberg et al. (2008); (33) Metchev & Hillenbrand (2009); (34) Fuhrmann et al. (2005); (35) Wilson et al. (2001); (36) Burgasser et al. (2005); (37) Hawley et al. (1996); (38) Gizis et al. (2000); (39) Kirkpatrick et al. (2001); (40) Bouy et al. (2003); (41) Ball et al. (2005); (42) ten Brummelaar et al. (2000); (43) The *Hipparcos* catalog; (44) Pourbaix et al. (2002); (45) Wertheimer & Laughlin (2006); (46) Potter et al. (2002); (47) Lu et al. (2001); (48) Reid et al. (1995); (49) Eisenbeiss et al. (2007); (50) Gatewood et al. (2001); (51) Mazeh et al. (2002); (52) Looper et al. (2007); (53) Raghavan et al. (2009); (54) Heintz (1990); (55) This work; (56) Liu et al. (2002); (57) Torres et al. (2002); (58) Makarov et al. (2007).

[a]Estimated using mass-sum of Aa+Ab from DM91 and mass of Aa from Nordström et al. (2004)